\documentclass[usegraphicx,usenatbib,useAMS]{mn2e}\usepackage{times}
  \let\url\relax
\def\apj{{ApJ}}
\def\apjs{{ApJS}} 
\def\apjl{{ApJL}}
\def\aap{{\em A.\&A}}
\def\aj{{AJ}}
\def\mnras{{MNRAS}}

\def\prd{{Phys Rev D}}
\def\nat{{Nature}}
\def\physrep{{Physics Reports}}
\newcommand{\be}{\begin{equation}}
\newcommand{\ba}{\begin{eqnarray}}
\newcommand{\ee}{\end{equation}}
\newcommand{\ea}{\end{eqnarray}}  

\def\lesssim{\mathrel{\hbox{\rlap{\hbox{\lower4pt\hbox{$\sim$}}}\hbox{$<$}}}}
\def\gtrsim{\mathrel{\hbox{\rlap{\hbox{\lower4pt\hbox{$\sim$}}}\hbox{$>$}}}}

\def\gtsima{$\; \buildrel > \over \sim \;$}
\def\ltsima{$\; \buildrel < \over \sim \;$}
\def\gsim{\lower.5ex\hbox{\gtsima}}
\def\lsim{\lower.5ex\hbox{\ltsima}}
\def\simgt{\lower.5ex\hbox{\gtsima}}
\def\simlt{\lower.5ex\hbox{\ltsima}}
\def\simpr{\lower.5ex\hbox{\prosima}}
\def\la{\lsim}
\def\ga{\gsim}

 \def\msun{{M_\odot}}

\def\simless{\mathbin{\lower 3pt\hbox
   {$\rlap{\raise 5pt\hbox{$\char'074$}}\mathchar''7218$}}}   % < or of order
\def\simgreat{\mathbin{\lower 3pt\hbox
   {$\rlap{\raise 5pt\hbox{$\char'076$}}\mathchar''7218$}}}   % > or of order

\begin{document}

\title  [Reionization sources and redshifted 21-cm]{Can 21cm observations 
discriminate between high-mass and low-mass galaxies as reionization sources?}
%What can redshifted 21-cm
%observations tell us about the reionization sources?}
 
\author[I. T. Iliev, et al.]{Ilian~T.~Iliev$^{1}$\thanks{e-mail: 
I.T.Iliev@sussex.ac.uk}, Garrelt Mellema$^{2}$, Paul R. Shapiro$^{3}$, 
Ue-Li Pen$^{4}$, Yi Mao$^{3}$, Jun Koda$^{3}$,\newauthor Kyungjin Ahn$^5$
\\
$^1$ Astronomy Centre, Department of Physics \& Astronomy, Pevensey II 
Building, University of Sussex, Falmer, Brighton BN1 9QH, United Kingdom\\    
$^2$ Department of Astronomy \& Oskar Klein Centre, Stockholm University, 
Albanova, SE-10691 Stockholm, Sweden\\
$^3$ Department of Astronomy, University of Texas, Austin, 
TX 78712-1083, U.S.A.\\
$^4$ Canadian Institute for Theoretical Astrophysics, University
  of Toronto, 60 St. George Street, Toronto, ON M5S 3H8, Canada\\
$^{5}$ Department of Earth Science Education, Chosun University, 
Gwangju 501-759, Korea%\\
}
\date{\today} \pubyear{2011} \volume{000}
\pagerange{1} \twocolumn \maketitle
\label{firstpage}

\begin{abstract}
The prospect of detecting the first galaxies by observing their 
impact on the intergalactic medium (IGM) as they reionized it
during the first billion years leads us to ask whether such 
indirect observations are capable of diagnosing which types of 
galaxies were most responsible for reionization. We attempt 
to answer this by extending the galaxy mass range of our first 
generation of large-scale radiative transfer simulations of
reionization in volumes large enough, $\geq(100/h Mpc)^3$, to 
make statistically meaningful predictions of observable signatures, 
downward from those above $10^9\,M_\odot$ (high-mass, atomic-cooling 
halos, or "HMACHs") to include those between $10^8$ and $10^9\,M_\odot$ 
(low-mass, atomic-cooling halos, or "LMACHs"), as well. Previously, 
we simulated the effects of both HMACHs and LMACHs but only by 
reducing the box size to $35/h$~Mpc, too small to apply those 
simulations to predict observables like the 21-cm background 
fluctuations.  Those simulations showed that LMACHS can make a 
difference, however. While LMACHs are even more abundant,  
photoheating suppresses their ability to form stars if they are 
located inside ionized regions of the IGM, so their contribution 
starts reionization earlier but tends to saturate over time 
before it ends. To predict the observables in that case while 
explicitly tracking the radiation from this full mass range of 
galaxies, we have had to advance both our N-body and radiative 
transfer methods substantially, as described here. With this, 
we perform new simulations in a box 163~Mpc on a side, and 
focus here on predictions of the 21cm background, to see if 
upcoming observations are capable of distinguishing a universe
ionized primarily by HMACHS from one in which both HMACHs and 
LMACHs are responsible, and to see how these results depend 
upon the uncertain source efficiencies.  

We find that 21-cm fluctuation power spectra observed by the first 
generation EoR/21cm radio interferometer arrays should be able to
distinguish the case of reionization by HMACHs alone from that 
by both HMACHs and LMACHs, together. Some reionization scenarios, 
e.g. one with abundant low-efficiency sources vs. one with 
self-regulation, yield very similar power spectra and rms evolution 
and thus can only be discriminated by their different mean 
reionization history and 21-cm PDF distributions. We find that the 
skewness of the 21-cm PDF distribution smoothed over LOFAR-like window 
shows a clear feature correlated with the rise of the rms due to 
patchiness. This is independent on the reionization scenario and thus 
provides a new approach for detecting the rise of large-scale patchiness 
and an independent check on other measurements, regardless of the 
detailed properties of the sources. The peak epoch of the 21-cm 
rms fluctuations depends significantly on the beam and bandwidth 
smoothing size as well as on the reionization scenario and can occur 
for ionized fractions as low as 30\% and as high as 70\%. Measurements 
of the mean photoionization rates are sensitive to the average density 
of the regions being studied and therefore could be strongly skewed in
certain cases. Finally, the simulation volume employed has very modest 
effects on the results during the early and intermediate stages of 
reionization, but late-time signatures could be significantly affected.
\end{abstract}


\section{Introduction}

Study of the Epoch of Reionization (EoR) has progressed in 
recent years in response to a number of new observational 
developments. The combination of the CMBR data from WMAP 
\citep{2011ApJS..192...18K,2011ApJS..192...16L}
and ever deeper ground-based observations of high-redshift 
QSOs, galaxies and GRBs \citep{2010ApJ...723..869O,
2011ApJ...734..119K,2011Natur.474..616M,2011ApJ...736....7C,
2011arXiv1106.6055K} clearly suggest that the reionization 
process started early and was quite extended in time. However,
observations of the effects of the EoR are just beginning to 
assemble constraints sufficient to diagnose the conditions 
which brought it about. Ongoing and upcoming observations are 
expected to put further, much more stringent constraints on 
the reionization history. The best constraints are likely to 
result from redshifted 21-cm experiments with the low-frequency 
radio interferometers GMRT\footnote{\url{http://gmrt.ncra.tifr.res.in/}}
 \citep{2011MNRAS.413.1174P}, LOFAR\footnote{\url{http://www.lofar.org/}} 
\citep[e.g.][]{2010MNRAS.405.2492H}, 
MWA \citep{2009IEEEP..97.1497L}\footnote{\url{http://www.mwatelescope.org/}} 
and PAPER \citep{2010AJ....139.1468P}. Additional information will
come from Planck satellite \citep{2011arXiv1101.2022P} and
measurements of the near infrared background with 
CIBER\footnote{\url{http://physics.ucsd.edu/~bkeating/CIBER.html}} 
and AKARI\footnote{\url{http://irsa.ipac.caltech.edu/Missions/akari.html}},
among others.

The understanding and correct interpretation of these observational 
results requires detailed modelling. Specific characteristics and 
features of the observable signatures provide information about 
different aspects of EoR. One of the central questions is what are
the nature, abundances and physical properties of the ionizing
sources. Our purpose in this work is to explore how observations 
of the EoR might diagnose the nature of the reionization sources.

Our first generation of simulations of inhomogeneous reionization 
combined cosmological N-body simulations of galaxy and large-scale 
structure formation and the intergalactic density and velocity 
fields with detailed radiative transfer calculations of the ionizing 
radiation from every galactic halo whose formation we resolved in a 
volume large enough to make statistically meaningful predictions of 
observable consequences of reionization \citep{2006MNRAS.369.1625I,
2006MNRAS.372..679M,2008MNRAS.384..863I,kSZ,cmbpol,2008MNRAS.391...63I,
2009MNRAS.393.1449H,2010MNRAS.406.2521I,2010ApJ...710.1089F}.  These 
simulations were in comoving boxes of size 100$/h$ ($=143$ for $h=0.7$ 
henceforth) Mpc on a 
side. Previous simulations had been limited to much smaller volumes, 
too small to serve this purpose. Earlier simulations of smaller 
volumes, for example, underestimated the width of the time interval 
for the global transition of the intergalactic medium (IGM) from 
neutral to ionized 
\citep{2000ApJ...535..530G,2002ApJ...575...33R,2003MNRAS.344..607S,
2003MNRAS.344L...7C}, as well as the amplitude of the kSZ fluctuations 
in the temperature of the CMB from the EOR \citep{2001ApJ...551....3G,
2005MNRAS.360.1063S}. The characteristic size of the intergalactic H~II 
regions during the EOR is expected to reach 10's of Mpc 
\citep{2004ApJ...613....1F,2006MNRAS.365..115F,2011MNRAS.413.1353F} 
before they grow large enough to overlap. Any fluctuations introduced 
by this ``patchiness'' scale, therefore, require simulation volumes at 
least this large to model reionization. Moreover, since the first 
generation of radio observations seeking to detect fluctuations in the 
brightness temperature of the 21cm background from the EoR have angular 
resolution of a few arcminutes, a simulation box size in excess of 
$\sim100$~Mpc is necessary to characterize the power spectrum, at 
wavenumbers as small as $\sim 0.1\,{\rm Mpc}^{-1}$ where the initial 
sensitivity will peak. 

Our simulations of a comoving volume 100$/h$ (= 143) Mpc on a side
were limited, however, by the mass resolution of the N-body
simulations, to the direct simulation of galactic halo sources more
massive than $\sim 2\times10^9\,\msun$.  Halo sources are also
possible at lower mass if halo gas can radiatively cool below the halo
virial temperature to make star formation possible.  This includes
halos above about $10^8\,\msun$, for which collisional excitation of H
atoms can radiatively cool the primordial-composition halo gas since
the virial temperature is above $10^4$ K.  We shall refer to these
halos between about $10^8$ and $10^9\,\msun$ as low-mass
atomic-cooling halos (``LMACHs''), to distinguish them from the halos
above $10^9\,\msun$, which we shall call high-mass, atomic-cooling
halos (``HMACHs''). LMACHs are more numerous than HMACHs at these
epochs, so it might be thought that they would dominate
reionization. However, unlike the HMACHs, the LMACHs are vulnerable to
the negative feedback effects of photo-heating if they form inside a
pre-exisiting H~II region of the IGM, since the pressure of the IGM
would then prevent the intergalactic gas from collapsing
gravitationally into their dark matter host halos
\citep{1992MNRAS.256P..43E,1994ApJ...427...25S,1996MNRAS.278L..49Q,
  1997ApJ...478...13N,2004ApJ...610L...5S,2008MNRAS.390..920O,
  2008MNRAS.390.1071M}.  As first emphasized by
\citet{1994ApJ...427...25S}, this limits their contribution to
reionization. Halos of even smaller mass than LMACHs would be even
more vulnerable to the negative feedback effects of photoionization
heating, since they would, in addition to being prevented from
capturing intergalactic gas, photoevaporate whatever interstellar gas
they had already accumulated before they were engulfed by reionization
\citep{2004MNRAS.348..753S,2005MNRAS...361..405I}.  Before that
reionization, however, these minihalos with masses below about
$10^8\,\msun$, nevertheless could have formed stars if enough H$_2$
molecules were present in them.  Since their virial temperatures are
below $10^4$ K, their gas is too cold for collisional excitation and
radiative cooling by H atoms, but cooling through the collisionally
exciting rotational-vibrational levels of H$_2$ is possible. However,
these molecules were easily dissociated by the rising UV
background of starlight at energies below the ionization threshold of
H atoms, an inevitable by-product of the same stars that
contributed to reionization. This tends to limit the contribution of
minihalos to reionization early in the EoR
\citep{2000ApJ...534...11H, 2009ApJ...695.1430A}. Previous estimates of
the minihalo contribution, as a result, assume that this contribution
is smaller than that of the LMACHs and HMACHS, so for now, we shall
neglect it, although patchiness in the UV background may make them
important than one might naively think \citep{2009ApJ...695.1430A}.
% However, we will re-address that contribution in our next
%paper in this series.

To investigate the impact of the LMACHs on reionization and its 
observable properties by direct radiative transfer simulation of 
reionization, our first generation of simulations boosted the halo 
mass resolution in order to resolve all the halos of mass $10^8\,\msun$ 
and above, but sacrificed volume by simulating in a box of size 
37$/h$ (= 53 Mpc) on a side \citep{2007MNRAS.376..534I}. These 
simulations demonstrated explicitly that reionization in the 
presence of the LMACHs and their suppression if they formed inside 
pre-existing H~II regions during the EoR was ``self-regulated''.  
The more LMACHs that formed, the more volume and mass of the 
IGM was ionized, but as this ionized fraction 
grew, so did the fraction of the total LMACH halo population that 
formed inside the ionized regions and was suppressed as sources of 
reionization.   This meant that, although the LMACHs dominated the 
early phase of reionization, their contribution to reionization 
eventually saturated, and reionization was finished by the HMACHs, 
whose abundance rose exponentially over time.  An observational 
consequence of importance that is a possible signature of this 
process was an early onset but late finish for the EoR, as 
required to explain the high values of electron scattering optical 
depth reported from WMAP observations of the large-angle fluctuations 
in the polarization of the CMB. To make further predictions of 
observable consequences, however, we need to enlarge the volume 
of these simulations to greater than 100/h Mpc on a side, without 
losing this enhanced mass resolution necessary to resolve the LMACHs.  
That is the purpose of the new developments we report in this paper.  

To accomplish this goal, it was first necessary to improve and 
advance both our N-body and radiative transfer methods in order 
to be able to simulate halo formation with much higher mass 
resolution and to transfer the ionizing radiation from a much 
larger number of sources. Toward this end, both codes had to 
become massively parallel, running on thousands of computing 
cores, as well as more efficient. These numerical developments
are discussed in more detail in \S~\ref{sim:sect} below and in
\citet{2008arXiv0806.2887I}. 

While the work described here follows naturally from our own
previous work as described above, it also differs substantially 
from other work in the literature to-date involving large-scale 
radiative transfer simulations of reionization. A full account 
of that literature is well-beyond the scope of this paper, but 
we will mention a few points to distinguish the current work.  
Our N-body simulations resolve all galactic halo sources of 
$10^8$ $\msun$ and above, in a comoving box as large as $114/h=163$~Mpc 
on a side. We post-process the density field of the IGM and the 
galactic halo source populations derived from these N-body 
simulations by performing a detailed, ray-tracing calculation 
on a grid of $256^3$ cells.  The simulations described in 
\citet{2007MNRAS.377.1043M} were based upon post-processing 
N-body simulations with halo mass resolution above $10^9$ $\msun$, 
in a box $65.6/h$ Mpc on a side, a volume which is five times 
smaller than ours, also on a grid of $256^3$ cells. 
\citet{2007MNRAS.377.1043M}  were, thus, unable to treat explicitly 
the halo mass range below $10^9$ $\msun$, which is subject to 
suppression by the negative feedback effects of reionization, but 
they did include a semi-analytical, ``subgrid'' approximation for 
the contribution from the unresolved, smaller-mass halos, including 
some feedback effects. \citet{2007ApJ...671....1T} considered a 
$50/h$ Mpc box, hence, an order of magnitude smaller volume than ours, 
and $180^3$ cells for radiative transfer, but their halo mass resolution 
was similar to ours.  They did not consider the feedback effects of 
reionization on the small-mass galactic halo sources as we do here.  
\citet{2008ApJ...681..756S} subsequently applied this method to 
simulate a volume $100/h$ Mpc on a side ($2/3$ of our volume), with 
similar halo mass resolution, on a radiative transfer grid of $360^3$ 
cells, to study the structure of the patchy ionization field during 
the EOR, again without considering feedback and doing only a single 
simulation, without considering any variation of the (highly uncertain)
reionization parameters. Recently, the codes of 
\citet{2007MNRAS.377.1043M} and \citet{2007ApJ...671....1T} were 
compared with each other in \citet{2011MNRAS.414..727Z}, by comparing 
results for the ionization fields and 21-cm brightness temperature 
fluctuation statistics, for reionizaton simulations advanced part-way 
through the epoch of reionization, to the 72\% ionized point, based 
upon post-processing a previously-simulated input density field in a 
box $100/h$ Mpc on a side, smoothed to a radiative transfer grid with 
$256^3$ cells, with halo mass resolution of $10^8$ solar masses, without 
any feedback or back-reaction on the halo sources. They also considered 
a single reionization scenario, with no parameter variation. Finally, 
\citet{2010ApJ...724..244A} simulated radiative transfer in several 
different boxes, as large as $100/h$ Mpc, by post-processing a density 
field and galaxy population derived from separate simulations that 
combined N-body dynamics and hydrodynamics on a $1024^3$ grid, but with 
minimum resolved galaxy masses in that case as large as $8\times 10^9$ 
$\msun$. No feedback from reionization on galactic sources was considered. 
While there are other distinctions of detail both between these other 
simulations and ours and of one from another, we have listed these above 
to make it clear that our current paper will describe simulations and 
results which are new, {\it both} because they are based upon different 
methodology, applied in greater depth than previously to predict 
observables like the 21cm background from the EOR, {\it and} because 
they are on an unprecedented scale.   

The rest of paper is organized as follows. In \S~\ref{sim:sect} 
we present our codes, numerical methods and simulations. In
\S~\ref{results:sect} we discuss our results on the formation
of early cosmic structures. In \S~\ref{basic_results:sect} we
present our results on the basic reionization features, reionization
history, integrated electron-scattering optical depth and geometry
of ionized patches. The observational signatures derived from 
our simulations are presented and discussed in \S~\ref{observ:sect}. 
Our conclusions are summarized in \S~\ref{summary:sect}.
Finally, in Appendix~\ref{appendixA} we present the (physically less
realistic) cases of reionization by rare, massive sources, while in 
Appendix~\ref{appendixB} we discuss the more technical point of
the dependence of our results on the Jeans suppression threshold
for low-mass sources.

\section{Simulations}
\label{sim:sect}
Our basic methodology has been previously described 
in \citet{2006MNRAS.369.1625I,2006MNRAS.372..679M} and 
\citet{2007MNRAS.376..534I}. Due to the much larger scale
of our current simulations compared to our previous ones, 
both our structure formation and our radiative transfer 
code had to be significantly developed and re-designed,
in particular to allow their massive paralellization on
distributed-memory machines. In this section we present 
our new set of simulations, along with a summary of our 
methods and parallel code scaling to large number of 
computing cores.

\subsection{N-body simulations}
\begin{table*}
\caption{N-body simulation parameters. Background cosmology 
is based on the WMAP 5-year results. 
}
\label{summary_N-body_table}
\begin{center}
\begin{tabular}{@{}llllll}
\hline
boxsize & $N_{part}$   & mesh   & spatial resolution & $m_{particle}$ & $M_{halo,min}$
\\[2mm]\hline
37 $\,h^{-1}$Mpc & $1024^3$ & $2048^3$ & $1.81\, kpc/h$ & $5.05\times10^6\,M_\odot$ & $1.01\times10^8\,M_\odot$
\\[2mm]
%56 $\,h^{-1}$Mpc & $1500^3$ & $3000^3$ & $1.87\, kpc/h$ & $5.57\times10^6\,M_\odot$ & $1.11\times10^8\,M_\odot$
%\\[2mm]
64 $\,h^{-1}$Mpc & $1728^3$ & $3456^3$ & $1.85\, kpc/h$ & $5.44\times10^6\,M_\odot$ & $1.09\times10^8\,M_\odot$
\\[2mm]
74 $\,h^{-1}$Mpc & $2048^3$ & $4096^3$ & $1.81\, kpc/h$ & $5.05\times10^6\,M_\odot$ & $1.01\times10^8\,M_\odot$
\\[2mm]
114 $\,h^{-1}$Mpc & $3072^3$ & $6144^3$ & $1.86\, kpc/h$ & $5.47\times10^6\,M_\odot$ & $1.09\times10^8\,M_\odot$
\\[2mm]
\hline
\end{tabular}
\end{center}
\end{table*}

\begin{figure*}
  \begin{center}
    \includegraphics[width=3in]{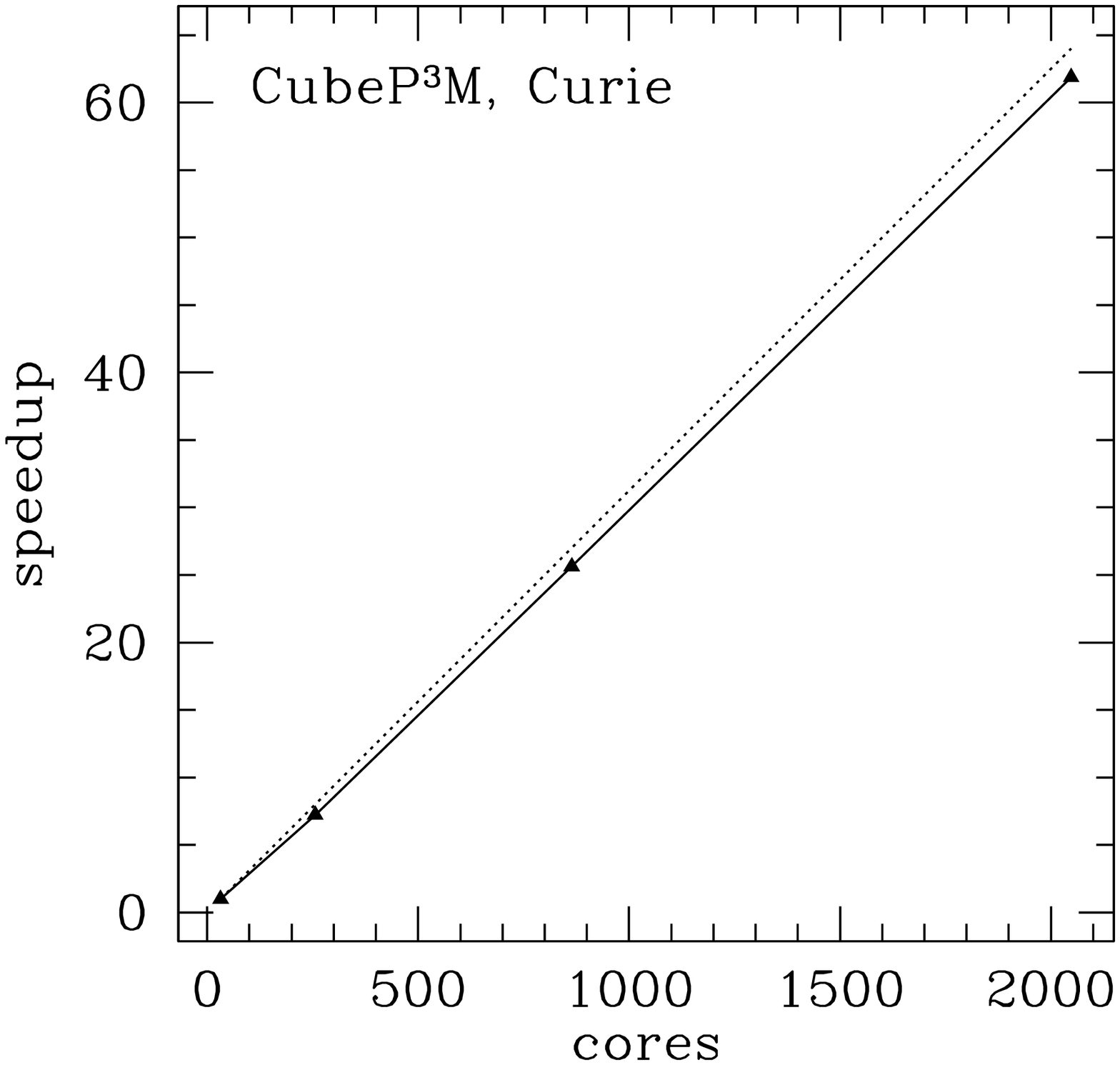}
    \includegraphics[width=3in]{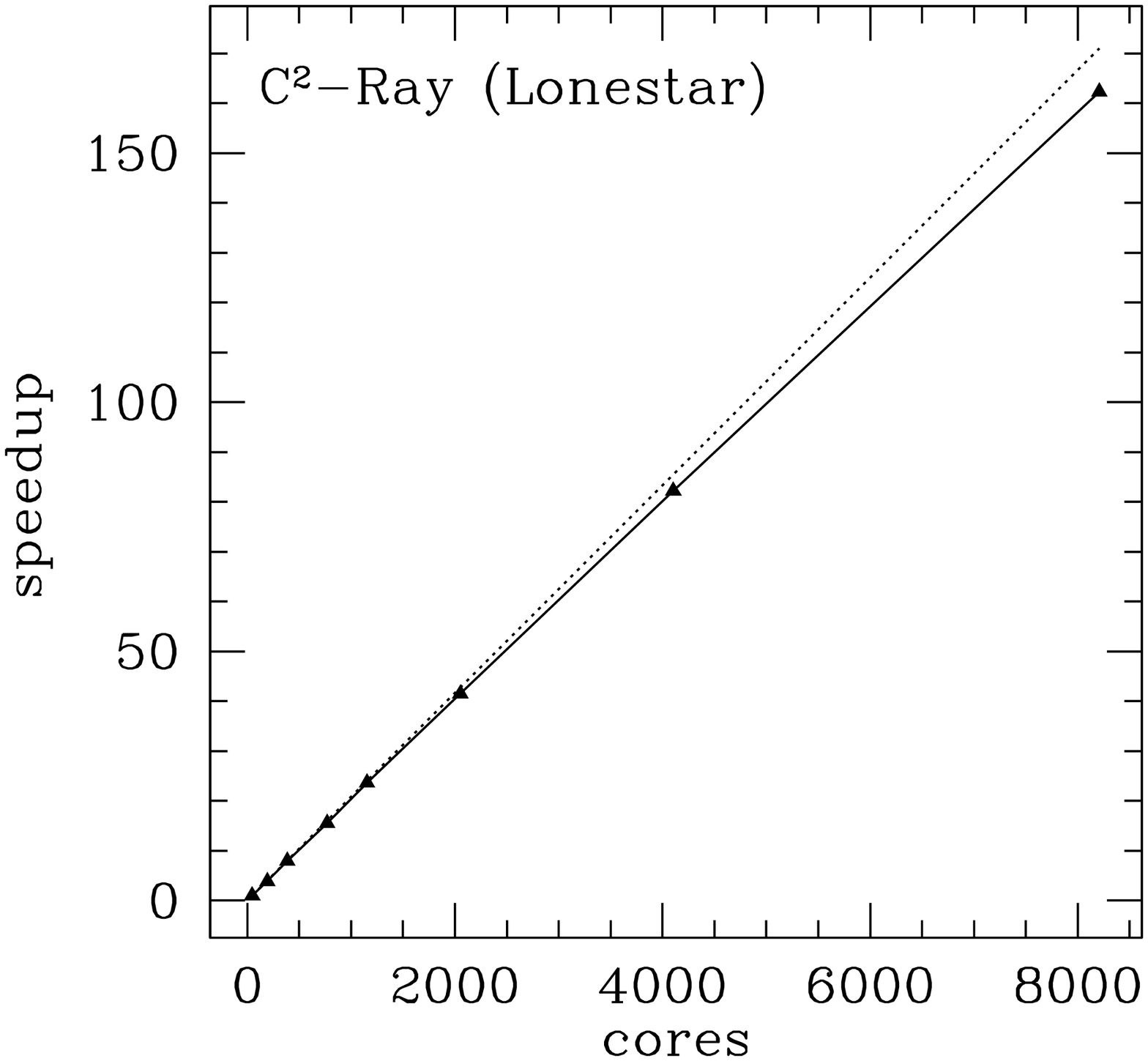}
  \end{center}
%  \vskip -0.5cm 
  \caption{(a)(left) Scaling of the CubeP$^3$M code. Plotted are 
    the code speedup vs. the number of computational cores used. 
    Both quantities are normalized to the smallest run in each 
    case. (b)(right) Scaling of the C$^2$-Ray code. Plotted are 
    the code speedup %, (number of sources)$\times$(number of 
    %cells in grid)/(wall-clock time), 
    vs. the number of computational cores used,
    again normalized to the smallest of the three runs compared. 
    Dashed line indicates the ideal weak scaling for each case.
    \label{scaling}}
\end{figure*}

We start by performing very high resolution N-body simulations of 
the formation of high-redshift structures. We use the CubeP$^3$M 
N-body code\footnote{
\url{http://www.cita.utoronto.ca/mediawiki/index.php/CubePM}, 
 for description of the code see also \citep{2008arXiv0806.2887I}.}
which evolved from the particle-mesh (PM) code PMFAST 
\citet{2005NewA...10..393M}. 
In CubeP$^3$M several important new features were introduced in 
comparison with these previous codes. The first one is the 
addition of a short-range direct particle-particle force, making 
it a $P^3M$ (particle-particle-particle-mesh) code. This 
significantly improves its spatial resolution and accuracy at 
small scales compared to PM codes. A second important new 
development is that CubeP$^3$M is now a massively parallel code 
which can run efficiently on either distributed- or shared-memory
 machines. This is achieved through cubical equal-volume domain 
decomposition and a hybrid MPI and OpenMP approach (see 
\citep{2008arXiv0806.2887I} for more details). CubeP$^3$M scales 
well (with 'weak' scaling, whereby the execution time rises 
proportionally to the problem size) up to thousands of processors, 
as shown in Fig.~\ref{scaling} (left) and has to date been run on 
up to 21,976 computing cores, following up to $5488^3$ particles 
\citep{2010arXiv1005.2502I}. These scaling tests were run on the 
Texas Advanced Computing Center (TACC) computer, {\it Lonestar}, 
and on the currently-available portion of the European Petascale 
computer under development in France, {\it Curie}, at CEA, part 
of the Partnership for Advanced Computing in Europe (PRACE). Our 
results show almost perfect scaling of CubeP$^3$M, within 3\% of 
the ideal one (dashed line), for up to 2,048 cores.

We performed a series of N-body cosmic structure formation 
simulations (summarized in Table~\ref{summary_N-body_table}) with 
increasing simulation box size, from $37\, h^{-1}$Mpc ($53$~Mpc) up 
to $114\, h^{-1}$Mpc ($163$~Mpc), but all with a fixed spatial and 
mass resolution. These N-body simulations were run on a range of 
core number from 256 ($1024^3$ particles) up to 2048 cores ($3072^3$ 
particles). The force smoothing length is fixed to $1/20$ of 
the mean inter-particle spacing, or $1.8\,h^{-1}$~kpc. The largest
of these simulations follows a volume which is 50\% larger than
the largest structure formation simulation performed previously
at similar resolution. 

The N-body simulations required between 4,100 (for $1024^3$ particles) 
and 159,000 (for $3072^3$ particles) computing hours (computing cores 
$\times$ wall-clock hours) on the TACC computer {\it Ranger} 
(SunBlade x6420 with AMD x86 64 Opteron Quad Core, 2.3 GHz, 9.2 
GFlops per core “Barcelona” processors and Infiniband networking). 
The particle mass is $5\times10^6M_\odot$, which guarantees that all 
atomically-cooling halos ($M>10^8M_\odot$) are resolved with at least 
20 particles. We use a spherical overdensity halo finder with 
overdensity parameter fixed to 178 of mean density (Note that 
$\bar{\rho}_M \ne \rho_{\rm crit}$ except for very high redshift. 
The series of N-body simulations with an increasing size allowed 
us to test the convergence of our results with computational box 
size.

The background cosmology is based on WMAP 5-year data combined
with constraints from baryonic acoustic oscillations and 
high-redshift supernovae 
%WMAP+BAO+SN  ******
($\Omega_M = 0.27, \Omega_\Lambda=0.73, h=0.7, \Omega_b=0.044, 
\sigma_8 =0.8, n=0.96$). The linear power spectrum of density 
fluctuations was calculated with the code CAMB \citep{Lewis:1999bs}.
Initial conditions were generated using the Zel'dovich approximation 
at sufficiently high redshift ($z_i=300$) to ensure against numerical 
artifacts \citep{2006MNRAS.373..369C}.

\subsection{Radiative transfer simulations}
\label{rt_sims_sect}

\begin{table*}
\caption{Reionization simulation parameters and global 
reionization history results. All runs use background cosmology 
based on the WMAP 5-year results.}
\label{summary_table}
\begin{center}
\begin{tabular}{@{}llllllllllllll}\hline
\hline
label & run & boxsize & $g_{\gamma}(f_{\gamma})
\,^3$ & $g_{\gamma}(f_{\gamma})$ &mesh & supp. & min source & min unsupp. & $\tau_{\rm es}$ & $z_{10\%}$&$z_{50\%}$&$z_{90\%}$&$z_{\rm ov}$ \\
&      &[cMpc]  & HMACH & LMACH & & & [$M_\odot$] & halo [$M_\odot$] & & &&&
\\[2mm]
\hline
L1&163Mpc\_g8.7\_130S   &163 & 8.7 (10) & 130 (150) &$256^3$& yes & $10^8   $& $10^9   $ &0.080&13.3&9.4&8.6  &8.3
\\[2mm]
L2&163Mpc\_g1.7\_8.7S   &163 & 1.7 (2)  & 8.7 (10)  &$256^3$& yes & $10^8   $& $10^9   $ &0.058&9.9&7.6&6.9&6.7
\\[2mm]
L3&163Mpc\_g21.7\_0  &163 & 21.7 (25) & 0 (0)&$256^3$& no & 2.2$\times$$10^9 $& 2.2$\times$$10^9 $ & 0.070 &10.3& 9.1&8.6&8.4
\\[2mm]
S1&53Mpc\_g8.7\_130S    &53  & 8.7 (10) & 130 (150) &$256^3$& yes & $10^8   $& $10^9   $ &0.084&13.6&9.8&8.9  &8.5
\\[2mm] 
S2&53Mpc\_g1.7\_8.7S    &53  & 1.7 (2)  & 8.7 (10)  &$256^3$& yes & $10^8   $& $10^9   $ &0.059&10.0&7.7&6.9  &6.7
\\[2mm]
S3&53Mpc\_g8.7\_130     &53  & 8.7 (10) & 130 (150) &$256^3$& no & $10^8   $ & $10^8   $ &0.131&15.6 & 13.9 & 13.2 & 12.9
\\[2mm]
S4&53Mpc\_g0.4\_5.3     &53  & 0.35 (0.4)& 5.3 (6) &$256^3$& no & $10^8   $ & $10^8   $    &0.078&11.7&9.7&8.9  &8.6
\\[2mm]
S5&53Mpc\_g10.4\_0      &53  & 10.4 (12) & 0 &$256^3$& no & $10^9   $ & $10^9   $        &0.071&10.5&9.1&8.5  &8.3
\\
\hline
S6&53Mpc\_g8.7\_130S9   &53  & 8.7 (10) & 130 (150) &$256^3$& yes$^{2}$ & $10^8   $& $10^9   $ &0.111&14.9&12.6&10.7  &9.5
\\[2mm]
S7&53Mpc\_g8.7\_130S5   &53  & 8.7 (10) & 130 (150) &$256^3$& yes$^{2}$ & $10^8   $& $10^9   $ &0.089&13.9&10.1&9.0  &8.6
\\
\hline
S8&53Mpc\_uvS\_1e9      &53  & variable$^{1}$ & 0 &$256^3$& no & $10^9  $& $10^9   $  &0.084&13.7&9.7&8.9  &8.5
\\[2mm] 
S9&53Mpc\_uvS\_1e10     &53  & variable$^{1}$ & 0 &$256^3$& no & $10^{10}  $& $10^{10}  $&0.080 &12.2&9.8&8.9&8.5
%\\[2mm]
%\hline
\end{tabular}
\end{center}
$^1$ see Figure~\ref{g_eff} and discussion in Appendix~\ref{appendixA}.  
$^2$ employing a different suppression criterion, 
see Appendix~\ref{appendixB}.  \\
$^3$ $f_\gamma$ is related to $g_\gamma$ by Eqn.~(\ref{eqn:g_gamma_convention}) with $\Delta t = 11.53$ Myrs. 
\end{table*}

The radiative transfer simulations are performed with our code 
C$^2$-Ray (Conservative Causal Ray-Tracing) \citep{methodpaper}. 
The method is explicitly photon-conserving in both space and time 
for individual sources and approximately (to a good approximation) 
photon-conserving for multiple sources, which ensures correct 
tracking of ionization fronts without loss of accuracy, independent 
of the spatial and time resolution, with corresponding great gains 
in efficiency. The code has been tested in detail against a number 
of exact analytical solutions \citep{methodpaper}, as well as in 
direct comparison with a number of other independent radiative 
transfer methods on a standardized set of benchmark problems 
\citep{comparison1,comparison2}. The ionizing radiation is 
ray-traced from every source to every grid cell using the short 
characteristics method, whereby the neutral column density 
between the source and a given cell is given by interpolation of 
the column densities of the previous cells which lie closer to 
the source, in addition to the neutral 
column density through the cell itself. The contribution of each 
source to the local photoionization rate of a given cell is first 
calculated independently, after which all contributions are added 
together and a nonequilibrium chemistry solver is used to 
calculate the resulting ionization state. Ordinarily, multiple 
sources contribute to the local photoionization rate of each cell. 
Changes in the rate modify the neutral fraction and thus the 
neutral column density, which in turn changes the photoionization 
rates themselves (since either more or less radiation reaches the 
cell). An iteration procedure is thus called for in order to 
converge to the correct, self-consistent solution. While our basic 
methodology remains essentially as described in \citet{methodpaper}, 
our C$^2$-Ray code has been thoroughly re-written in Fortran 90, 
made more flexible and modular and parallelized for distributed-memory 
machines. In terms of parallelization strategy, due to the causal 
nature of the ray-tracing procedure (i.e. the state of each cell 
can be calculated only after all previous cells, closer to the source 
are done) it is not possible to employ domain decomposition (except 
for a limited one, into octants, see below), although other approaches 
exist which seek ways to overcome this limitation 
\citep[]{2001MNRAS.321..593N,2006A&A...452..907R}. Instead, the 
main code loop over the sources of ionizing radiation is done in 
massively parallel fashion. Each MPI node has a copy of the density 
field and receives a number of sources whose radiation is to be traced 
through the grid. For the large-scale cosmological reionization 
problem there are typically hundreds of thousands to millions of 
sources, thus our code scales well up to tens of thousands of cores 
at least (see next section). For problems with (relatively) low 
number of ionizing sources such parallelization strategy would 
be inefficient, but such problems are not sufficiently 
computationally-intensive to require such massive parallelization 
and could, instead, be solved on a smaller number of nodes, or even 
in serial.  A similar situation occurs for the initial steps of the 
simulations presented below, when the cosmological structure 
formation is not yet much advanced, thus only a few to few tens of 
halos form. However, their number increases exponentially over time, 
quickly reaching thousands, and then tens and hundreds of thousands. 
We therefore start our simulations on a small number of cores 
(typically 32), raising to thousands of cores as more sources form.

As mentioned above, a limited domain decomposition onto octants is 
possible for our method, since those are independent of each other 
within the short-characteristic ray-tracing framework. We use this 
to (optionally) improve the memory efficiency of the code by doing 
the grid octants in parallel within each MPI node using OpenMP 
multi-threading. This way each MPI node needs only one copy of the 
grid, which is shared amongst the cores within the node.

The radiative transfer problem size scales proportionally to both 
the grid size and the number of sources. Results, shown in 
Figure~\ref{scaling} (right) demonstrate almost perfect scaling, 
within $\sim10\%$ from the ideal one, for up to 8,192 cores.

The N-body simulations discussed above provide us with the spatial 
distribution of cosmological structures and their evolution in 
time. We then use this information as input to a full 3D radiative 
transfer simulations of the reionization history, as follows. We 
saved series of time-slices, both particle lists and halo 
catalogues from redshift 50 down to 6, uniformly spaced in time, 
every $\Delta t=11.53$~Myr, a total of 76 slices. Based on the 
particle distribution at each redshift we used SPH-style smoothing 
scheme using the nearest neighbours (to be described in detail in 
a companion paper, in prep.) to produce regular-grid density and 
bulk velocity fields at the radiative transfer resolution of 
$256^3$ cells. 

All identified halos are potential sources of ionizing radiation, 
with a photon production rate per Myr, $\dot{N}_\gamma$, 
proportional to their mass, $M$:
\be
\dot{N}_\gamma=\frac{f_\gamma M\Omega_b}{\Delta t\, \mu \Omega_0 m_p}\,, 
\ee
where $m_p$ is the proton mass and $f_\gamma = f_{\rm esc} f_\star N_\star$ 
is an ionizing photon production efficiency parameter which
includes the efficiency of converting gas into stars, $f_*$,
the ionizing photon escape fraction from the halo into the IGM,
$f_{\rm esc}$ and the number of ionizing photons produced per
stellar atom, $N_\star$. The latter parameter depends on the
assumed IMF for the stellar population and varies between 4,000
and $\sim 100,000$. %(Pop III vs. Pop II, etc.
Halos were assigned different luminosities according to whether 
their mass was above (``large sources'') or below (``small sources'') 
$10^9 M_\odot$ (but above $10^8 M_\odot$, the minimum resolved halo 
mass). Low-mass sources are assumed to be suppressed within 
ionized regions (for ionization fraction higher than 10\%), through 
Jeans-mass filtering, as discussed in \citet{2007MNRAS.376..534I}.

We note that while previously we used the 
factor, $f_\gamma$, to characterize the source efficiencies, 
here we define a slightly different factor, $g_\gamma$, that is 
given by
\be
g_\gamma$=$f_\gamma\left(\frac{10 \;\mathrm{Myr}}{\Delta t}\right)\,
\label{eqn:g_gamma_convention}
\ee
where ${\Delta t}$ is the time between two snapshots from the 
N-body simulation. The new factor $g_\gamma$ has the advantage 
that it is independent of the length of the time interval 
between the density slices, and as such it allows a direct 
comparison between runs with different $\Delta t$. For reader's 
convenience we listed the values of both parameters in 
Table~\ref{summary_table}. We also note that the specific 
numerical values of the efficiency parameters are strongly 
dependent on the background cosmology adopted and the minimum 
source halo mass. Therefore, parameter values for simulations 
based on different underlying cosmology and resolution should 
not be compared directly, but would require a cosmology and 
resolution-dependent conversion coefficients to achieve the 
same reionization history. 

Our full simulation notation reads $Lbox\_gI\_J(S)(K)$ (the 
bracketed quantities are listed only when needed), where $'Lbox'$
is the simulation box size in Mpc, $'I'$ and $'J'$ are the values 
of the $g_{\rm gamma}$ factor for HMACHs and LMACHs, respectively,
the symbol `S' means that the small sources are suppressed within 
already-ionized regions and 'K' indicated the ionized fraction 
threshold for a given radiative transfer cell above which this 
suppression occurs for halos residing in that cell, which is 0.1 
if not listed explicitly and raised to 0.9 or 0.5 for cases $K=9$ 
and $K=5$, respectively (see below for details). For example, 
53Mpc\_g8.7\_130S indicates that large sources have an efficiency 
$g_\gamma=8.7$, while small sources have an efficiency $g_\gamma=130$ 
and are suppressed in ionized regions. 

We have performed series of radiative transfer simulations with 
varying underlying assumptions about the source efficiencies and
the suppression conditions imposed on the low-mass sources, as 
summarized in Table~\ref{summary_table}. For our radiative 
transfer simulations we use the data from the largest N-body box, 
$114/h=163$~Mpc, and the smallest one of these, $37/h=53$~Mpc. 
The former volume is sufficiently large to faithfully represent 
the reionization observables, while the latter one affords much 
faster and computationally cheaper simulations, which allows us 
to explore a wider parameter space. These two very different 
computational volumes also allow us to investigate resolution 
effects and evaluate which features of reionization and observable 
signatures are sensitive to the box size and which are less so. 
We label all runs by a short label (listed in the first column 
of Table~\ref{summary_table}) for more compact notation. Large-box 
runs are labelled L1-L3, while small-box ones are labelled S1-S9.

Our fiducial runs, to which all the others will be compared 
are 163Mpc\_g8.7\_130S (L1) and the companion small-box one
with same source efficiencies, 53Mpc\_g8.7\_130S (S1). These 
parameters yield a relatively early overlap and high electron 
scattering optical depth. The second set of simulations, 
163Mpc\_g1.7\_8.7S (L2) and 53Mpc\_g1.7\_8.7S (S2) are in the 
opposite limit, which assumes considerably lower efficiencies 
for both types of sources and as a consequence serves as a model 
for a late-overlap, extended, more photon-poor reionization 
scenario. These two cases are designed to roughly bracket the 
range of observationally-allowed reionization scenarios.

Our third large-box simulation, 163Mpc\_g21.7\_0 (L3) is 
equivalent to our previous large-box simulations without 
self-regulation presented in \citep{2008MNRAS.384..863I}, 
except for the updated background cosmology and the source 
photon production efficies, adjusted here to yield the same 
overlap epoch as our fiducial case L1. Therefore, L1 and L3
share the underlying density structures and sources (apart 
from the different minimum mass cutoffs) and hence a 
head-to-head comparison yields the effects of the presence 
of low-mass sources and Jeans-mass filtering.

The rest of our cases, S3 to S9, test various aspects of the 
reionization source modelling. Simulation S3 is an extreme 
case which has the same source efficiencies as our fiducial 
case L1, but assumes no suppression occurs. Naturally, this 
results in a very early reionization and very high integrated 
optical depth, $\tau_{\rm es}=0.131$, which is well outside the 
WMAP5 $1-\sigma$ range of $\tau_{\rm es}=0.084\pm0.016$. Simulation
53Mpc\_g0.4\_5.3 (S4) is again without suppression, but here we 
tuned down the efficiencies of both types of sources so as to 
achieve approximately the same overlap epoch as in our fiducial 
case L1. In simulation 53Mpc\_g10.4\_0 (S5) we assume that there 
are no low-mass, $M<10^9M_\odot$ sources at all and we adjust the 
photon efficiency of the remaining sources to again reach overlap 
at roughly the same epoch as in the fiducial case. 

\begin{figure*}
\begin{center}  
\includegraphics[width=3.2in]{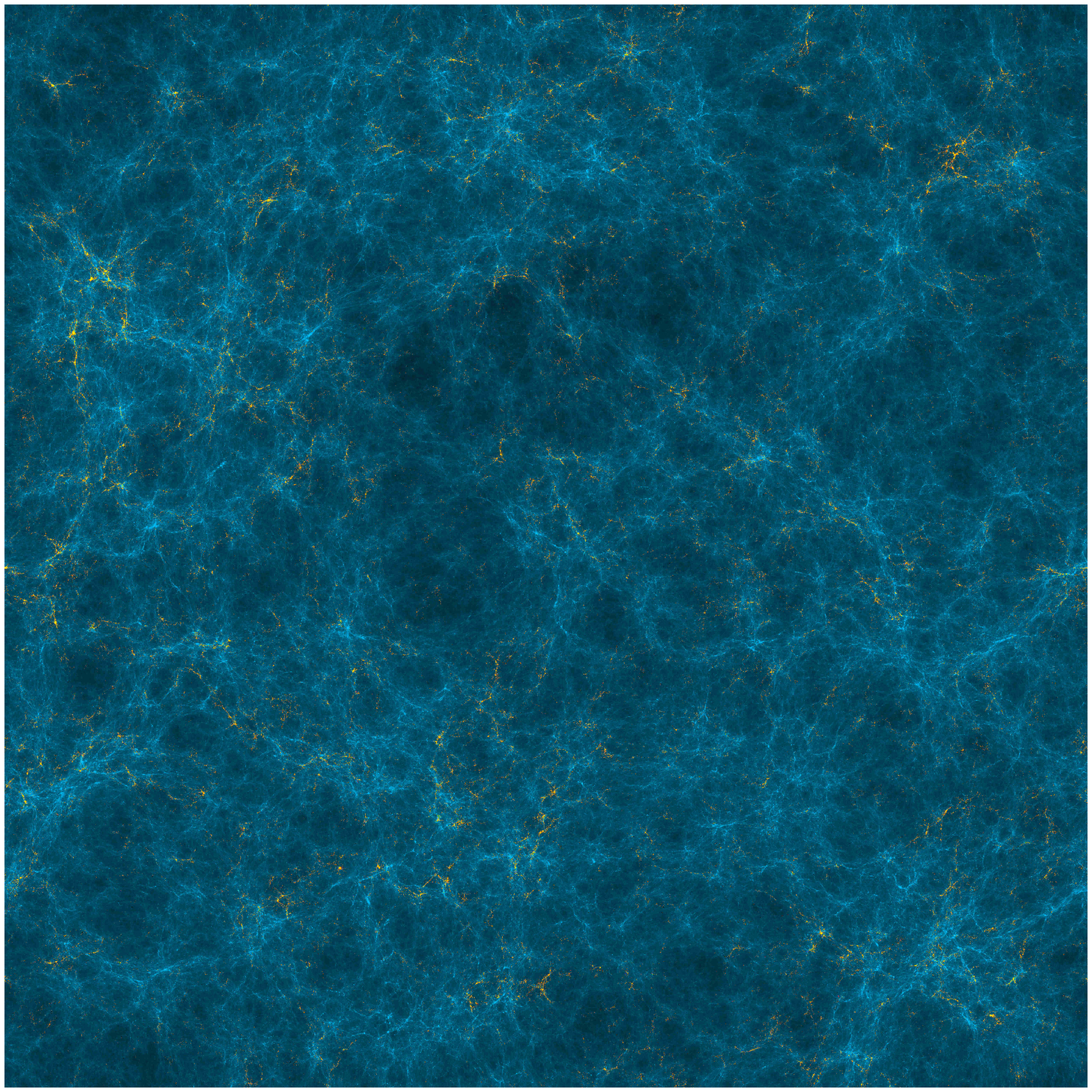} 
\includegraphics[width=3.2in]{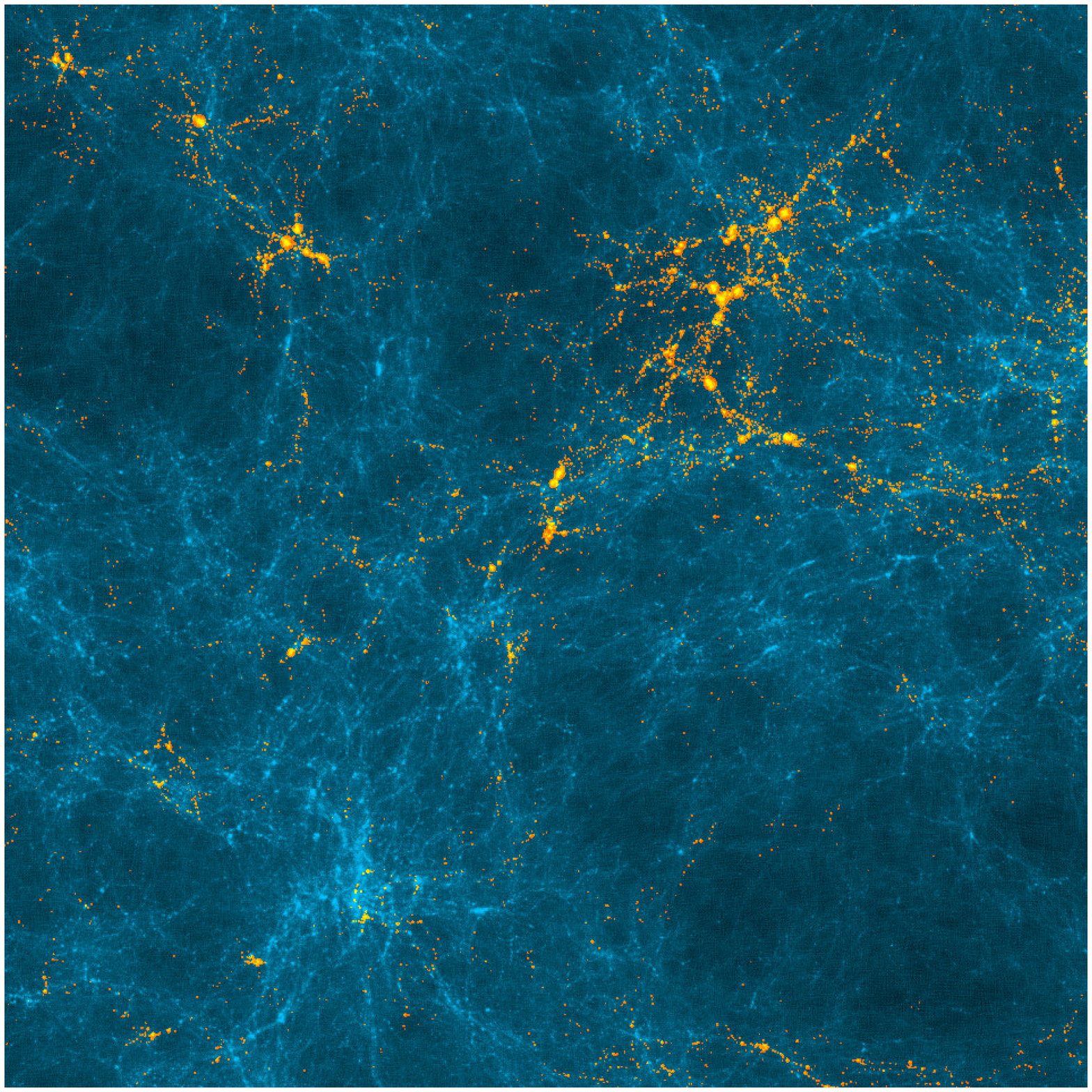} 
%\vspace{-1in}
\caption{
\label{z6_114Mpc_image}
(left) Slice of the Cosmic Web at redshift $z=6$ from our CubeP$^3$M 
simulation with $3072^3$ particles (29 billion) on a $6144^3$ fine 
grid in a comoving volume of $163$~Mpc on a side. Shown are the dark 
matter density (blue) and halos (in actual size; yellow). Image 
resolution is $6144\times6144$, the slice is $1/h$ Mpc thick. (right) 
Zoomed-up region (25.76$\times$25.76~Mpc) of the same image.
}
%\vspace{-0.5cm}
\end{center}
\end{figure*}

Additionally, we consider two scenarios which have exactly the
same time-dependent ionizing photon emissivity (and therefore
almost identical reionization history) as our fiducial case S1,
but with higher minimum source mass of $10^9M_\odot$ 
(53Mpc\_uvS\_1e9; S8) and $10^{10}M_\odot$ (53Mpc\_uvS\_1e10; S9).
The fixed ionizing photon emissivity results in unphysically 
high early source luminosities and we consider them primarily 
in order to illustrate the effect of re-distributing the full 
luminosity of all sources over the massive ones only, similar
to the models adopted in some recent work 
\citep{2009MNRAS.393...32T,2009A&A...495..389B}. These simulations
and some illustrative results from them are discussed in 
Appendix~\ref{appendixA}. 

Finally, in order to evaluate the rubistness of our source 
suppression model, we consider two more scenarios, 
53Mpc\_g8.7\_130S9 (S6) and 53Mpc\_g8.7\_130S5 (S7), whereby 
we raise the ionization threshold for low-mass source suppression 
to $x_{\rm threshold}=0.9$ and 0.5, respectively, from our fiducial 
threshold of $x_{\rm threshold}=0.1$. Since this is a more technical 
study we present its results separately, in Appendix~\ref{appendixB}.

The radiative transfer simulations presented in this work 
typically required $\sim0.5-1$ million computing hours ($163$~Mpc 
boxes) and $\sim10-30$ thousand computing hours ($53$~Mpc boxes), 
depending on the specific set of source parameters we adopted.

\section{Results: Early structure formation}
\label{results:sect}

\begin{figure*}
\begin{center}  
\includegraphics[width=3.2in]{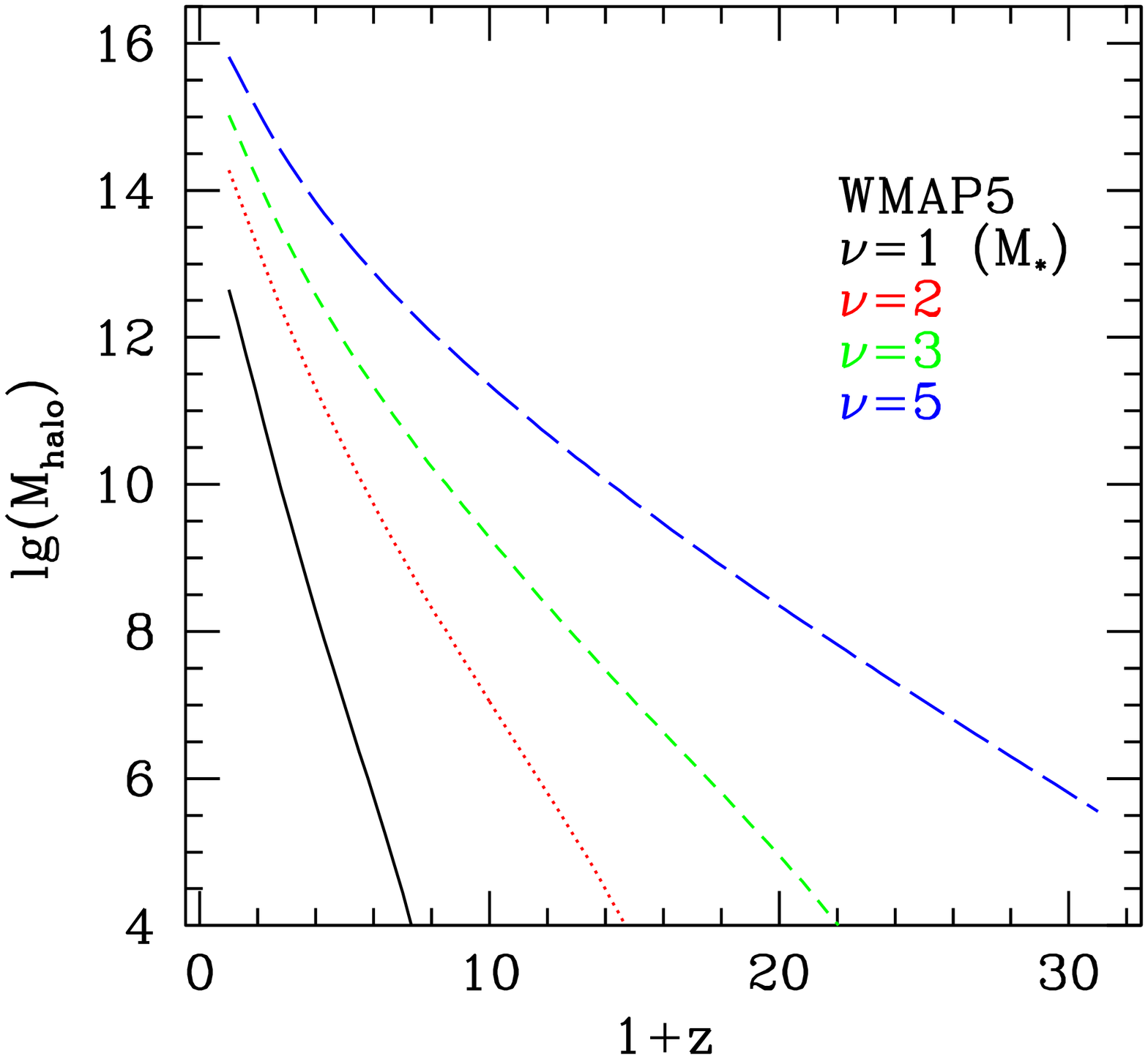} 
\includegraphics[width=3.2in]{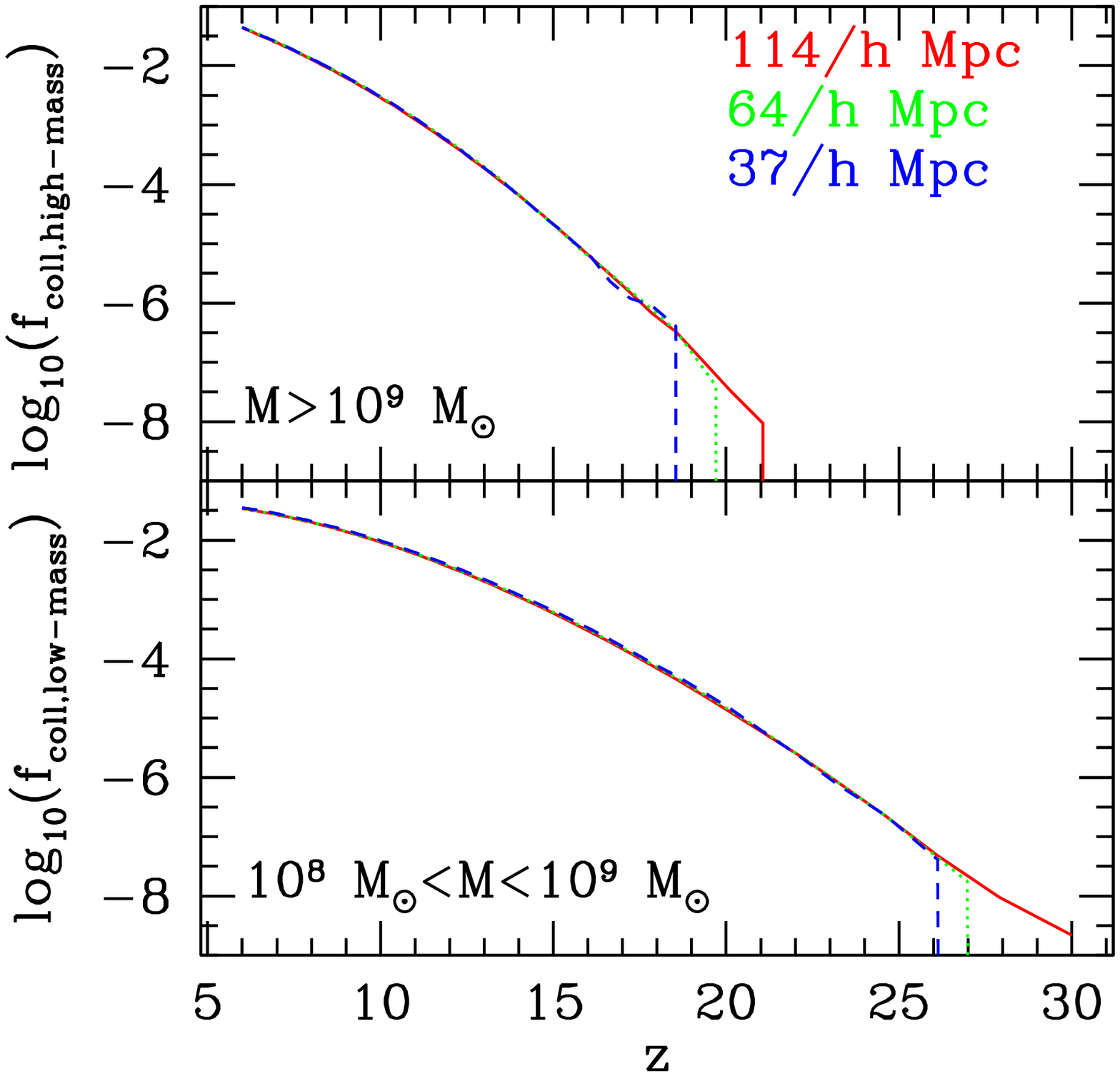} 
%\vspace{-1in}
\caption{
\label{fcoll_fig}
(left) Halo abundances for $\nu=\delta_cD_+/\sigma(0,M)$=1 ($M_*$;
 black, solid), 2 (red, dotted), 3 (green, short-dashed), and 5 
(blue, long-dashed). (right) Collapsed fraction of HMACHs (top) 
and LMACHs (bottom) halos.} 
%\vspace{-0.5cm}
\end{center}
\end{figure*}

\begin{figure*}
\begin{center}  
\includegraphics[width=2.3in]{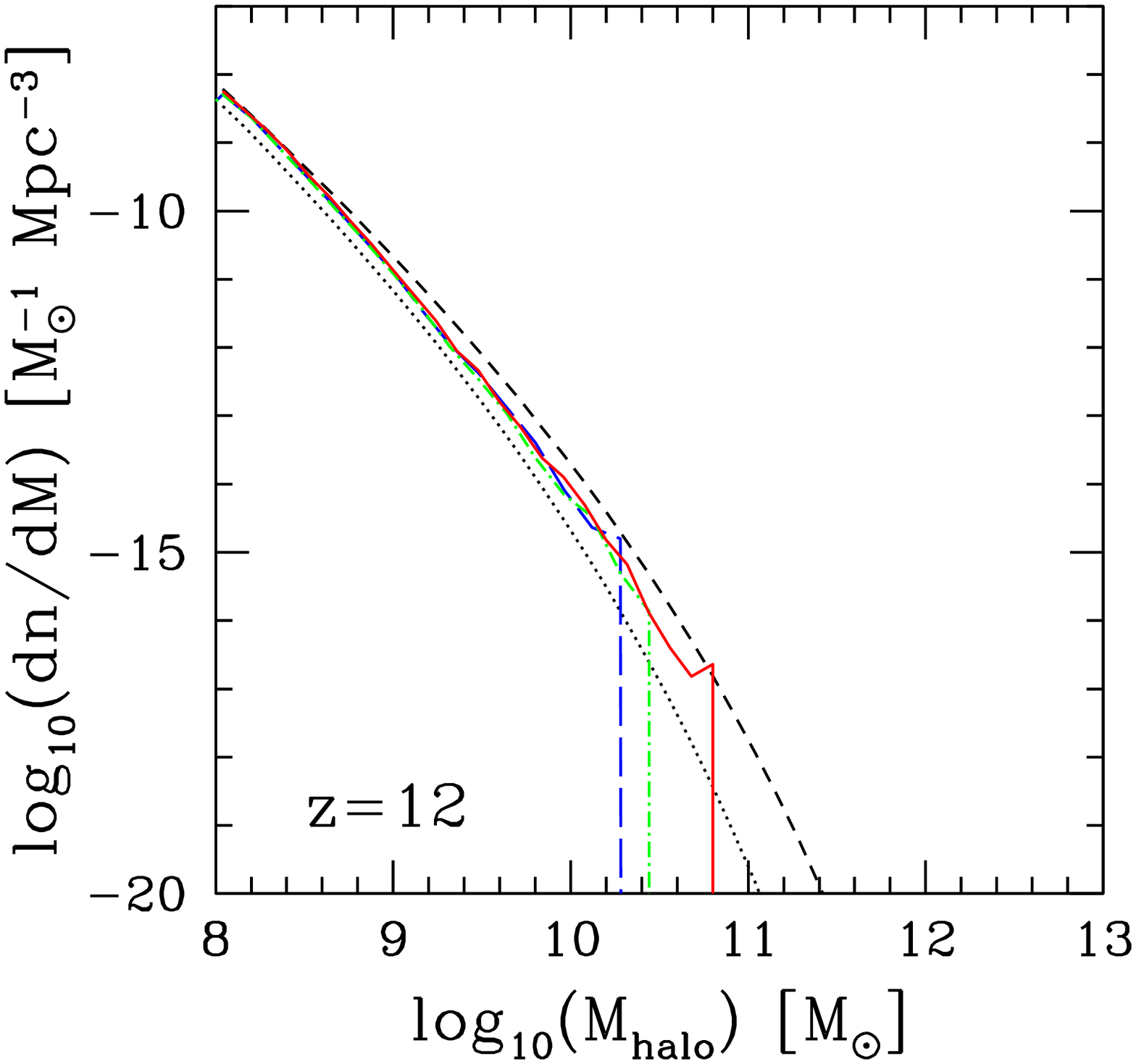} 
\includegraphics[width=2.3in]{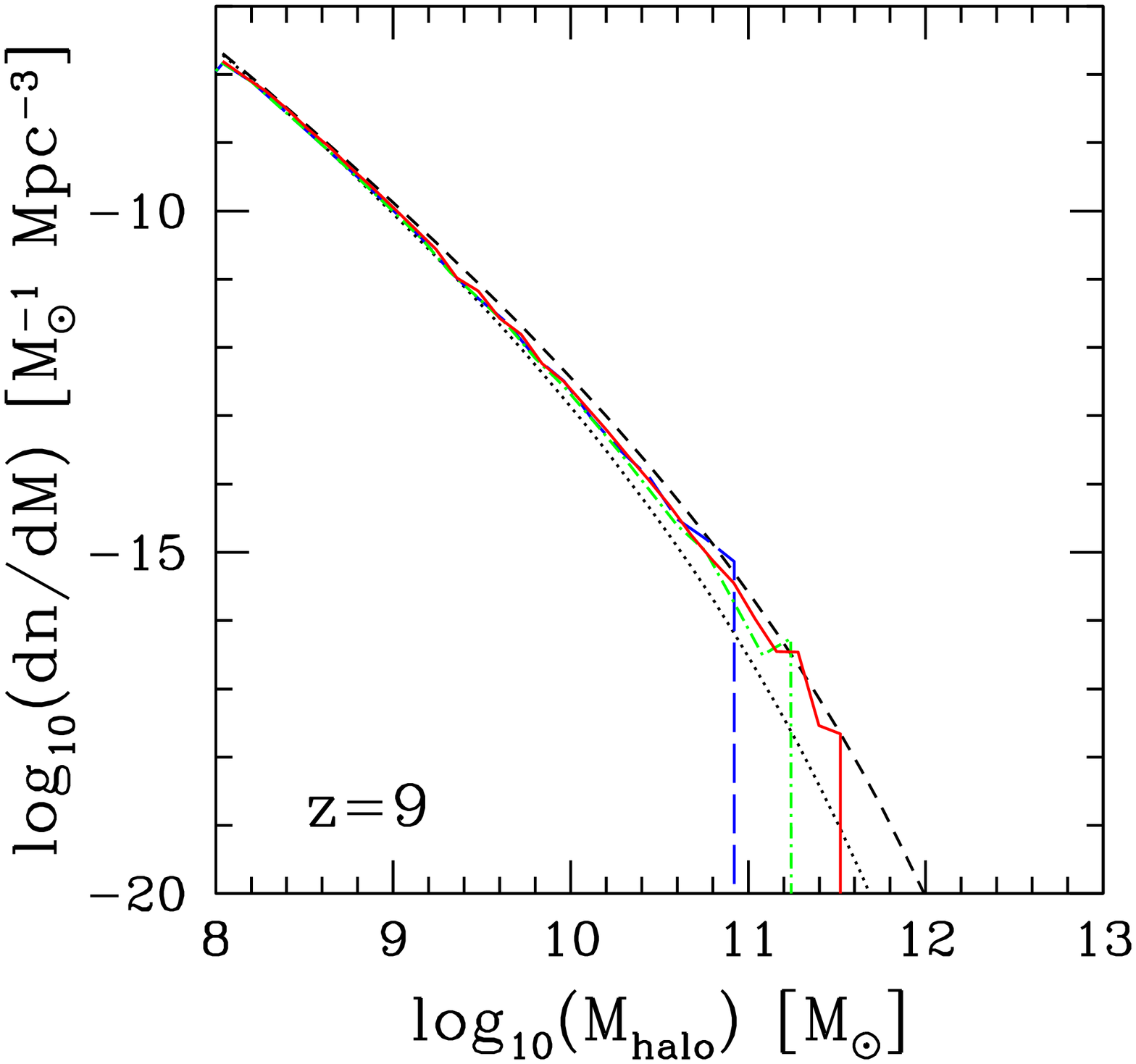} 
\includegraphics[width=2.3in]{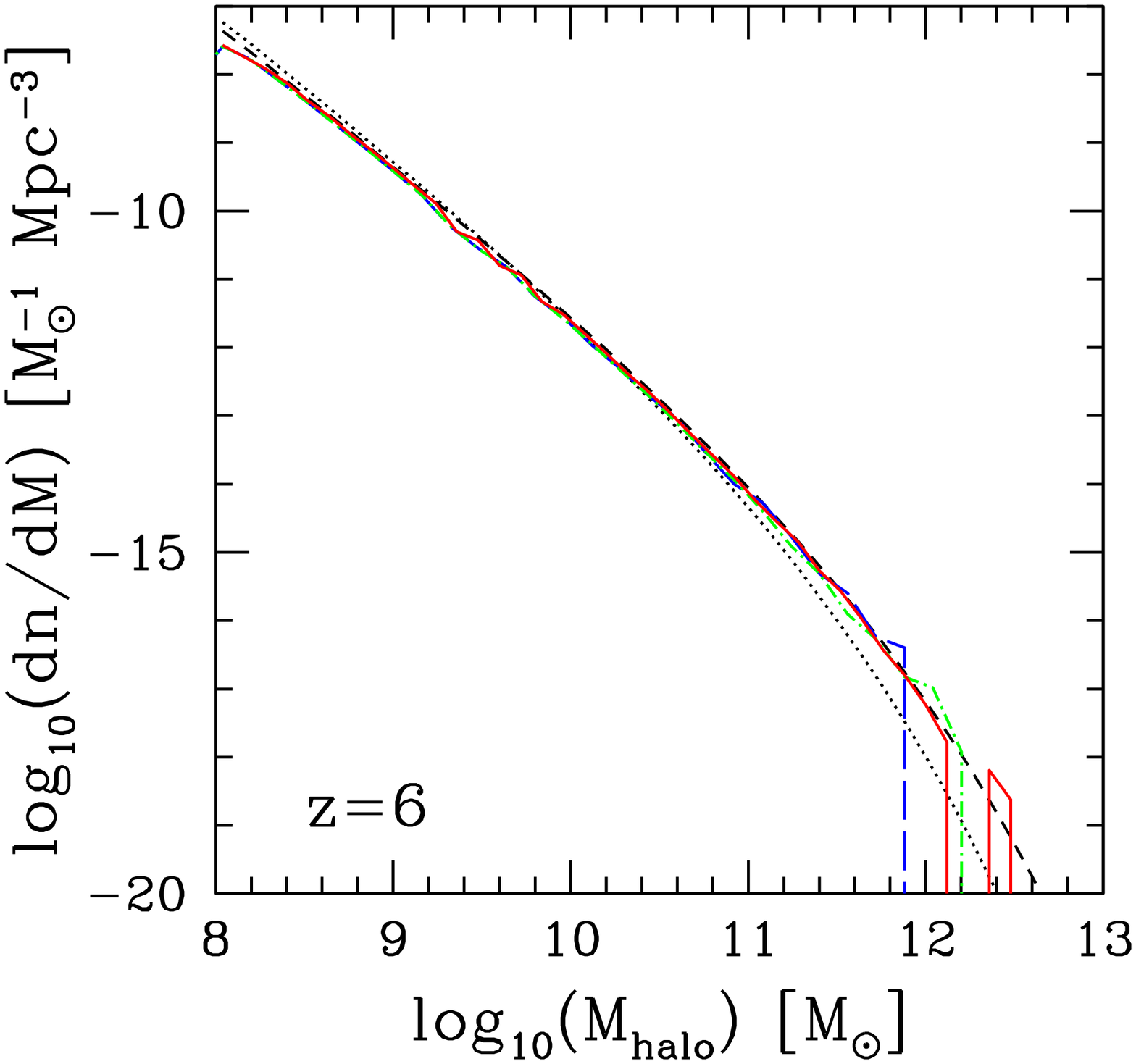} 
%\vspace{-1in}
\caption{
\label{mf_fig}
Simulated halo mass function at high-z derived from 114/h Mpc 
box (solid, red), 74/h Mpc box (dot-dashed, green), 37/h Mpc 
box (long-dashed, blue) at (left to right) $z=12, 9$ and 6. Also 
shown are the Press-Schechter (dotted, black) and Sheth-Tormen 
(short-dashed, black) analytical mass functions. 
}
\end{center}
\end{figure*}

\begin{figure*}
\begin{center}  
\includegraphics[width=3.2in]{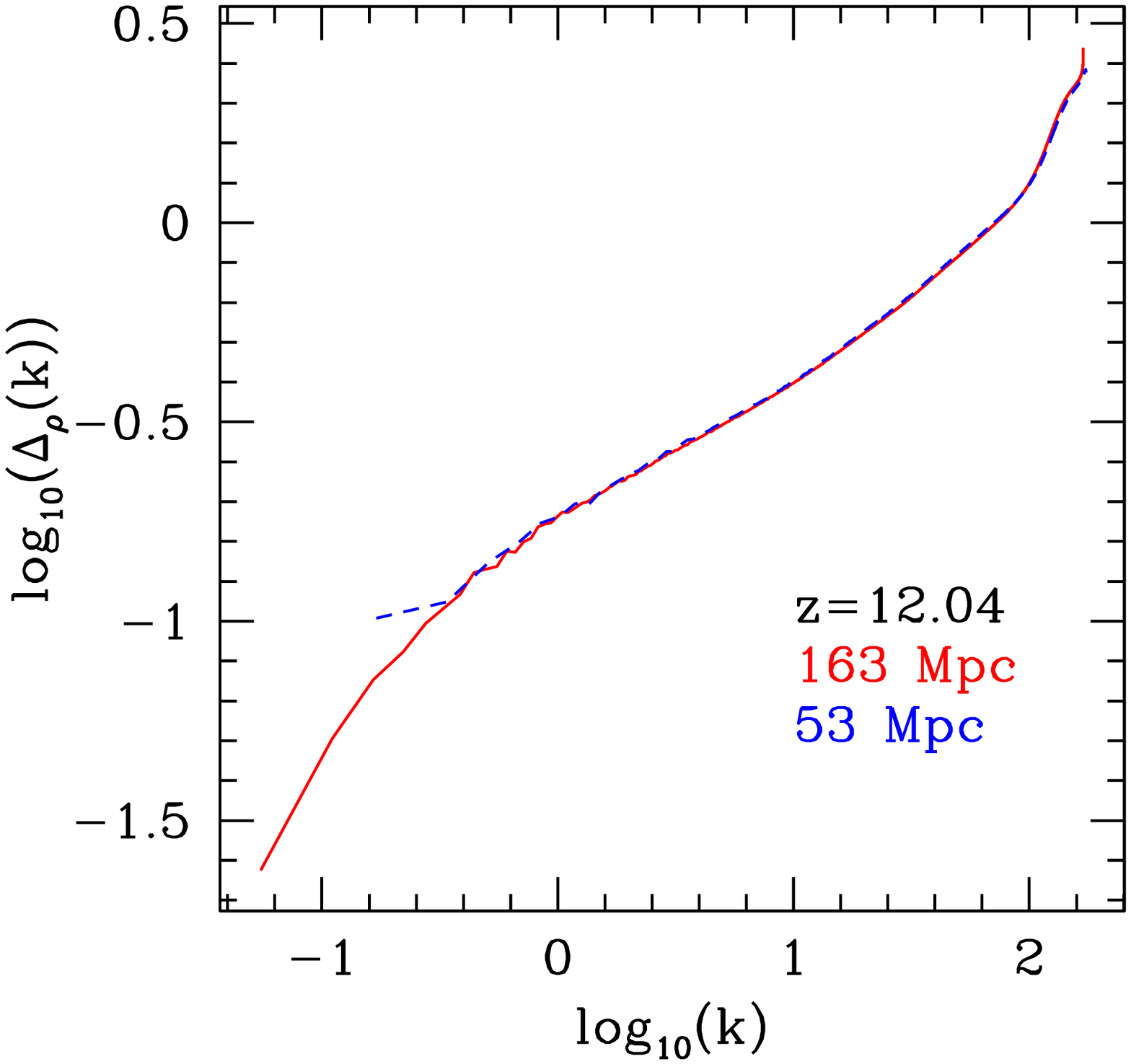} 
\includegraphics[width=3.2in]{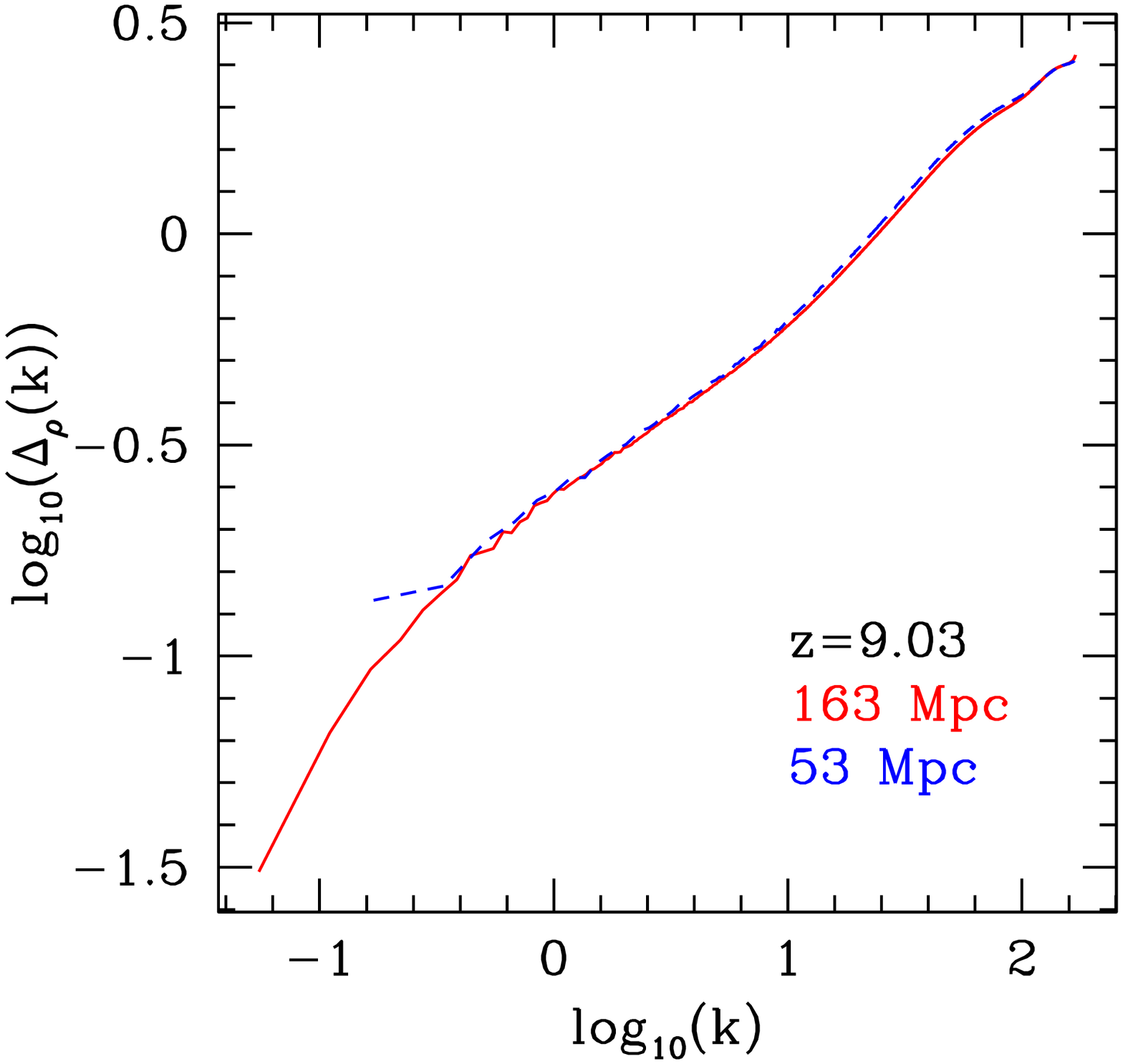} 
%\vspace{-1in}
\caption{
\label{dm_power_fig}
The power spectra of the density fields, 
$\Delta_{\rho}=(k^3P(k)/2\pi^2)^{1/2}$, for the $114/h=163$~Mpc 
box run (red, solid) and the $37/h=53$~Mpc box run (blue, 
dashed) at redshifts $z=12$ (left) and $z=9$ (right).} 
%\vspace{-0.5cm}
\end{center}
\end{figure*}

\begin{figure*}
\begin{center}  
\includegraphics[width=3.2in]{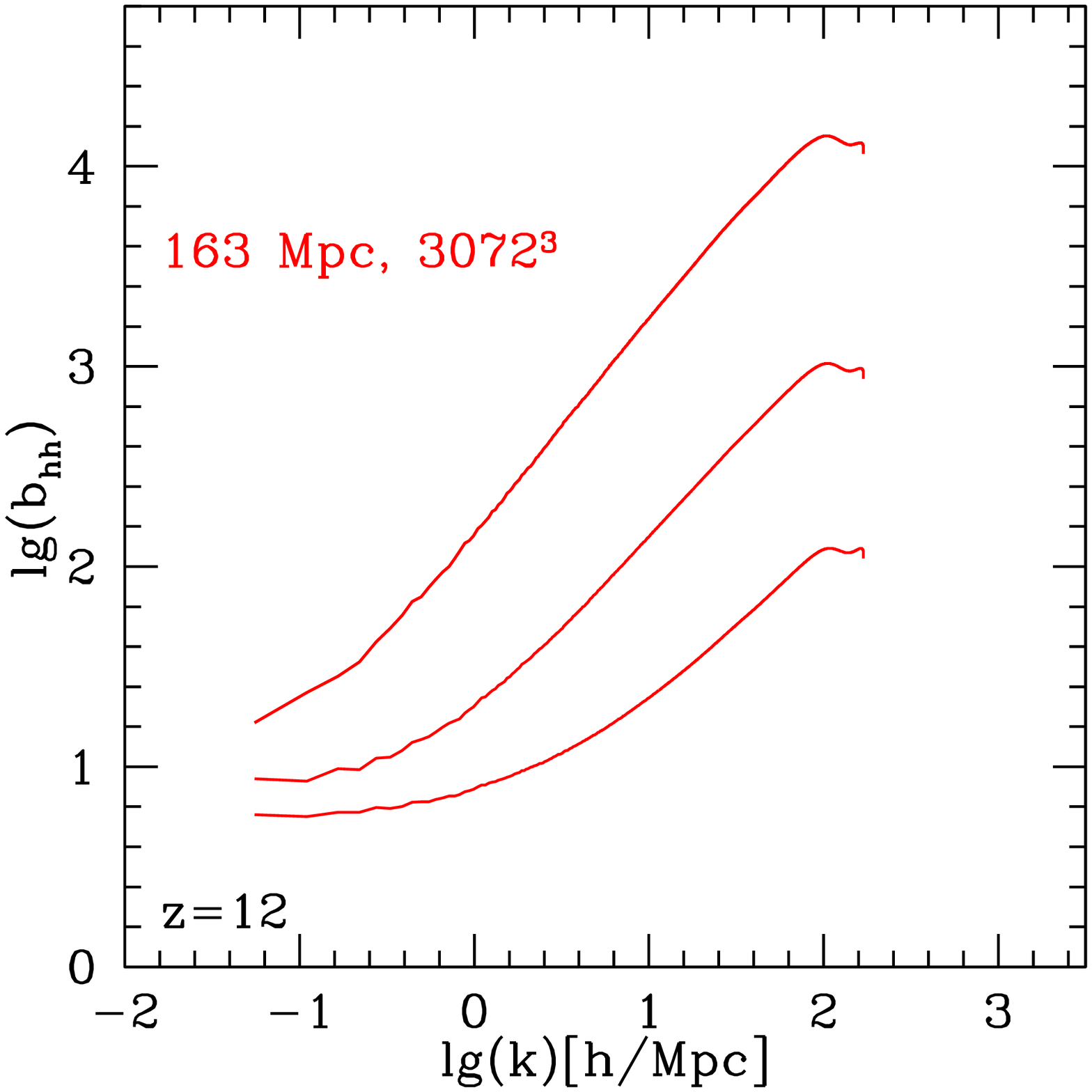} 
\includegraphics[width=3.2in]{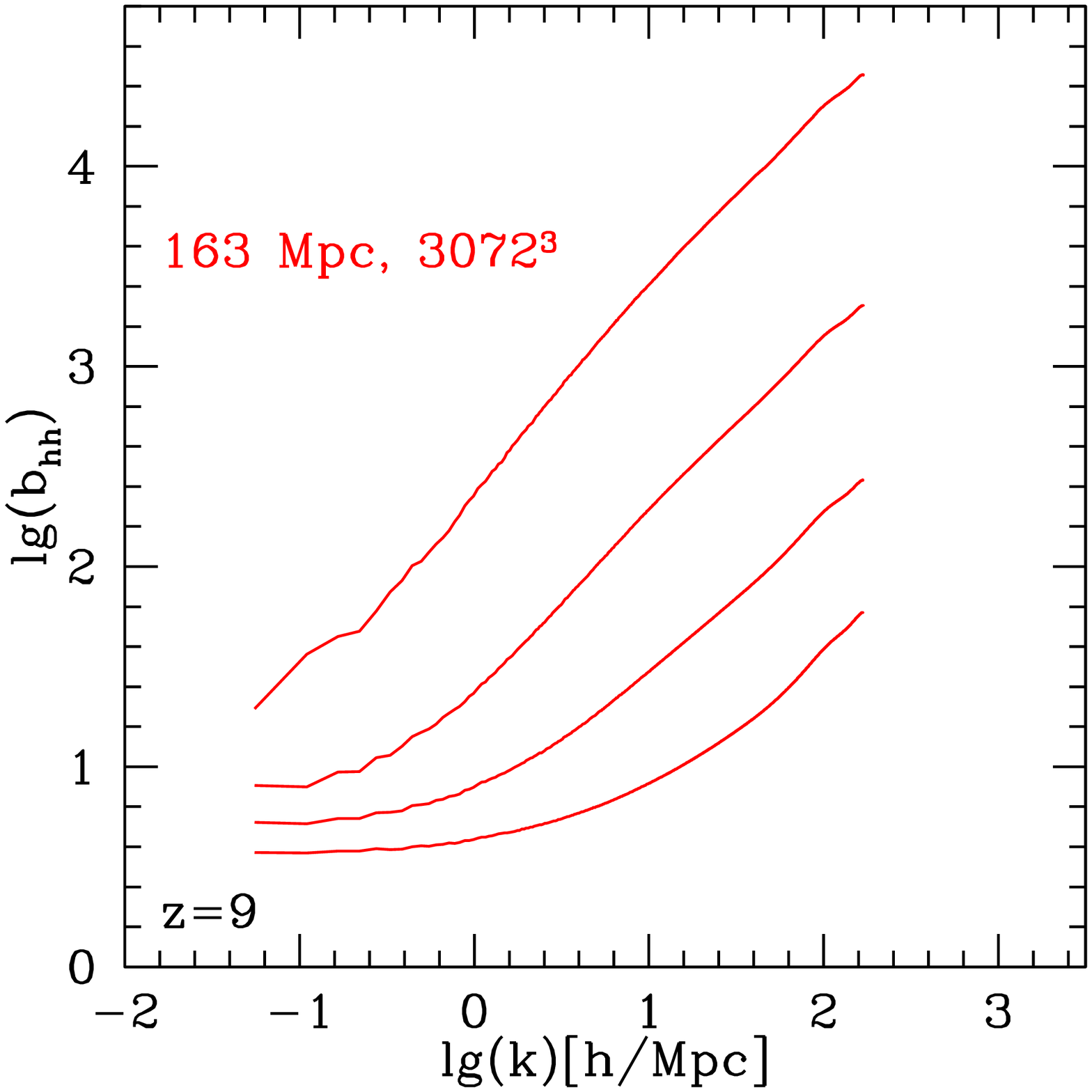} 
%\vspace{-1in}
\caption{
\label{halo_bias_fig}
The halo bias, $b_{\rm hh}=\Delta_{\rm hh}/\Delta_\rho$, for the 
$114/h=163$~Mpc box run (red, solid) at redshifts $z=12.04$ (left) 
and $z=9.03$ (right). Lines are for halos binned by decades of 
mass (bottom to top curve) $10^8M_\odot<M_{\rm halo}<10^9M_\odot$, 
$10^9M_\odot<M_{\rm halo}<10^{10}M_\odot$, 
$10^{10}M_\odot<M_{\rm halo}<10^{11}M_\odot$,
and $10^{11}M_\odot<M_{\rm halo}<10^{12}M_\odot$.} 
%\vspace{-0.5cm}
\end{center}
\end{figure*}

In Figure~\ref{z6_114Mpc_image} (left: full box, right: zoomed 
sub-volume) we show a slice of the density field and halos at 
redshift $z=6$ from our $114\,h^{-1}$~Mpc (163~Mpc), 
$3072^3$-particle N-body simulation. The structure formation is 
already well-advanced and strongly nonlinear at sub-Mpc scales. 
The very first resolved ($M_{\rm halo}>10^8M_\odot$) halo in this 
volume forms at $z=31$, while the first insupppressible halo
($M_{\rm halo}>10^9M_\odot$) forms at $z=21$. By $z=6$ there are 
over 20.5 million collapsed halos, of which $\sim18.7$ million 
low-mass ($M_{\rm halo}<10^9M_\odot$) halos and $\sim2$ million 
high-mass ($M_{\rm halo}>10^9M_\odot$) halos. The halos are 
strongly clustered at all times, more so going to higher redshifts, 
when they are ever rarer. The halo abundances are usually 
described in terms of $\nu=\delta_cD_+/\sigma(0,M)$, where 
$\delta_c$ is the linear density contrast corresponding to the 
moment of collapse of a top-hat density perturbation, 
$\sigma(0,M)$ is the present variance of the density fluctuations 
corresponding to the mass scale $M$, and $D_+$ is the growth 
factor of the density fluctuations. The halo masses corresponding 
to $\nu=1$ (most common, $M_*$, halos), $\nu=2$, $3$, (rare halos)
 and $\nu=5$ (extremely rare halos) are shown in 
Figure~\ref{fcoll_fig} (left). Clearly, before $\sim6$, within the
redshift range of interest here, there are no $1-\sigma$ halos at
all and all halos are rare. Atomically-cooling halos are 
$2-3-\sigma$ at the low end of the redshift interval and as rare
as $5-\sigma$ at early times. The evolution of the collapsed 
fractions in high-mass and low-mass halos is shown in 
Fig~\ref{fcoll_fig} (right). The collapsed fractions start very 
low and rise exponentially at early times when the halos are very 
rare. The collapsed fraction in low-mass halos reaches $10^{-3}$ 
and $1\%$ at $z=14$ and $z=9.8$, respectively. After that point 
it starts to level off as low-mass halos start to become less 
rare ($2-\sigma$ or less) and their collapsed fraction reaches 
3.4\% by $z=6$. The collapsed fraction in high-mass halos rises 
steeply all the way to $z=6$, eventually reaching 4.25\%. There 
is only a modest departure from the exponential growth, 
reflecting the fact that they remain quite rare throughout this 
period. The simulation volume has essentially no effect on the 
derived collapsed fractions, indicating numerical convergence 
on that quantity at fixed mass resolution. The only exception 
to this is at very early times ($z>26$ for the low-mass halos 
and $z>17$ for the high-mass ones) the corresponding halo 
populations are so rare ($\sim$5-$\sigma$ in each case) that 
Poisson noise (i.e. cosmic variance, due to the smaller volume) 
affects the results. For example, the first resolved halos form 
at $z=31$ in the $114\,h^{-1}$Mpc box, but only at $z=26$ in the 
$37\,h^{-1}$Mpc box. As soon as there is sufficient statistics 
for any given volume the collapsed fractions converge.

The halo mass functions derived from our simulations are shown 
in Figure~\ref{mf_fig} for a range of redsifts, $z=12-6$, along 
with the Press-Schechter \citep[PS;][]{1974ApJ...187..425P} and 
Sheth-Tormen \citep[ST;][]{2002MNRAS.329...61S} analytical mass 
functions. The halo abundancies at all redshifts fall between 
those two analytical predictions. PS always under-predicts the 
abundances of massive, rare halos, while ST over-predicts them. 
With time ST becomes a better match to the numerical results. 
This broadly agrees with previous results on the high-redshift 
mass functions \citep{2006MNRAS.369.1625I,2007MNRAS.374....2R,
2007ApJ...671.1160L}.

\begin{figure*}
\begin{center}  
\includegraphics[width=3.2in]{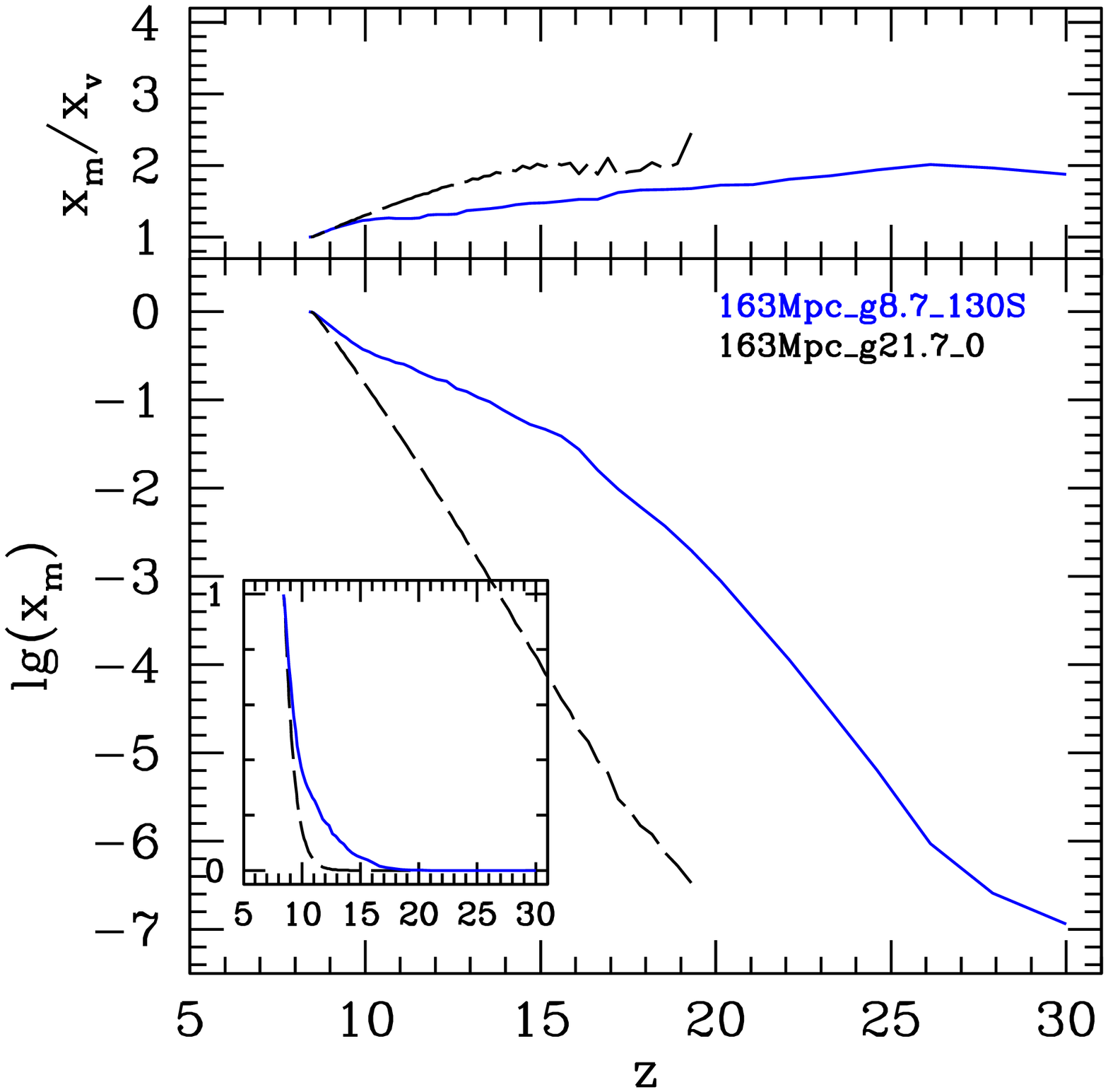}  
\includegraphics[width=3.2in]{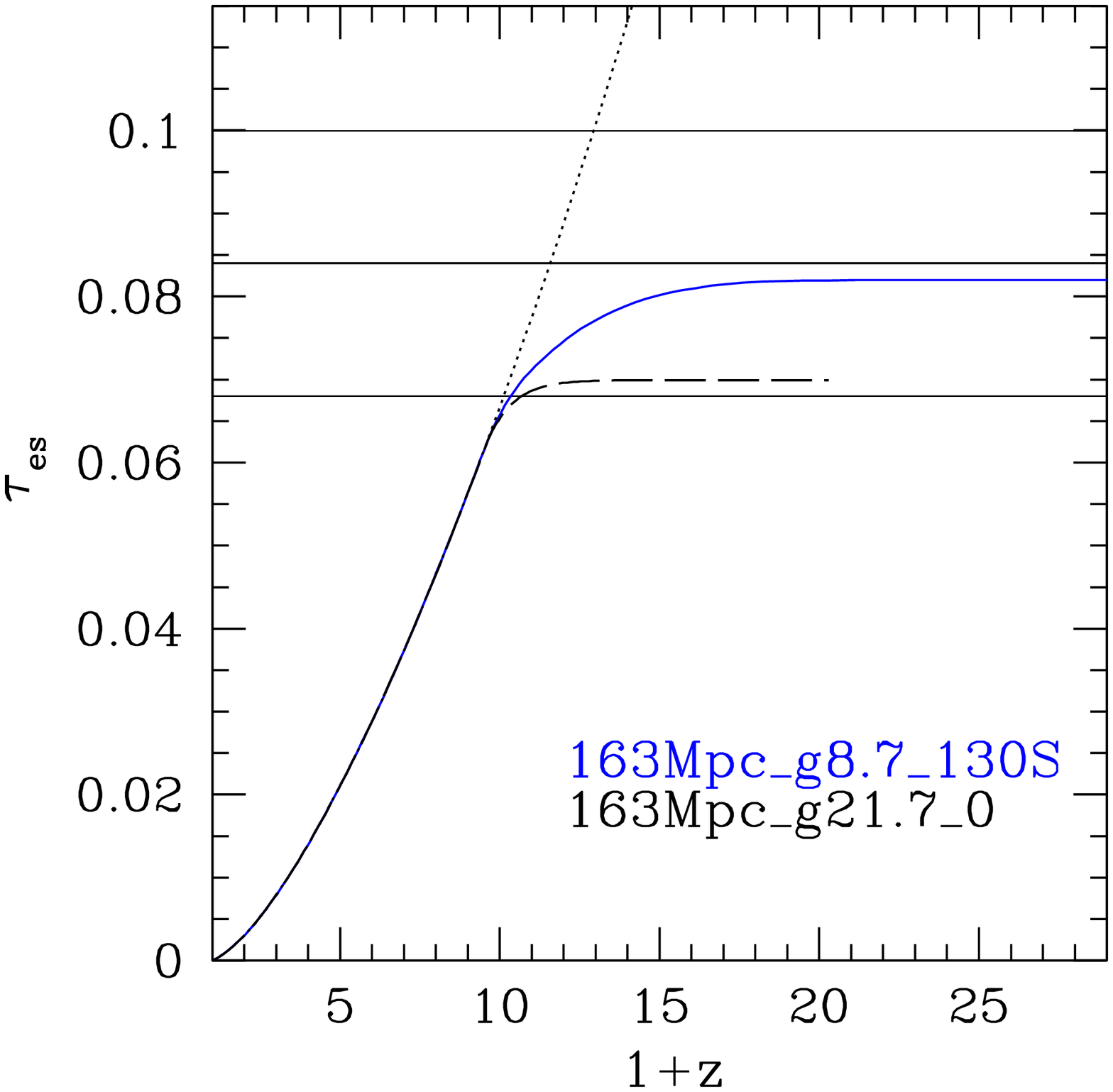}
\caption{
\label{reion_hist_selfreg_fig}
The effect of self-regulation on the reionization history and 
integrated electron-scattering optical depth: (left) 
Mass-weighted reionization histories (bottom) and the ratio of 
the mean mass-weighted and volume ionized fractions, $x_m/x_v$ 
(top) for our fiducial self-regulated case, L1 (blue, solid) 
and the corresponding non-selfregulated case with same overlap 
epoch, L3 (black, long-dashed). The computational box size is 
$163$~Mpc in both cases. Inset shows the same reionization 
histories in linear scale. (right) The corresponding electron 
scattering optical depth, $\tau_{\rm es}(z)$ integrated from 
redshift 0 to redshift $z$ for the same two simulations. 
Horizontal lines indicate the mean and 1-$\sigma$ band derived 
from the WMAP 5-year data, while the dotted line shows the 
value of $\tau_{\rm es}$ for a fully-ionized universe.}
%\vspace{-0.5cm}
\end{center}
\end{figure*}

In Fig.~\ref{dm_power_fig} we plot the total matter density 
field power spectra $\Delta_{\rho}=(k^3P(k)/2\pi^2)^{1/2}$ at 
two representative redshifts for our largest ($114$~Mpc$/h$) 
and smallest ($37$~Mpc$/h$) boxes. Power spectra were 
calculated by interpolating the N-body particles 
using a cloud-in-cell scheme onto the fine grid of the CubeP$^3$M 
code, with $6144^3$ and $2048^3$ cells, respectively. There is a 
close agreement between the two cases, apart from the expected 
variance at scales close to the box size. This shows that there 
is no missing density fluctuation power in the small box, except 
for the scales at or above the box size.

Finally, in Figure~\ref{halo_bias_fig} we show the halo-halo bias, 
calculated as the ratio of the halo autocorrelation power specrum
divided by the density field one, i.e. $b_{\rm hh}=\Delta_{\rm hh}/\Delta_\rho$. 
This measures the clustering of the dark matter halos with respect 
to the underlying matter density 
field at redshifts $z=12$ and 9, roughly corresponding to early and 
advanced stages of reionization. Because of the relative rarity of 
all halos studied here, their clustring is quite strong, particularly 
at the smallest resolved scales ($k\sim100\,h/$Mpc), where it is of 
order $100$ for the lowest mass halos and is as high as $\sim10^4$ 
for the most massive halos. At these small scales the bias is 
strongly nonlinear. At large scales the bias factors asymptote to a 
(mass-dependent) constant - the large-scale linear bias 
\citep{1996MNRAS.282..347M} %{\bf -- can we plot it too?}. 
The transition between the large-scale 
linear bias and the nonlinear one occurs around $k\sim1$ for the 
lowest-mass halos at $z=9$, rising to $k\sim0.1-0.5$ for the larger 
halos and/or higher redshifts. The largest, rarest halos 
($M>10^{10}M_\odot$ at $z=12$,$M>10^{11}M_\odot$ at $z=9$) show a 
roughly linear log-log $b(k)$ relation and never asymptote to the 
linear bias value within the k-range covered in our simulation. 

\section{Results: Basic Features of the Self-regulated 
reionization}
\label{basic_results:sect}

\subsection{The effects of self-regulation}

\begin{figure}
\begin{center}  
\includegraphics[width=3.2in]{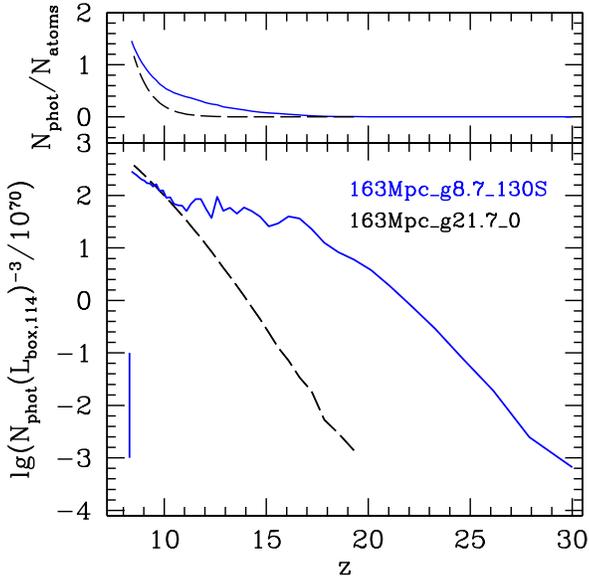} 
%\vspace{-1in}
\caption{
\label{lum_evol_selfreg_fig}
(bottom) Number of ionizing photons emitted by all active 
sources in the computational volume per timestep and (top) 
cumulative number of photons per total gas atom released 
into the IGM. Notation is the same as in 
Fig.~\ref{reion_hist_selfreg_fig}. The vertical line marks 
the overlap redshift ($z(x_m=0.99)$) for each case.}
%\vspace{-0.5cm}
\end{center}
\end{figure}

\begin{figure*}
\begin{center}  
\includegraphics[width=3.2in]{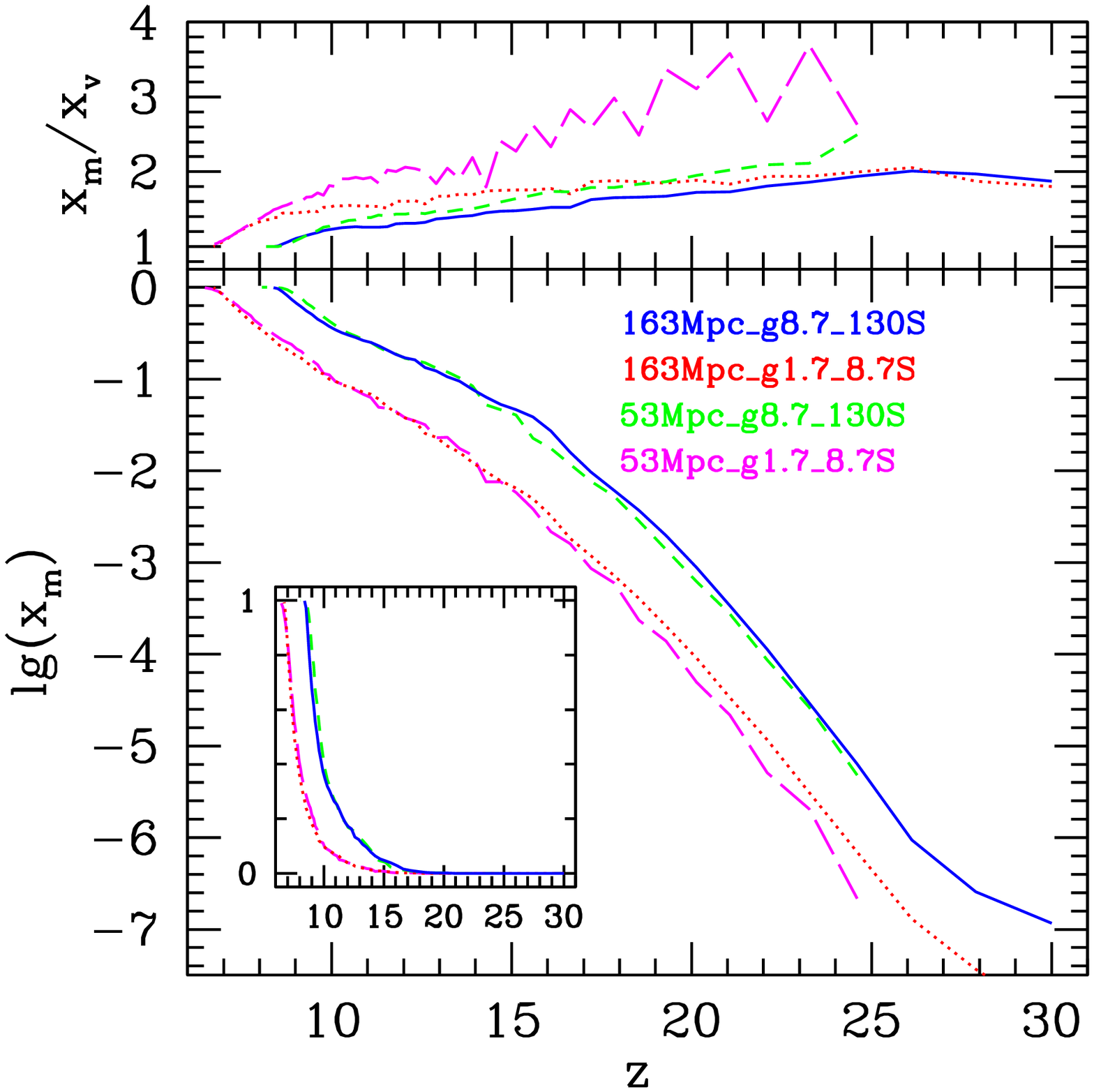} 
\includegraphics[width=3.2in]{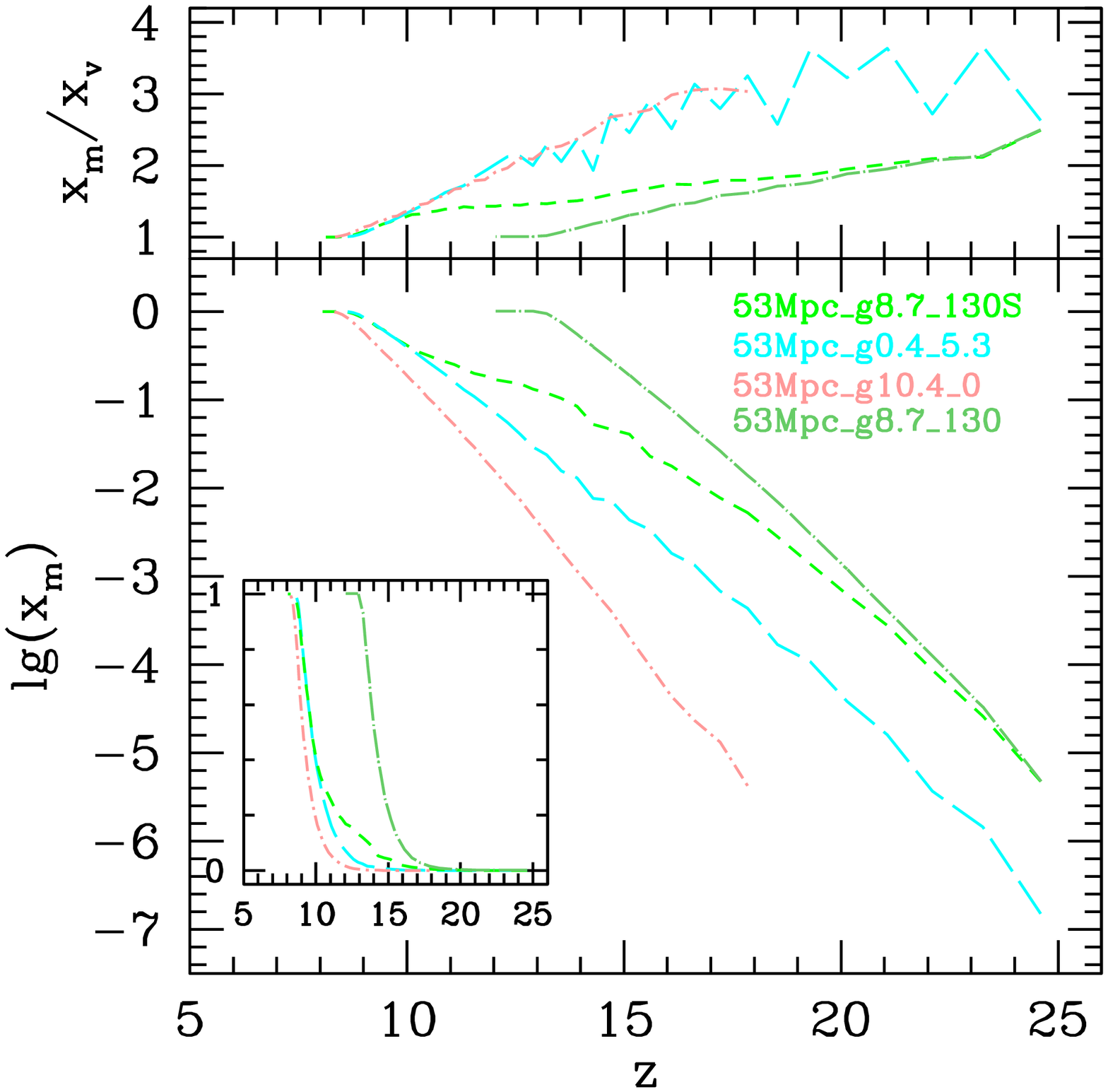} 
\caption{
\label{reion_hist_fig}
(bottom panels) Mass-weighted reionization histories for our fiducial 
self-regulated cases (left) and with varying assumptions about the 
ionizing sources and their suppression (right). (top panels) Ratio
of the mean mass-weighted and volume ionized fractions, $x_m/x_v$.
All cases are labelled by color and line-type, as follows: (left) 
L1 (blue, solid), S1 (green, short-dashed), L2 (magenta, long-dashed), 
and S2 (red, dotted), (right) S4 (cyan, long dashed), S5 (light red, 
dot-short dashed), and S3 (light green, dot-long dashed). For ease of 
comparison we show the fiducial case S1 on both plots.
}
%\vspace{-0.5cm}
\end{center}
\end{figure*}
We start by comparing our fiducial simulation L1 against the 
simulation L3, equivalent to our previous large-box simulations 
without LMACHs and their self-regulation \citep{2008MNRAS.384..863I}. 
The resulting reionization histories and integrated 
electron-scattering optical depths are shown in 
Figure~\ref{reion_hist_selfreg_fig}. As expected, the 
self-regulated model yields a much more gradual and extended 
reionization history, which starts with the formation of the 
first atomically-cooling halo sources at $z\sim31$ vs. the 
much later start, at $z\sim20$ in L3, due to the much larger, 
rarer sources in the latter case, which accordingly form later. 
The exponential rise in the numbers of the high-mass sources 
yields a steep, power-law like reionization history (reasonably 
well-fit by $\lg(x_m)\approx-1.226z+10.41$ for $x_m>0.03$, rising 
somewhat steeper than this earlier on) when it is driven solely by 
those sources, while in the self-regulated case the steep initial rise 
becomes much more gradual when the self-regulation first kicks in, 
around $z\sim16$, when the efficient low-mass sources are massively 
suppressed and thereby gradually give way to the less efficient 
high-mass ones which come to dominate at the latter stages of 
reionization. However, we note that even with self-regulation the 
reionization history remains monotonic and no plateaus, let alone 
double reionization, ever occur. The mass-weighted over volume 
ionized fraction (upper panel) is always lower in the self-regulated 
case, indicating that reionization has less pronounced inside-out 
character, i.e. ionized regions are less correlated with the highest 
density peaks in this case since reionization is driven by wider range
of sources, including low-mass, less biased ones. The integrated 
electron-scattering optical depth (Figure~\ref{reion_hist_selfreg_fig},
 right) is significantly boosted by the presence of low-mass 
sources, by about $0.01$ overall, most of it due to the early 
stages of reionization. For the particular source efficiencies 
we have chosen here both optical depths fall within the 1-$\sigma$ 
interval given by the WMAP 5-year data, albeit the value for 
the self-regulated fiducial case is very close to the central 
value, while the $\tau_{\rm es}$ for the L3 case is at the low 
1-$\sigma$ limit.

These reionization histories are a direct consequence of the 
overall number of ionizing photons emitted by all active sources, 
shown in Figure~\ref{lum_evol_selfreg_fig}. In the case L3 where 
no source suppression occurs the number of photons emitted per 
timestep simply rises proportionally to the halo collapsed 
fraction, roughly exponentially. In contrast, in the fiducial 
self-regulated case L1 the initial exponential rise is halted 
around redshift $z\sim16$ and rises very slowly (and moderately 
non-monotonically) until $z\sim11$, at which point sufficient 
number of high-mass, non-suppressible sources form to allow them 
to take over the evolution, while the low-mass sources become 
highly suppressed. Therefore, similarly to our earlier results 
in \citet{2007MNRAS.376..534I} the late phase of reionization 
and overlap epoch, $z_{\rm ov}$, are dominated by HMACHs, while 
the LMACHs dominate the early phase of reionization and provide 
a significant boost to the electron-scattering optical depth, 
$\tau_{\rm es}$. Ultimately, by overlap in both simulations L1 and 
L3 there are 1.2-1.6 ionizing photon per atom emitted, slightly 
more in the self-regulated case due to its more extended 
reionization history which yields more recombinations per atom.

\subsection{The effects of source efficiencies and box size}

The reionization histories derived from our suite of simulations 
are shown in Fig.~\ref{reion_hist_fig}. The reionization history 
is monotonic in all cases, although due to the self-regulation
the slope of the curves can vary significantly and in particular
can become almost horizontal for short periods of time when the
Jeans mass filtering compensates for the rise in source numbers.
The exact redshifts at which certain reionization milestones, 
10\%, 50\% and 90\% by mass, are reached are listed in 
Table~\ref{summary_table}. We also list there the epochs when
final overlap, which we define as the time when $x_m=0.99$, i.e.
at least 99\% of the mass is ionized, is reached in each case.

Our large-volume, self-regulated simulations L1 and L2 
(Fig.~\ref{reion_hist_fig}, left panels) have reionization 
histories which are very similar to each other, but offset 
by $\Delta z\sim2$. Overlap is reached at $z=8.3$ (6.7) in 
L1 (L2), corresponding to early (extended) reionization scenarios. 
The integrated electron scattering optical depth for L1 is 
$\tau_{\rm es}=0.080$ for the early reionization case, well within 
the current WMAP5 1-$\sigma$ constraints. The corresponding value 
for the low-efficiency, extended reionization scenario L2 is 0.058, 
which is outside the 1-$\sigma$ range, but still within 2-$\sigma$. 
On the other hand, the simulation volume (163 vs. 53 Mpc) has little 
effect on the global reionization histories. This is in agreement 
with the results in \citet{2006MNRAS.369.1625I}, which were derived 
by sub-dividing a 100$\,h^{-1}$Mpc volume into smaller ones, which
indicated that $\sim20-30\,h^{-1}$Mpc box is sufficient to reliably 
derive the global mean reionization history. Most variations between 
the corresponding large and small box simulations result solely from 
the different random realizations in the two cases. At early times 
($x_m<0.01$) there are also departures due to cosmic variance - 
unlike the larger, 163 Mpc, volume the more limited 53 Mpc one does 
not contain any sources at $z>25$ as those are statistically too 
rare to occur. Even when the very first halos appear in the 53 Mpc 
volume, they are initially so few that they are subject to very high 
shot noise fluctuations. Once there are statistically-significant 
numbers of sources in each size box the reionization histories 
converge and any fluctuations thereafter are simply due to the 
different random realizations. There is also some effect from the 
higher resolution of the small-box simulations, due to the 
better-resolved density field in those cases, which yields slightly 
increased recombinations. This effect is rather minor here however, 
because the relatively small difference in resolution results in 
only a marginal increase of the the recombination rates.

\begin{figure*}
\begin{center}  
\includegraphics[width=3.2in]{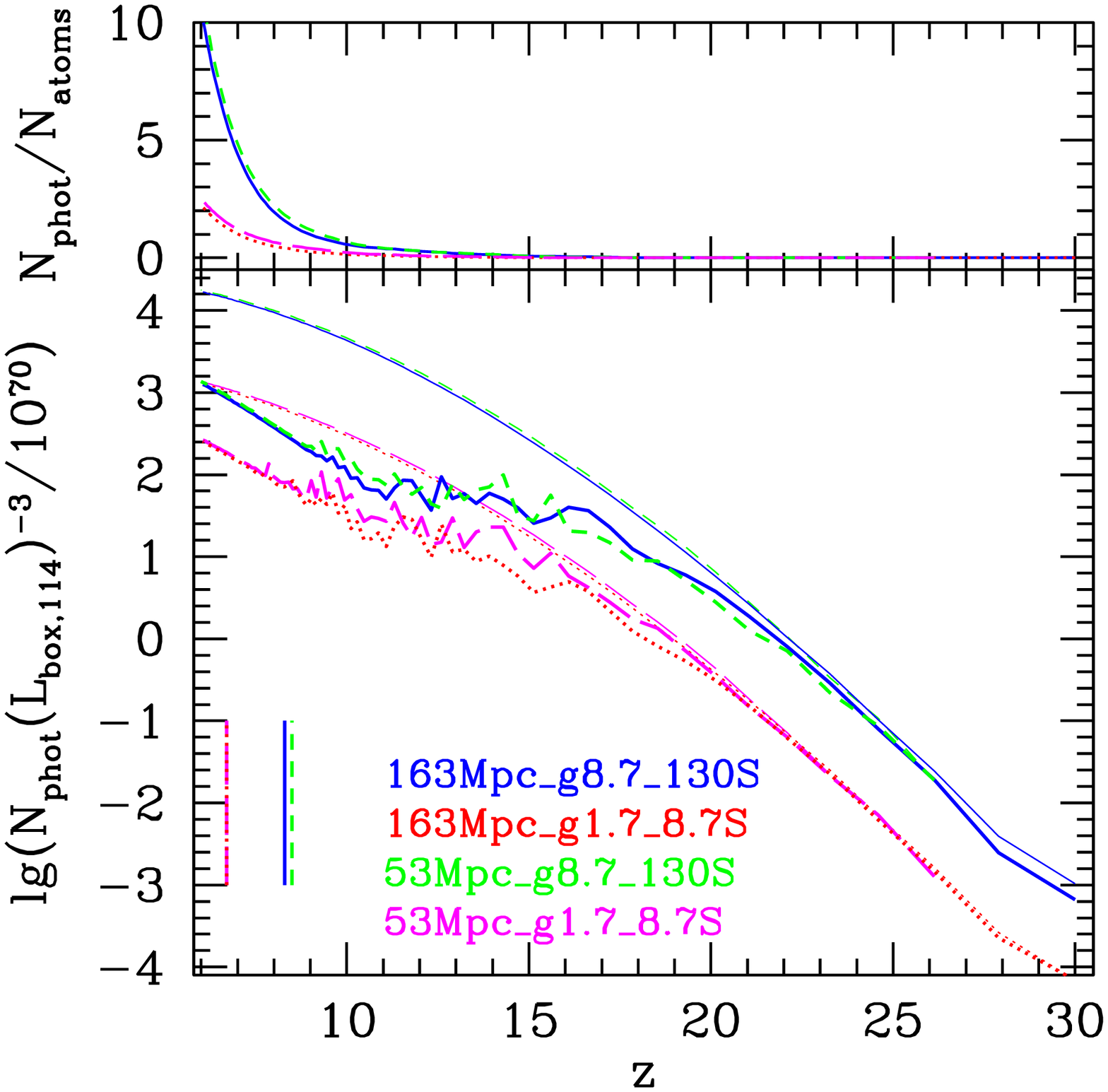} 
\includegraphics[width=3.2in]{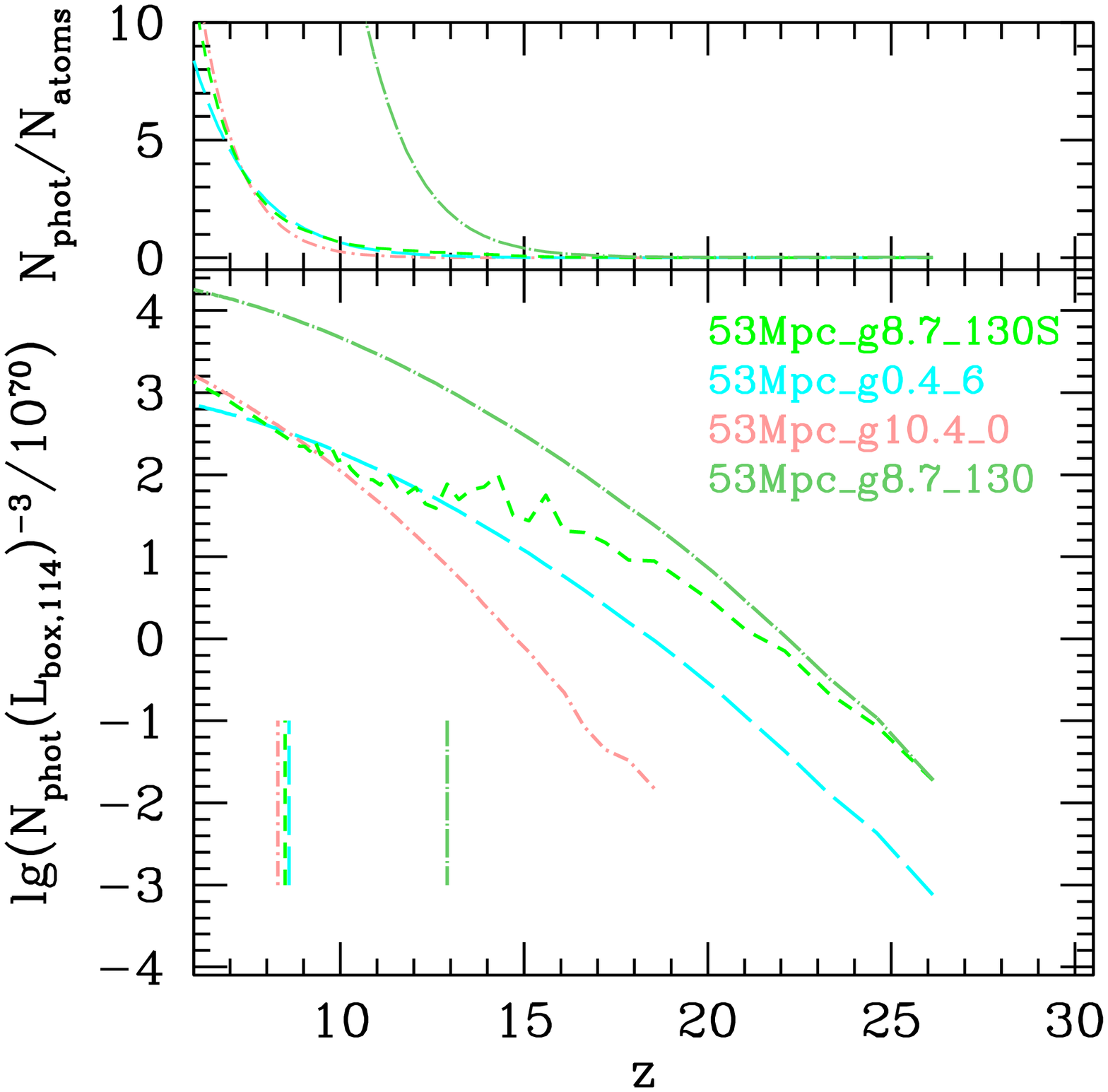} 
%\vspace{-1in}
\caption{
\label{lum_evol_fig}
(bottom panels) Number of ionizing photons emitted by all sources
(i.e. if there were no suppression; thin lines) and all active 
sources (thick lines) in the computational volume per timestep and 
(top panels) cumulative number of photons per total gas atom released 
into the IGM. Notation is the same as in Fig.~\ref{reion_hist_fig}.
Vertical lines with same colors and linetypes mark the overlap redshift 
in each case. 
}
\end{center}
\end{figure*}

In Fig.~\ref{reion_hist_fig} (top left panel) we show the ratio of 
the ionized fraction by mass, $x_m$ and by volume, $x_v$, which is 
equal to the average density of the ionized regions in units of the 
mean \citep{2006MNRAS.369.1625I}. These ratios start at about 2 
and remain above unity at all times, indicating that the reionization 
proceeds in an inside-out manner, with the high density peaks (where 
the first sources preferentially form) being ionized on average earlier 
than the mean and low-density ones. On average the ionized regions are 
denser in the low-efficiency cases. This behaviour could be expected 
based on the typically smaller H~II region sizes in those cases. They 
therefore stay in the immediate vicinity of the density peaks and do 
not propagate as much into the voids. The higher spatial resolution
of the 53 Mpc cases also yields somewhat higher mean density of the 
ionized regions compared to the corresponding 163 Mpc box cases.

\subsection{The effects of the source model: photon production 
efficiencies and minimum source mass}

In Fig.~\ref{reion_hist_fig} (right panels) we show the corresponding 
reionization history results when the source models are varied. We 
also replotted one of our fiducial cases, S1, for facilitating direct 
comparison with the self-regulated cases. All reionization histories 
remain largely monotonic throughout the evolution, which therefore 
is a fairly robust feature, independent of the particular ionizing 
source properties assigned. However, a wide range of overlap epochs 
- from as early as $z=12.9$ in the no suppression case S3 to $z=8.3$ 
in large-source-only case S5, and a wide range of slopes of the 
reionization history evolution are observed. 

Our fiducial case, S1, has the most extended reionization history 
of all, which starts with the formation of the first $10^8M_\odot$ 
halos at $z\sim26$ and reaches overlap at $z=8.9$. In comparison 
when all sources have the same efficiencies, but none are ever 
suppressed (case S3) the reionization history is very steep, 
roughly exponential, tracking the exponential rise of the collapsed 
fraction in halos (cf. Figure~\ref{fcoll_fig}). The no-suppression, 
low-efficiency case S4 also produces a very extended history since 
it also starts with the formation of the first $10^8M_\odot$ halos 
and, by design, reaches overlap at roughly the same time as S1. 
However, the sources are necessarily much weaker in that case 
compared to S1 and S3, and therefore the ionized fraction starts 
much lower compared to the fiducial simulation and only catches 
up with the self-regulated case at late times ($x_m\gtrsim0.25$). 
Finally, in case S5 only the massive sources are active, and 
therefore the reionization starts late, at $z\sim18$, but $x_m$ 
rises exponentially, in proportion of the collapsed fraction in 
those massive halos, reaching (again by design) overlap at the 
same time, $z=8.3$ as the fiducial case. The lack of low-mass 
halos therefore delays reionization considerably and naturally 
yields much lower intergated electron scattering optical depth 
(0.071 compared to 0.084 for S1, and 0.078 for S4). 

\begin{figure*}
  \begin{center}
   \includegraphics[width=2.2in]{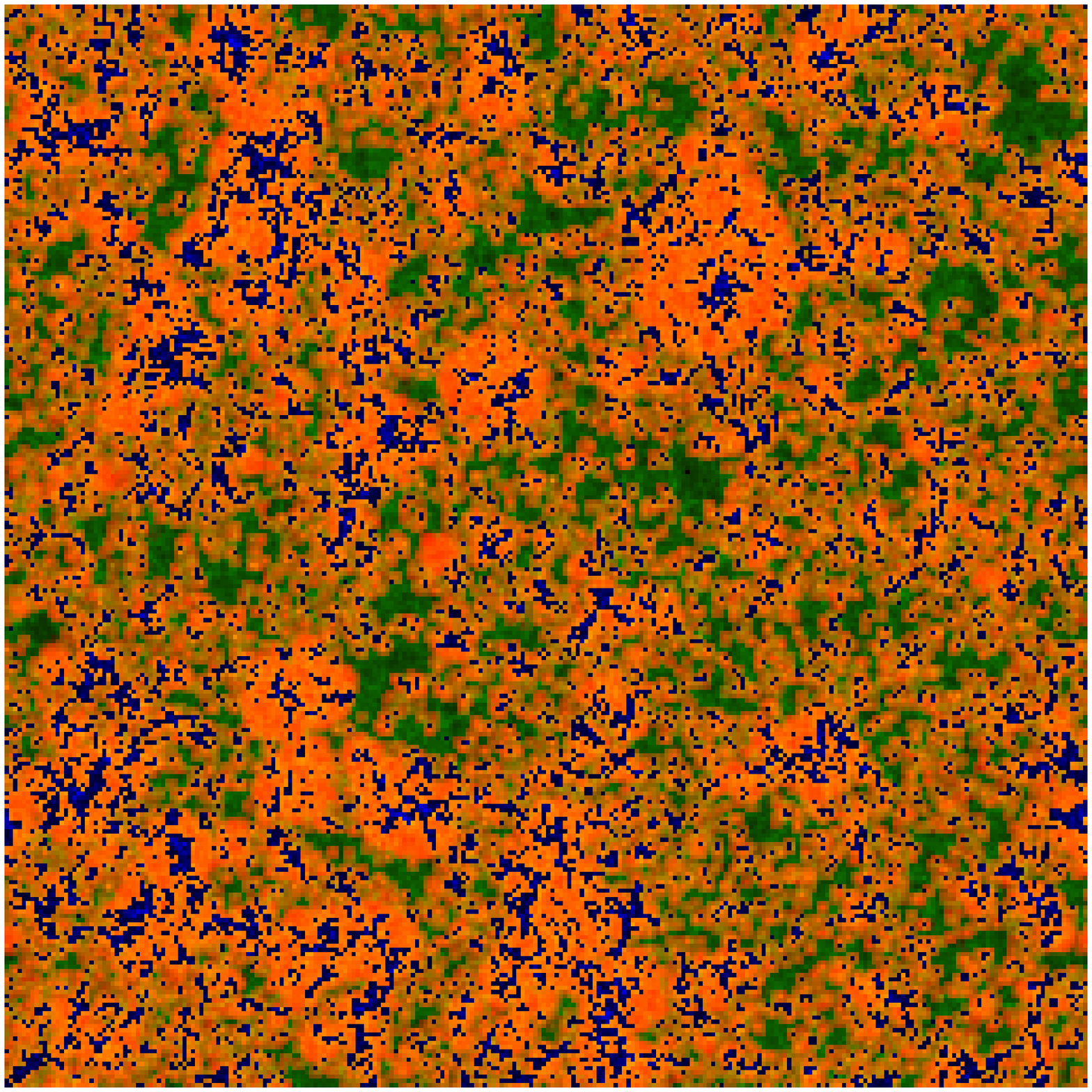}
    %\hskip 1.5cm
    \includegraphics[width=2.2in]{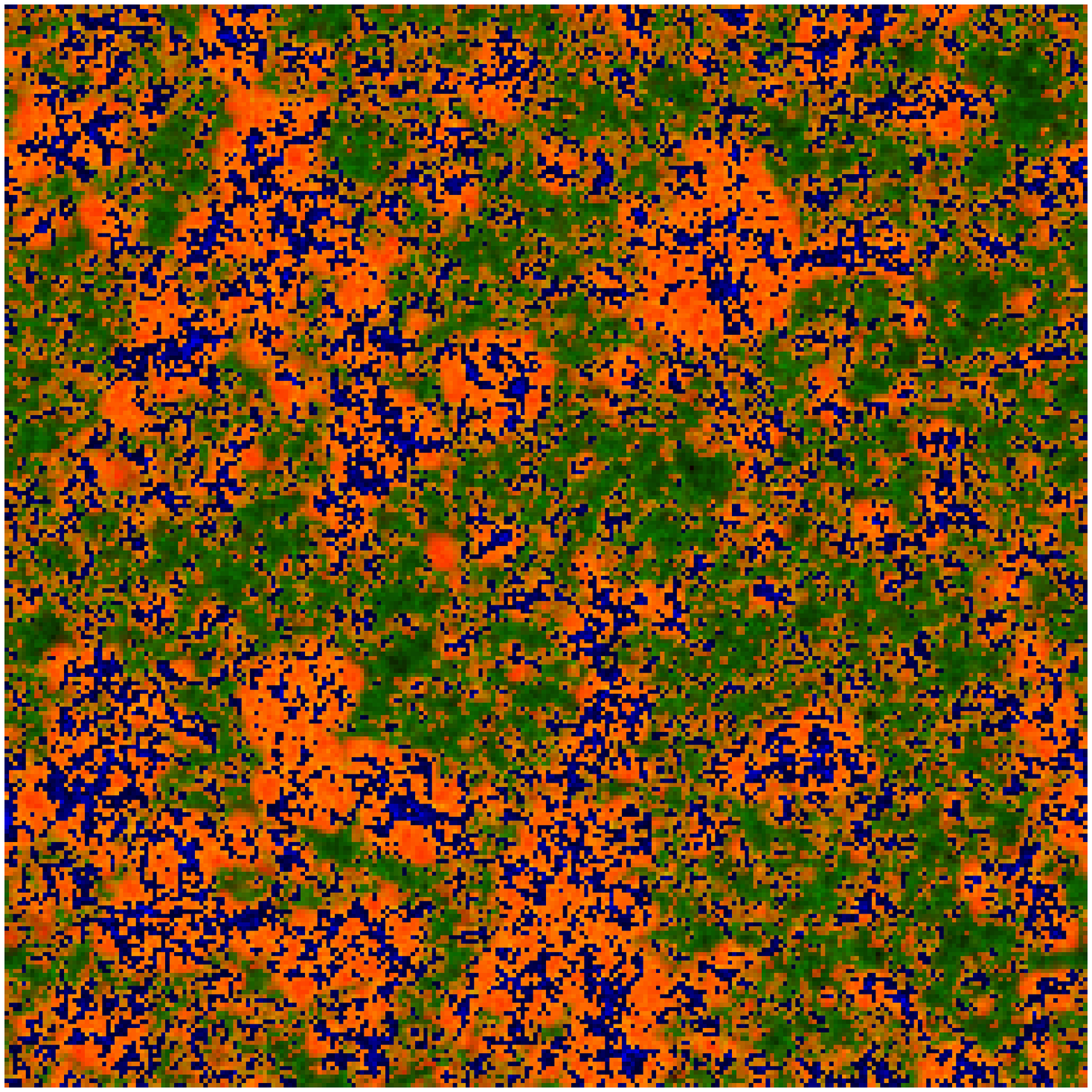}
    \includegraphics[width=2.2in]{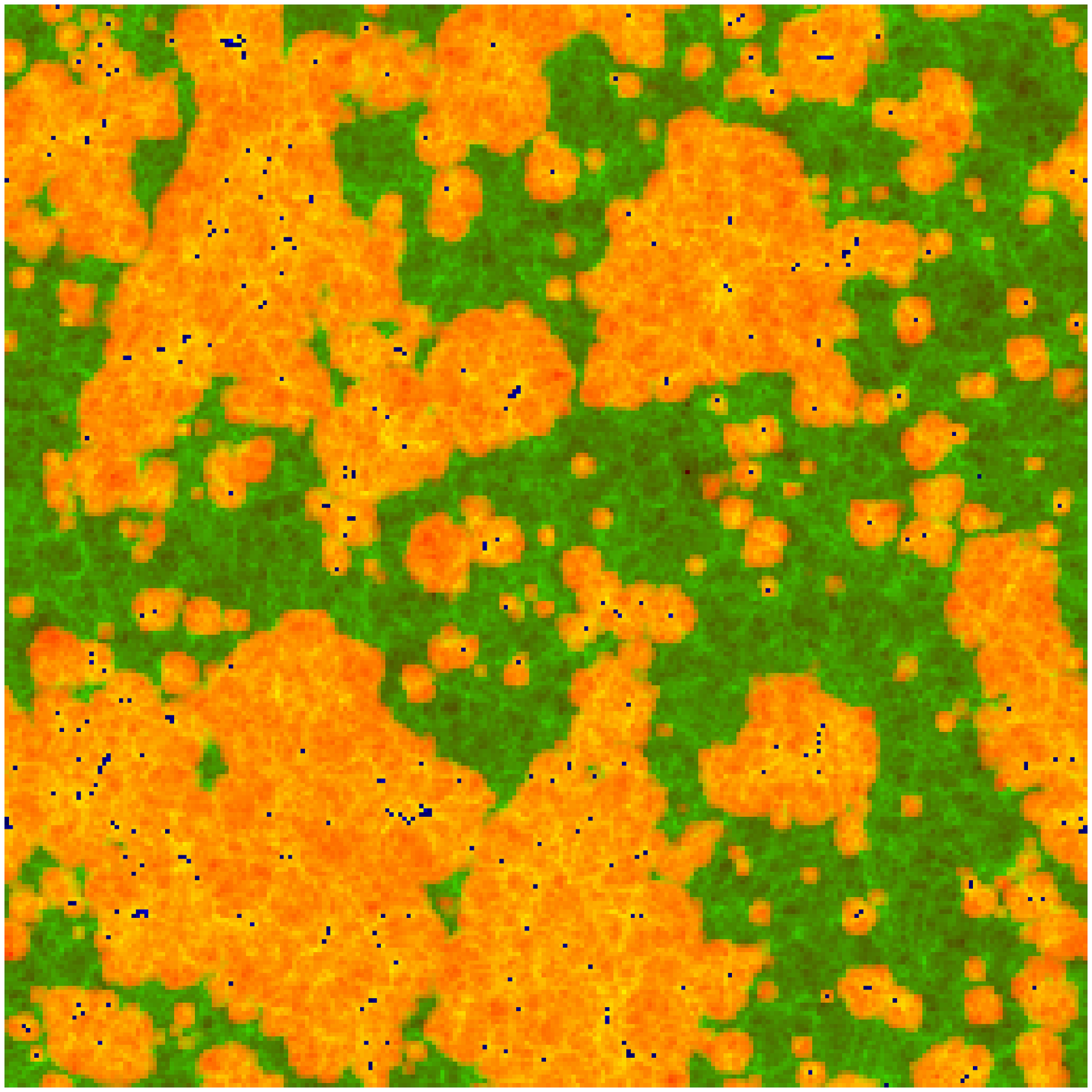}
    \vspace{-0.3cm}
  \end{center}
  \caption{Spatial slices of the ionized and neutral gas density 
    from our radiative transfer simulations with boxsize 
    $163$~Mpc: (a)(left) L1 (b)(middle) L2, and (c)(right) L3, 
    all at box-averaged ionized fraction by mass of $x_m\sim0.50$. 
    Shown are the density field (green) overlayed with the ionized 
    fraction (red/orange/yellow) and the cells containing active 
    sources (dark/blue). 
    \label{images}}
\end{figure*}

The mean overdensity of the ionized regions, $x_m/x_v$ 
(Fig.~\ref{reion_hist_fig}, top right panel) is above unity for 
all cases and at all times, demonstrating the robustness of the 
inside-out nature of reionization, in agreement with our original 
findings \citep{2006MNRAS.369.1625I}. Compared to our fiducial 
simulation, S1, the ratio $x_m/x_v$ is significantly higher for 
cases S4 and S5. This is due to the fact that the H~II regions 
are more tightly correlated with the density peaks in those cases, 
because the number of sources, which form at the density peaks, 
rises exponentially in these cases. Finally, simulation S3 show 
an intermediate behaviour, similar to the fiducial case, S1, but 
with somewhat faster decrease of the mean overdensity of the 
ionized regions, due to higher ionization of the low-density 
regions in this case.

\begin{figure*}
  \begin{center}
    \includegraphics[width=2.2in]{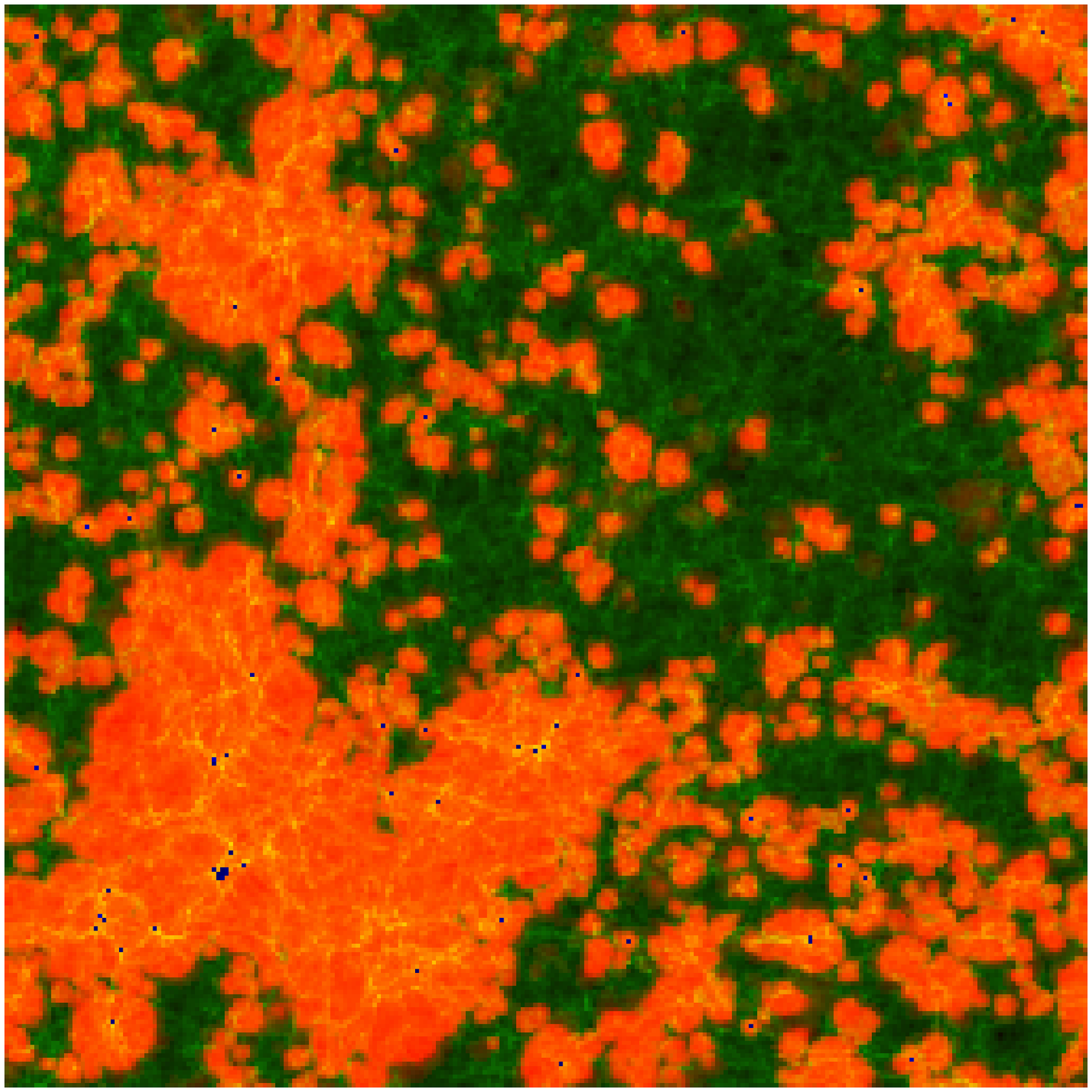}
    \includegraphics[width=2.2in]{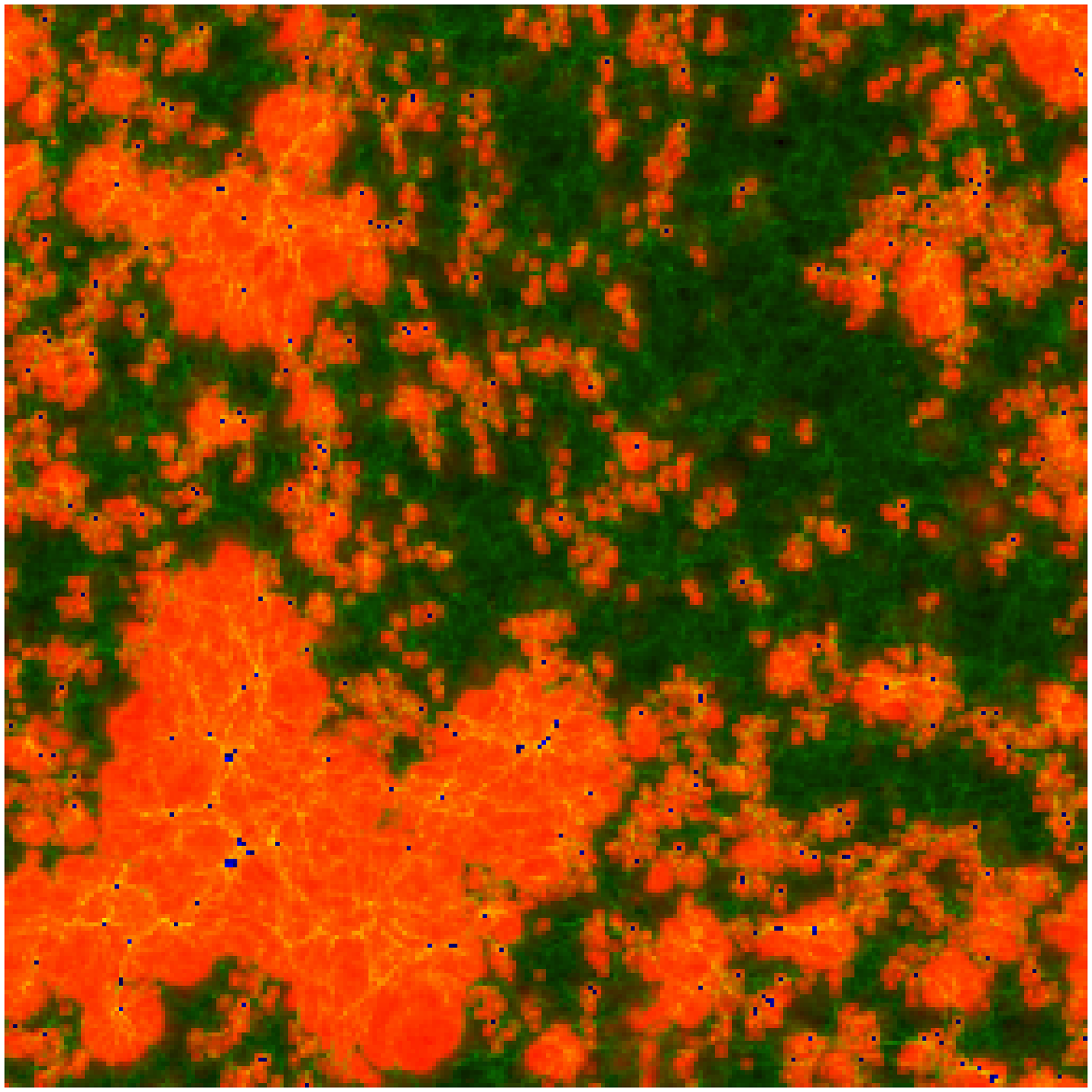}
    \includegraphics[width=2.2in]{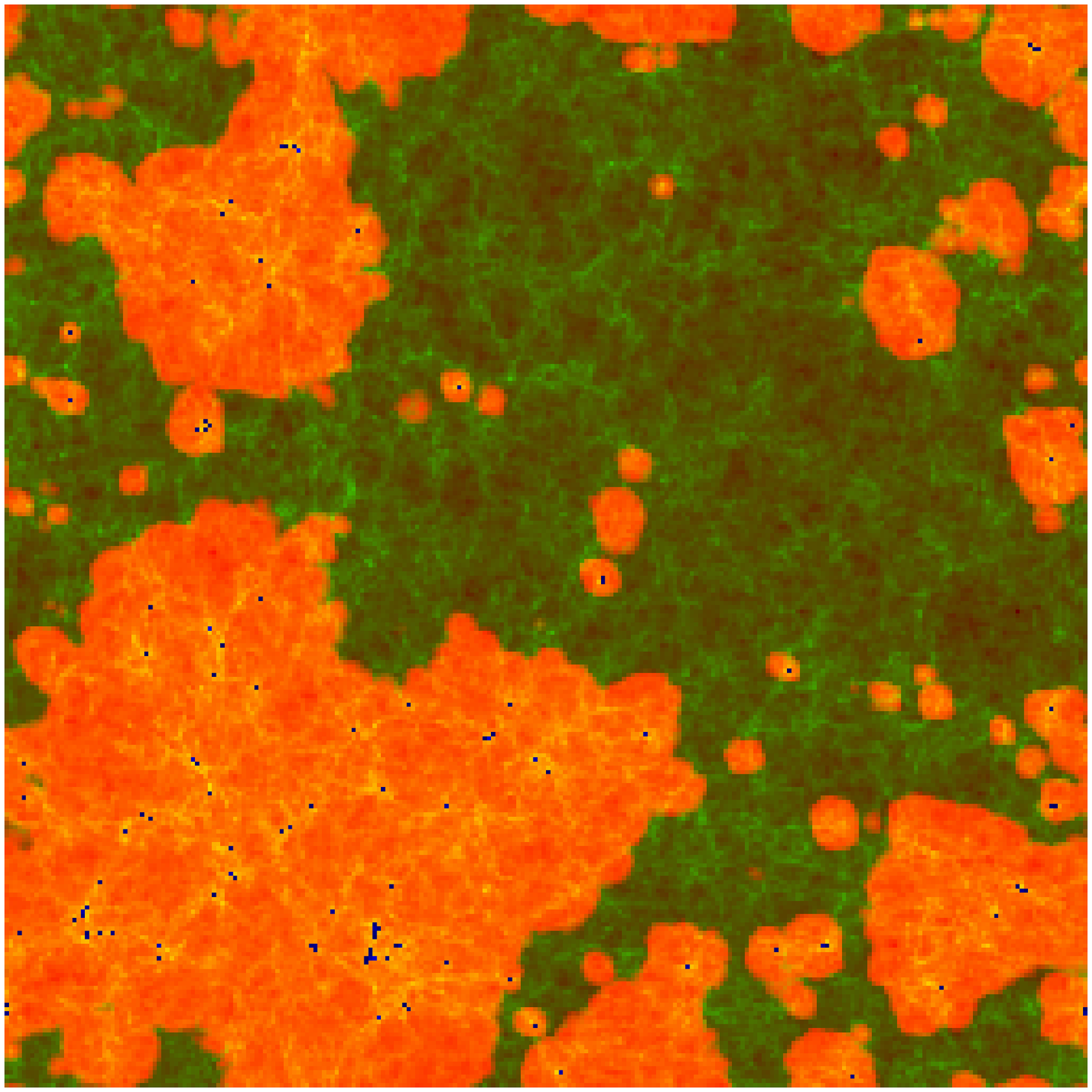}
    \includegraphics[width=2.2in]{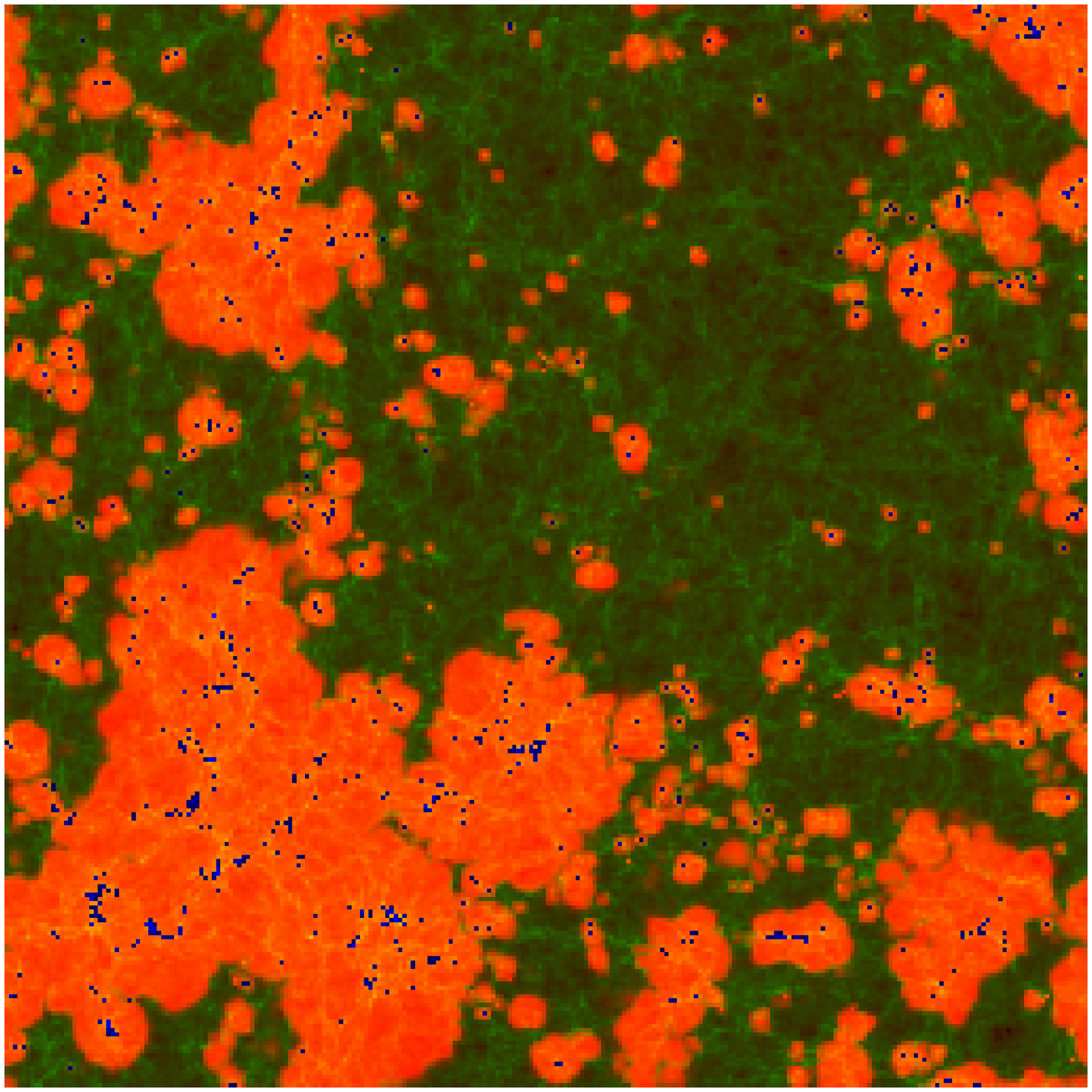}
    \includegraphics[width=2.2in]{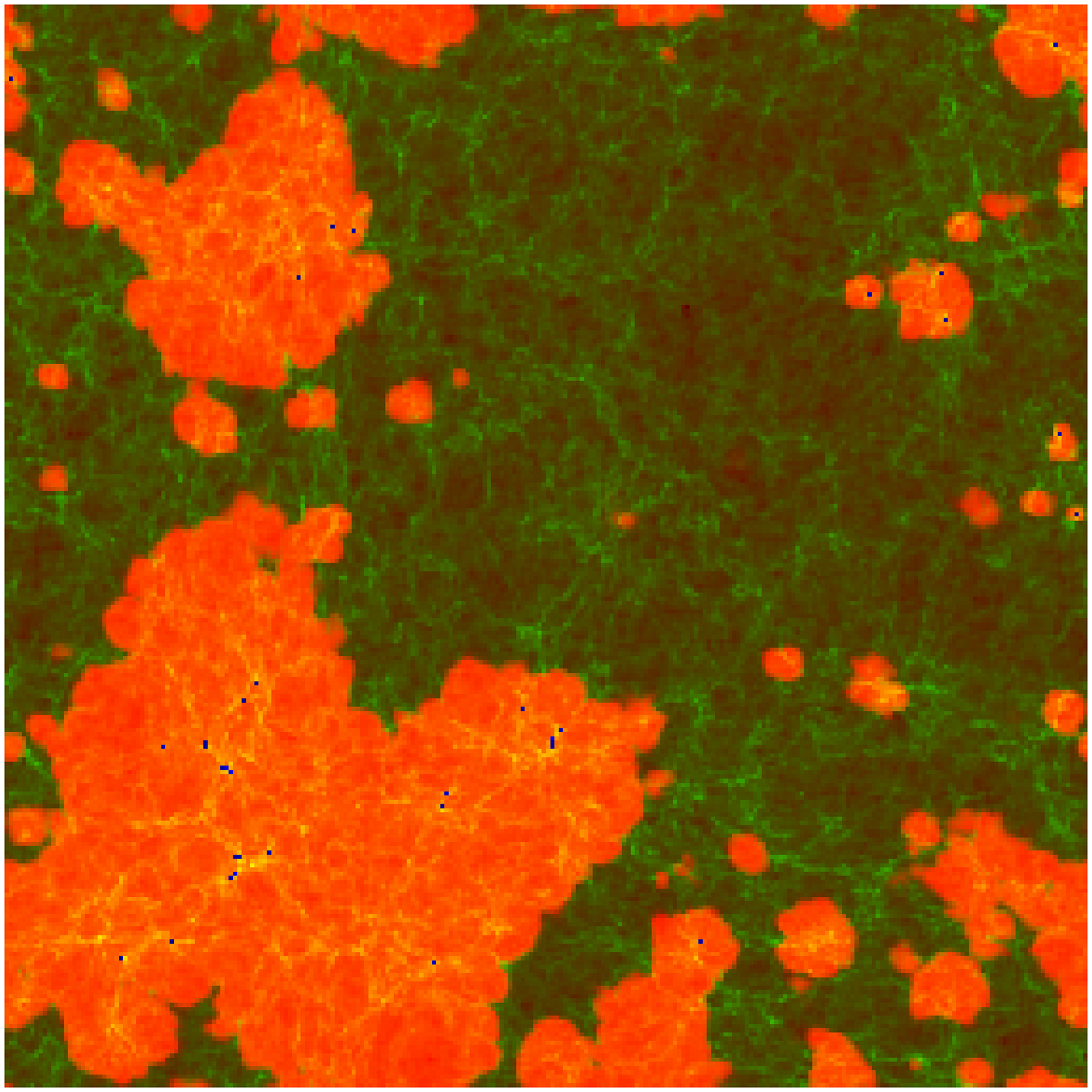}
    \vspace{-0.3cm}
  \end{center}
  \caption{Spatial slices of the ionized and neutral gas density 
    from our radiative transfer simulations with boxsize 53~Mpc, 
    all at box-averaged ionized fraction by mass $x_m\sim0.50$. 
    Shown are the density field (green) overlayed with the ionized 
    fraction (red/orange/yellow) and the cells containing sources 
    (dark/blue). Shown are (left to right and top to bottom) cases 
    S1, S2, S3, S4, and S5.
    \label{images_37Mpc}}
\end{figure*}

The corresponding evolution curves of the cumulative number of 
ionizing photons emitted within each simulation are shown in 
Fig.~\ref{lum_evol_fig}. Indicated are also the number of photons 
which would have been emitted if no self-regulation has taken place 
(thin lines) and the overlap redshifts (vertical marks). Starting 
with our self-regulated cases (Fig.~\ref{lum_evol_fig}, left) we 
see that the low-mass source suppression does not have a significant 
impact until $z\sim20-22$, but after that has a great impact, 
reducing the overall number of emitted photons by up to a factor of 
$\sim30$ for highly-efficient low-mass sources and about a factor of 
10 for less efficient ones. The number of ionizing photons emitted 
per timestep also becomes variable and non-monotonic function of 
redshift due to the complex interplay of suppression and new source 
formation during the self-regulation process. Eventually, by overlap 
(indicated for each case by the vertical lines) all low-mass sources 
are suppressed and the number of photons continues to rise smoothly 
as ever more high-mass sources form. The simulation boxsize makes 
little difference in the ionizing photon production, apart from 
modest variations due to the different random realization in each 
case. In terms of the cumulative numbers of emitted photons per 
atom (top panel), by redshift $z=6$ up to 10 photons per atom are
produced in the efficient-sources case and up to 2 photons per atom
in the low-efficiency cases. However, at their respective overlap 
redshifts approximately the same number are produced, about 1.5, 
i.e. on average only about one recombination per every 2 atoms 
occurs during the evolution. Therefore, recombinations are relatively 
unimportant in these runs. The reason for this is that much of the 
density fluctuations are at very small scales, well below our 
radiative transfer grid resolution. This additional small-scale power 
can be added as sub-grid clumping of the gas, calculated based on 
much higher resolution simulations. We will consider the effects 
of sub-grid clumping in a companion paper (Koda et al., in prep.). 

Turning our attention to the set of cases with different source 
efficiency models (Figure~\ref{lum_evol_fig}, right), we first 
note that the three models which by construction have very similar 
overlap epochs (our fiducial case, S1, and cases S4 and S5) also 
have almost identical photon production numbers at overlap. All 
three reach this point in a very different manner, however. In 
run S4 all sources are quite weak, but no sources are ever 
suppressed, and the emissivity per timestep reaches the fiducial 
case once most low-mass sources are suppressed in the latter case. 
In case S5 only the massive sources are present and consequently 
its emissivity lags significantly at early times until eventually 
the exponential rise of those sources allows it to join the other 
two cases at $z\lesssim11$. We also note that after overlap the 
photon emissivity in the no-suppression case S4 lags behind the 
others because its high-mass sources are very inefficient and the 
collapsed fraction in low-mass sources by this point does not rise as 
fast as the one for the high-mass sources. Finally, the number of 
photons produced in the no suppression, high source efficiency case, 
S3, simply follows the total collapsed fraction in all sources and 
therefore rises almost exponentially, roughly parallel to the curve 
for S4, eventually surpassing 10 photons per atom before $z=10$. 
However, at its own (very early) overlap epoch even this case 
produces the same number of photons as the others, about two per IGM 
atom. 

In Figures~\ref{images} and \ref{images_37Mpc} we show slices through 
our simulation volume showing the geometry of the H~II regions at 
$x_m\sim0.5$, overlayed on the corresponding density field for all 
our simulations. We also mark the cells containing active sources 
(blue/dark). Comparing first the large, $163$~box cases 
(Fig.~\ref{images}), we note that, as could be expected in the 
high-efficiency fiducial case L1 there are many more active sources 
in the low-efficiency one, L2. In contrast, in simulation L3 there 
are many fewer sources dure to its higher mass cutoff, which only 
leaves the high-mass, rare sources present. The large-scale 
structures are quite similar in size and shape in the two 
self-regulated simulations, but the fiducial case L1 yields much 
more small-scale ionized patches even though it reaches half-ionized 
state noticeably earlier, at which point there are many fewer, and 
more clustered, sources. The reason for this apparently 
counter-intuitive behaviour is that the much weaker sources in case 
L2 have difficulties fully ionizing their own cells (which, at 445 
kpc/h linear size, are relatively large), and therefore large number 
of sources are needed to produce a sizeable fully-ionized patch. In 
contrast, the much more efficient sources in our fiducial case L1 
easily ionize their own cell, resulting in many small-size H~II 
regions, instead of the more scattered, partially-ionized cells in 
L2. On the other hand, in the non-self-regulated case L3 we find 
many fewer, larger ionized regions, in agreement with our previous 
results in \citep{2008MNRAS.384..863I}. The large-scale structures 
have some similarities to the ones found in the self-regulated cases, 
as could be expected given that all simulations share the underlying 
large-scale cosmic structures. However, the ionized regions are in a 
more advanced stage where they start merging together, and there is 
far less small-scale structure due to the absence of low-mass, weaker 
sources. In that case there are also no partially-ionized regions, 
since there is no low-mass source suppression Jeans-mass filtering 
(i.e. sources do not die), and all sources are sufficiently luminous 
to completely ionize their own local volume. 

The corresponding images from our small-box simulations at the same 
ionized fraction of $x_m\sim0.5$, are shown in Fig.~\ref{images_37Mpc}.
 Once again, the large-scale structures, which tend to strongly 
corellate with the underlying distribution of density and clustered 
halos, are generally quite similar. In contrast, ionized patches 
produced by multiple, less biased sources whose distribution does 
follow the knots and filaments of the Cosmic Web are much more 
irregularly-shaped. There are significant differences in the 
smaller-scale structures among the range of simulations. The 
self-regulated cases, S1, S2 and to a lesser extend case S4 have 
the most small-scale structure, including both small H~II regions 
and rough, irregularly-shaped large H~II region boundaries. In 
contrast, S5, which does not include the low-mass sources and S3, 
which includes efficient, unsuppressible low-mass sources both 
yield many fewer ionized patches with smoother boundaries, which 
reflects the rarity and highly clustered nature of their active 
sources. We have presented a more detailed discussion of the H~II 
region geometry, size distribution and topological characteristics 
in a recent companion paper based on a subset of the current suite
of simulations \citep{2011MNRAS.413.1353F}.

\section{Observational signatures}
\label{observ:sect}

We now turn our attention to the reionization observables and 
specifically how are they related to the assumed source 
populations and their efficiencies. A better understaning of 
these dependencies should allow us in turn to use the observational 
data to constrain the properties of the reionization sources.
Our main focus will be on the redshifted 21-cm signatures, 
although we also briefly discuss the photoionization rates
in the IGM, related to the measurements of
the Gunn-Peterson effect and the gas temperature.  

\subsection{Photoionization rates} 

\begin{figure}
  \begin{center}
    \includegraphics[width=3.3in]{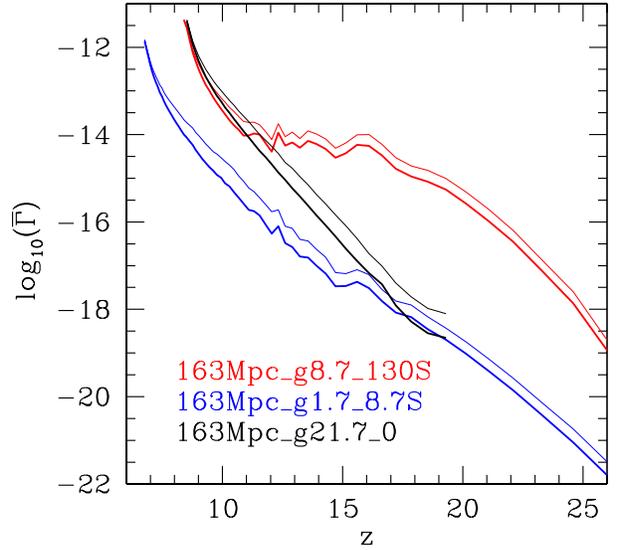}
    %\hskip 1.5cm
    \vspace{-0.3cm}
  \end{center}
  \caption{Evolution of the mean mass-weighted (thin lines) 
  and volume-weighted (thick lines) photoionization rates in 
  our computational volume for simulations  L1 (red), L2 (blue) 
  and L3 (black). 
    \label{gamma_means}}
\end{figure}

\begin{figure}
  \begin{center}
    \includegraphics[width=3.3in]{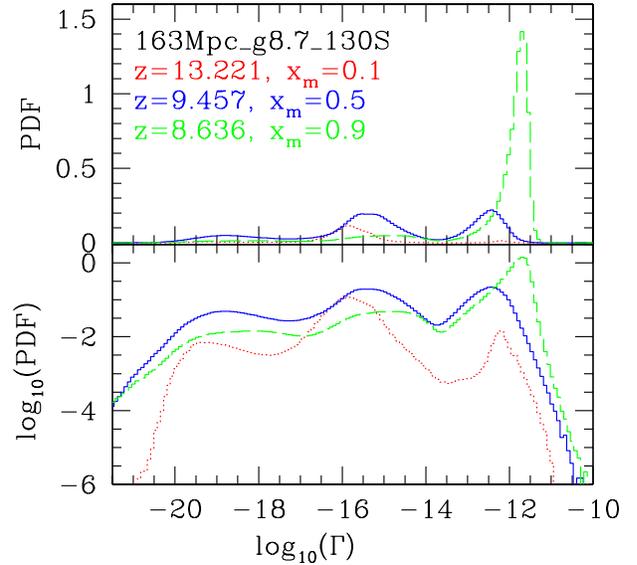}
    %\hskip 1.5cm
    \vspace{-0.3cm}
  \end{center}
  \caption{Photoionization rate PDF's for our fiducial case L1 
    for epochs when the ionized fraction by mass is $x_m=0.1$ (red), 
    $x_m=0.5$ (blue), $x_m=0.9$ (green) plotted in linear (top) and 
    log (bottom) scales. 
    \label{gamma_PDFs}}
\end{figure}

\begin{figure*}
  \begin{center}
   \includegraphics[width=2.3in]{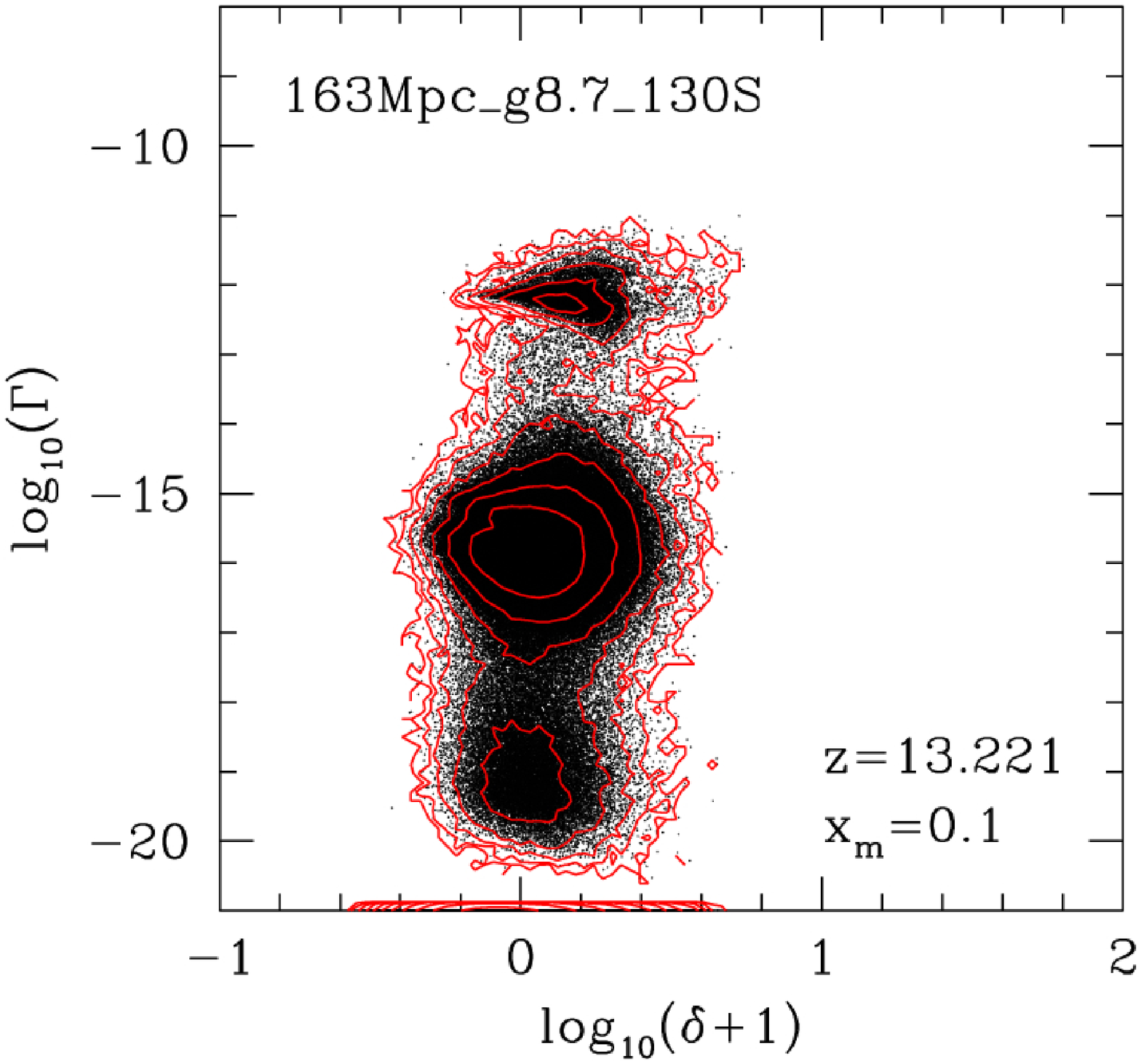}
    %\hskip 1.5cm
   \includegraphics[width=2.3in]{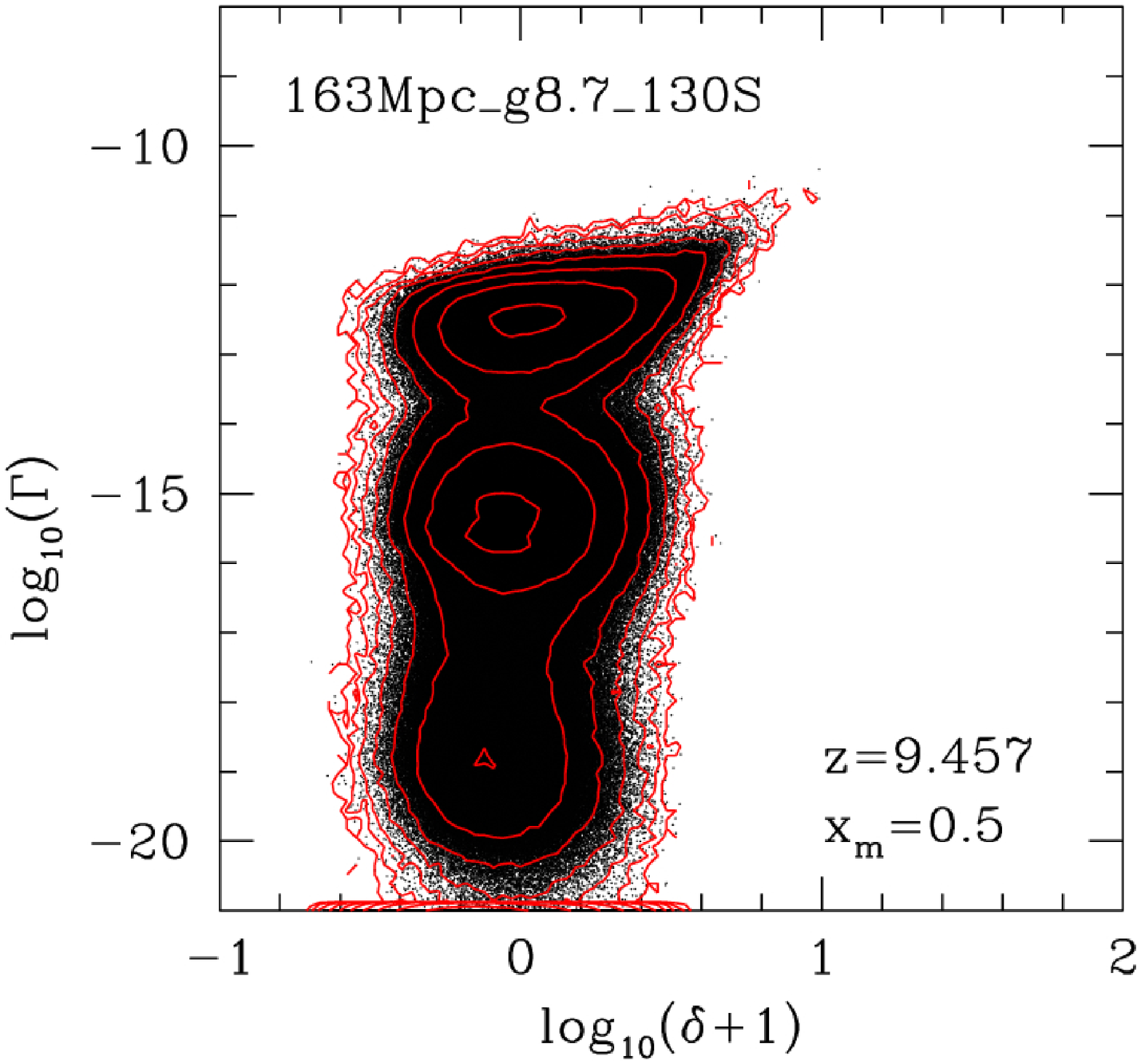}
   \includegraphics[width=2.3in]{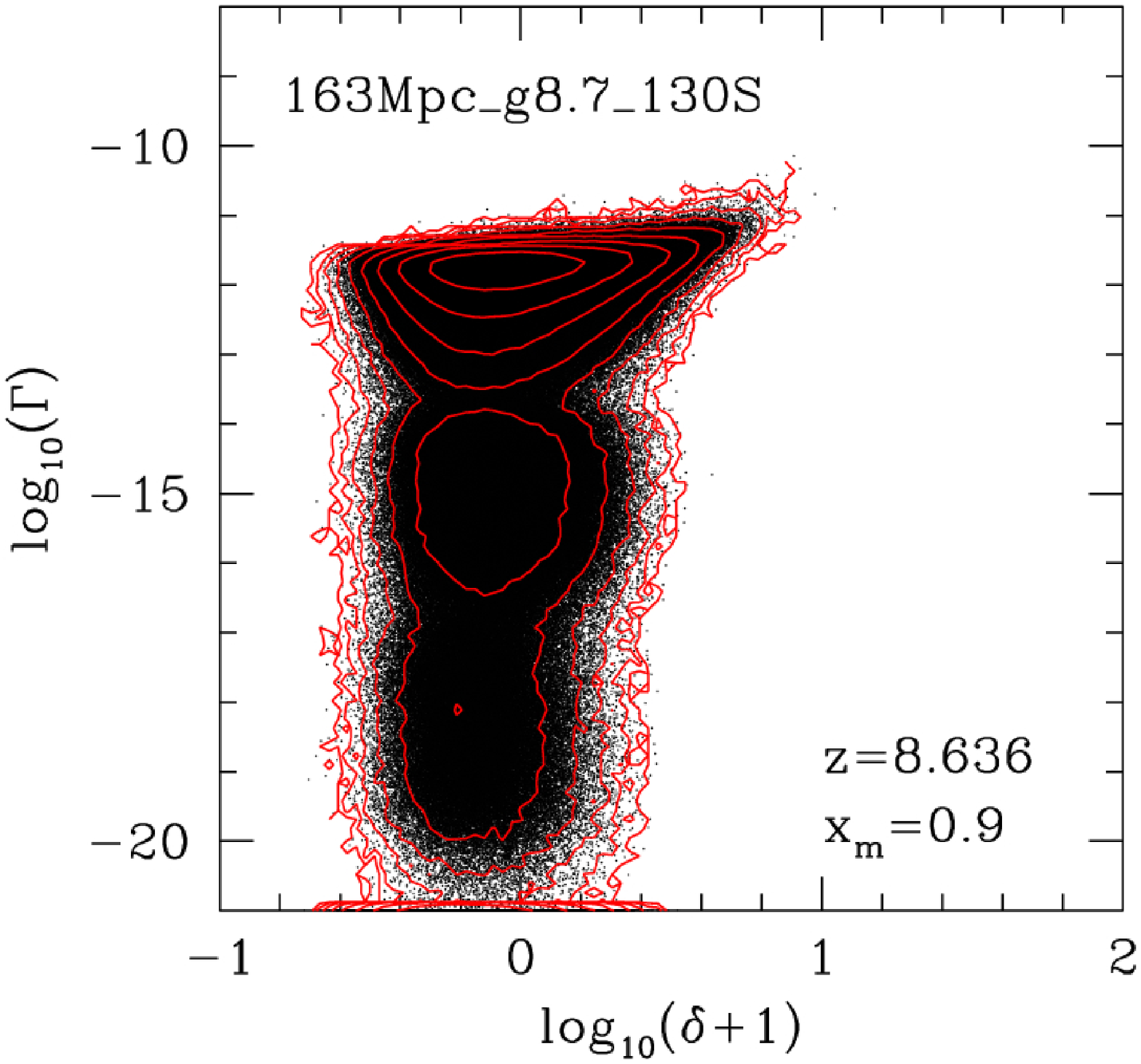}
    \vspace{-0.3cm}
  \end{center}
  \caption{Cell-by-cell photoionization rate - overdensity correlation 
   scatter plot at $x_m=0.1$ (left), $x_m=0.5$ (middle), and  $x_m=0.9$ 
    (right). Contours are logarithmic, from 10 cells up every 0.5 dex.
  \label{gamma_vs_delta}}
\end{figure*}

An important, if indirect, observable signature of reionization
is the mean photoionization rate in the IGM. At present this 
quantity has only been measured for the post-reionization IGM 
at $z<6$, derived based on the small residual neutral fraction 
and its corresponding Ly-$\alpha$ optical depth. It therefore 
typically characterizes only the final EoR stages, around and 
after overlap.  

The redshift evolution of the mean photoionization rates, $\Gamma$, 
averaged over our simulation volume for our large-box simulations 
L1-L3 are shown in Figure~\ref{gamma_means}. In overall curve shape 
and timing the evolution roughly mirrors the reionization histories 
for these three cases. This is natural, since the mean photoionization 
rate is average of the fraction from the ionized regions, where the
$\Gamma$ values are high and fairly uniform at $\sim10^{-12}\,s^{-1}$ 
(see also Fig.~\ref{gamma_PDFs}), and the neutral regions, where 
$\Gamma\sim0$. The mean photionization rate in our fiducial case 
L1 initially rise roughly exponentially, until self-regulation 
becomes wide-spread at $\sim16$, at which point the stall at around 
$10^{-14}$ until the high-mass sources become sufficiently abundant 
to dominate the evolution, which occurs around $z\sim11$. At this 
time the mean rates resume their steep rise, reaching a peak of 
$\Gamma\sim$few $\times 10^{-12}$ by overlap. The situation is 
somewhat different for the low photon production efficiency case, 
L2. The mean $\Gamma$ values are much lower in this case, by 2-4 
order of magnitude, and are mostly rising monotonically throughout 
the evolution, as the Jeans-mass suppression effects are milder in 
this scenario. The peak value reached in this case is about 
$10^{-12}$, in rough agreement with the measured one at $z\sim6$. 
However, we should note that any direct comparisons to 
observationally-derived values are at best approximate since our
simulations at present do not take into account the Lyman-limit 
systems (LLS), which are likely to limit the growth of the mean 
free path of the photons and thus limit $\Gamma$, as well. Before 
overlap the mean free path is dictated by the remaining neutral 
regions and the (still fairly high) residual neutral gas fraction 
within the ionized regions, and therefore the LLS are unimportant 
and do not affect our simulation results. The same is probably 
not true after overlap and we will study the effects of LLS in 
future work. For our current purposes the lack of LLS means that 
we cannot yet make a firm conclusion that the low efficiency, 
late-overlap case, L2, fits the observations better than case L1.

The mean photoionization rate for the non-self-regulated case L3 
is intermediate between L1 and L2. It starts from very low values, 
$\sim10^{-19}$, when there are still only a very few high-mass sources, 
around $z\sim 17$, but then rise sharply, roughly exponentially, and 
converges (by construction, since efficiencies were picked so they 
overlap at the same time) to the values for L1 at later times.

The mass-weighted photoionization rates (thin lines) are 
significantly higher, by factors of up to 2-3 than the 
volume-weighted ones (thick lines) at all times and for 
all simulations. This is easy to understand given the 
inside-out nature of reionization, whereby the ionizing 
sources are found in dense regions, which pushes the 
mass-weighted means higher. Such large differences are 
interesting, however, since they can possibly skew the 
observationally-derived values. Probes of the mean, low-density 
IGM will therefore yield considerably lower values for 
$\Gamma$ than any measurements which are more sensitive 
to denser regions, e.g. around sources.  

Several illustrative PDF's (at cell size, here $445$~kpc/h) of 
the photoionization rates for our fiducial simulation L1 are 
shown in Figure~\ref{gamma_PDFs}. Plotted are the PDF's at 
early ($x_m=0.1$), intermediate ($x_m=0.5$) and late ($x_m=0.9$) 
stages of the evolution. These can be compared to the 
no low-mass sources data we presented previously in 
\citep{2008MNRAS.391...63I}. At all times the PDFs show a 
characteristic, three-peaked profile. The rightmost peak, 
at $\Gamma_{-12}\sim1$, is formed by the cells inside the 
H~II regions, while the other two peaks correspond to 
partially-ionized cells, predominantly at the expanding 
I-fronts and relic (i.e. recombining) H~II regions. As 
we have shown in \citep{2008MNRAS.391...63I}, the ionization 
state is close to or at equilibrium deep inside the ionized 
regions, but far from equilibrium at the I-fronts. This holds
true for the current simulations, as well. However, compared 
to our previous simulations the low-mass source suppression in 
the current runs yields a significant fraction of volume in 
relic H~II regions and partially-ionized cells and thus higher 
peaks at lower $\Gamma$ values than was observed in the 
simulations without suppression.

In Figure~\ref{gamma_vs_delta} we show scatter plots and the 
corresponding contour levels of the local photoionization rates, 
$\Gamma$ vs. density in units of the mean, 
$1+\delta\equiv{\rho_{\rm cell}}/\bar{\rho}$ for our fiducial case, 
L1 and $x_m=0.1,0.5$ and 0.9. Overall, there is a clear positive 
correlation between the density and the photoionization rate. 
This could be expected, given that sources, around which the 
photorates peak form preferentially in high-density regions. 
However, the relationship between the two is complex, the 
correlation is weak and the scatter significant. Similarly to the 
PDF's discussed above, three peaks are observed. The high-$\Gamma$ 
peak ($\Gamma\sim10^{-12}\,$s$^{-1}$) consists of the ionized 
cells, which are typically denser than average. The middle 
peak, at $\Gamma\sim10^{-15}\,$s$^{-1}$, corresponds to I-fronts 
and other partially-ionized regions, while the cells with still
lower values ($\Gamma\sim0$) correspond to the still-neutral 
regions. At early times ($x_m=0.1$) the majority of cells is either 
neutral or partially-ionized and the correlation with the local density
is very weak. When the process advances ($x_m=0.5$) a large population 
of fully-ionized, high-$\Gamma$ cells develops and within the H~II
regions the photoionization rate is fairly well correlated with the 
density, albeit still with a large scatter. On the other hand, for
the photoionization rates in the partially-ionized regions shows 
essentially no correlation with the density. At late times ($x_m=0.9$)
these trends become even more pronounced and a quite tight correlation
develops for the highest-density regions. The overall behaviour is
consistent with what we previously observed when no source suppression
were present \citep{2008MNRAS.391...63I}, albeit with some minor
quantitative differences.

\subsection{Redshifted 21-cm}

The differential brightness temperature of the redshifted 21-cm 
emission with respect to the CMB is determined by the density of 
neutral hydrogen, $\rho_{\rm HI}$, and its spin temperature, 
$T_{\rm S}$ and is given by 
\ba
 \delta T_b&=&\frac{T_{\rm S} - T_{\rm CMB}}{1+z}(1-e^{-\tau})\nonumber\\
&\approx&
\frac{T_{\rm S} - T_{\rm CMB}}{1+z}
\frac{3\lambda_0^3A_{10}T_*n_{HI}(z)}{32\pi T_S H(z)}\label{temp21cm}
\\
&=&{28.5\,\rm mK}\left(\frac{1+z}{10}\right)^{1/2}(1+\delta)
\left(\frac{\Omega_b}{0.042}\frac{h}{0.73}\right)
\left(\frac{0.24}{\Omega_m}\right)^{1/2}
\nonumber
\ea
\citep{1959ApJ...129..536F}, where $z$ is the redshift, $T_{\rm CMB}$ 
is the temperature of the CMB radiation at that redshift, $\tau$ is 
the corresponding 21-cm optical depth, assumed to be small when 
writing equation~\ref{temp21cm}, $\lambda_0=21.16$~cm is the 
rest-frame wavelength of the line, $A_{10}=2.85\times10^{-15}\,\rm s^{-1}$
 is the Einstein A-coefficient, $T_*=0.068$~K corresponds to the 
energy difference between the two levels, 
$1+\delta={n_{\rm HI}}/{ \langle n_H \rangle}$ is the mean number 
density of neutral hydrogen in units of the mean number density 
of hydrogen at redshift $z$, 
\ba
\langle n_H \rangle(z)&=&
\frac{\Omega_b\rho_{\rm crit,0}}{\mu_Hm_p}(1+z)^3\nonumber\\
&=&1.909\times10^{-7}\rm cm^{-3}\left(\frac{\Omega_b}{0.042}\right)
(1+z)^3,
\ea
with $\mu_H=1.32$ the corresponding mean molecular weight 
(assuming 24\% He abundance), and $H(z)$ is the redshift-dependent 
Hubble constant,
\ba
  H(z)&=&
H_0[\Omega_{\rm m}(1+z)^3+\Omega_{\rm k}(1+z)^2+\Omega_\Lambda]^{1/2}
                      \nonumber\\  
&=&H_0E(z)\approx H_0\Omega_{\rm m}^{1/2}(1+z)^{3/2},
\ea
where $H_0$ is its value at present, and the last approximation 
is valid for $z\gg 1$. Throughout this work we assume that 
$T_{\rm S} \gg T_{\rm CMB}$ i.e. that all of the neutral IGM gas 
is Ly-$\alpha$-pumped by the background of UV below 13.6~eV 
from early sources and heated well above the CMB temperature 
(due to e.g. a small amount of X-ray heating), and thus the 
21-cm line is seen in emission. These assumptions are generally 
well-justified, except possibly at the earliest times 
\citep[see e.g.][and references therein]{2006PhR...433..181F}. 

\begin{figure*}
\begin{center} \vspace{-0.5in}  
\includegraphics[height=3.2in]{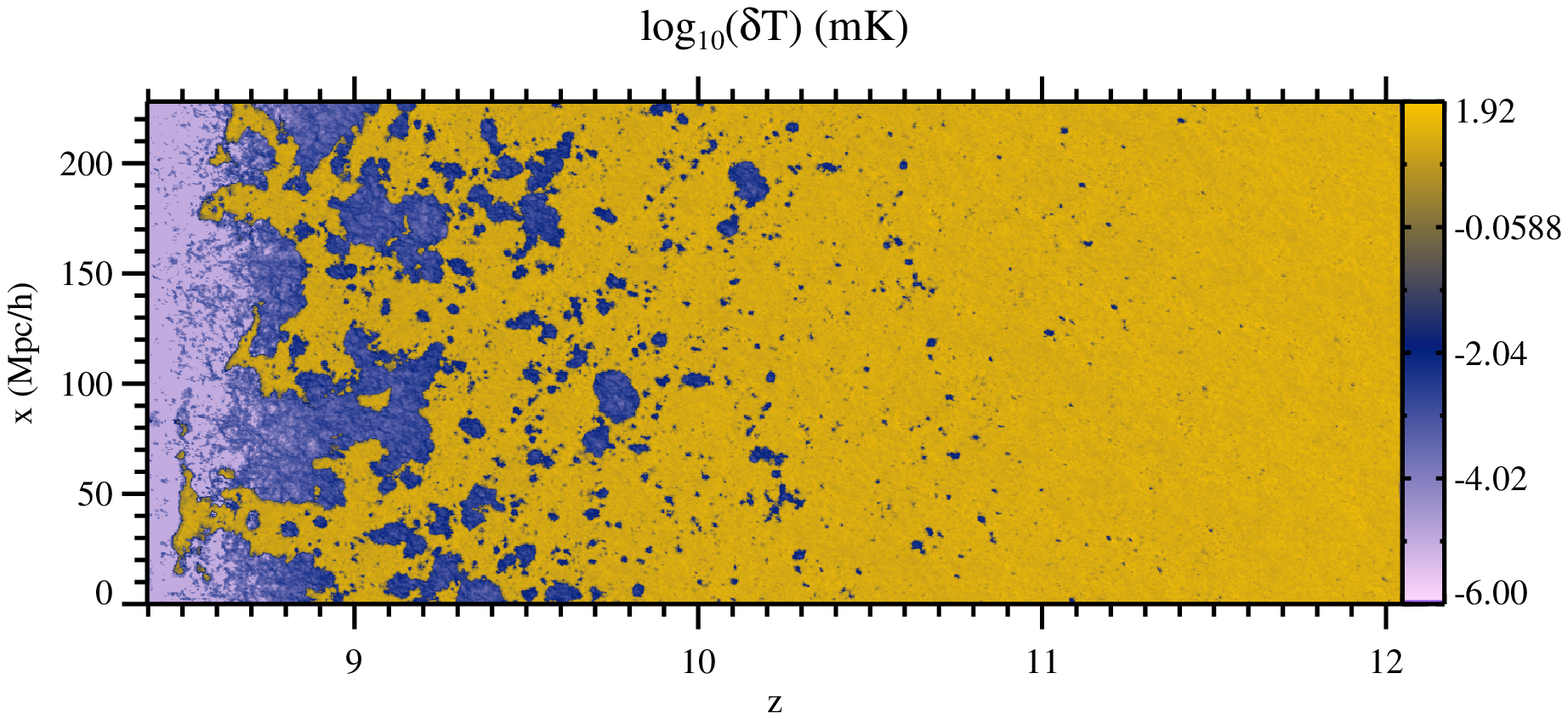}
\vspace{-0.45in} 
\includegraphics[height=3.2in]{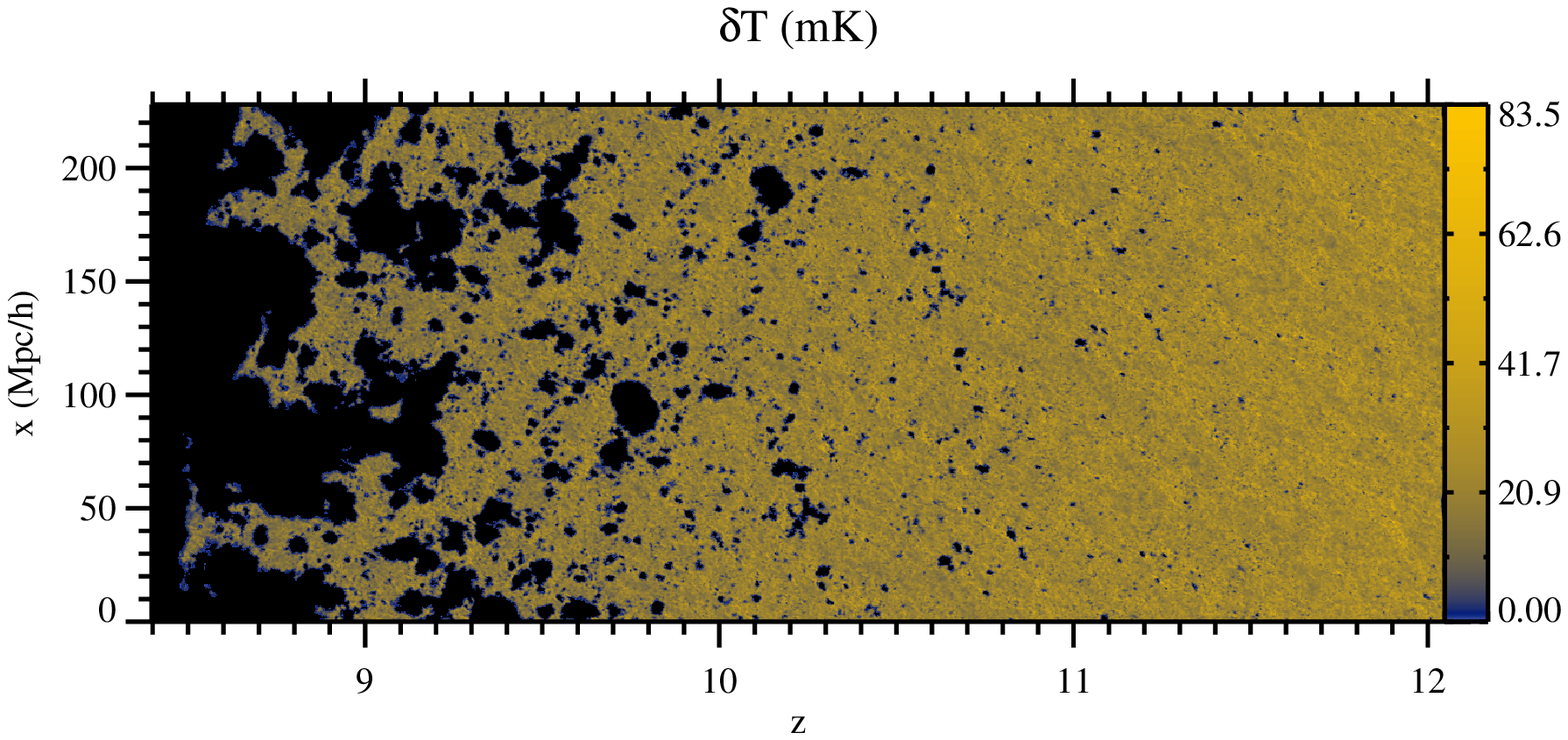} 
%\vspace{-0.9in} 
\includegraphics[height=3.2in]{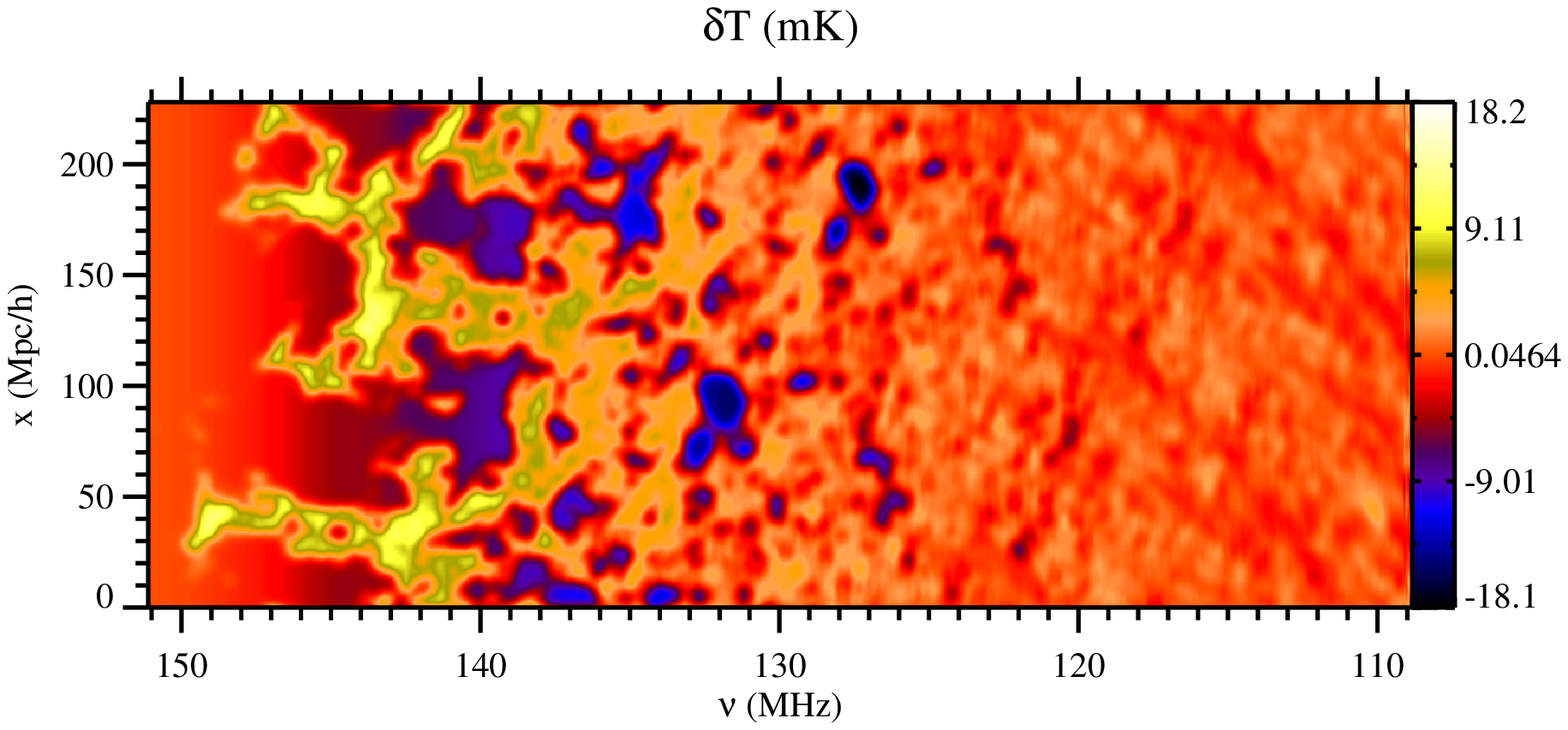} 
\vspace{-0.7cm}
\caption{
\label{21cm_image_fig}
Position-redshift and position-frequency slices from our fiducial
simulation L1. These slices illustrate the large-scale geometry of
reionization and the significant local variations in reionization
history as seen at redshifted 21-cm line.  Observationally they
correspond to slices through an image-frequency volume of a radio
array. The top and middle images shows the differential brightness
temperature at the full grid resolution in decimal log and linear
scale, respectively.  The bottom image shows the same $\delta T_b$
data, but smoothed with a Gaussian beam of 3' and (tophat) bandwidth
of 0.45 MHz, roughly corresponding to the expected parameters for the
LOFAR EoR observations. In order to mimic the behaviour of an
interferometer the mean signal has been subtracted for every frequency
slice. The spatial scale is given in comoving Mpc and we note that for
visualization purposes we have doubled (periodically) the box size in
the spatial direction. The redshift-space distortions due to the
peculiar velocities are also included.  }
%\vspace{-0.5cm}
\end{center}
\end{figure*}

\begin{figure*}
\begin{center}  
\includegraphics[width=3.2in]{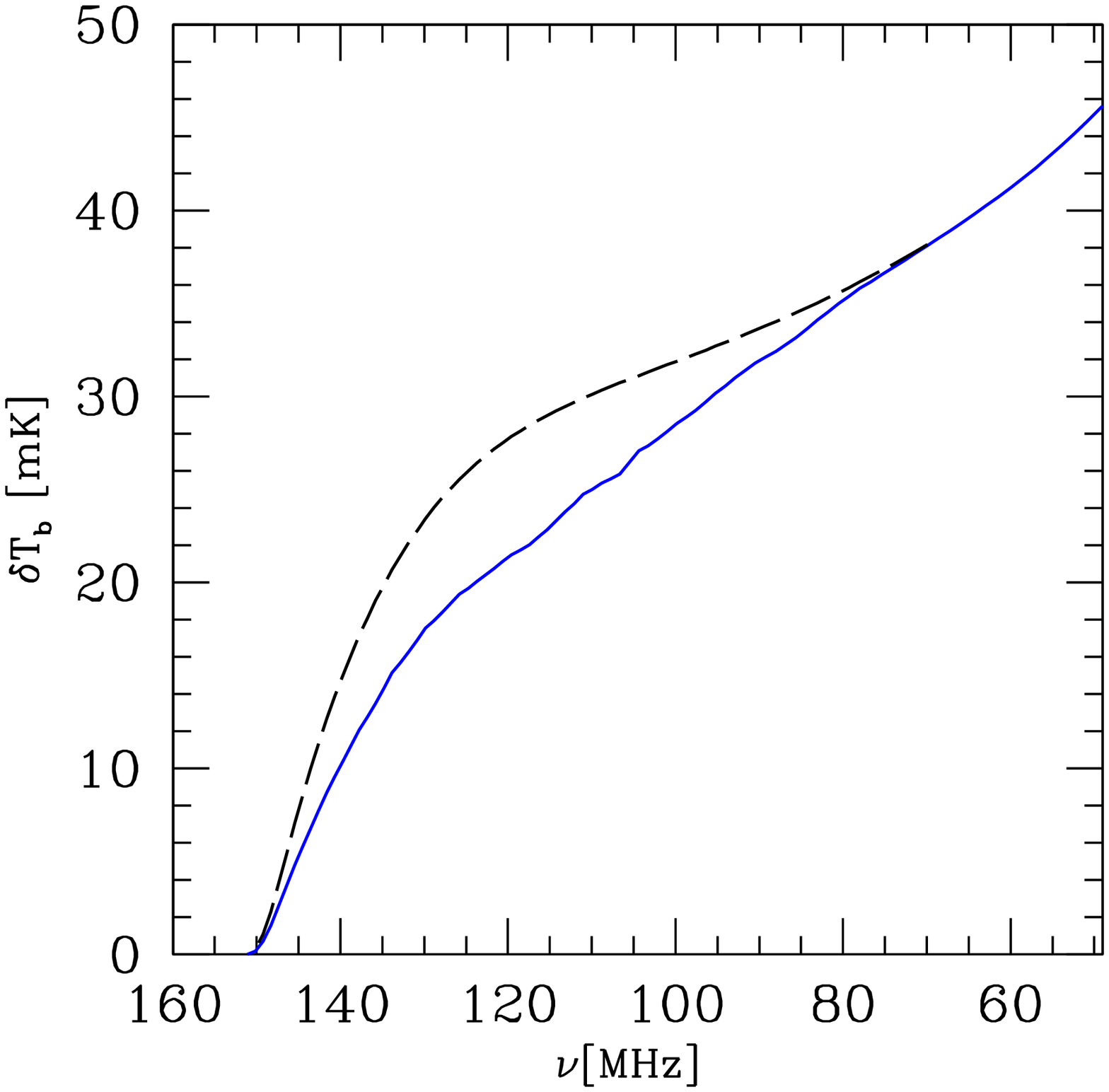} 
\includegraphics[width=3.2in]{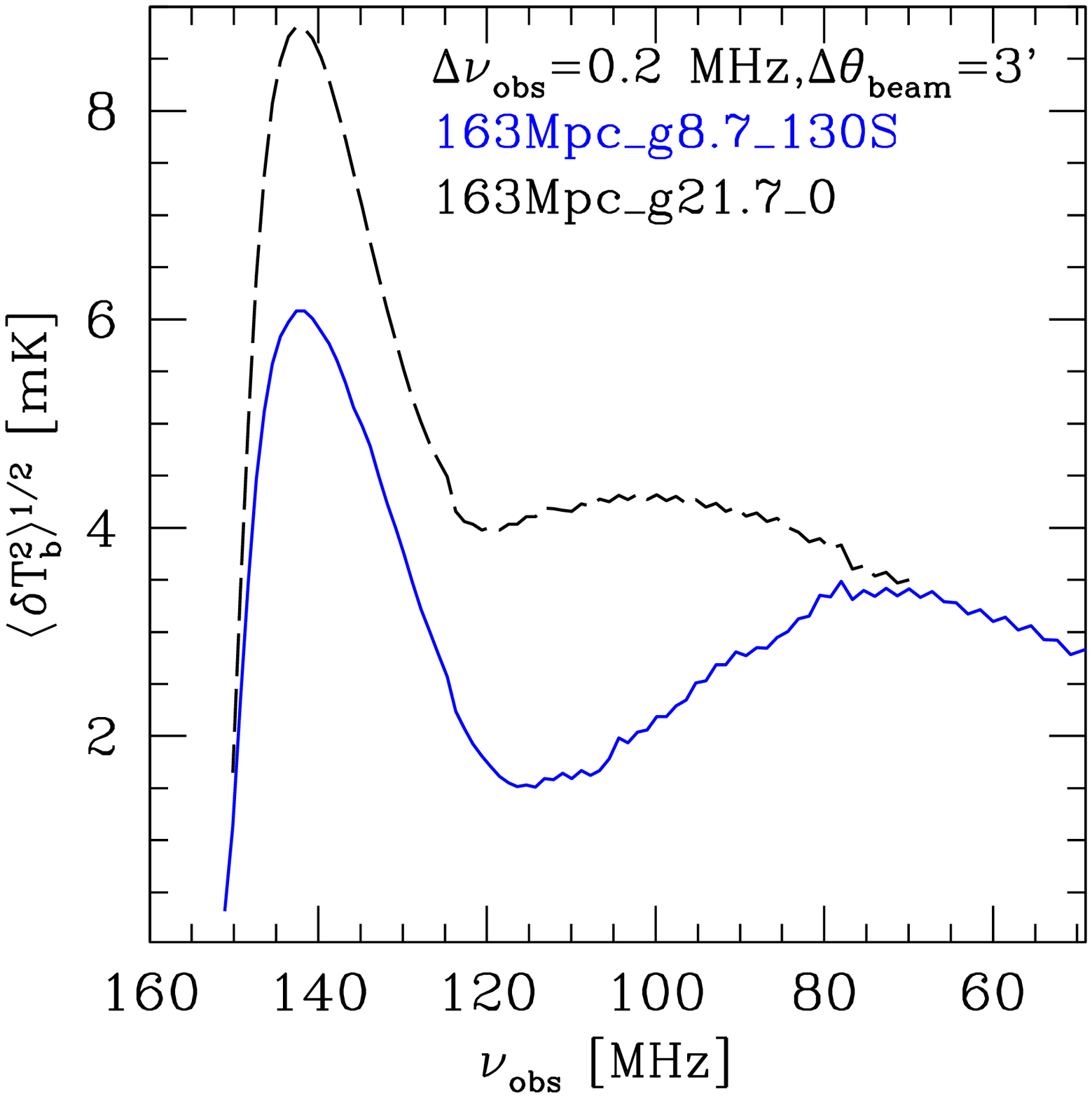}
\caption{The evolution of the mean 21-cm background (left) and 
its rms fluctuations for Gaussian beamsize $3'$ and bandwidth 
$0.2$~MHz and boxcar frequency filter (right) vs. observed 
21-cm frequency. Shown are simulations L1 (blue, solid) and L3 
(black, long-dashed). 
\label{21cm_selfreg_fig}
}
%\vspace{-0.5cm}
\end{center}
\end{figure*}  

\subsubsection{Evolution of the patchiness}

In Figure~\ref{21cm_image_fig} we show space-redshift (space-frequency)
slices cut through the simulated image cube as a radio array would see 
it (ignoring foregrounds). The spatial dimension is on the vertical, 
where we have duplicated the computational volume for visualization 
purposes, while the redshift/frequency is along the horizontal. Images 
are of the 21-cm emission differential brightness temperature signal 
extracted from our fiducial simulation L1, continuously-interpolated 
in redshift/frequency including redshift-space distortions due to 
peculiar velocities. The volume is cut at an oblique angle in order 
to minimize artificial repetition of structures along any line of 
sight. The top and middle panel show images (in log, which shows 
better the residual H~I fraction in the ionized regions and linear 
scale, which shows better the neutral structures) at the full 
simulation resolution, which is much higher than what current 
experiments will achieve given the sensitivity constraints. The 
bottom panel shows same data, but smoothed with a Gaussian beam 
and an integrated bandwidth which both roughly correspond to the 
values adopted in the LOFAR EoR experiment. To mimic the fact
that an interferometer such as LOFAR is insensitive to the global
signal, the mean signal at every frequency slice has been subtracted.

At high redshift, here $z>11$ all H~II regions are small and 
largely isolated. Smoothing the data to the LOFAR resolution 
(the ones for MWA and GMRT are even lower) renders such small 
structures undetectable. One needs at least $\sim1'$ or better 
resolution for potentially observing them, making this regime
a potential target for future, more sensitive experiments e.g.
SKA. However, at intermediate redshifts (here $z\sim10$) the 
ionized regions quickly grow by merging and remain clearly 
visible also after beam- and bandwidth smoothing. Even though 
some detail is lost, the large-scale structure of the ionization 
field remains visible all the way to the overlap epoch, here 
$z=8.4$. As was noted above, compared to the simulations with 
no self-regulation \citep[e.g.][]{2008MNRAS.391...63I}, the 
suppression of low-mass sources introduces much more small-scale 
structure and many, mostly small partially-ionized and relic H~II 
regions. However, the smoothing to the radio array resolution 
largely eliminates this fine-scale structure and the result is, 
at least visually, not dramatically different from the case with 
no self-regulation. The minimum and maximum values of the 
differential temperature are also similar. We consider more 
quantitative measures of the 21-cm signal next. 

\subsubsection{21-cm background: mean and rms}

\begin{figure*}
\begin{center}  
\includegraphics[width=3.2in]{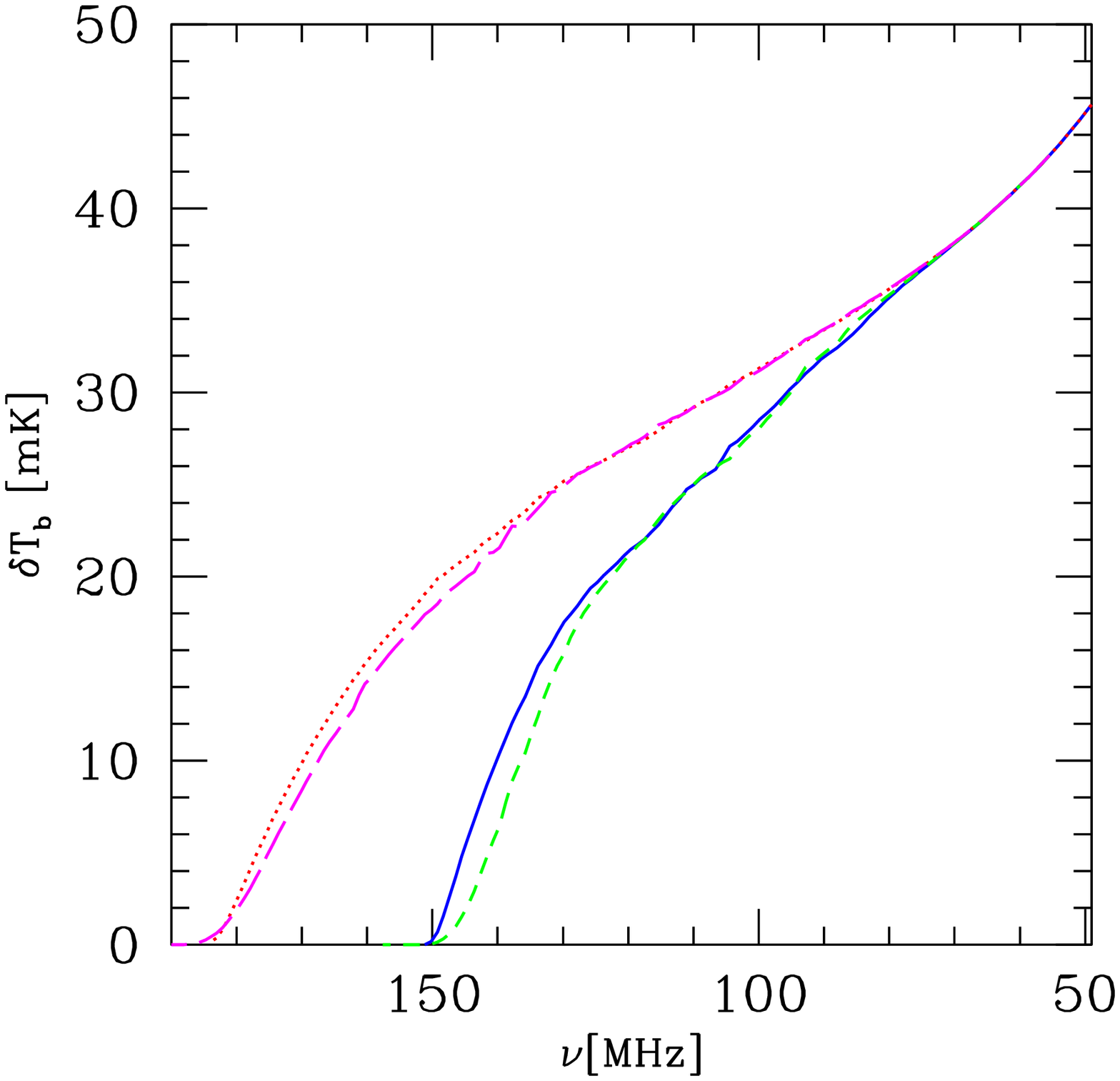} 
\includegraphics[width=3.2in]{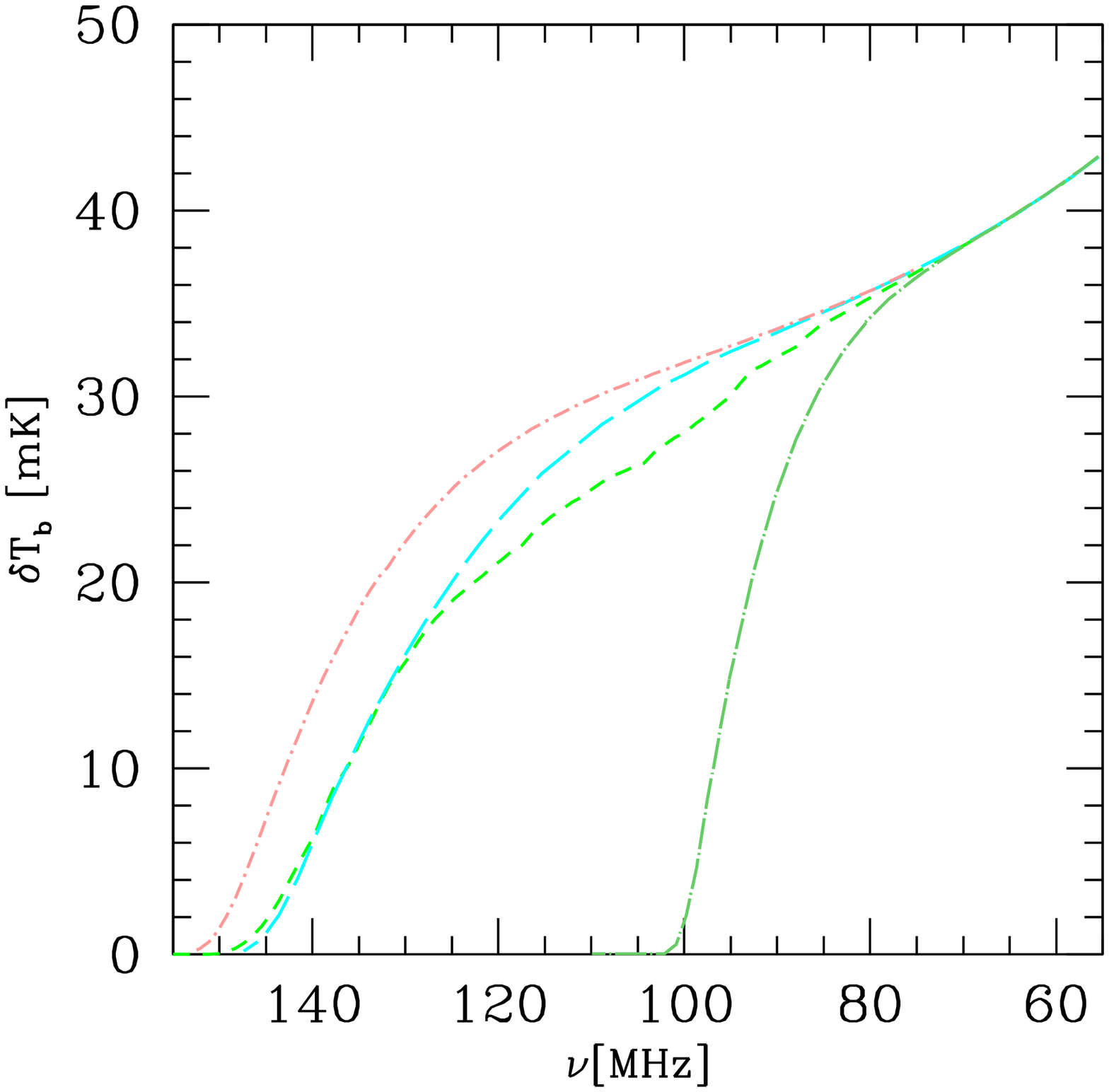} 
%\vspace{-1in}
\caption{
\label{21cm_bg_fig}
The evolution of the mean 21-cm background for our fiducial cases 
(left) and varying the source model (right). All cases are labelled 
by color and line-type, as follows: (left) L1 (blue, solid), S1 
(green, short-dashed), L2 (magenta, long-dashed), and S2 (red, 
dotted), (right) S4 (cyan, long dashed), S5 (light red, dot-short 
dashed), and S3 (light green, dot-long dashed)} 
%\vspace{-0.5cm}
\end{center}
\end{figure*} 
The evolution of the mean differential brightness temperature and 
its rms fluctuations for our fiducial case L1 and L3, which 
corresponds to our previous simulations with no self-regulation in 
\citep{2008MNRAS.391...63I} are shown in Figure~\ref{21cm_selfreg_fig}.
 The presence of low-mass sources and Jeans mass filtering yields 
initially a steeper decline of the mean 21-cm emission starting 
from $\nu\sim80$~MHz ($z\sim17$), at which point the low-mass 
sources start forming in larger numbers (becoming $\nu\sim3$ halos, 
cf. Figure~\ref{fcoll_fig}), while the high-mass sources are still 
very rare. At $\nu\sim130$~MHz ($z\sim10$) the high-mass sources in 
turn become $3-\sigma$ halos, i.e. relatively more common and the 
mean $\delta T_b$ evolution for case L3 steepens, eventially 
reaching the same overlap epoch (by construction). In terms of 
detectability in experiments looking for looking for rapid changes 
in the 21cm signal as the Universe reionizes 
\citep{1999A&A...345..380S,2010Natur.468..796B}, this behaviour 
means that the case of self-regulation is even more difficult to 
detect than the one without. While without self-regulation the 
global signal drops fast by about 25 mK between 130 and 150 MHz, 
with it the drop at the higher frequencies is more gradual. The 
decrease with self-regulation is somewhat steeper at lower 
frequencies, $\nu=80-120$~MHz, but it is still fairly gradual and 
more difficult to detect.

Comparing the rms fluctuations averaged over LOFAR-like beam and 
bandwidth (Figure~\ref{21cm_selfreg_fig}, right) we see that the 
overall evolution follows similar paths in both cases. Early-on 
very little of the gas is ionized and the fluctuations therefore 
simply follow the density ones. Only when a significant ionized 
fraction develops do the fluctuations depart from the underlying 
density. For simulation L3 this occurs fairly late, at $\nu>110$~MHz, 
compared to much earlier, $\nu>80$~MHz, for the fiducial simulation. 
At this point the rms fluctuations slightly dip, as the highest 
density peaks are ionized, which diminishes the mean $\delta T_b$ 
but does not boost the fluctuations since the H~II regions are still 
smaller than the smoothing size. As the H~II regions grow, the 
fluctuations increase again, reaching a peak before the signal 
dips again as the IGM becomes highly ionized. The peak position 
(at 142~MHz) remains the same in both cases and is thus not 
affected by self-regulation. The rms fluctuations are lower with 
self-regulation, by about 1/3 at the peak. The reason for this is 
the lower mean differential brightness temperature in that case, 
as can be seen in the bottom panel. When the fluctuations are 
normalized by the mean they become indentical in the two cases 
once they surpass the density fluctuations ($\nu>127$~MHz). Before 
that point the fluctuations in case L3 closely follow the density 
ones, while in case L1 they are lower because the highest-density
peaks have already been ionized.

\begin{figure*}
\begin{center}  
\includegraphics[width=3.2in]{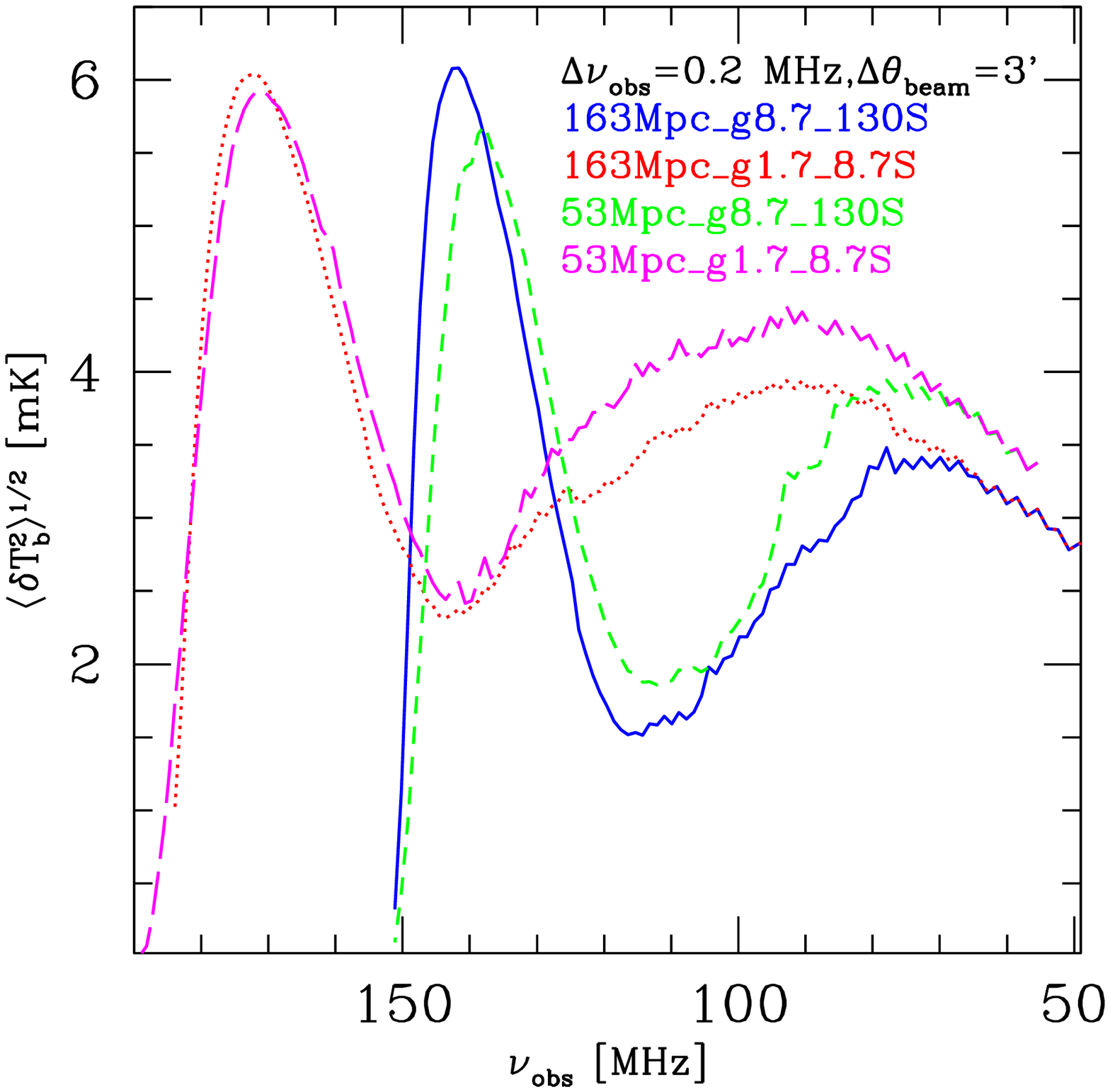} 
\includegraphics[width=3.2in]{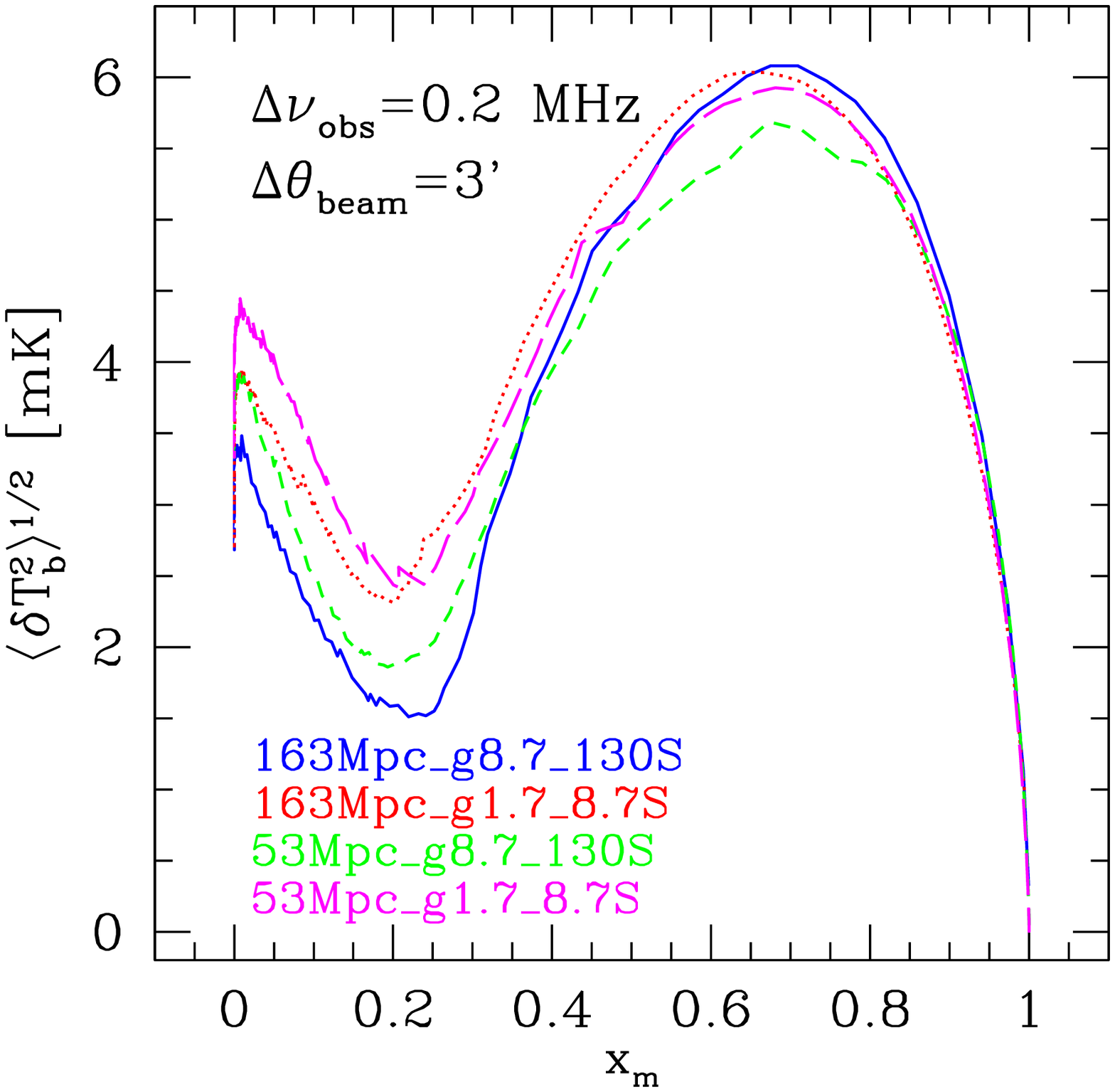} 
%\vspace{-1in}
\caption{
\label{21cm_fluct_fig}
The evolution of the rms fluctuations of the 21-cm background, 
for beamsize $3'$ and bandwidth $0.2$~MHz and boxcar filter vs. 
redshift (left) and vs. mean mass-weighted ionized fraction (right). 
Shown are our fiducial simulations L1 (blue, solid), S1 (green, 
short-dashed), L2 (magenta, long-dashed), and S2 (red, dotted).}
%\vspace{-0.5cm}
\end{center}
\end{figure*}

In Figure~\ref{21cm_bg_fig} we show the evolution of the mean 
redshifted 21-cm differential brightness temperature for high 
vs. low source efficiencies and different box sizes (left) and 
for varying source models (right). The lower photon efficiencies 
(simulation L2) predictably yield a more gradual transition of 
the global IGM from neutral to ionized state compared to our 
fiducial case L1. E.g. the evolution from 25~mK to $\sim0$~mK 
occurs over $\sim50$~MHz, from 130 to 180~MHz. Such an evolution 
would make detection of the 'global step' even more difficult 
compared to the fiducial case. The computational volume adopted 
for the simulation makes very little difference to the predicted
mean 21-cm signal, demonstrating again that $37$~Mpc$/h$ box is 
sufficiently large to faithfully represent the mean reionization 
history. Most cases with varying UV-source models 
(Figure~\ref{21cm_bg_fig}, right) yield mean 21-cm histories 
which are quite similar to each other, a consequence of their 
analogous reionization histories. The only noticeable differences 
are at intermediate frequencies, between 90 and 130 MHz, where 
case S4 gives higher $\delta T_b$ by up to 5 mK. The only 
significantly different evolutions are provided by cases S5 and 
S3. Those scenarios exhibit sharper 21-cm step due to their 
faster, exponential rise of the ionizing photon emissivity due 
to the weak or no suppression in those cases.

\begin{figure*}
\begin{center}  
\includegraphics[width=3.3in]{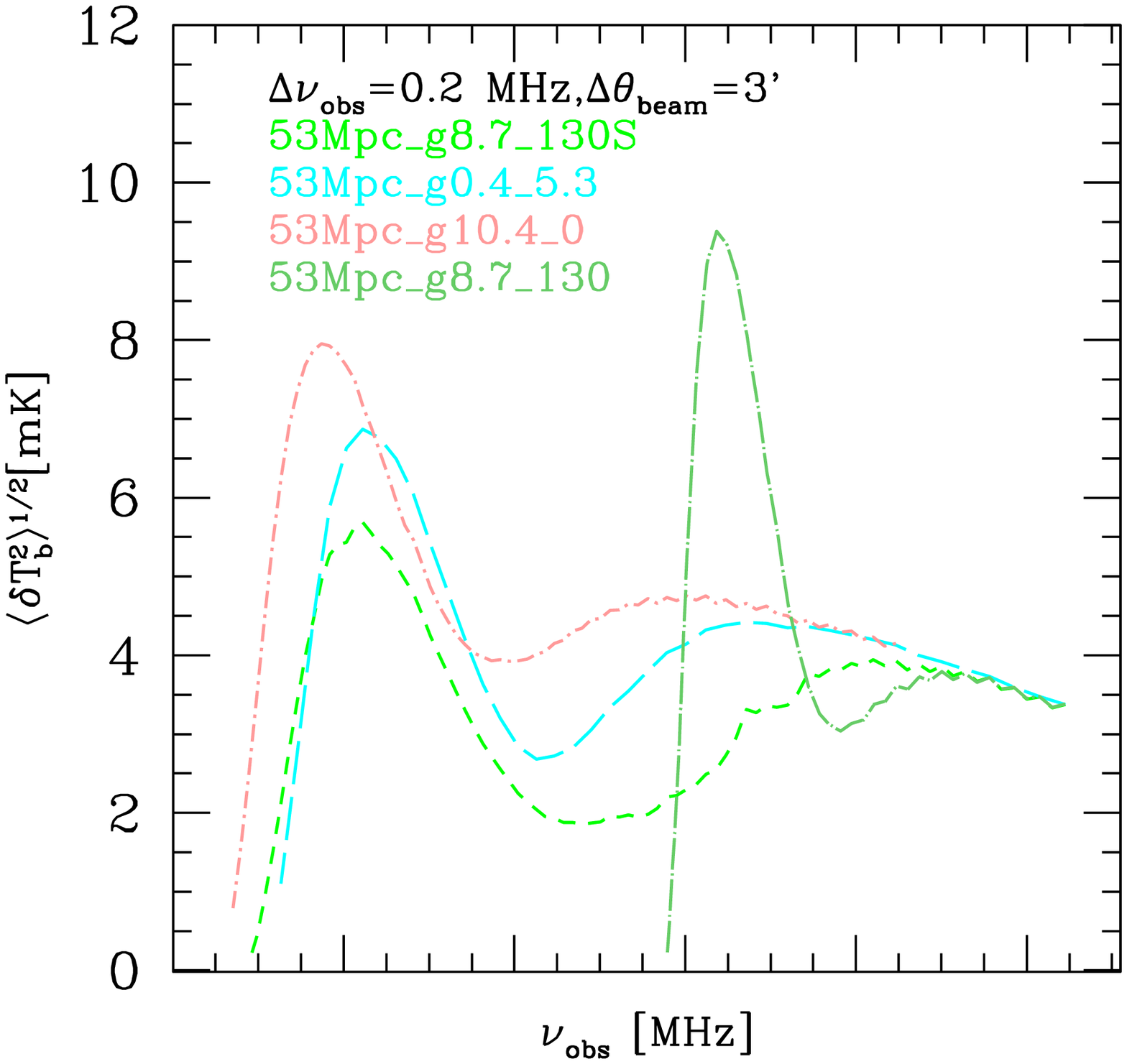} 
\includegraphics[width=3.2in]{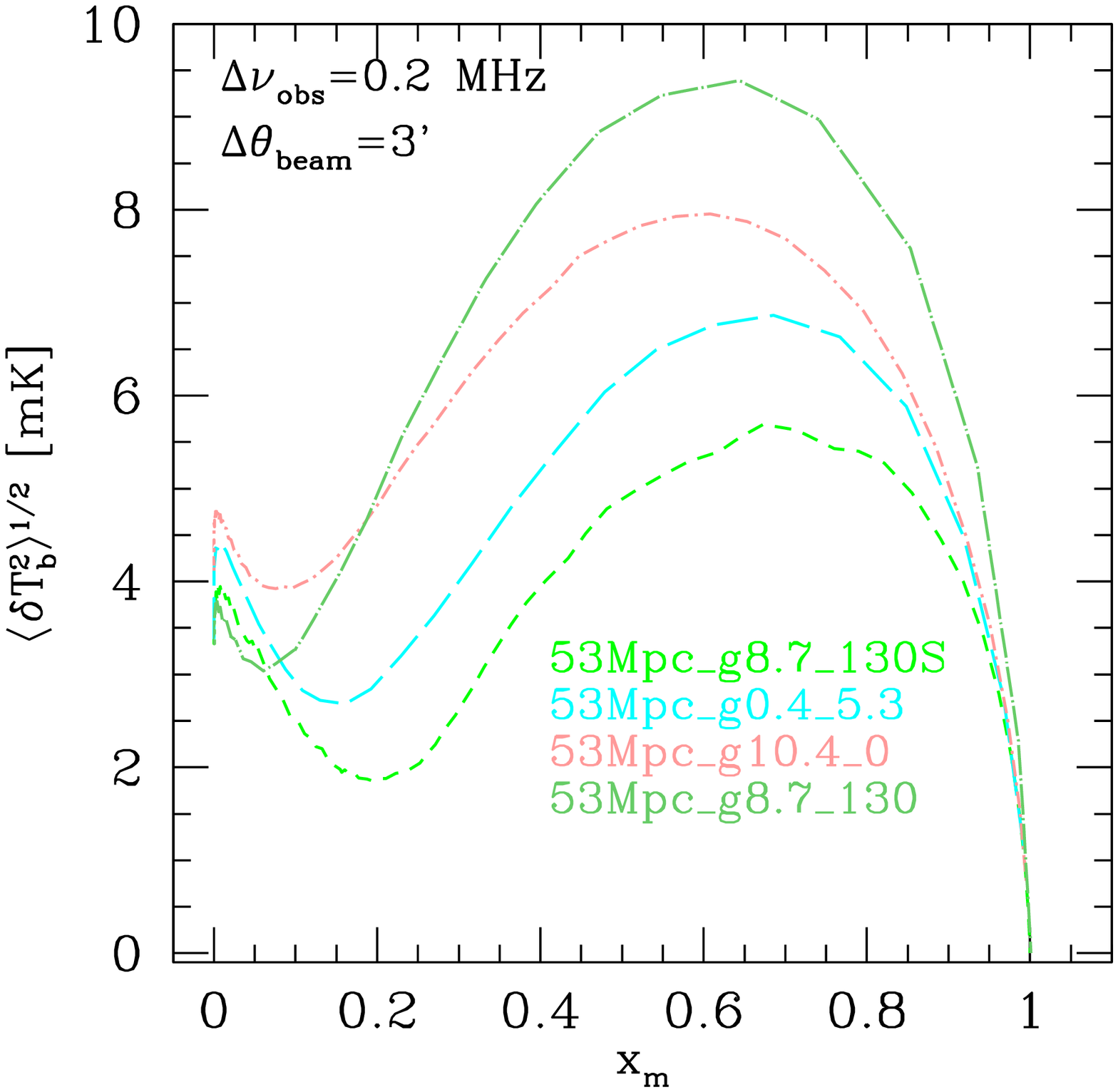} 
%\vspace{-1in}
\caption{
\label{21cm_fluct_uv_fig} Same as Fig.~\ref{21cm_fluct_fig},
but for varying UV source models: S1 (green, short-dashed), 
S4 (cyan, long dashed), S5 (light red, dot-short dashed), 
and S3 (light green, dot-long dashed).
}
%\vspace{-0.5cm}
\end{center}
\end{figure*}

The evolution of the rms 21-cm emission fluctuations for 
LOFAR-like beam and bandwidths corresponding to the same sets 
of simulations as in Figure~\ref{21cm_bg_fig} are shown in 
Figures~\ref{21cm_fluct_fig} (high vs. low ionizing efficiencies 
and varying boxsize) and \ref{21cm_fluct_uv_fig} (different UV 
source models). We show the same data vs. observed frequency 
(left) and ionized fraction (right). The latter takes away the 
reionization timing and allows comparison at the same stages of 
each reionization history regardless of when they actually occur 
in time.

In all cases the rms evolution roughly follows the same path, 
with an initial rise tracking the underlying density fluctuations 
when the IGM is mostly neutral, with a subsequent decrease when 
the first H~II regions appear followed by a second, higher peak 
of the fluctuations at later times when the initially small 
ionizing regions grow, overlap locally and as a result match 
better the interferometer beam and bandwidth resolution, and a 
final decline when most hydrogen is ionized. However, despite 
this recurring pattern, there are significant, interesting, and 
often instructive differences among the models. Varying the 
ionizing photon emission efficiencies primarily changes the 
timing of the peak of the fluctuations (Figure~\ref{21cm_fluct_fig}, 
left), from 142~MHz ($z=9$) for the fiducial case L1 to 172~MHz 
($z=7.24$) for the low-efficiency case L2. However, as seen in 
Figure~\ref{21cm_fluct_fig} (right) this shift to later times 
is fully explained by the delayed reionization in the 
low-efficiency model and both curves peak at mass-weighted 
ionized fraction $x_m\sim0.7$. We note that the latter value 
is dependent, apart from the reionization parameters, also on 
the beam and bandwidth considered, as the peak is reached when 
the typical H~II region size is best matched to the radio array 
resolution, therefore the peak occurs earlier in the reionization 
history for higher resolution and later for lower one. The 
fluctuation dip due to the earliest H~II regions also occurs at 
the same point of the reionization history ($x_m\sim0.2-0.25$) 
in both cases, but the lowest rms values reached differ 
significantly at 1.5 mK for the fiducial model vs. 2.3~mK for the 
low-efficiency one. The reason for this is that in the former case 
the bottom occurs earlier, when the sources responsible are rarer, 
more biased and therefore ionize the highest density peaks, which
results in a larger decrease in the fluctuations. The simulation 
volume (and, correspondingly, radiative transfer grid resolution) 
has moderate, but appreciable effect on the 21-cm fluctuations in 
our fiducial case. The peak height is decreased by 7\%, from 6.1~mK 
to 5.7~mK, but it is also shifted to earlier time/lower frequency 
(to 138~MHz). Interestingly, while the dip of the fluctuations is 
also shifted to lower frequency for the smaller box simulation, 
the lowest rms value is in fact higher. At first sight this appears 
counter-intuitive, since naively we might expect that the higher grid 
resolution in the smaller volume to yield a larger rms decrease (since 
the density peaks where the first sources form are resolved better in 
this case). What actually occurs here is more complicated, however. 
Statistically, there are fewer (and lower) high-density peaks in the 
smaller volume, which diminishes the effect of the very first sources
on the fluctuations. Furthermore, the very first sources form later in
the smaller box, again due to its much smaller volume, which means that
the 21-cm rms fluctuations track the density ones for somewhat longer.
On the other hand the effects of box size and resolution for the 
low-efficiency cases are small and manifest themselves solely through 
the higher underlying density fluctuations in the small-box, 
high-resolution simulation. Both the peak and dip reach the same 
value for the two cases and occur at the same frequencies.

\begin{figure*}
\begin{center}  
\includegraphics[width=2.3in]{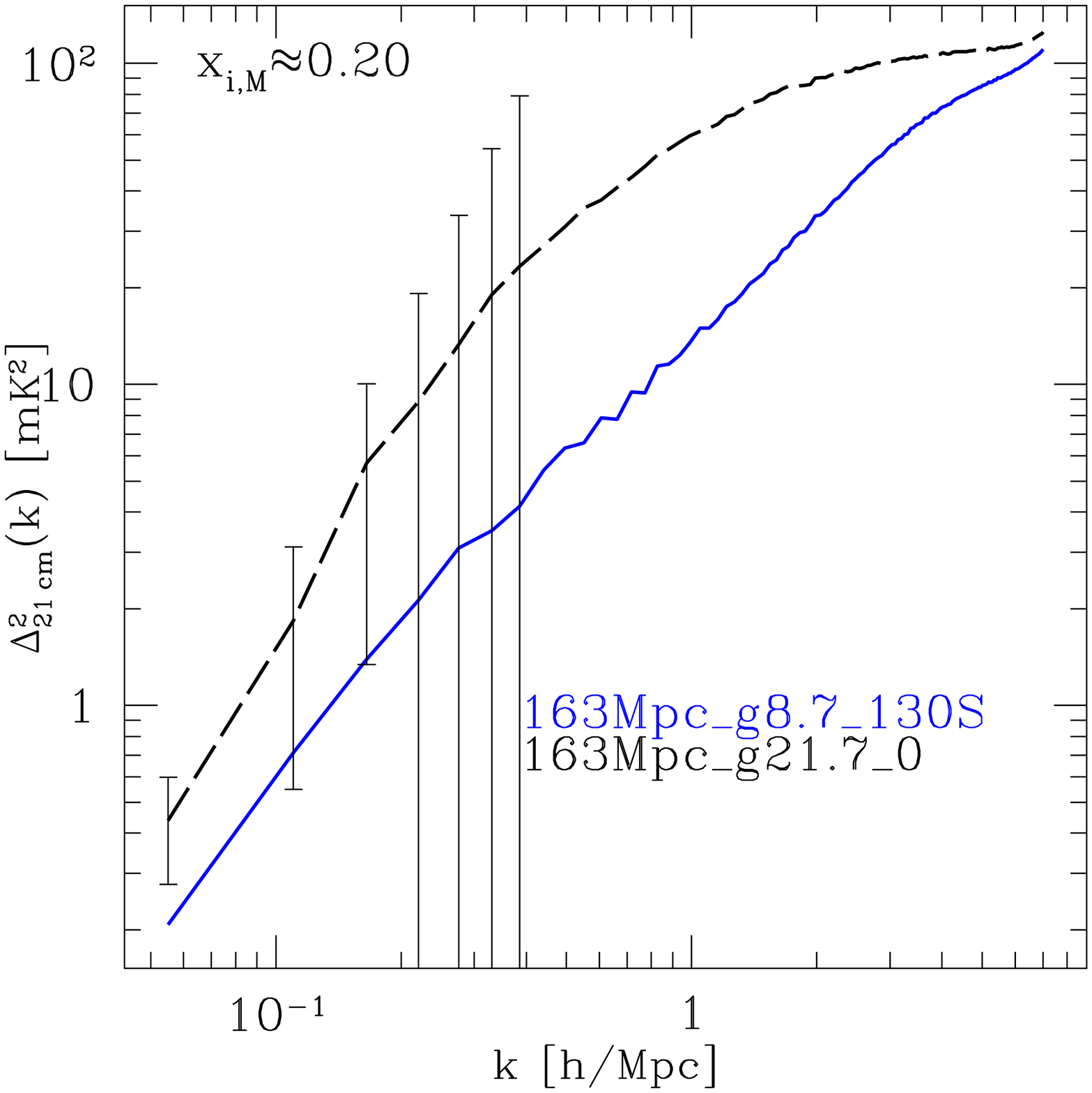} 
\includegraphics[width=2.3in]{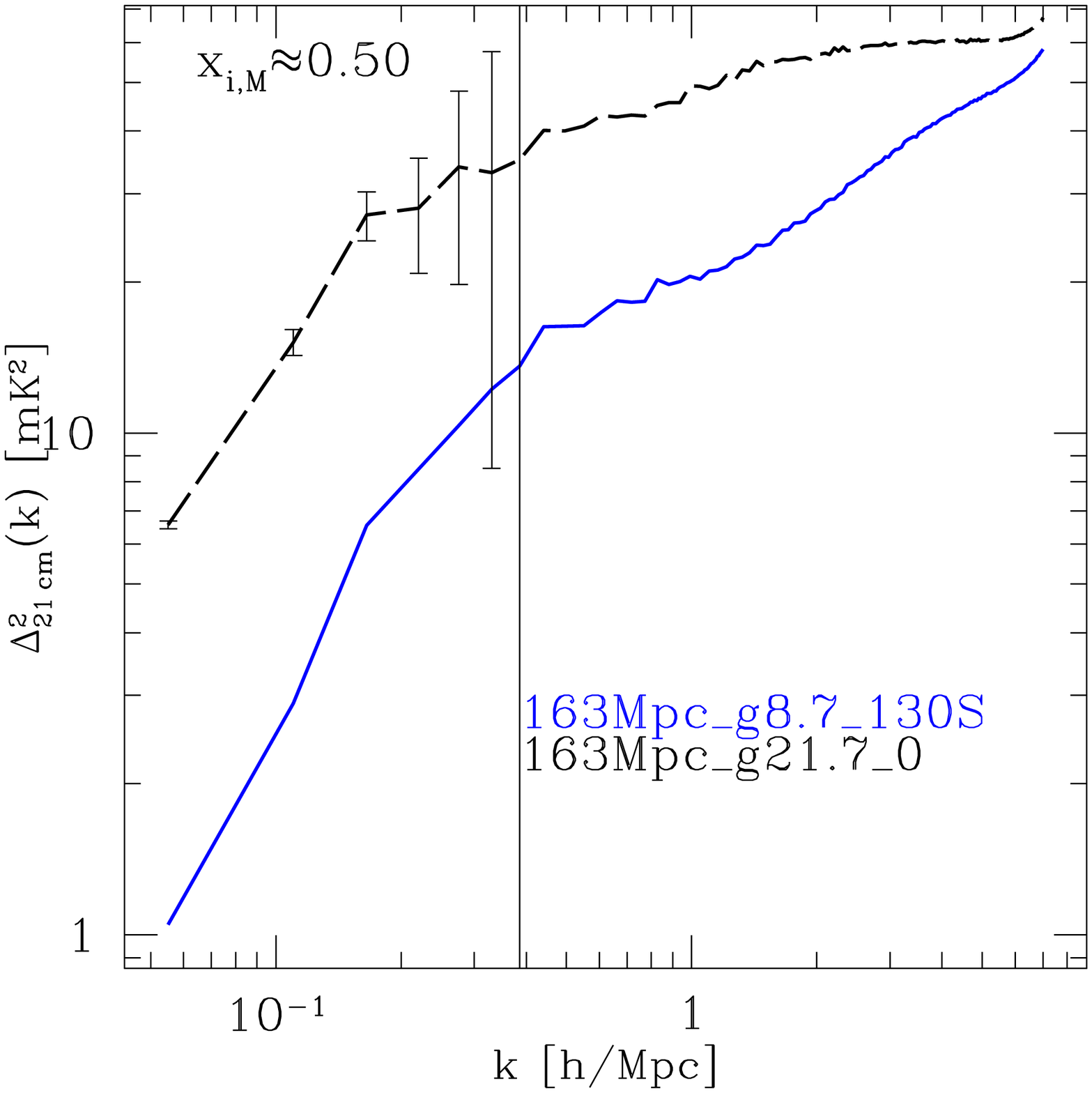} 
\includegraphics[width=2.3in]{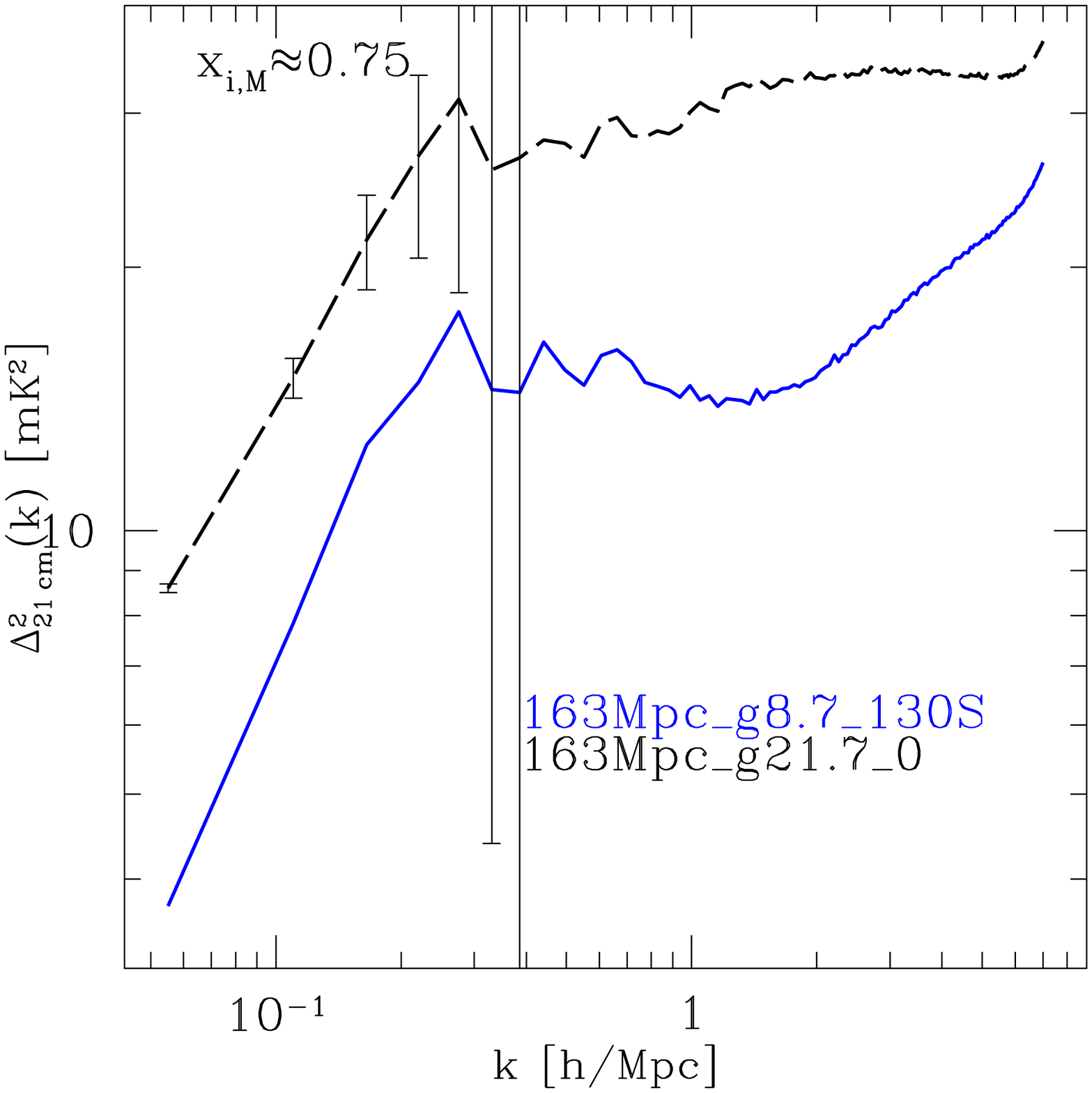} 
%\vspace{-1in}
\caption{
\label{21cm_selfreg_power_fig} The effect of self-regulation on the 
21-cm differential brightness temperature fluctuation power spectra. 
Shown are the epochs at which the ionized fractions are (left) 
$x_m=0.2$, (middle)  $x_m=0.5$ and (right) $x_m=0.75$ for our 
fiducial self-regulated case, L1 (blue, solid) and the corresponding 
non-selfregulated case with same overlap epoch, L3 (black, long-dashed).}
%\vspace{-0.5cm}
\end{center}
\end{figure*}

On the other hand, variations of the ionizing source model, yield 
a wider variety of 21-cm rms evolutions 
(Figure~\ref{21cm_fluct_uv_fig}). Interestingly, all models 
exhibit the basic evolution features seen in our fiducial 
simulations - the initial dip of the rms value when the first H~II 
regions appear, followed by a (relatively narrower) peak at later 
times when the process is sufficiently advanced for the typical 
patch size to roughly match the radio beam and bandwidth (though
we note that more extreme, and unrealistic, source models can
produce rms fluctuations with a very different shape, see 
Appendix~\ref{appendixA}). There are significant variations in 
the details of the evolution, however. The simulations with no 
low-mass source suppression but same overlap as in our fiducial 
case (S4 and S5) yield rms fluctuations peaks which are at 
roughly the same frequency as the fiducial case ($\nu\sim140$~MHz, 
more specifically), but the $\langle \delta T_b^2\rangle^{1/2}$ 
peak values are up to 50\% higher ($\sim7-8$~mK). The early 
reionization, no suppression case, S3, gives a still higher peak 
value, reaching almost 10~mK, and a narrower peak.

A different way to consider the same data is to plot the 
differential brightness temperature evolution in terms of its 
own reionization history, i.e. against the mass-weighted 
ionized fraction, $x_m$, (Figure~\ref{21cm_fluct_uv_fig}, left), 
which removes the dependence on the absolute timing of 
reionization. There is only a modest variation in the reionization 
stage (i.e. ionized fraction, $x_m$) at which the rms peak is 
reached, which ranges $x_m\sim0.6-0.7$ (i.e. relatively late in 
the reionization history). However, as we noted above, the peak 
value itself varies significantly between the simulations. It is 
highest for the high-efficiency, no suppression model, S2 
($\langle \delta T_b^2\rangle^{1/2}=9.4\,$mK 
compared to 3.8 mK for our fiducial case S1). Lower efficiency 
and no suppression (S4) results in a significantly lower peak 
(6.9~mK), while when only massive halos are present (L5) the 
peak is again quite high ($\langle \delta T_b^2\rangle^{1/2}=8.0\,$mK). 
Therefore, in general more abrupt reionization scenarios (S3, S5) 
result in higher fluctuations at the peak, while more extended 
ones (due to self-regulation or lower efficiencies) give lower 
rms peak values.

\subsubsection{21-cm background fluctuations: power spectra}
 
\begin{figure*}
\begin{center}  
\includegraphics[width=2.3in]{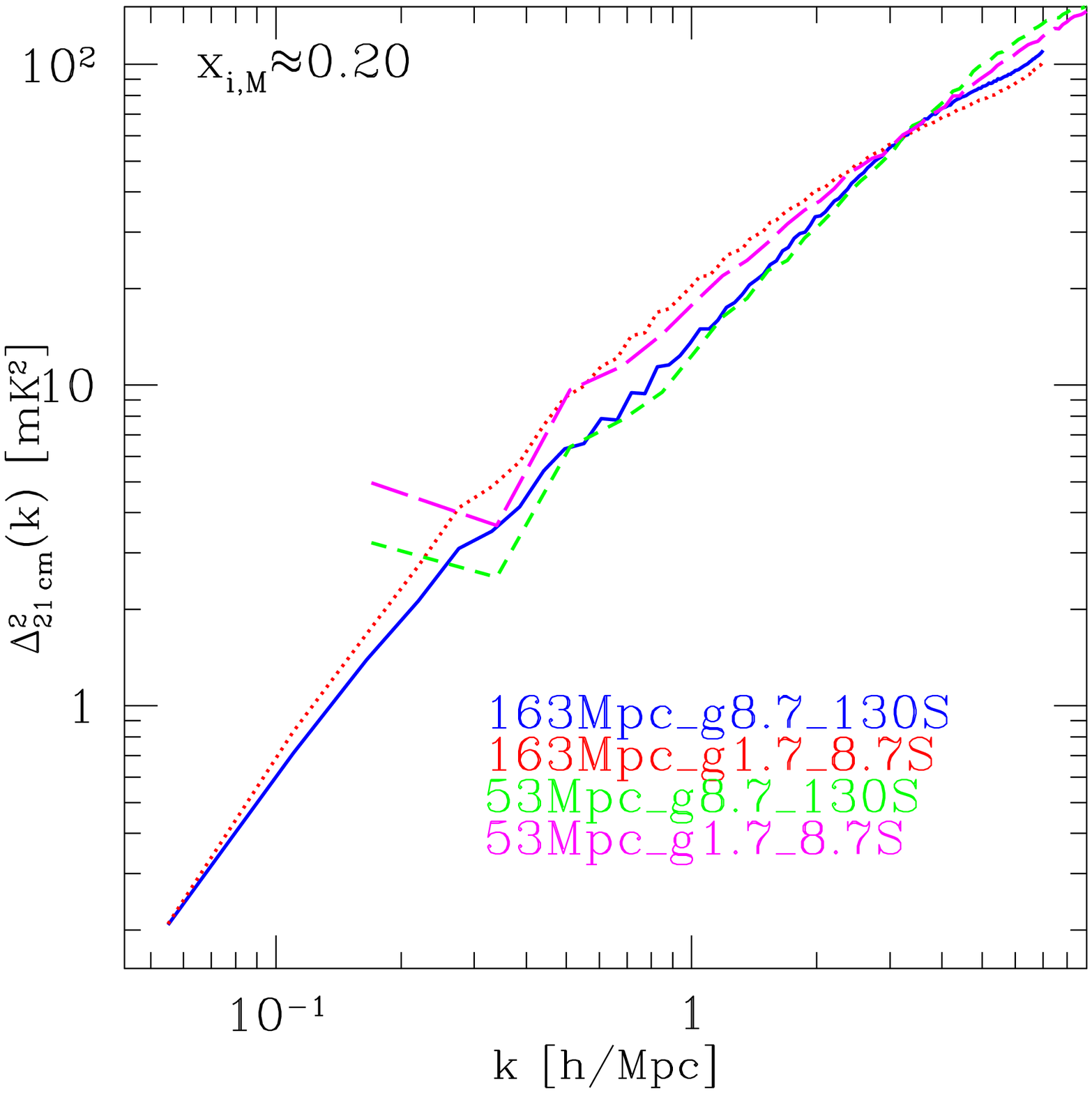} 
\includegraphics[width=2.3in]{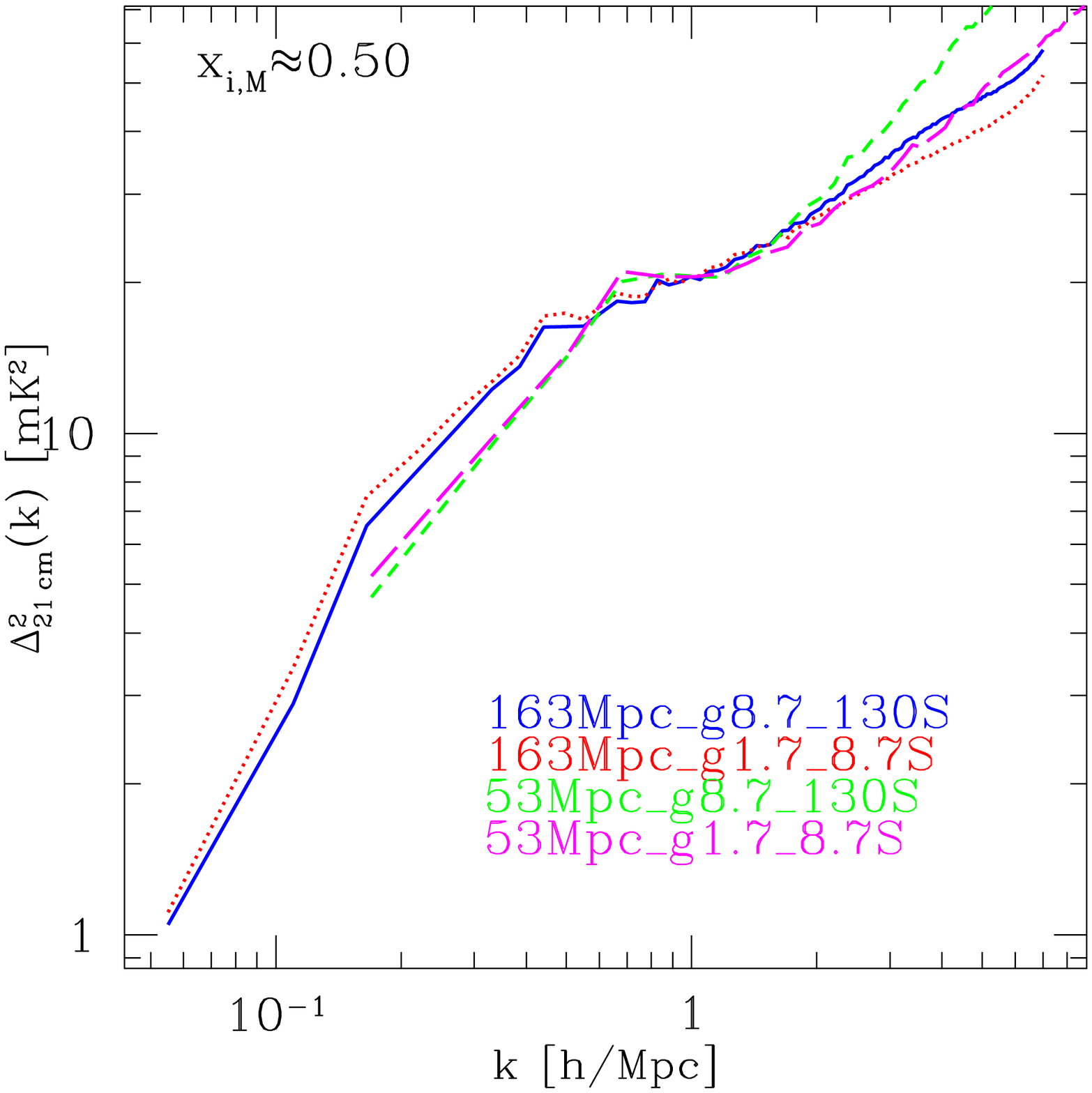} 
\includegraphics[width=2.3in]{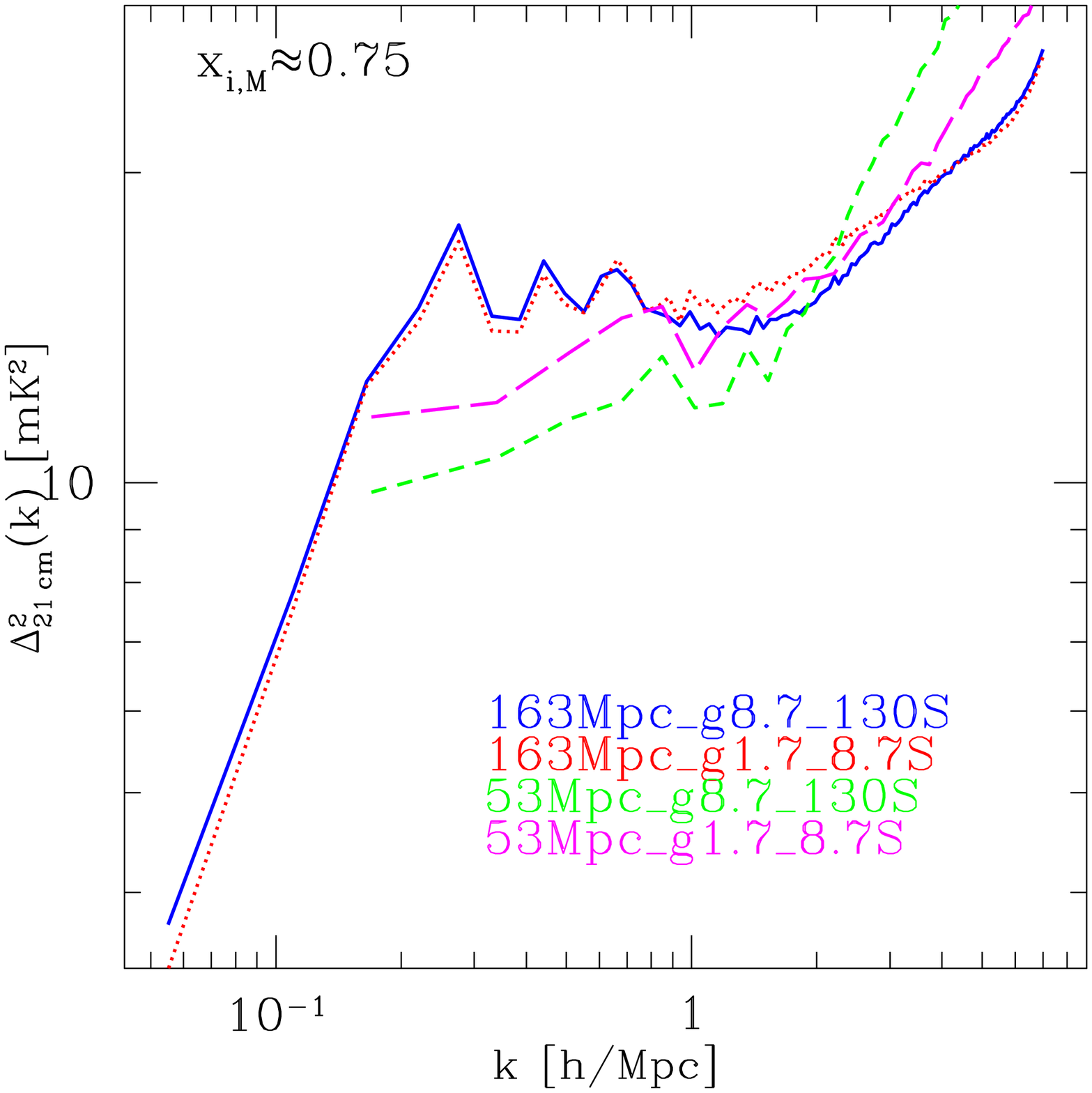} 
%\vspace{-1in}
\caption{
\label{21cm_power_fig} 21-cm differential brightness temperature
fluctuation power spectra. Shown are the epochs at which the 
ionized fractions are (left) $x_m=0.2$, (middle)  $x_m=0.5$ and 
(right) $x_m=0.75$. All cases are labelled by color and line-type, 
as follows: L1 (blue, solid), S1 (green, short-dashed), L2 
(magenta, long-dashed), and S2 (red, dotted).}
%\vspace{-0.5cm}
\end{center}
\end{figure*}

We now turn our attention to the (3D) power spectra of the 21-cm 
emission derived from our simulations. We construct the brightness 
temperature datacube in the redshift space using what we term the 
{\it PPM-RRM} scheme, as follows. We first develop an adaptive-kernel, 
SPH-like approach to compute the bulk-flow velocity of the IGM at any 
position, directly from N-body particle data. We paint the particle 
mass by the hydrogenic neutral fraction of the RT grid that the 
particle resides. Then N-body particles are Doppler-shifted to 
their apparent locations by LOS bulk-flow velocity, new smoothing 
kernel lengths are computed using the new particle positions in 
redshift-space, and halo-excluded particle data (i.e., H~I mass) 
are again smoothed onto a regular, redshift-space grid at RT grid 
resolution. Then we compute the redshift-space HI density 
fluctuation, and 21-cm brightness temperature measured in redshift 
space by 
\be
\delta T_b^s ({\bf s}) = \widehat{\delta T}_b (z_{\rm cos}) \, 
\left[1+\delta^s_{\rho_{\rm HI}}({\bf s})\right],
\ee
where $\widehat{\delta T}_b$ was defined in equation~\ref{temp21cm}. 
We compute the redshift-space power spectrum using FFT. We refer the 
reader to \citet{2011arXiv1104.2094M} for a detailed discussion of 
this methodology.

In Figure~\ref{21cm_selfreg_power_fig} we compare the 21-cm power 
spectra for models L3 (equivalent to our previous simulations with 
no self-regulation) and our fiducial self-regulated case L1 for
several representative stages of reionization. In both cases the
21-cm power spectra are initially close to a power law, with no
characteristic scale, and with only the L3 power flattening out 
at small scales due to its lack of very small-scale structures 
and smooth, large H~II regions. Overall, there is always less 
power in our fiducial case, by factor of 50\% (in $\Delta_k$), 
or more. As ionized regions continue to expand a characteristic 
scale starts to emerge, which for these particular simulations is 
around wavenumbers $k\sim0.2-0.8\,h/$Mpc. Interestingly, this 
feature shows up at the same scales regardless of the presence of 
low-mass sources, indicating that the characteristic H~II region 
size is caused by the clustering of the high-mass, unsupressible 
sources.

On the other hand, the assumed ionizing source efficiencies 
have at most only a minor effect on the power spectra once 
the self-regulation is included (Figure~\ref{21cm_power_fig}).
The characteristic H~II region scale is the same in the two 
cases and arises at the same point in the reionization history. 
The modest differences in the power spectra at the early stages 
of reionization arise due to the preferential ionization of the 
high density peaks (where the first sources form). At the same
ionized fraction $x_m$ there are many more sources in the 
low-efficiency case, forming very small H~II regions, vs. fewer, 
larger ones in the high-efficiency case. As a result, the 
low-efficiency model yields less power at very small scales 
($k>4$), but a little more power at intermediate scales 
($k=0.2-2$). At the middle and late stages of reionization the 
two power spectra at the same $x_m$ are largely identical, with 
only small differences due to the different amount of small-scale 
structure. The simulation volume utilized also has little effect 
on the derived power spectra at early times, essentially just 
shifting the range of $k$ over which the results obtained are 
reliable. However, as the H~II regions grow at intermediate and 
late stages of reionization, their sizes become comparable to 
the simulation volume and as a result the $37/h$~Mpc volume 
cannot represent the bubble size distribution correctly and the 
calculated power spectra completely miss the characteristic H~II 
region scale. On the other hand, the smaller volumes yield more 
fluctuation power at small scales, due to their superior spatial
grid resolution. These results argue for a strong caution when 
using small (sub-100$/h$~Mpc) boxes for predicting any EoR 
observables at late times. 
\begin{figure*}
\begin{center}  
\includegraphics[width=2.3in]{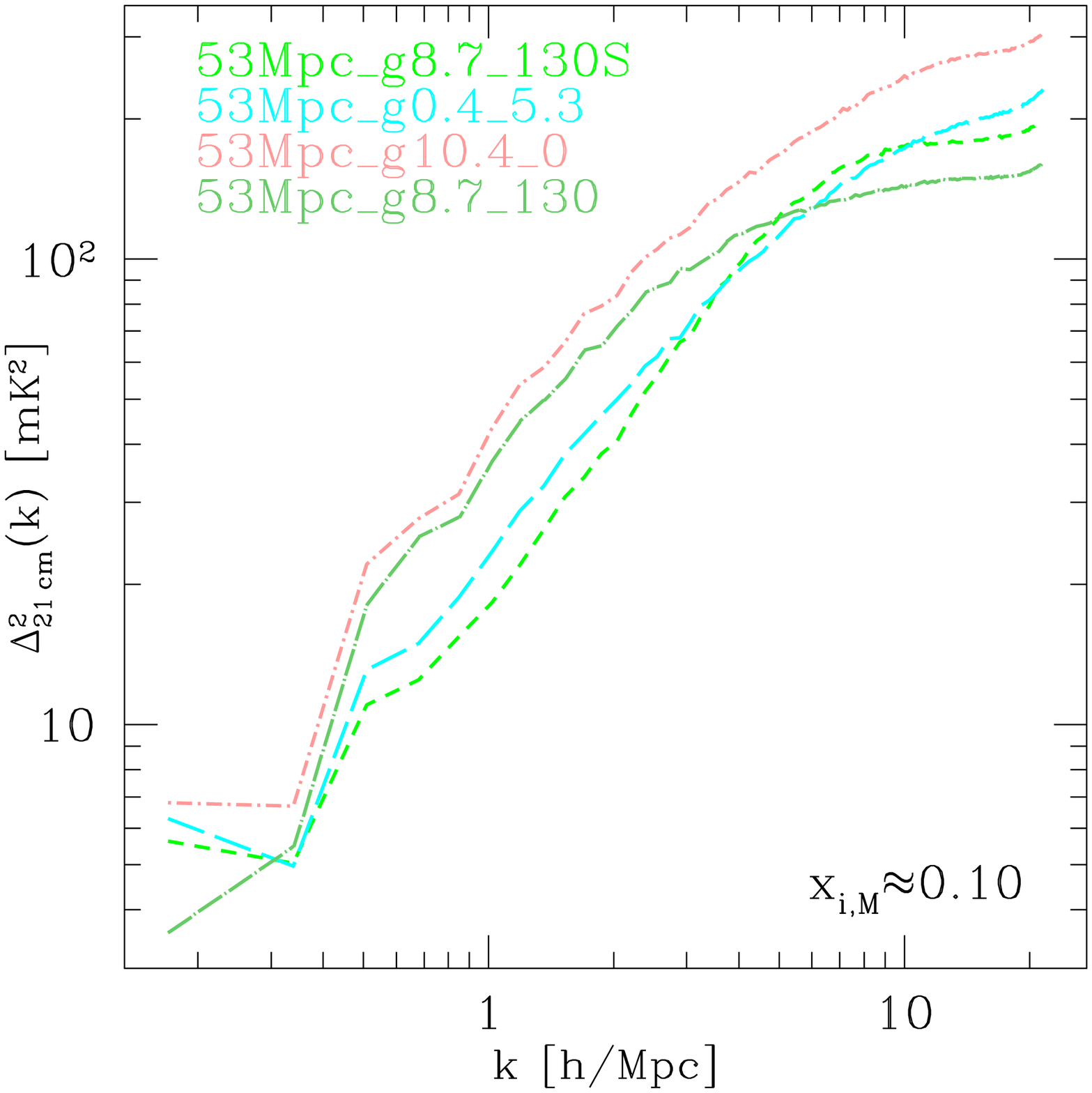} 
\includegraphics[width=2.3in]{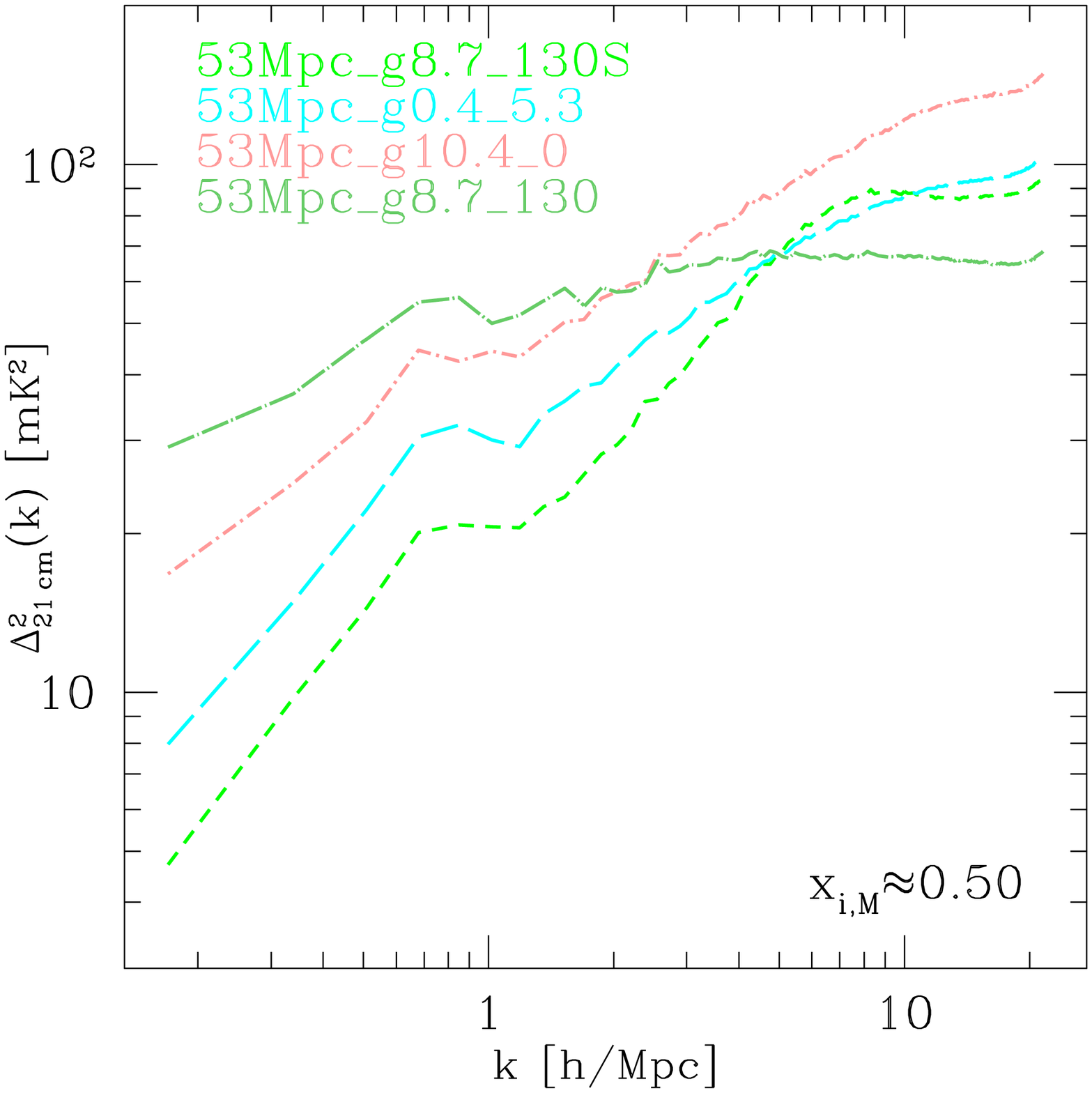} 
\includegraphics[width=2.3in]{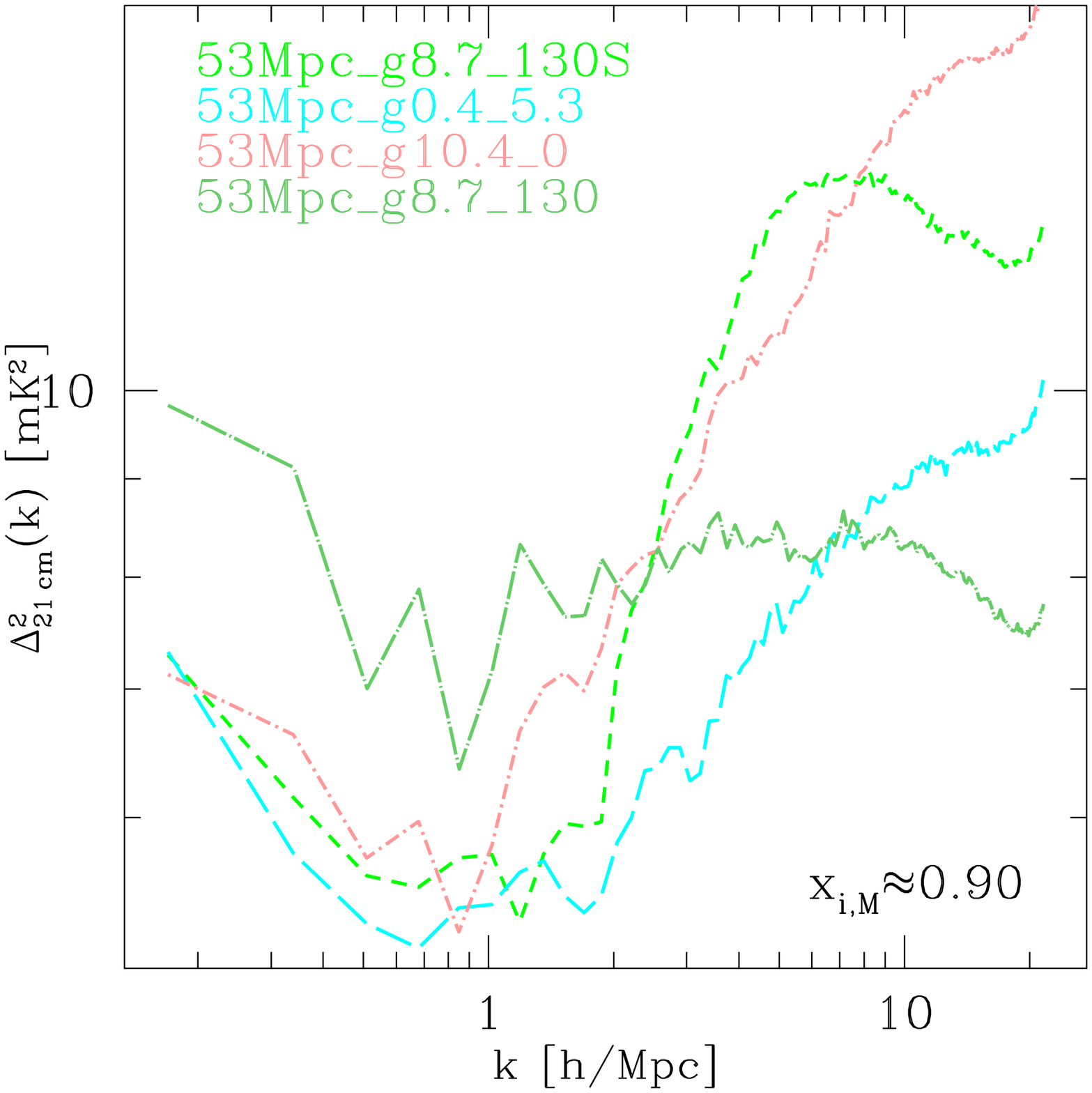} 
%\vspace{-1in}
\caption{
\label{21cm_power_misc_fig} 21-cm differential brightness 
temperature fluctuation power spectra for varying source 
models. Shown are the epochs at which the ionized fractions 
are (left) $x_m=0.1$, (middle) $x_m=0.5$ and (right) $x_m=0.9$. 
All cases are labelled by color and line-type, as follows: S1 
(green, short-dashed), S4 (cyan, long dashed), S5 (light red, 
dot-short dashed), and S3 (light green, dot-long dashed) 
}
%\vspace{-0.5cm}
\end{center}
\end{figure*}

Turning our attention to the varying source models 
(Figure~\ref{21cm_power_misc_fig}), several trends become 
clear. The lack of Jeans mass filtering (case S3 vs. the 
fiducial S1) results in much more power at large scales 
($k<5\,h/$Mpc early, $k<2\,h/$Mpc at late times), but 
considerably less power on small scales, consequence of 
the large H~II regions with smooth boundaries in case S3 
produced by its luminous and strongly clustered sources. 
This also results in very flat power spectra for S3, with 
roughly constant power at all scales at intermediate and 
late times ($x_m=0.5$ and 0.9). On the other hand, if the 
low-mass sources are not present at all (case S5 vs. S3) 
there is more power on all scales during the early stages 
of reionization ($x_m=0.1$). However, at intermediate and 
late stages of reionization the power spectra for the 
large-source only case S5 remain steeper, with considerably 
more power at small scales, which results in cases S3 and 
S5 power spectra crossing at $k\sim2\,h/$Mpc. The reason 
for this somewhat counterintuitive behaviour is that the 
same reionization stage is reached in S5 much later than 
in S3, by which time there are many more and less clustered
high-mass sources, which in turn results in more power at 
small scales. Finally, the low-efficiency case with no 
self-regulation (S4) yields very similar power spectra to 
our fiducial simulation S1 throughout the evolution, with 
only slightly more power at large scales during the early 
and intermediate stages, and slightly less small-scale 
power at the late stages. This suggests that the S1 and S4 
scenarios might be difficult to distinguish solely through 
power spectra measurements.

\subsubsection{21-cm background fluctuations: PDFs and non-Gaussianity}

\begin{figure*}
\begin{center}  
\includegraphics[width=2.3in]{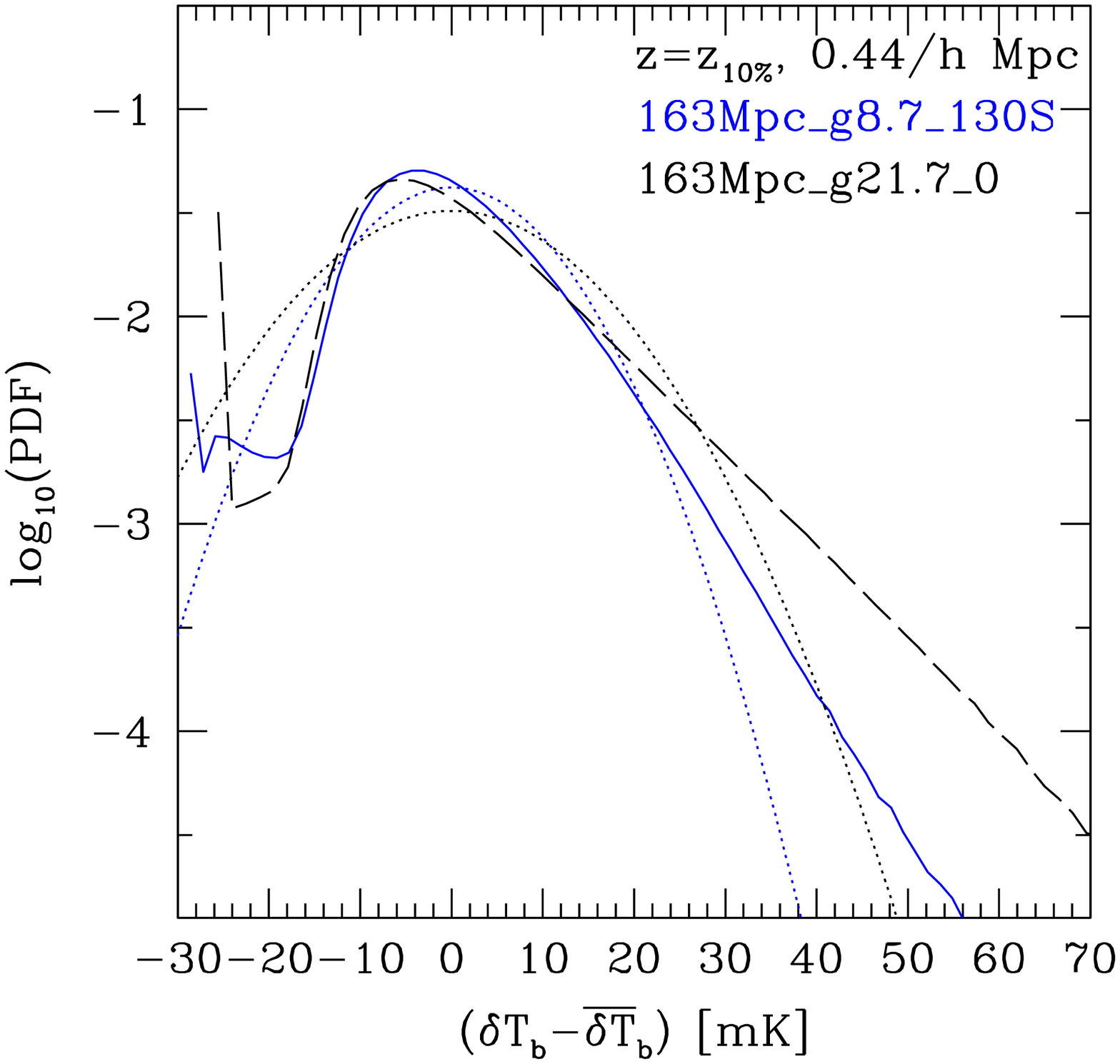} 
\includegraphics[width=2.3in]{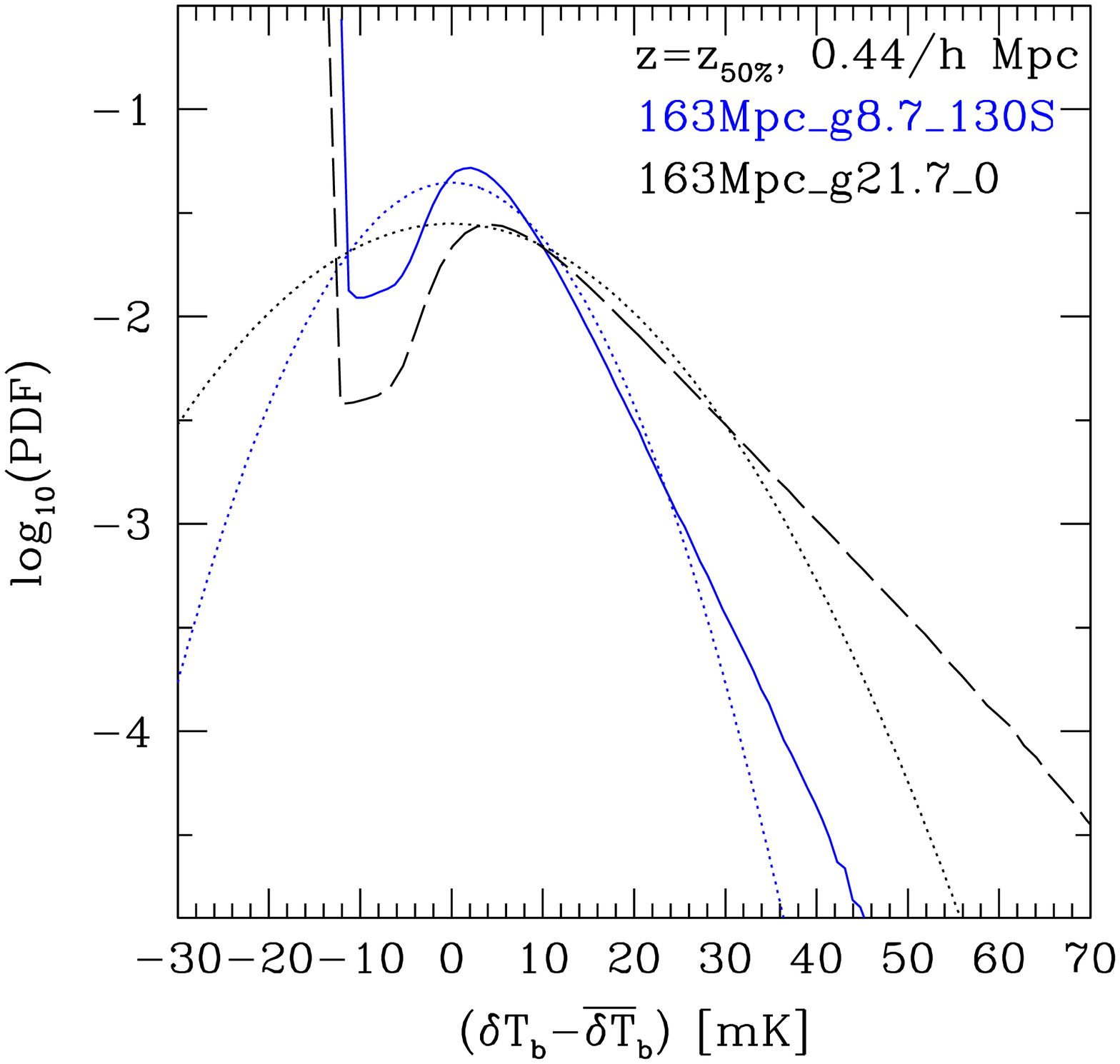}
\includegraphics[width=2.3in]{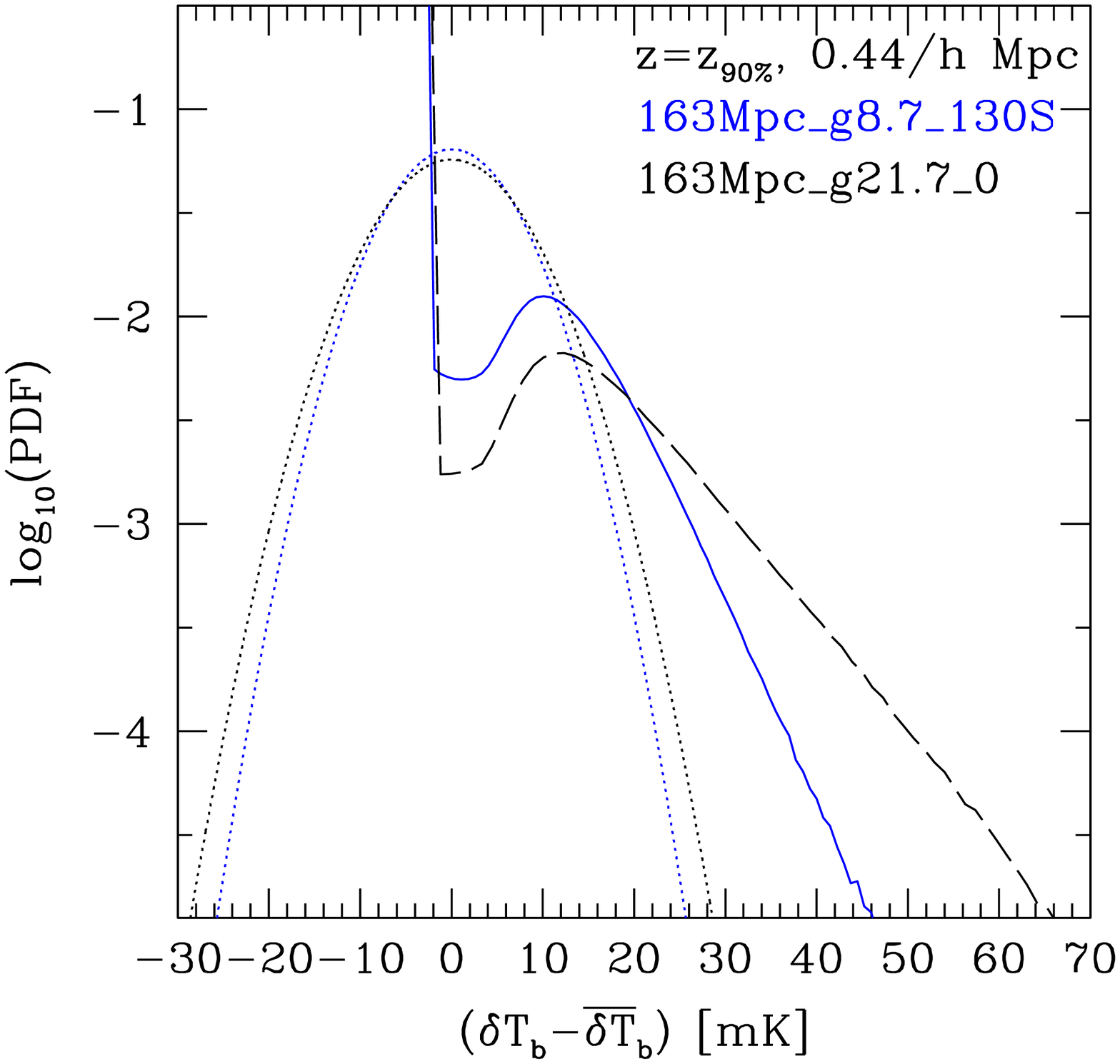} 
%\vspace{-1in}
\caption{The effect of self-regulation on the PDF distribution 
of the 21-cm signal. Shown are the epochs at which the ionized 
fractions are $x_m=0.1$ (left), $x_m=0.5$ (middle) and $x_m=0.9$ 
(right) for our fiducial self-regulated case, L1 (blue, solid) 
and the corresponding non-selfregulated case with same overlap 
epoch, L3 (black, long-dashed). The PDFs are cell-by-cell 
% of sizes $0.445\,h^{−1}$~Mpc 
(i.e. no smoothing apart from the numerical grid resolution)
Also indicated are the Gaussian distributions with the same 
mean values and standard deviations (dotted lines, corresponding 
colours).
\label{21cm_PDF_selfreg_1cell_fig}
}
%\vspace{-0.5cm}
\end{center}
\end{figure*}

\begin{figure*}
\begin{center}  
\includegraphics[width=2.3in]{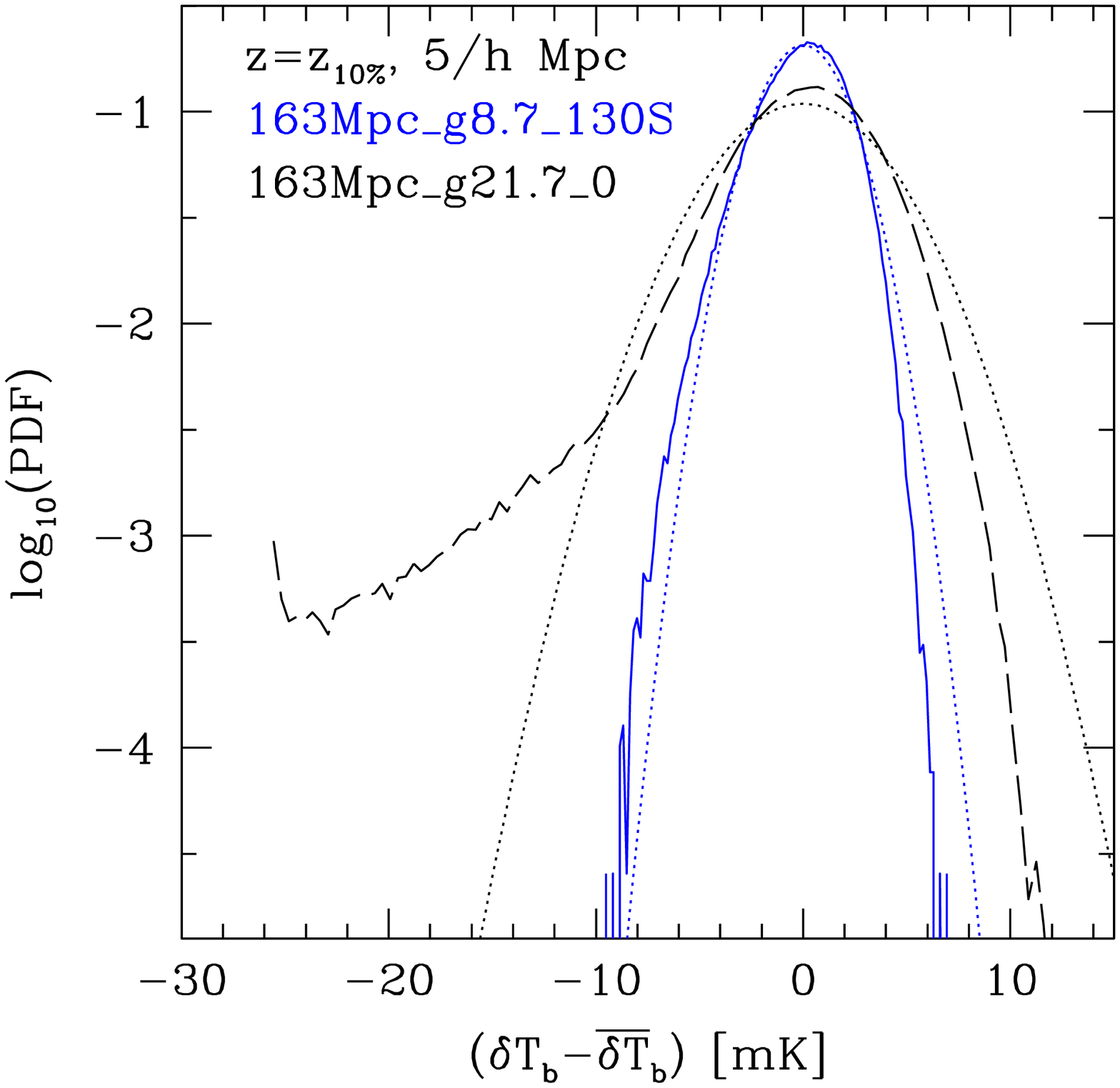} 
\includegraphics[width=2.3in]{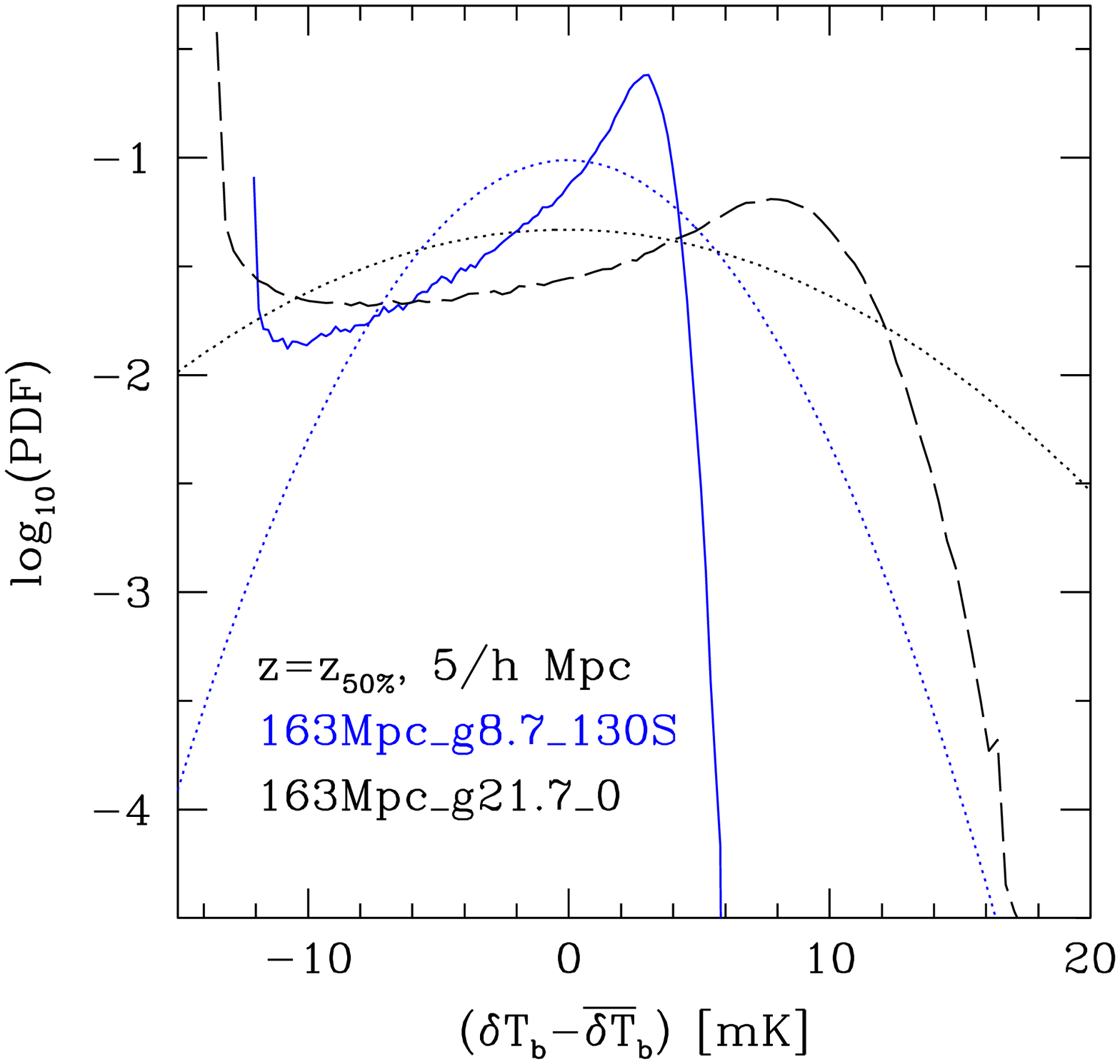}
\includegraphics[width=2.3in]{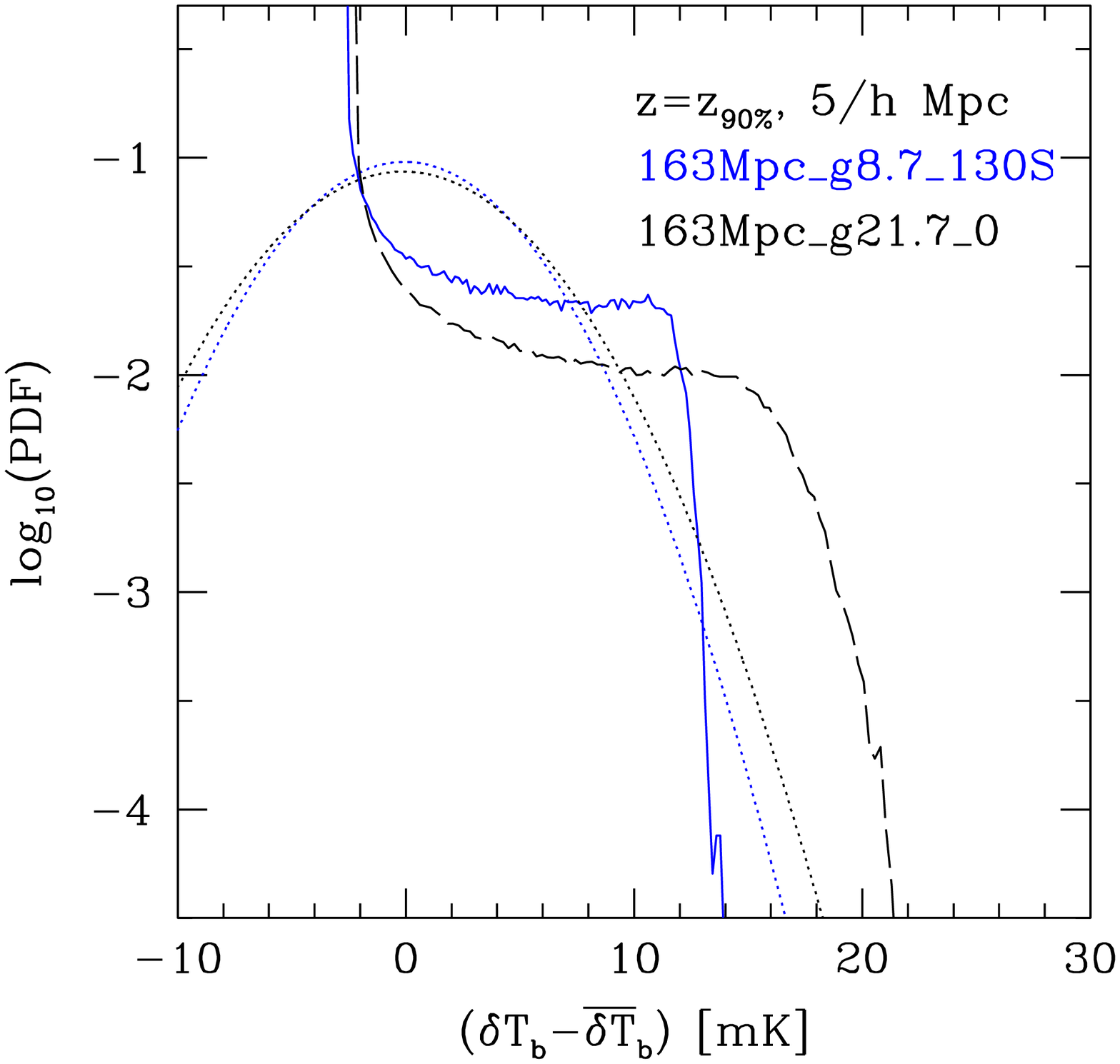} 
%\vspace{-1in}
\caption{Same as in Fig.~\ref{21cm_PDF_selfreg_1cell_fig}, but for
boxcar smoothing of $5\,h^{−1}$~Mpc. 
\label{21cm_PDF_selfreg_5Mpc_fig}
}
%\vspace{-0.5cm}
\end{center}
\end{figure*}

The 21-cm power spectra would fully characterize the emission 
field if the differential brightness distribution were purely 
Gaussian. However, generally that is not the case during 
reionization, as we have shown in previous work 
\citep{2006MNRAS.372..679M,2008MNRAS.384..863I,2009MNRAS.393.1449H}. 
The probability distribution functions (PDFs) of $\delta T_b$ 
could be significantly non-Gaussian, particularly at the later 
stages of reionization \citep{2006MNRAS.372..679M} and their 
measured skewness can be used to discriminate between different
reionization scenarios \citep{2009MNRAS.393.1449H}. The PDF's 
and their evolution could also be used to derive the reionization 
history of the IGM \citep{2010MNRAS.406.2521I,2010MNRAS.408.2373G}.

\begin{figure*}
\begin{center}  
\includegraphics[width=2.3in]{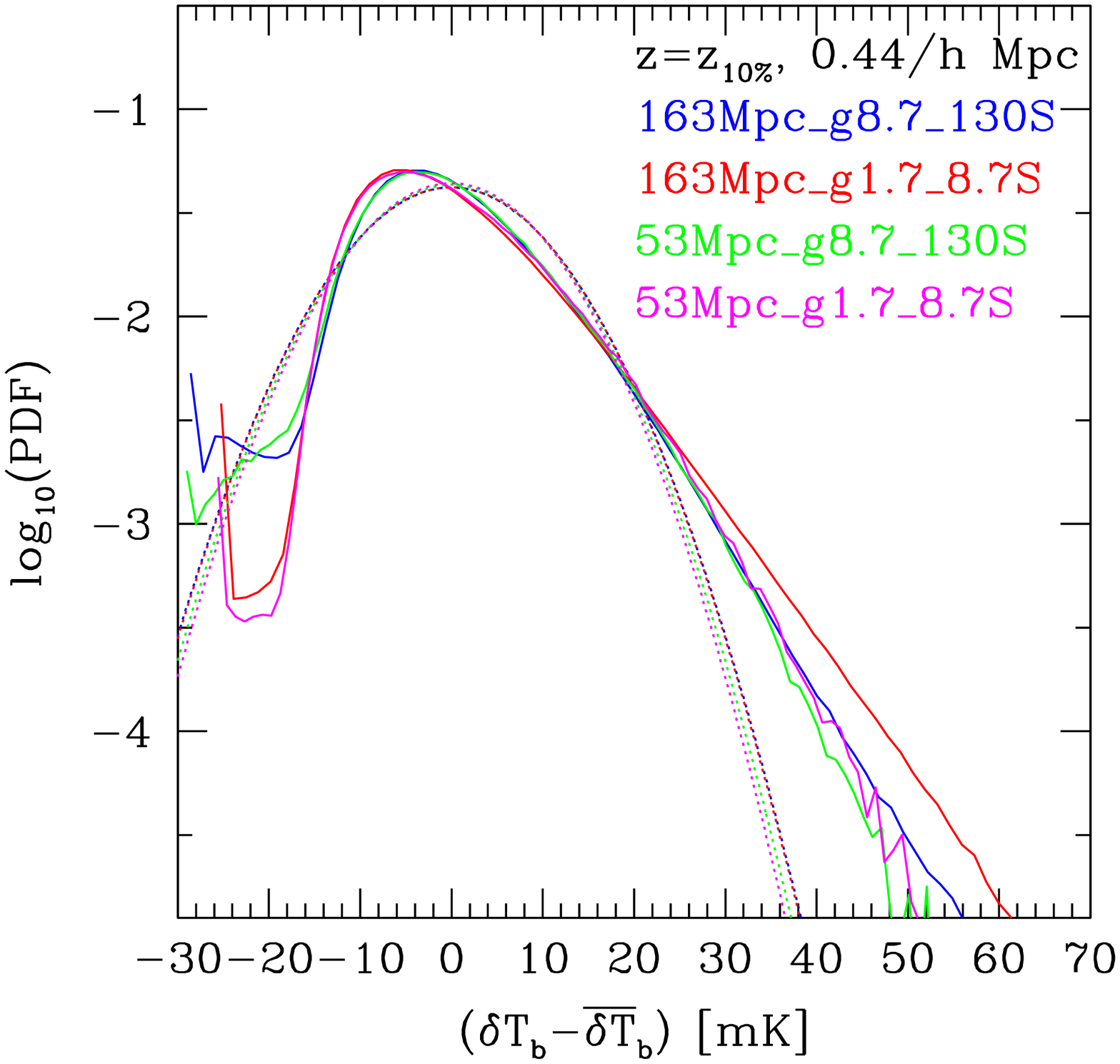} 
\includegraphics[width=2.3in]{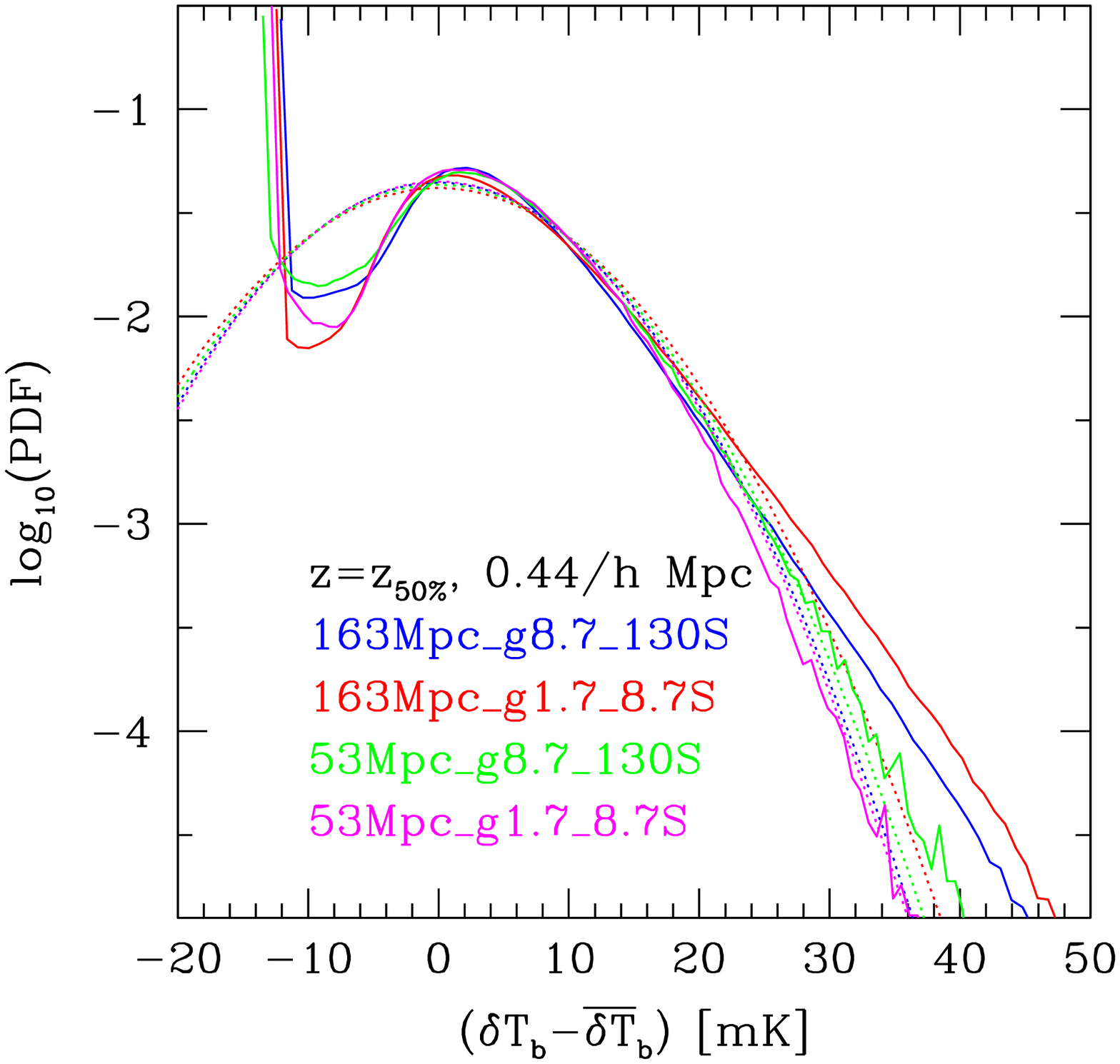}
\includegraphics[width=2.3in]{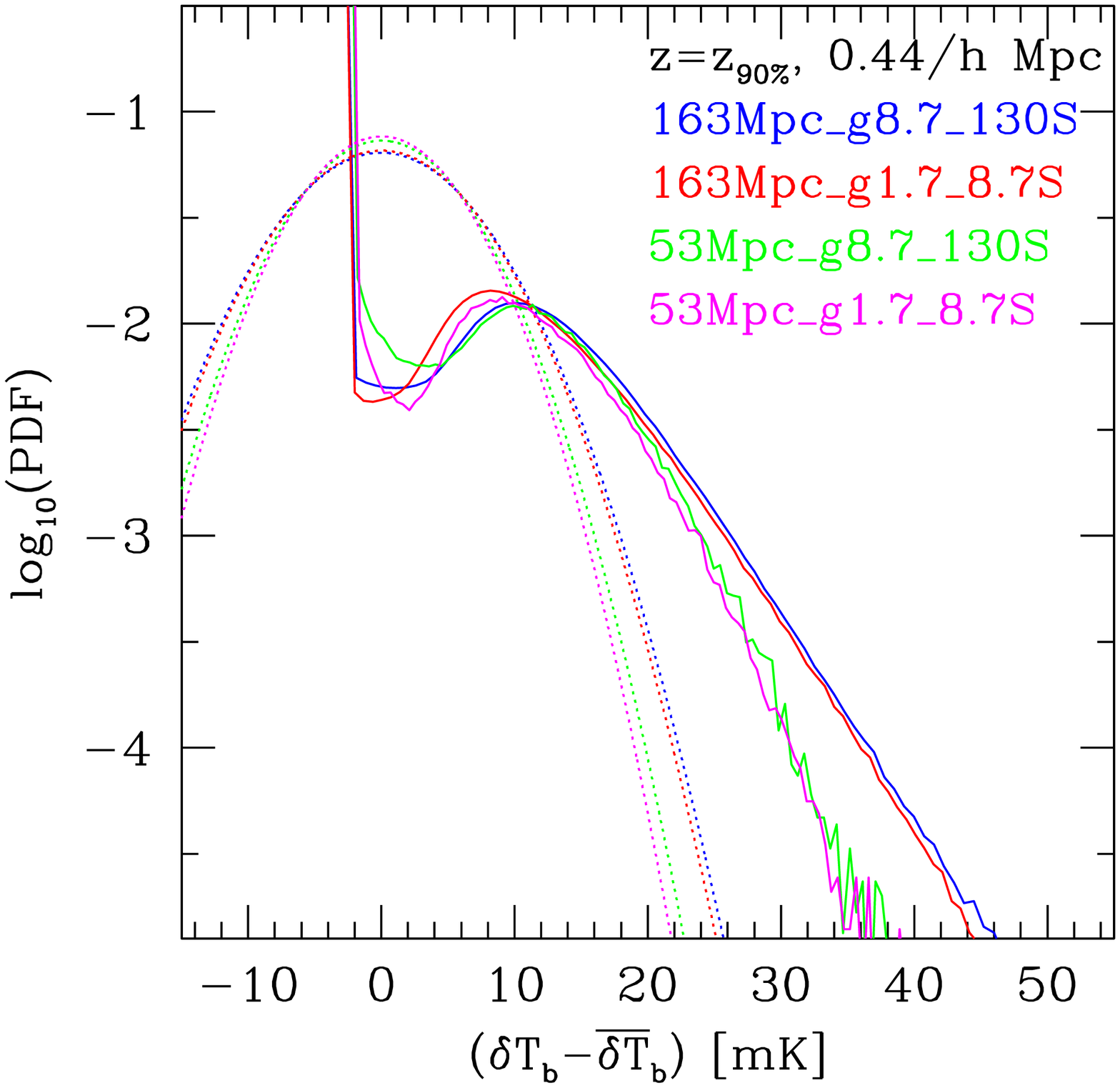} 
%\vspace{-1in}
\caption{The effect of the source efficiencies (high- vs. low 
efficiency) and box size on the PDF distribution of the 21-cm 
signal. Shown are the epochs at which the ionized fractions 
are $x_m=0.1$ (left), $x_m=0.5$ (middle) and $x_m=0.9$ (right) 
for our fiducial self-regulated cases, L1 (blue, solid), L2 
(red, solid), S1 (green, solid), S2 (magenta, solid). The PDFs 
are cell-by-cell (i.e. no smoothing apart from the numerical 
grid resolution). Also indicated are the Gaussian distributions 
with the same mean values and standard deviations (dotted lines, 
corresponding colours).
\label{21cm_PDF_1cell_fig}
}
%\vspace{-0.5cm}
\end{center}
\end{figure*}

\begin{figure*}
\begin{center}  
\includegraphics[width=2.3in]{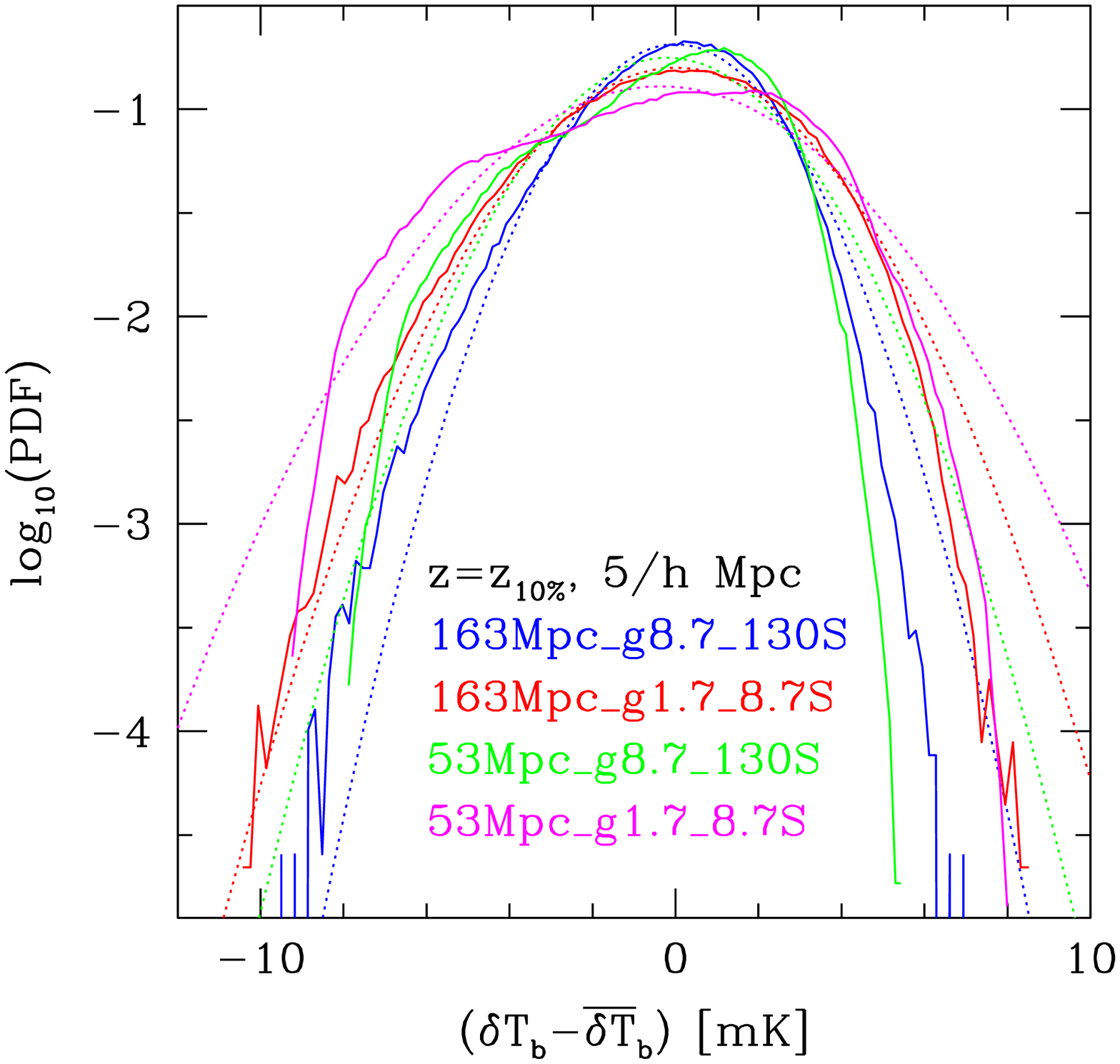} 
\includegraphics[width=2.3in]{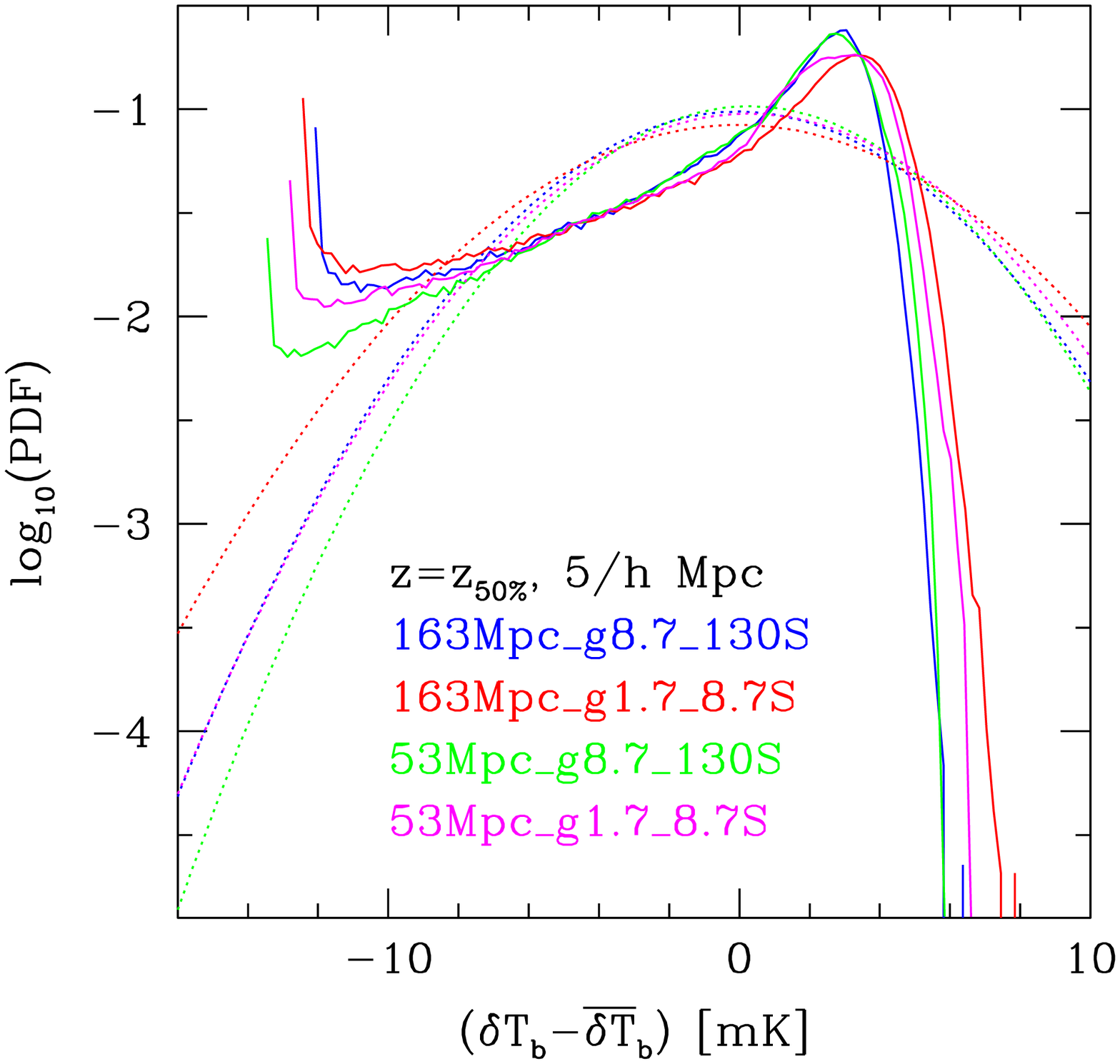}
\includegraphics[width=2.3in]{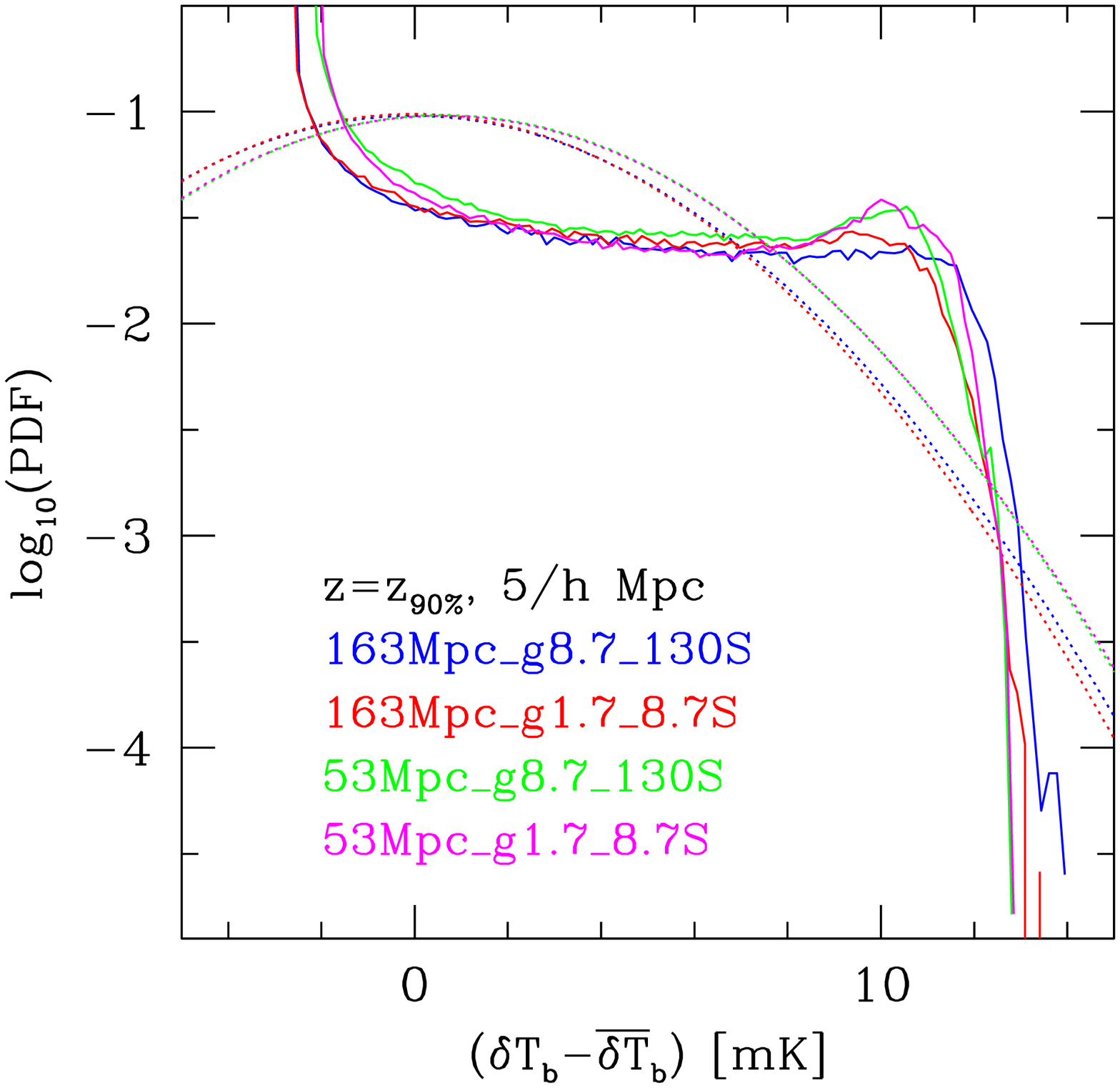} 
%\vspace{-1in}
\caption{Same as in Fig.~\ref{21cm_PDF_1cell_fig}, but for
boxcar smoothing of $5\,h^{−1}$~Mpc. 
\label{21cm_PDF_5Mpc_fig}
}
%\vspace{-0.5cm}
\end{center}
\end{figure*}

\begin{figure*}
\begin{center}  
\includegraphics[width=2.3in]{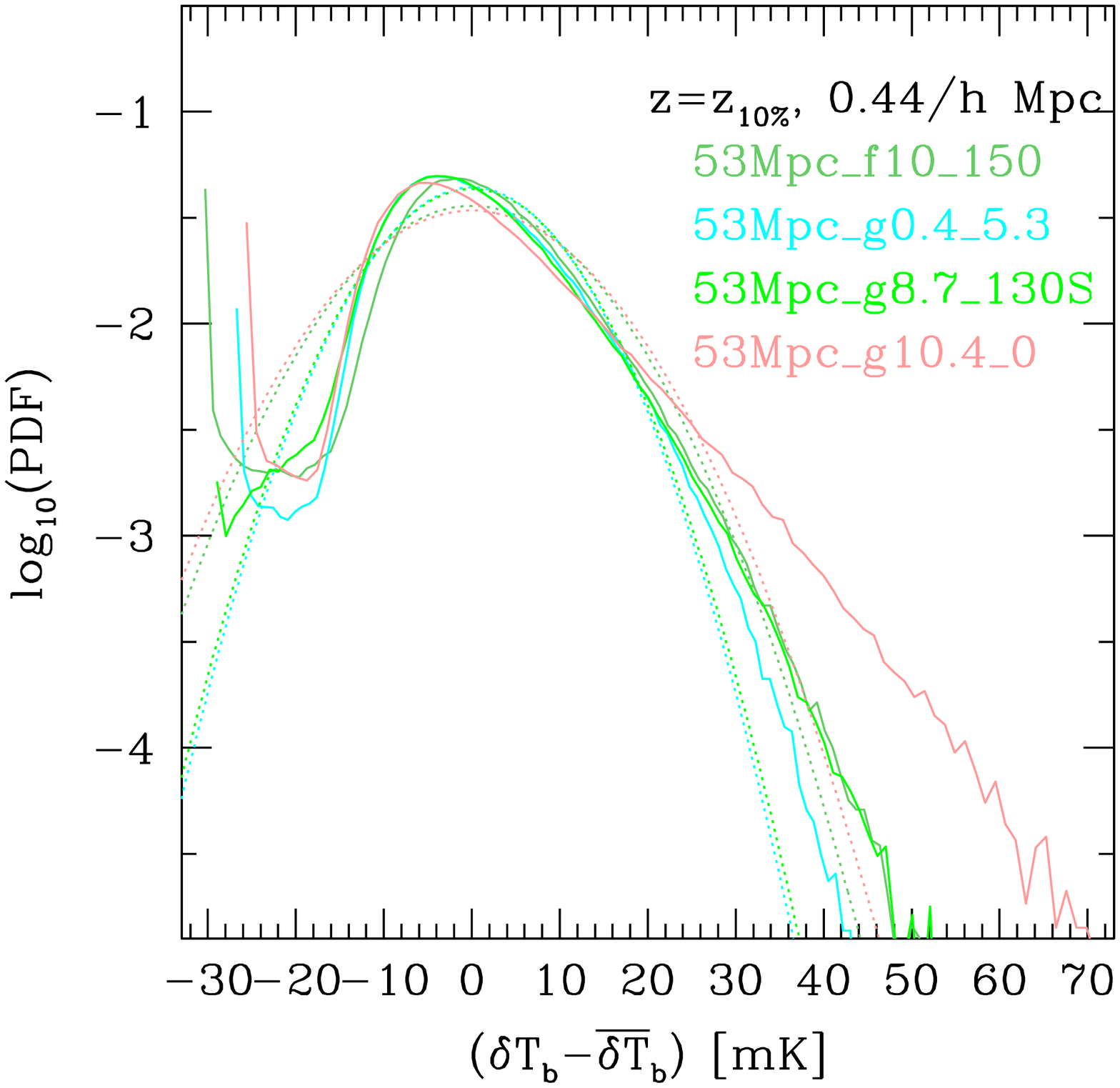} 
\includegraphics[width=2.3in]{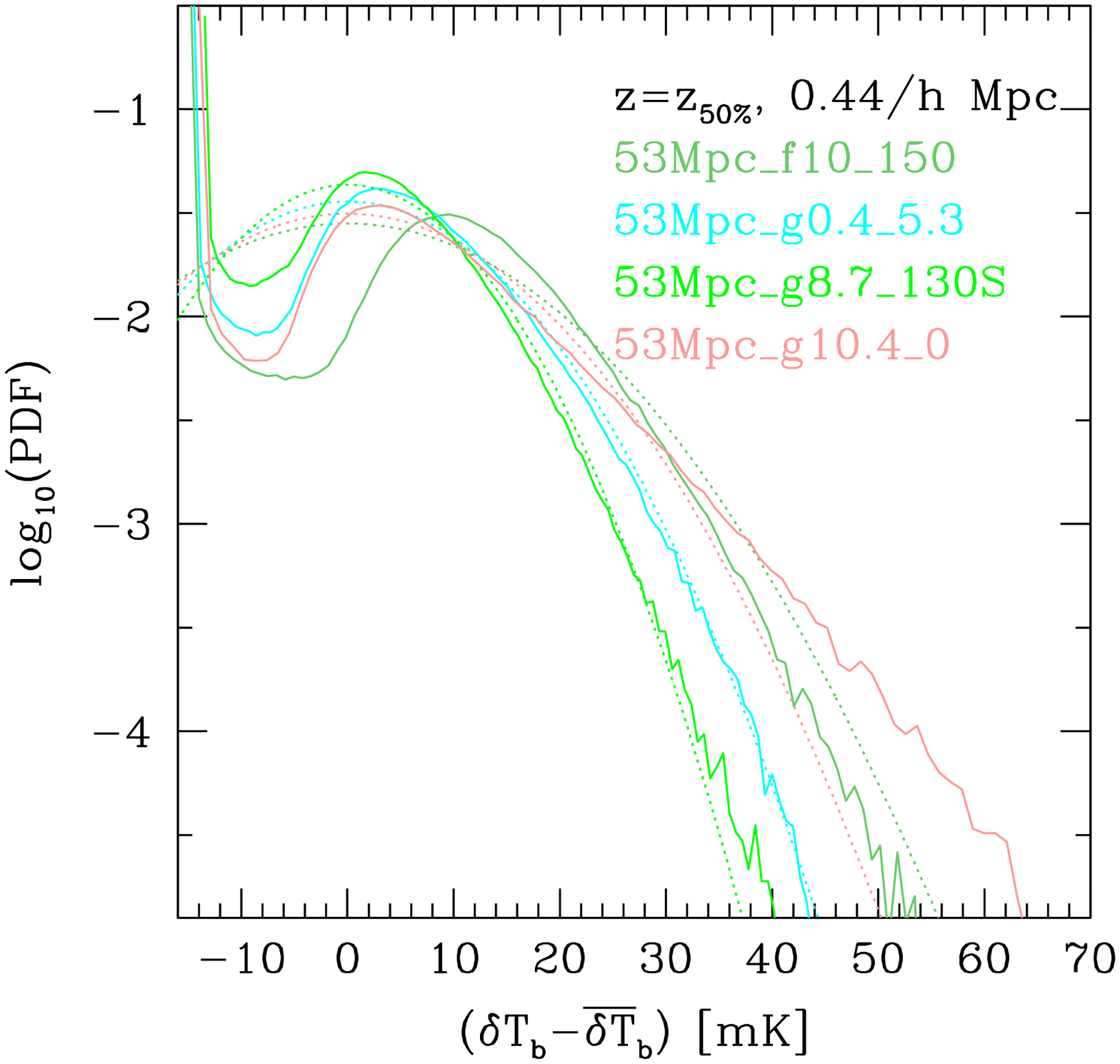}
\includegraphics[width=2.3in]{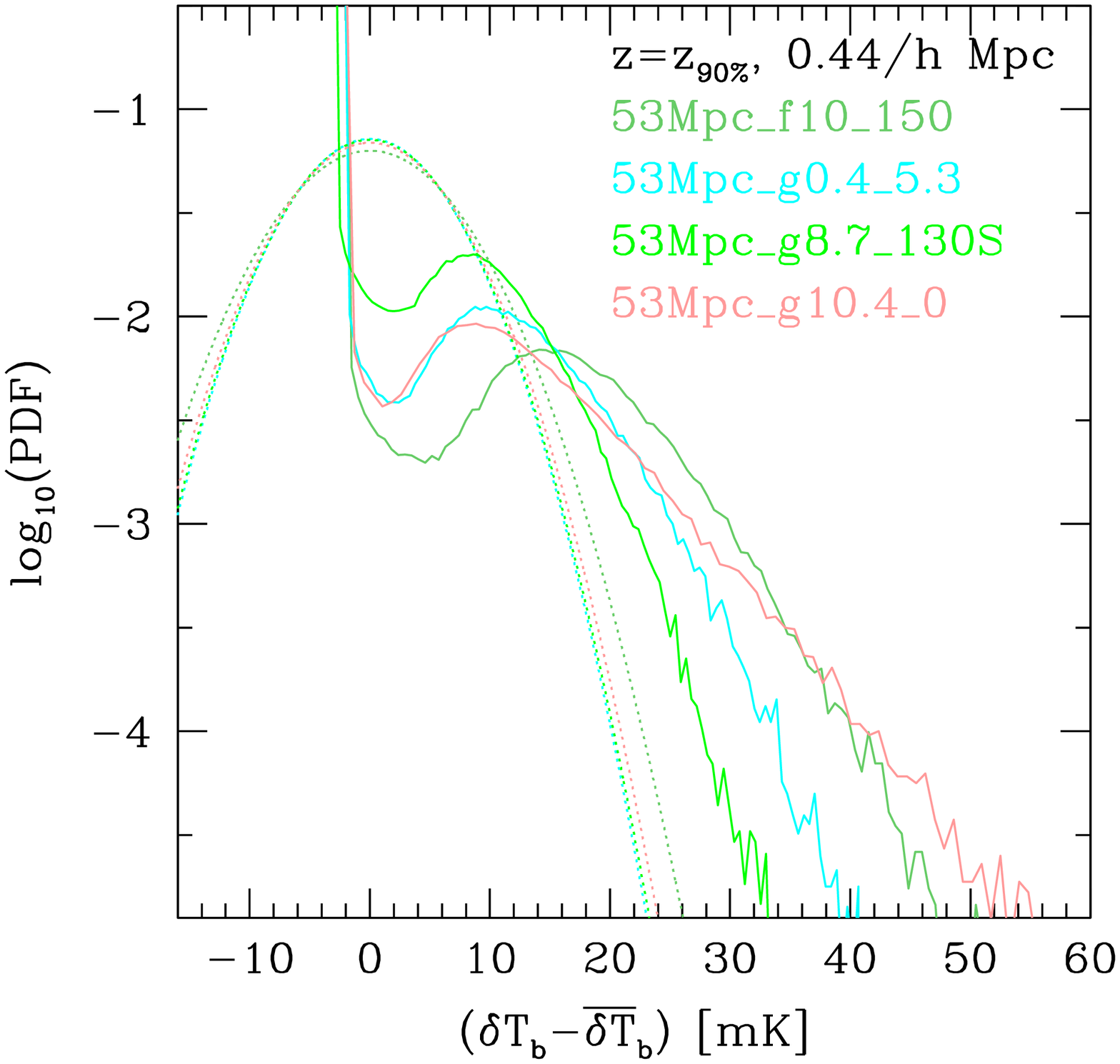} 
%\vspace{-1in}
\caption{The effect of varying source models on the PDF distribution 
of the 21-cm signal. Shown are the epochs at which the ionized 
fractions are $x_m=0.1$ (left), $x_m=0.5$ (middle) and $x_m=0.9$ 
(right) for our fiducial self-regulated case, S1 (green, solid), 
S4 (cyan, solid), and S5 (light red, solid). The PDFs are cell-by-cell
(i.e. no smoothing apart from the numerical grid resolution). Also 
indicated are the Gaussian distributions with the same mean values 
and standard deviations (dotted lines, corresponding colours). 
\label{21cm_PDF_uv_1cell_fig}
}
%\vspace{-0.5cm}
\end{center}
\end{figure*}

The 21-cm cell-by-cell PDFs with and without the presence of 
low-mass, suppressible sources at three representative stages 
of reionization ($x_m=0.1, 0.5$ and 0.9) are shown in 
Figure~\ref{21cm_PDF_1cell_fig}. Early on ($x_m=0.1$) the 
distributions are mostly following the underlying density 
field, and as a consequence are mostly Gaussian. There is a 
non-Gaussian tail for high differential brightness temperatures 
due to density nonlinearities. Reionization itself introduces 
some non-Gaussianity at low $\delta T_b$ as the first H~II 
regions form around the highest density peaks, moving the
corresponding cells into the extreme left of the distributions 
(i.e. holes in the neutral gas distribution, with 
$\delta T_b\sim0$). This slightly skews the distribution 
towards below-average (i.e. negative in $\delta T_b-\bar{\delta T_b}$) 
temperature values since the low-density regions remain more 
neutral, on average. Self-regulation mitigates those trends 
somewhat, as the low-mass sources are less clustered and more 
uniformly distributed throughout the volume, rather than being 
only at the highest density peaks (which are strongly clustered, 
as a consequence of the Gaussian density field statistics, see 
Figure~\ref{halo_bias_fig}). For the same reasons the PDF with 
self-regulation is also slightly less wide than without, as 
evidenced by the Gaussian distributions with the same mean and 
width as the actual PDFs. During the later stages of reionization 
the distribution becomes ever more non-Gaussian, with the most 
prominent feature due to the ionized regions 
($\delta T_b-\bar{\delta T_b}<0$). The remaining neutral regions 
are a mixture of voids (low, but positive 
$\delta T_b-\bar{\delta T_b}$) and a few remaining higher-density 
regions (e.g. filaments) which form the 21-cm bright non-Gaussian 
tail. Once again, both features are much reduced when low-mass 
source suppression is taken into account due to the weaker 
clustering of such sources. The width of the PDFs decreases over 
time, and does so faster when the low mass sources are not present. 
At late times the two distributions have almost the same means and 
widths, although the actual distributions remain significantly
different.

\begin{figure*}
\begin{center}  
\includegraphics[width=2.3in]{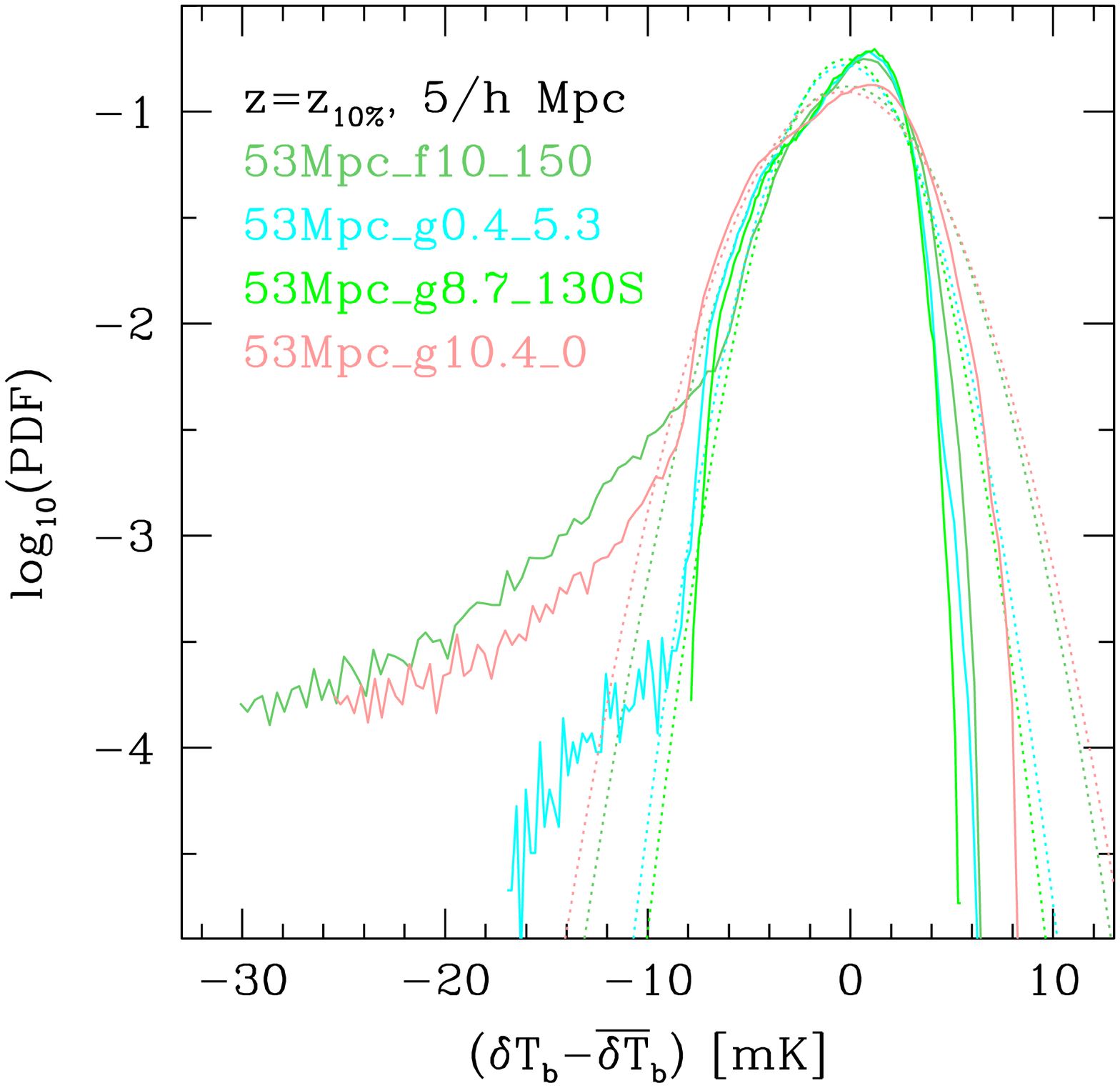} 
\includegraphics[width=2.3in]{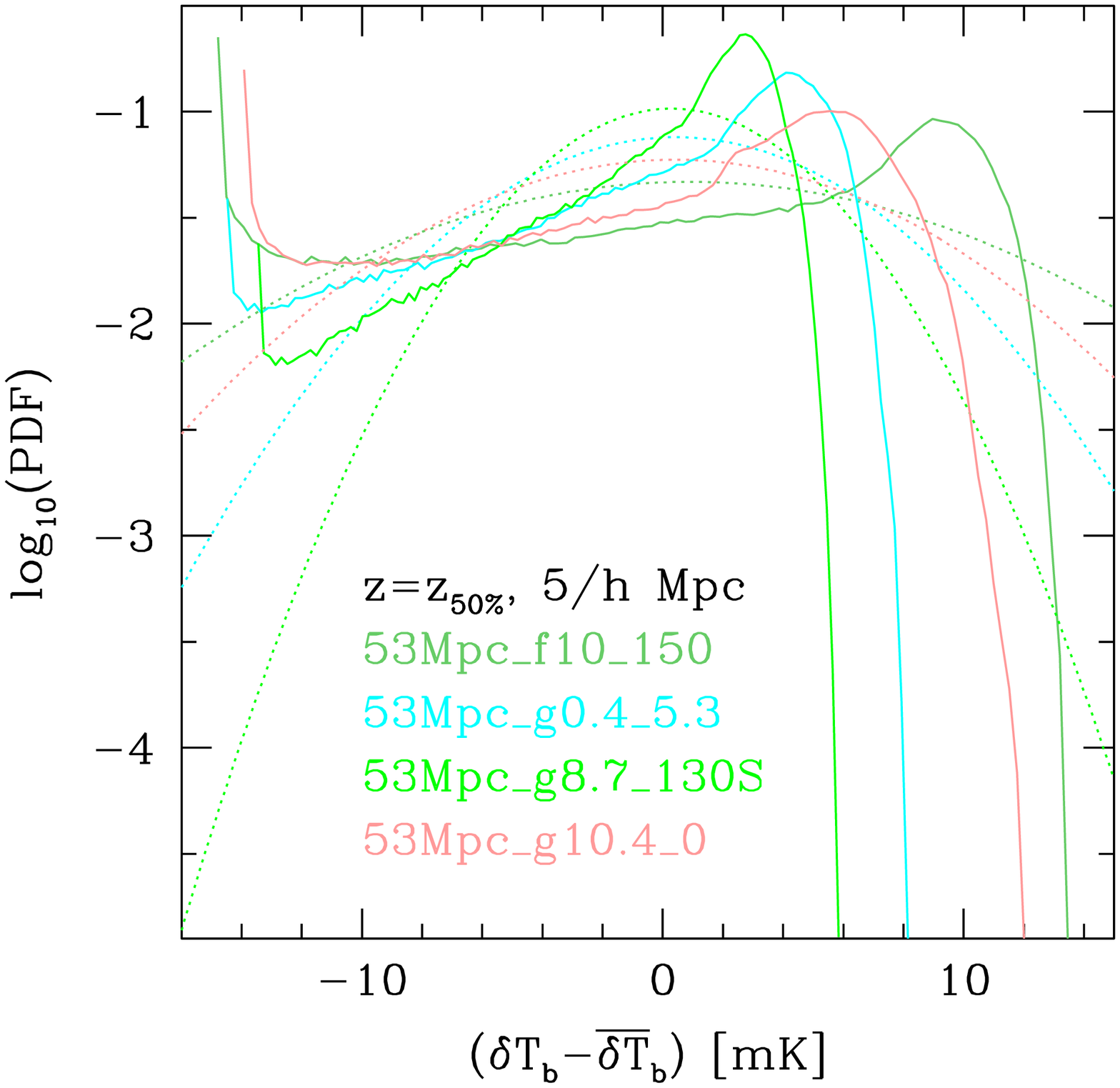}
\includegraphics[width=2.3in]{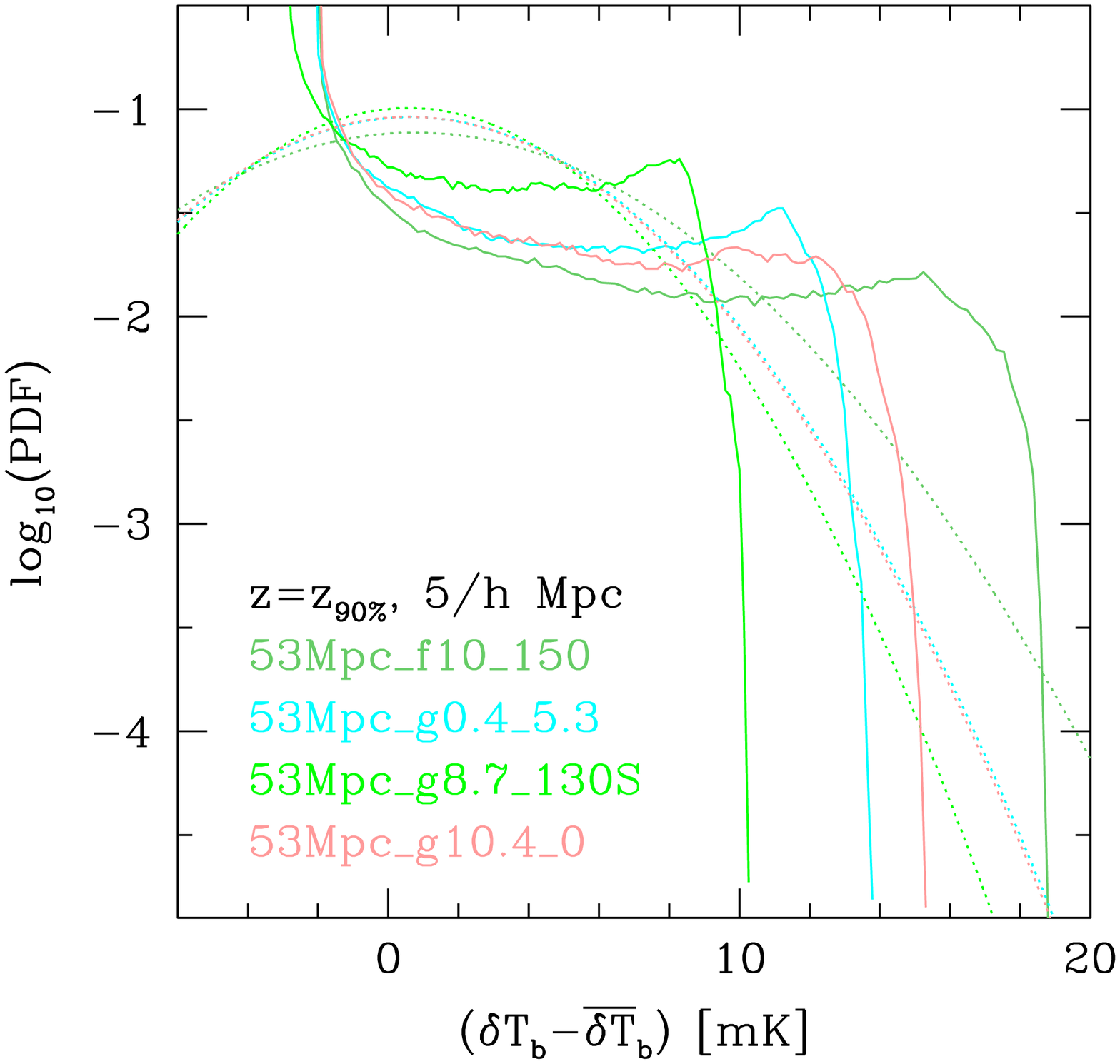} 
%\vspace{-1in}
\caption{Same as in Fig.~\ref{21cm_PDF_uv_1cell_fig}, but for
boxcar smoothing of $5\,h^{−1}$~Mpc. 
\label{21cm_PDF_uv_5Mpc_fig}
}
%\vspace{-0.5cm}
\end{center}
\end{figure*}

\begin{figure*}
\begin{center}  
\includegraphics[width=2.3in]{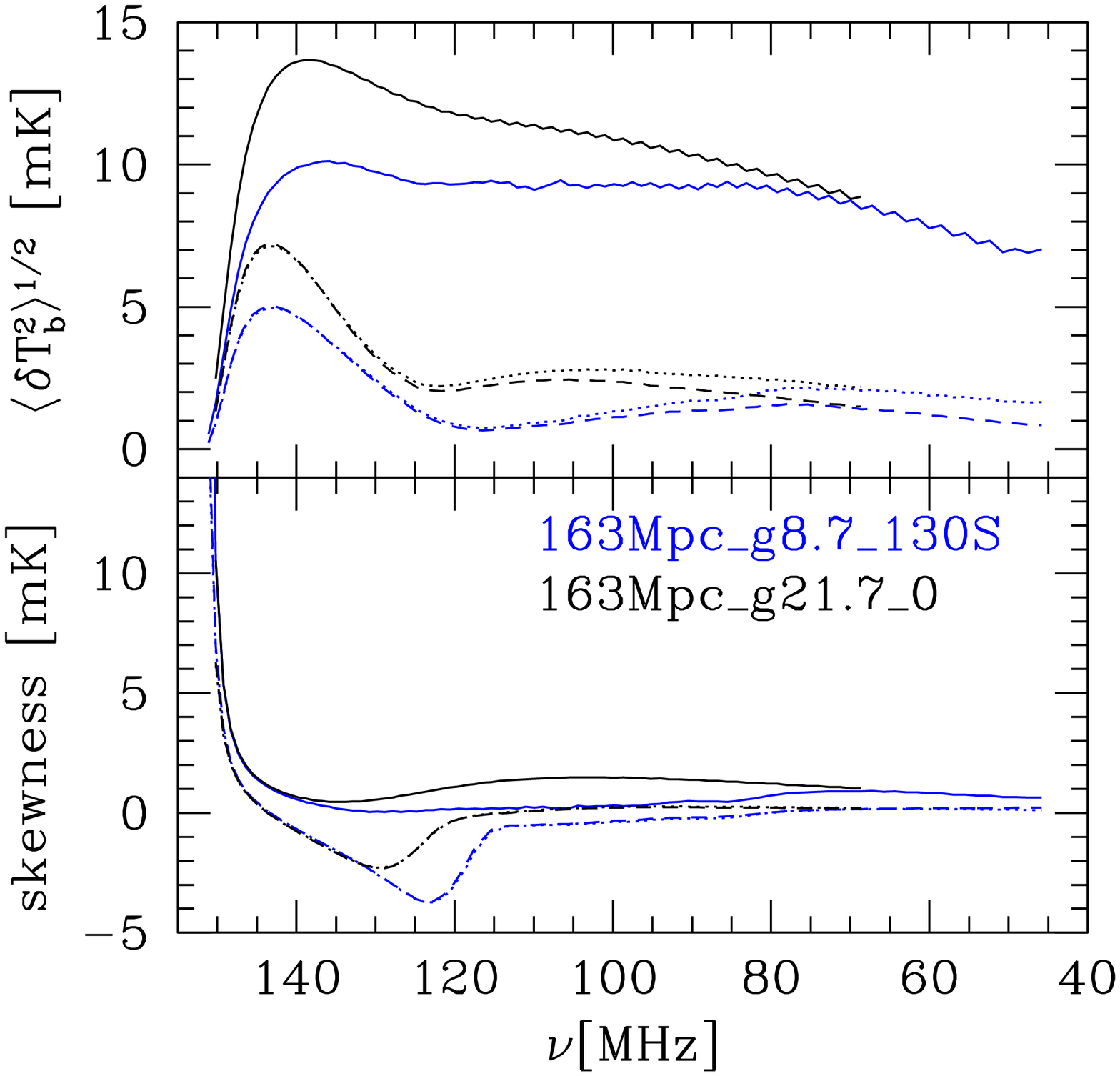} 
\includegraphics[width=2.3in]{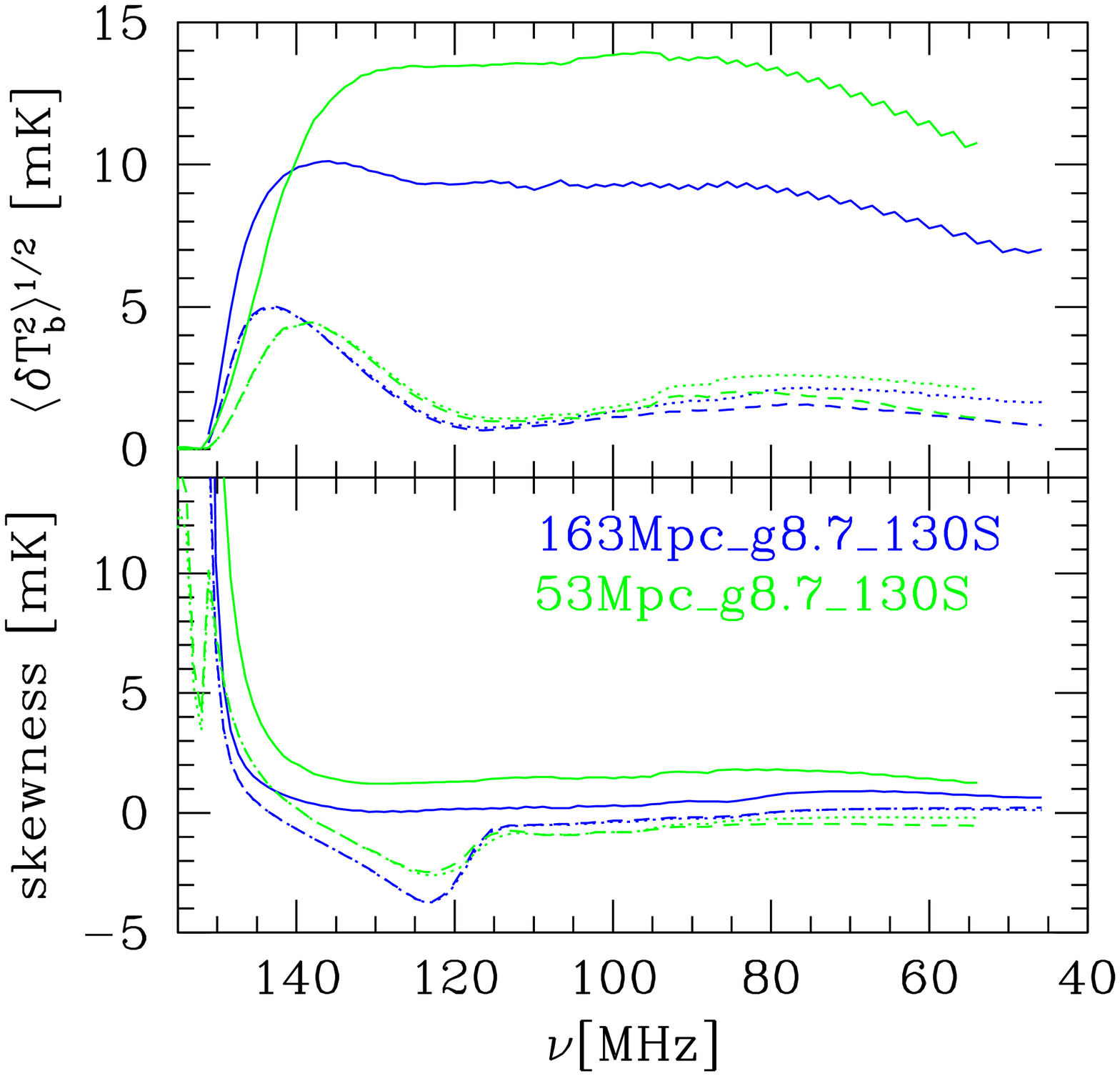} 
\includegraphics[width=2.3in]{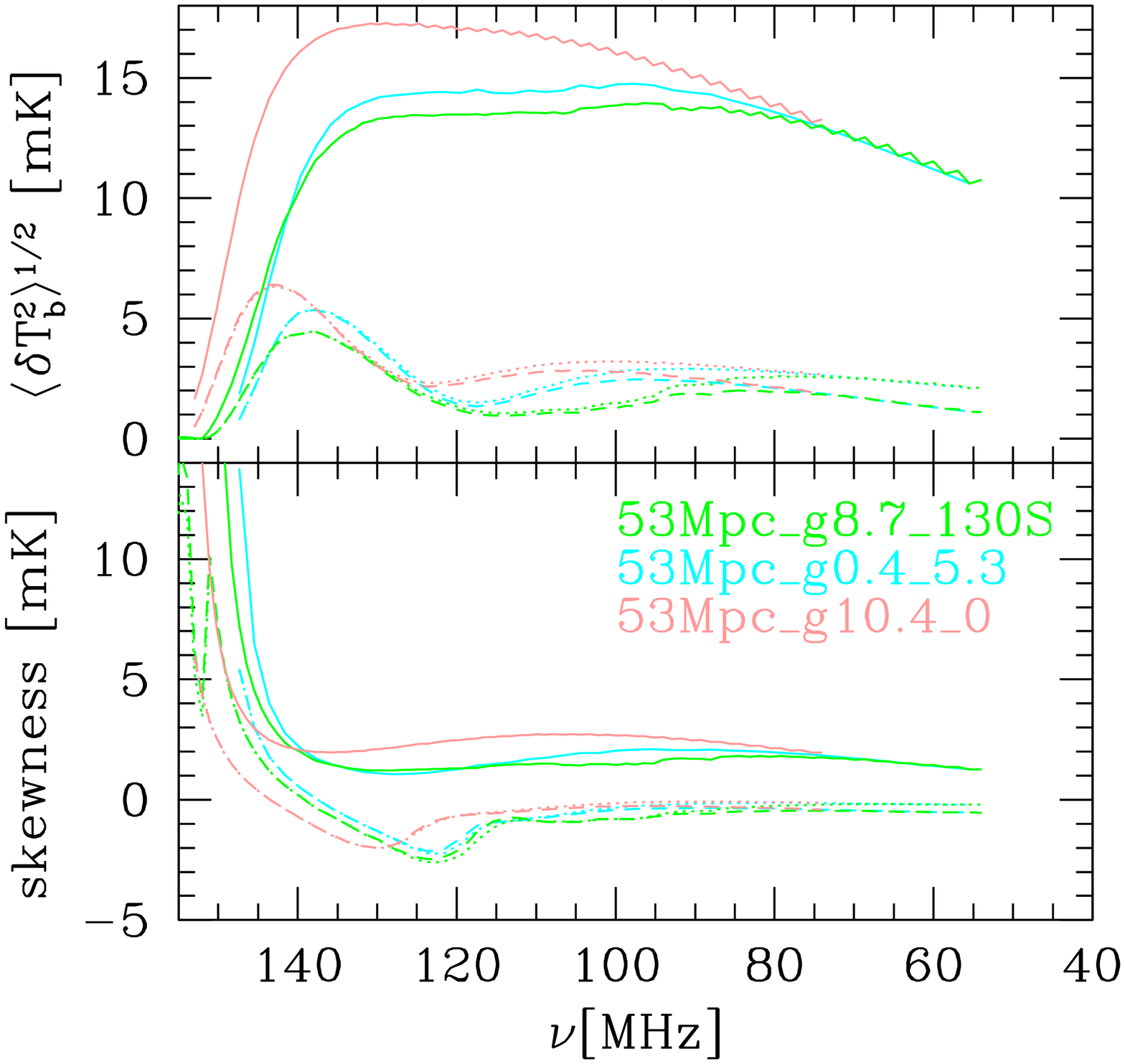} 
%\vspace{-1in}
\caption{
\label{21cm_skew_fig}
(top) The evolution of the rms of the 21-cm fluctuations.
and (bottom) evolution of the skewness of the 21-cm PDFs 
for (left panels) simulations L1 (blue), and L3 (black); 
(middle panels) simulations L1 (blue) and S1 (green); 
and (right panels) simulations S1 (green), S4 (cyan) and 
S5 (light red). Shown are the results at full simulation 
resolution (solid lines), smoothed with 3' Gaussian beam 
and 440 kHz bandwidth (dotted lines) and smoothed with a
Gaussian beam corresponding to a 2.5km maximum baseline 
and 440 kHz bandwidth (dashed lines).}
%\vspace{-0.5cm}
\end{center}
\end{figure*}

When the same PDFs are smoothed with a 5~Mpc$h^{-1}$ window 
(roughly similar in size to e.g. the LOFAR beam, albeit here 
we use a different window shape for simplicity of the 
calculations) the results become notably different 
(Figure~\ref{21cm_PDF_5Mpc_fig}). The smoothed PDFs at early 
times ($x_m=0.1$) become significantly more Gaussian, although 
some residual non-Gaussian tails remain at both high and low 
$\delta T_b$. As could be expected, the smoothed distributions 
also become much less wide compared to the unsmoothed ones, 
since the smoothing window averages the values, flattening the 
highest peaks and deepest valleys of the distributions. 

Interestingly, at the middle and late stages of reionization 
($x_m=0.5$, and 0.9) the opposite happens, namely that the 
smoothed PDF distributions become less Gaussian for any 
$\delta T_b$ value. The PDF distributions with and without 
self-regulation have similar shapes, but the presence of 
low-mass sources makes the distribution much less wide. For 
$x_m=0.5$ the very brightest peaks are fewer than a Gaussian 
would predict, but there are many more intermediate-brightness 
($5\, {\rm mK}<\delta T_b-\bar{\delta T_b}<12\,$mK with no 
self-regulation, $\delta T_b-\bar{\delta T_b}<5\,$mK with)
ones. At the late stages of reionization ($x_m=0.9$) both 
cases show many more bright peaks 
($10\, {\rm mK}<\delta T_b-\bar{\delta T_b}$) than a Gaussian 
would predict, although the self-regulated case yields fewer 
very bright peaks ($15\, {\rm mK}<\delta T_b-\bar{\delta T_b}$) 
than either the corresponding Gaussian or the non-self-regulated 
case. Finally, regardless of the above differences in the PDFs, 
their equivalent widths are very similar for the two simulations.

The PDFs for our fiducial self-regulated high- and low-efficiency 
cases L1, L2, S1 and S2 are shown in Figures~\ref{21cm_PDF_1cell_fig} 
and \ref{21cm_PDF_5Mpc_fig}. Unlike the presence and self-regulation 
of low-mass sources presented above, which influenced the PDFs 
significantly, neither the source efficiencies nor the box size 
have any dramatic effect of the resulting PDFs. The smaller boxes 
do not capture well the bright wing of the distribution because 
the highest density peaks are rare and the volume in these cases 
is too small to capture them. The effect of varying source 
efficiency manifests itself by yielding more bright peaks during 
the early stages of reionization and somewhat brighter peaks at
its middle stages.

Finally, the results with a varying source models are shown in 
Figures~\ref{21cm_PDF_uv_1cell_fig} and \ref{21cm_PDF_uv_5Mpc_fig}.
Here for clarity we just show a representative sub-sample of our 
full simulation suite. Upon inspection several general trends 
become clear. If only high-mass sources are present (model S5) 
the distributions are noticeably wider, with a long non-Gaussian 
tail at high differential brightness temperatures 
($\delta T_b-\bar{\delta T_b}>30-40$~mK) than in the fiducial 
self-regulation case, S1. Conversely, there are many fewer regions 
with low, but positive (i.e. still mostly neutral) differential 
brightness temperatures ($\delta T_b-\bar{\delta T_b}<15$~mK).  
The reason for this is that the massive sources form only at 
the highest density peaks, leaving neutral many other density 
peaks which have not yet collapsed. The high-density, neutral 
gas in those peaks is reflected in the non-Gaussian tail at 
high differential brightness temperatures. Lastly, model S4 
(low efficiency sources and no suppression) yields intermediate 
PDF between the fiducial run and the high-mass sources only 
runs. The PDF's are therefore mostly dependent on which 
population of sources is active (high or low mass), but are 
not very sensitive to the details of the reionization history 
(models S5 and S8, not shown here, which have the same source 
population active, but with different efficiencies over time 
and thus different reionization history yield very similar 
distributions). These trends are independent of the smoothing 
employed, as can be seen in Figure~\ref{21cm_PDF_uv_5Mpc_fig}, 
although naturally the range of differential brightness 
temperatures is much reduced by the smoothing. The only new 
feature found in the smoothed data is the non-Gaussian tails 
for negative $\delta T_b-\bar{\delta T_b}$  at early times 
($x_m=0.1$). These are result of the H~II regions 
($\delta T_b-\bar{\delta T_b}<0$) in the non-self-regulated 
cases growing relatively large quickly. Consequently, even at 
these early times their sizes become comparable to the smoothing 
window size, which results in the non-Gaussian tails. These were 
not present in the cell-wise PDF distribution, as the individual 
cells tend to be either fully ionized 
($\delta T_b-\bar{\delta T_b}<-25$~mK) or mostly neutral.

The level of non-Gaussianity of the PDF distributions can, to a 
first order, be characterized by their skewness, which in turn
can be used to distinguish and extract the reionization signals
\citep{2009MNRAS.393.1449H}. In Fig.~\ref{21cm_skew_fig} we show
the evolution of the skewness vs. frequency for selected models.
We show the skewness for the cell-wise PDF, as well as smoothed 
with 3' Gaussian beam and 440 kHz bandwidth and smoothed with 
Gaussian beam corresponding to a 2.5km maximum baseline and 440 
kHz bandwidth (bottom panels). Both sets of beam and bandwidth 
smoothing are roughly as expected for the LOFAR array. We also 
plotted the rms of the correspondingly-smoothed 21-cm differential 
brightness temperature fluctuations (top panels). We note that 
because of the different beam- and bandwidth smoothing employed
here the rms values are slightly different from the ones shown 
in Figs.~\ref{21cm_selfreg_fig}, \ref{21cm_fluct_fig} and 
\ref{21cm_fluct_uv_fig}. 

In all cases, regardless of the specific reionization scenario
the skewness evolution for the unsmoothed (1-cell) PDFs follows 
similar pattern. It says at a roughly constant, positive value 
throughout the evolution, until it shoots up just before overlap,
very similar to the behaviour observed in \citet{2009MNRAS.393.1449H}.
The the beam- and bandwidth-smoothing of the PDFs introduces a
significant feature in the skewness, whereby it becomes negative 
during the intermediate stages of the evolution. Interestingly, 
this dip of the skewness to negative values closely corresponds 
to the rise and peak of the differential brightness temperature
rms fluctuations, preceding it slightly in time. This feature 
is universal, observed for every reionization scenario and source 
model we consider here and suggests an interesting approach for
a detection and/or independent confirmation of the rise and peak 
of the 21-cm rms fluctuations during cosmic reionization.

On the other hand, the skewness of the 21-cm PDF distributions 
proves to be fairly insensitive to the source model or reionization
scenario, resulting in only slight changes in the values. The 
skewness of the smoothed PDFs also proves largely independent of
the box size and resolution and of the details of the interferometer
beam assumptions (i.e. if it is fixed in angular size or evolves with 
frequency). This suggests that while the feature in the evolution of 
the skewness is a good indicator of the rise in patchiness, it most
likely cannot be used for constraining the properties of the ionizing 
sources. 

\section{Summary}
\label{summary:sect}

In this work we have used a large set of cosmological structure
formation and reionization simulations in an attempt to gain 
insight into what can be learned about the properties of the 
reionization sources based on their observational signatures. 
In particular, we are interested in determining what 
observations could be used to discriminate between certain 
source models, thereby restricting the available parameter 
space. Here we primarily focused on the redshifted 21-cm 
signatures, as these can in principle probe the full 
reionization history and offer a wide range of different 
probes, from the mean history, through fluctuations measures 
like rms evolution and power spectra, to PDFs and higher-order 
statistics which can detect non-Gaussian features. 

The observable features of the epoch of reionization derive 
from the gradual mean transition of the IGM from neutral to 
highly ionized state, as well as from the patchiness of that 
transition. The mean transition depends largely on the overall 
number of ionizing photons emitted by the source per unit time, 
with some correction due to recombinations which is 
position-dependent due to density spatial variations. The 
patchiness, on the other hand depends, in a complicated way on the 
abundances, clustering and efficiencies of the ionizing sources.

Our structure formation simulations confirm previous results by
us and other groups that the high-redshift halo mass functions 
are inconsistent with either of the widely-used Press-Schechter 
and Sheth-Tormen analytical fits. In particular, the abundance 
of rare halos is strongly underestimated by PS, but over-estimated 
by ST. We find that the nonlinear halo bias is extremely high and 
very scale-dependent. Linear bias regime is only reached at very 
large scales, $k\gtrsim0.1$. For the rarest halos (3-$\sigma$ and 
above) linear bias regime is never reached even within our largest, 
163~Mpc volume. Therefore, proper account for the nonlinear bias 
of halos is important and any calculations assuming linear bias 
are underestimating the halo clustering significantly.

The Jeans mass filtering of low-mass halos in ionized regions and 
the related self-regulation of reionization results in significantly 
more extended reionization history and higher integrated electron 
scattering optical depth (by $\Delta\tau_{\rm es}\sim0.01$) compared
to the high-mass source-only scenario with the same overlap redshift, 
albeit both optical depths are still within the current constraints 
from WMAP. Even more significant are the changes in the reionization 
geometry, resulting in corresponding differences in the reionization 
observables. The 21-cm fluctuations are lower at all scales and their 
PDF distributions are somewhat more Gaussian, although significant 
non-Gaussianity remains. 
 
In all our simulations reionization occurs inside-out, with the 
high-density regions being reionized on average earlier than the
mean and low-density ones. This inside-out nature of reionization 
results in the mass-weighted IGM photoionization rates being 
considerably larger, by factor of a few, than volume-weighted ones. 
This should always be taken into account as it can easily skew 
observational measurements depending on the mean density of the 
regions being probed.

The skewness of the 21-cm PDF distribution smoothed over LOFAR-like 
window shows a clear feature correlated with the rise of the rms due 
to patchiness. This feature does not exist in the unsmoothed data,
indicating that it is related to the non-Gaussianity of the 
large-scale patchy distribution of 21cm emission. The feature exists 
for any reionization scenario and ionizing source properties and 
thereby provides a different approach for detecting the rise of 
large-scale patchiness and an independent check on other detections.

The peak position of the 21-cm rms fluctuations depends 
significantly on the beam and bandwidth smoothing size as 
well as on the reionization scenario. As a consequence, it 
does not always occur at 50\% ionization fraction as sometimes 
is claimed, but instead can happen for ionized fractions as 
low as 30\% and as high as 70\%. 

The simulation volume has only a modest effect on the results
as long as the typical size of the ionized patches is smaller
than the volume. However, the fluctuations at large scales 
(above approximately a fifth of the boxsize) are severely 
affected. This is especially important at late times, when the 
ionized patches grow very large. Therefore, at least $\sim100/h$~Mpc 
boxes are required to model the fluctuations during the late 
stages of reionization.

The ionizing source efficiencies and their correlation properties 
introduce clear signatures in the reionization observables. As a 
direct consequence of that, one cannot model low-mass, unresolved 
sources by simply assigning their emissivity to the resolved 
higher-mass sources as the latter have abundance which is highly 
variable over time and different clustering properties from 
lower-mass sources which provide the bulk of the ionizing photons 
(see Appendix~\ref{appendixA}). 

When self-regulation is present there are only minor differences
between the 21-cm observational signatures resulting from high- 
and low-efficiency ionizing sources, apart from an overall shift 
of the reionization history. The corresponding PDF distributions 
are also very similar, which suggests that the source efficiencies
in such models can only be constrained by the overall timing of 
the mean reionization history.

Scenarios where low-mass sources are completely absent, e.g. 
somehow rendered sterile, are relatively easily distinguishable
from the ones where they are present (even if strongly suppressed).
On the other hand, our results suggest that numerous low-efficiency 
sources (case S4) can mimic the effects of suppression (S1). Such 
scenarios therefore might be difficult to distinguish solely based
on power spectra and similar measurements. However, they might still 
be discriminated through 21-cm PDFs as the no suppression case creates 
many more high-brightness peaks. Similarly, a high-mass source only
scenario (S5) gives quite similar results to the high-efficiency,
no suppression case (S3) at the same stages of reionization (albeit
these cases overlap at different times for the parameters we have 
chosen), although they do differ in terms of power at large scales 
and in number of bright peaks. 

The results presented in this work should not be considered 
to be an ultimate, realistic prediction of the reionization 
signals. While the assumptions about source efficiencies and 
suppression we made for our fiducial cases are reasonable 
based on our current knowledge and likely bracket the realistic
range, the uncertainties are still substantial. As more
observational data becomes available over time it can be used 
to restrict the parameter space further and help us refine our 
theoretical models, which in turn will provide a valuable tool 
for interpreting the meaning of the observational results in 
terms of early structure formation, source efficiencies, 
suppression mechanisms, etc. In this framework our current 
study, which evaluates in a controlled way the effects of a 
set of widely different assumptions about the sources of 
ionizing radiation is a very useful step towards a more 
complete understanding of early galaxy formation and feedback.

\section*{Acknowledgments} ITI was supported by The Southeast Physics
Network (SEPNet) and the Science and Technology Facilities Council
grants ST/F002858/1 and ST/I000976/1.  This study was supported in
part by Swiss National Science Foundation grant 200021-116696/1,
Swedish Research Council grant 2009-4088, NSF grants AST-0708176 and
AST-1009799, NASA grants NNX07AH09G, NNG04G177G and NNX11AE09G, and
Chandra grant SAO TM8-9009X. The authors acknowledge the TeraGrid and
the Texas Advanced Computing Center (TACC) at The University of Texas
at Austin ( URL: http://www.tacc.utexas.edu), as well as the Swedish
National Infrastructure for Computing (SNIC) resources at HPC2N
(Ume\aa, Sweden) for providing HPC and visualization resources that
have contributed to the research results reported within this paper
and Partnership for Advanced Computing in Europe (PRACE) grant
2010PA0442 to ITI. KA is supported in part by Basic Science Research
Program through the National Research Foundation of Korea (NRF) funded
by the Ministry of Education, Science and Technology (MEST;
2009-0068141,2009-0076868) and by KICOS through
K20702020016-07E0200-01610 funded by MOST.

\appendix

\section{Reionization by rare, massive, variable-luminosity sources}
\label{appendixA}

\begin{figure}
  \begin{center}
    \includegraphics[width=3in]{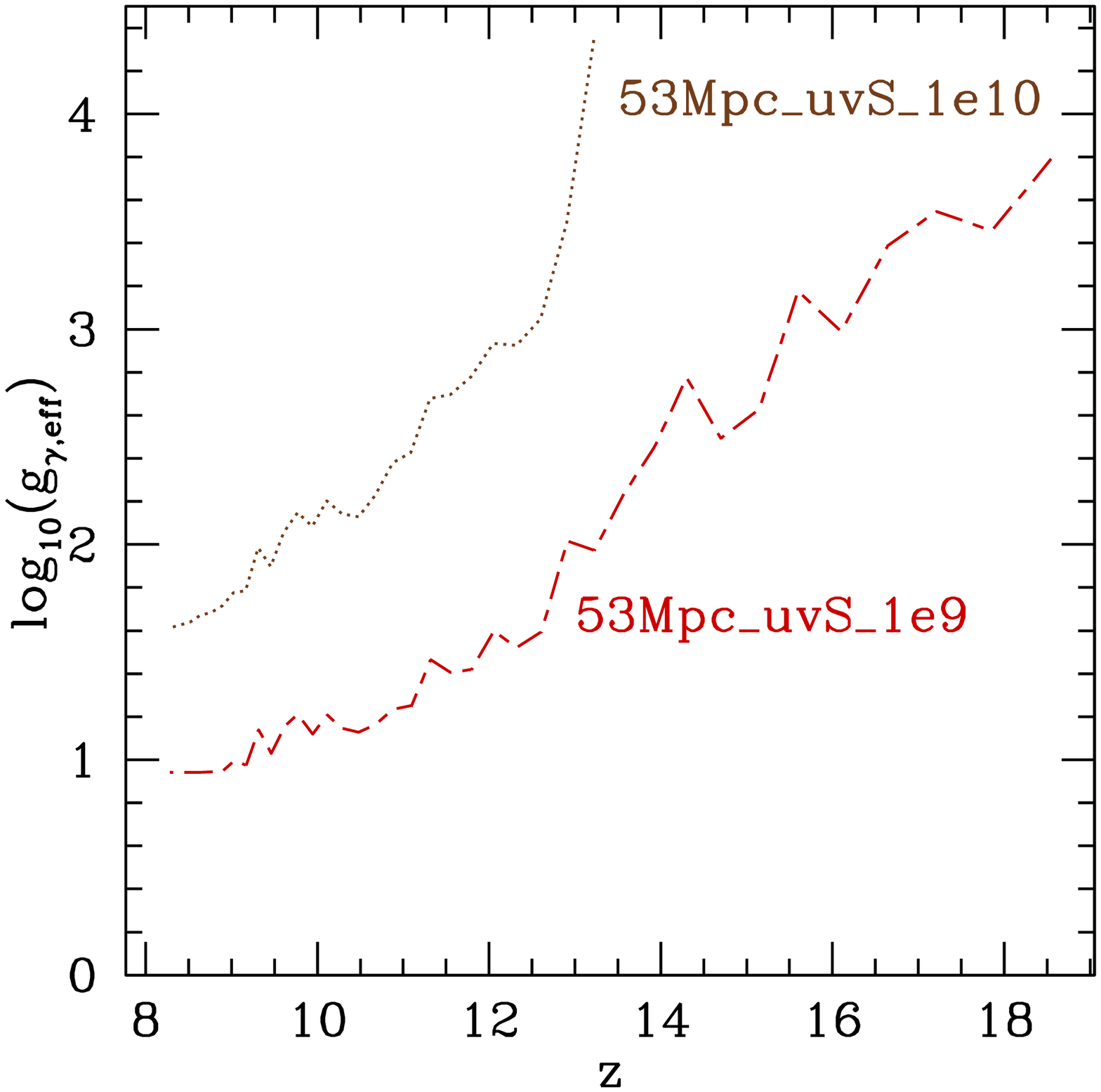}
  \end{center}
%  \vskip -0.5cm 
  \caption{Effective efficiency factors $g_\gamma$ vs. redshift for 
simulations S8 and S9, defined so as to ensure the same total number 
of ionizing photons emitted per atom as in our fiducial simulation, 
S1.\label{g_eff}}
\end{figure}

\begin{figure*}
  \begin{center}
   \includegraphics[width=2.2in]{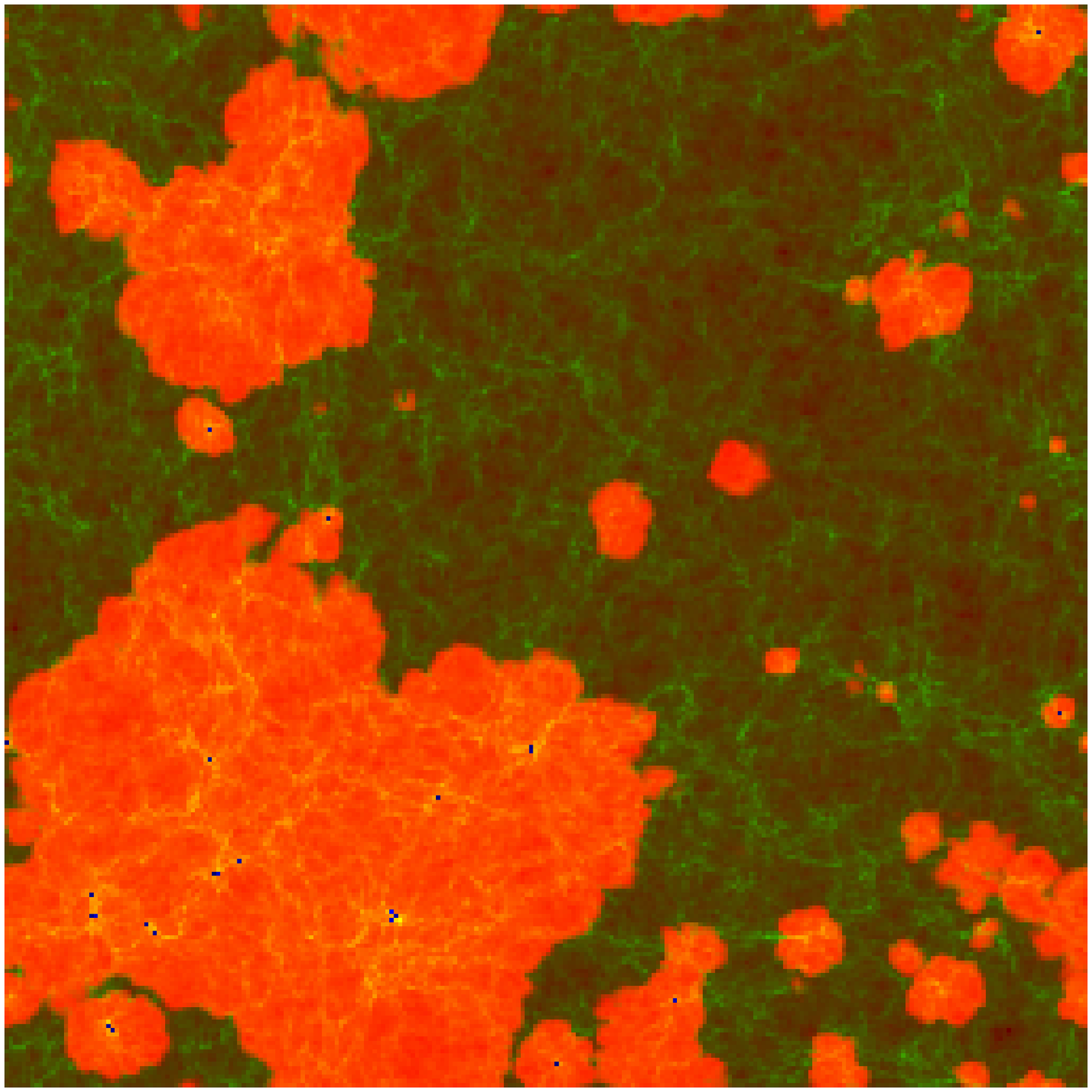}
   \includegraphics[width=2.2in]{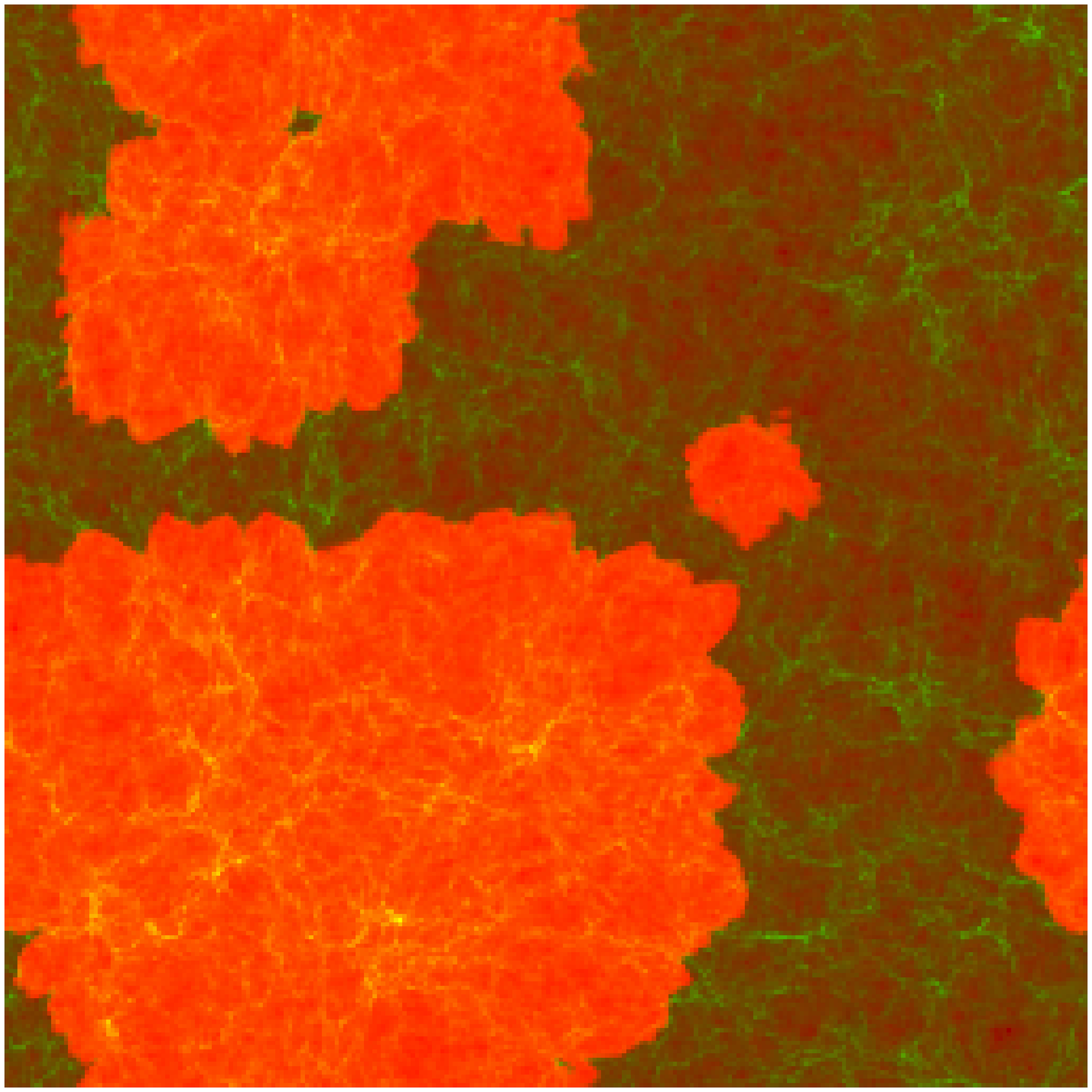}
    \vspace{-0.3cm}
  \end{center}
  \caption{Spatial slices of the ionized and neutral gas density 
    from our radiative transfer simulations with boxsize $53$~Mpc 
    at box-averaged ionized fraction by mass $x_m\sim0.50$. Shown 
    are the density field (green) overlayed with the ionized fraction 
    (red/orange/yellow) and the cells containing sources (dark/blue). 
    Shown are cases S8 and S9.
    \label{images_37Mpc_add}}
\end{figure*}

\begin{figure*}
\begin{center}  
\includegraphics[width=3.4in]{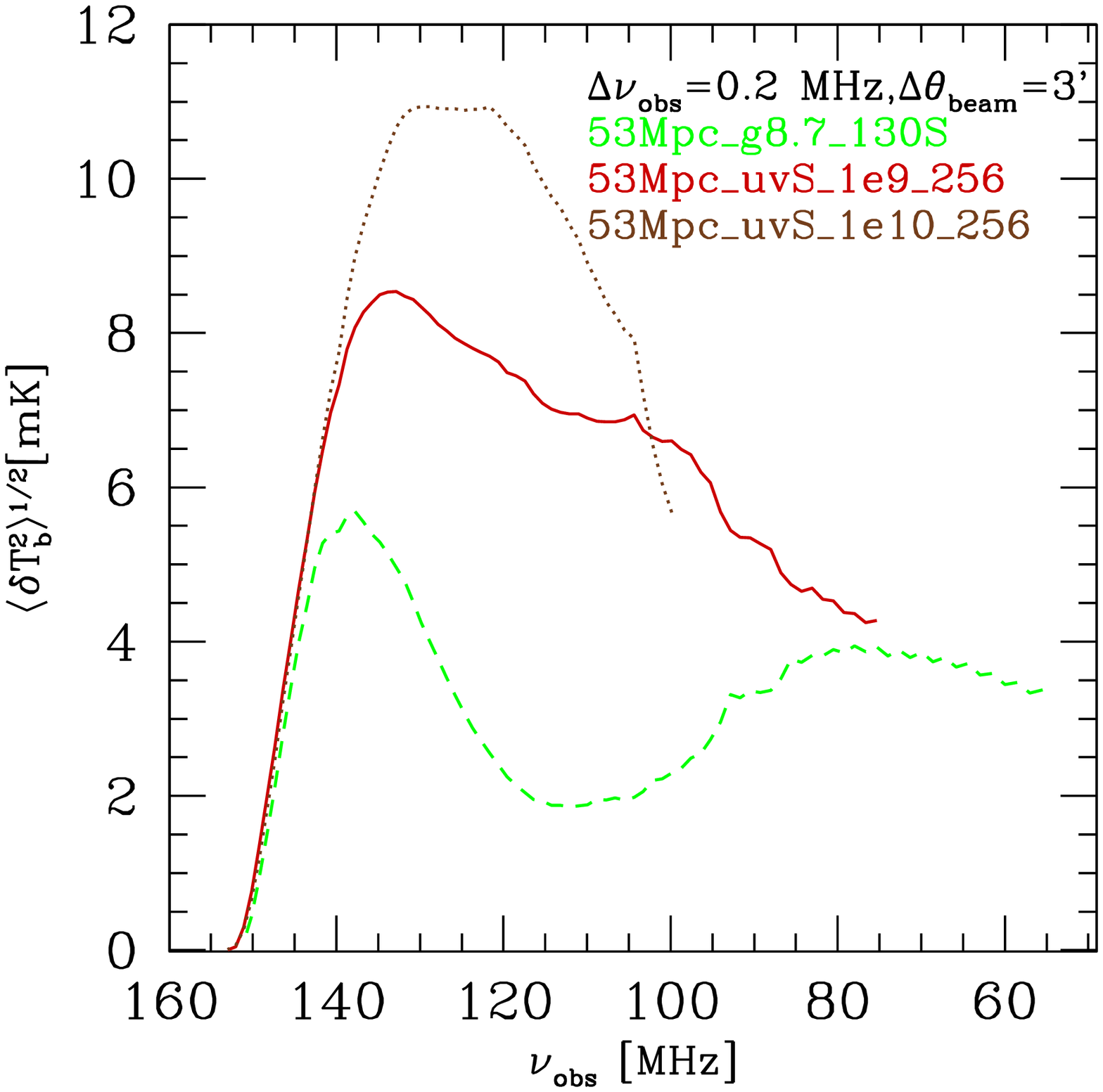} 
\includegraphics[width=3.3in]{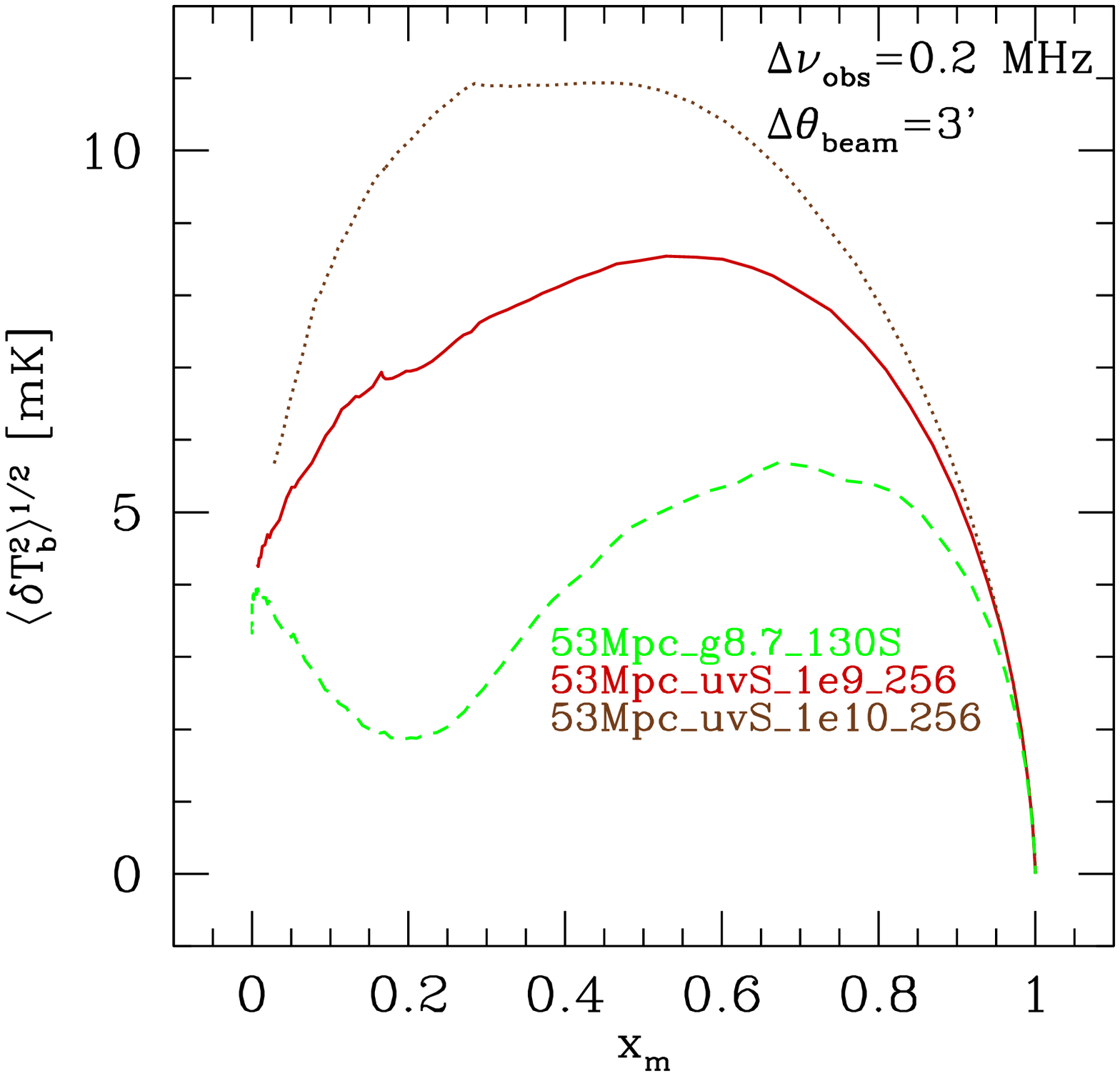}
%\vspace{-1in}
\caption{
\label{21cm_fluct_uvS_fig}
The evolution of the rms fluctuations of the 21-cm background, 
for beamsize $3'$ and bandwidth $0.2$~MHz and boxcar filter vs. 
frequency (left) and vs. mean mass-weighted ionized fraction 
(right). Shown are simulations S1 (green, short dashed), S8 
(dark red, long dash-short dash) and S9 (brown, dotted).}
%\vspace{-0.5cm}
\end{center}
\end{figure*}

Simulations 53Mpc\_uvS\_1e9 and 53Mpc\_uvS\_1e10 (S8 and S9 
in Table~\ref{summary_table}) are investigating the effects 
of keeping the global, volume-averaged emissivity of ionizing 
photons per unit time fixed at each redshift, while raising 
the minimum source mass, assumed to be $10^9M_\odot$ for S8 and 
$10^{10}M_\odot$ for S9. The overall number of photons emitted 
at each timestep are set to be exactly equal, at all times, to 
the one yielded by our fiducial case, S1. The resulting effective 
efficiencies $g_\gamma$ are shown in Figure~\ref{g_eff}. Simulation 
S8 has the same high-mass halo population as our fiducial simulation, 
but no active low-mass sources (LMACHs) at all. Therefore, the 
effective efficiencies start very high, at several thousand photons 
per atom, as the relatively few high-mass sources at early time have 
to 'compensate' for the more numerous low-mass sources present 
in the fiducial case which are missing here, as well as for all 
photons emitted before $z\sim19$ in our fiducial simulation, 
during which time there are no active sources larger than 
$10^9M_\odot$. However, as the number of high-mass sources rises 
exponentially, $g_{\gamma,eff}$ drops precipitously, to less than 
40 by $z=12.6$ and less than 20 by $z=11$. Towards overlap 
$g_{\gamma,eff}$ settles on $\sim8.7$, the value adopted in our 
fiducial case, as by then all low-mass sources are suppressed 
and the high-mass sources are identical to the ones in the 
fiducial case. Note that although in this case the sources 
belong to the same halos as in S5 and overlap is reached at a 
similar redshift, this S8 case is different in assuming the 
same step-by-step total emissivity as in our fiducial case, 
L1, which naturally makes them variable in time, unlike case 
L5 discussed before, which had a fixed photon emissivity per 
unit halo mass. 

\begin{figure*}
\begin{center}  
\includegraphics[width=2.3in]{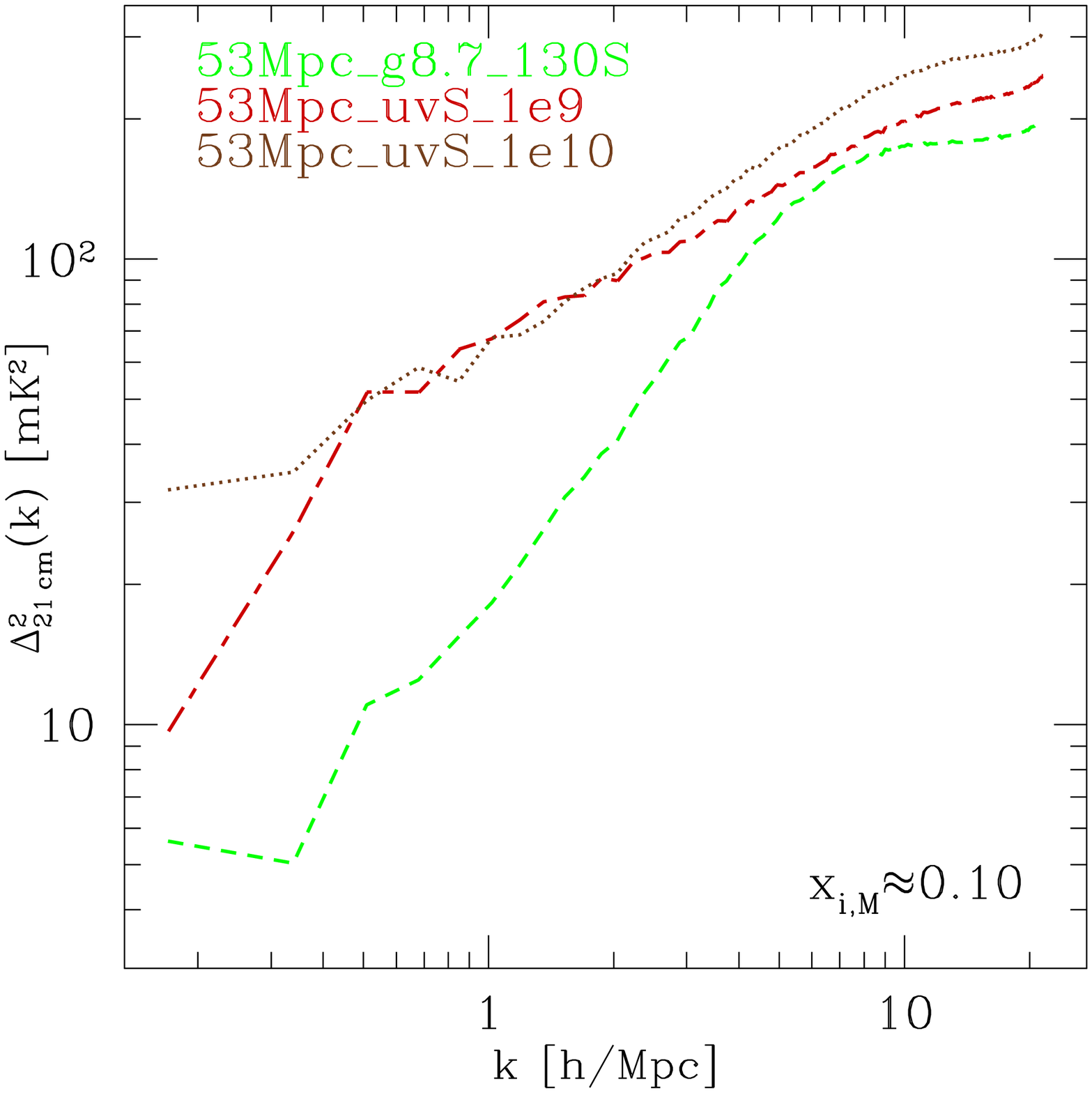} 
\includegraphics[width=2.3in]{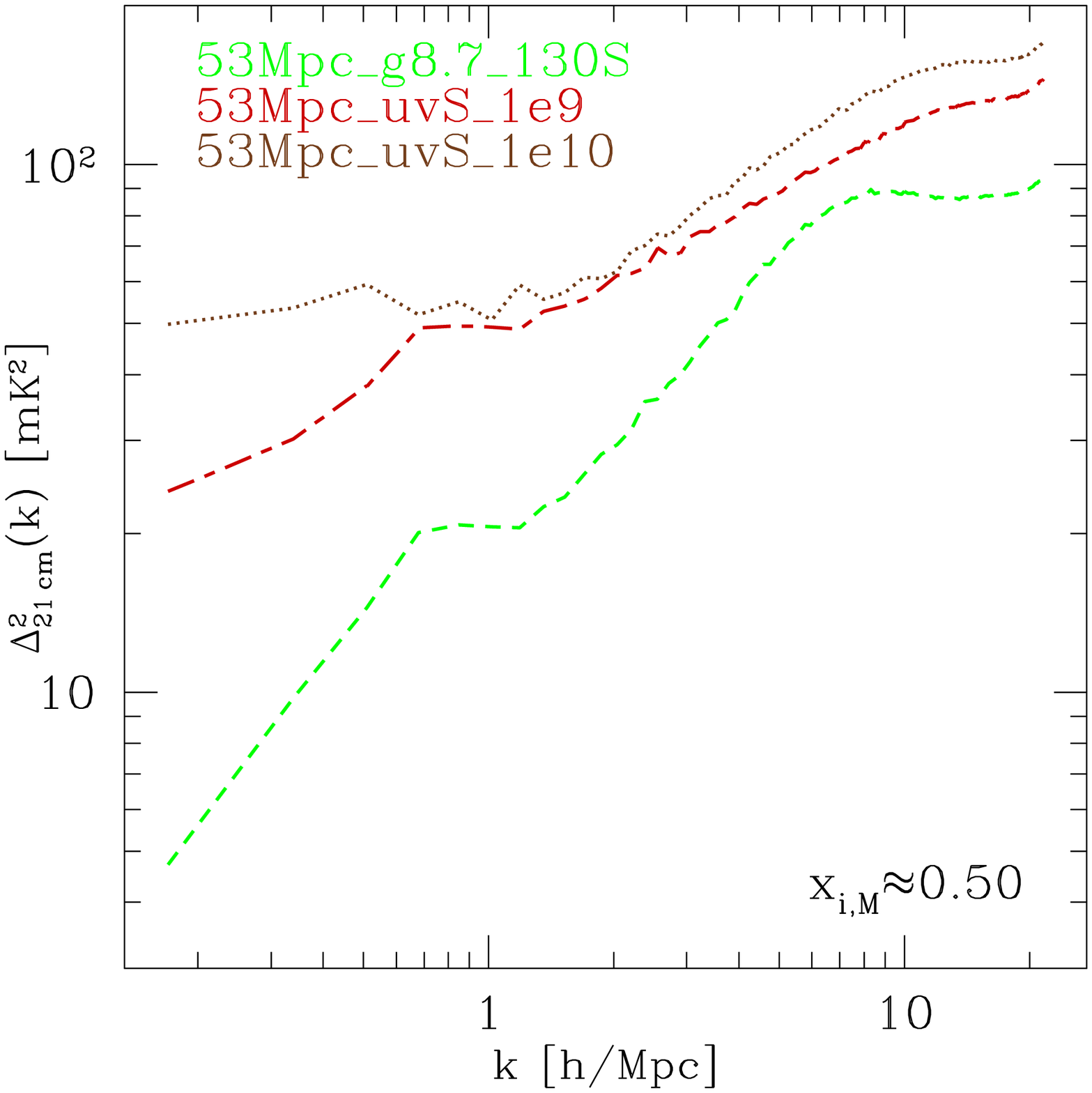} 
\includegraphics[width=2.3in]{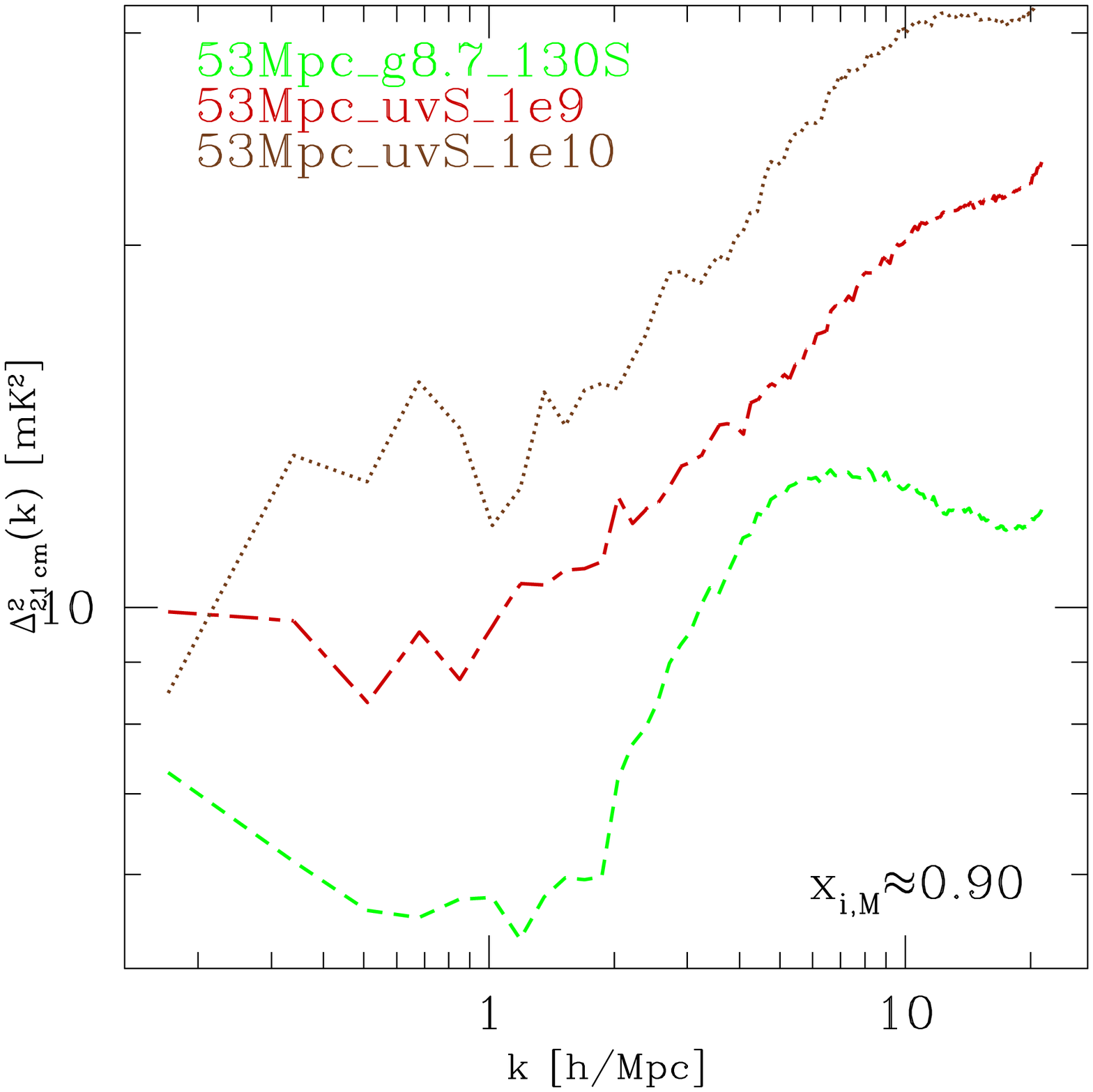} 
%\vspace{-1in}
\caption{
\label{21cm_power_uv_fig} 21-cm differential brightness 
temperature fluctuation power spectra for varying source 
models. Shown are the epochs at which the ionized fractions 
are (left) $x_m=0.1$, (middle) $x_m=0.5$ and (right) $x_m=0.9$. 
All cases are labelled by color and line-type, as follows: S1 
(green, short-dashed), S8 (dark red, long dash-short dash) and
S9 (brown, dotted).
}
%\vspace{-0.5cm}
\end{center}
\end{figure*}

Case S9 is still more extreme, since only quite massive halos, 
with masses above $10^{10}M_\odot$ are allowed to be active 
sources. The first such massive halos form in our simulation 
only at $z=13.2$ and they remain relatively rare ($\sim3-\sigma$) 
even at overlap ($z=8.2$). As a consequence, their effective 
efficiency is very high at all times, starting at over 20,000 
and reaching $\sim40$ at overlap. In order to avoid hyper-luminous 
sources during the first timestep, we distributed the photons that 
were emitted at $z>13.2$ in the fiducial case over the first several 
timesteps of run S9. Clearly, both S8 and S9 scenarios are not very 
realistic physically, given this vast range of change in the source
efficiencies.

Since in cases S8 and S9 we imposed the same global integrated
ionizing photon emissivities per timestep as in our fiducial 
case S1 (but higher minimum source mass), the averaged global 
reionization histories of those two cases closely follow the 
one of the fiducial simulation once the first haloes above the 
respective minimum cutoff form in our volume. The only remaining 
difference is that at early times ($z>11$) the ionized fraction 
in case S9 is a little lower than in the other two cases, as a 
consequence of our imposition of a bit more gradual initial 
release of photons in this case, in order to avoid hyper-luminous 
sources, as explained above. However, unlike S1, both simulations 
S8 and S9 yield $x_m/x_v\approx1$, since in those latter cases the 
ionized patches produced by the few, luminous sources present are 
far less correlated with the underlying density field 
(Fig.~\ref{21cm_fluct_uvS_fig}). This is a consequence of the 
I-fronts quickly escaping into the nearby voids, which compensates 
for the exponential rise of the number of ionizing sources forming 
at the high density peaks (although we note that even in this case 
reionization remains inside-out, as the ionized regions are still
over-dense on average). This results in H~II region distributions
which are clearly distinct from the rest. As the cutoff mass increases, 
there are exponentially fewer ionizing sources, which consequently 
are much more efficient (cf. Fig.~\ref{g_eff}). Hence, those 
hyper-luminous sources produce correspondingly large H~II regions,
which are less correlated with the underlying density field and are 
more spherical than in the other cases, as they are produced by few, 
but highly clustered sources.

This distinct H~II region geometry of cases S8 and S9 also yields 
very characteristic 21-cm signatures. The massive, rare, highly 
efficient sources quickly produce very large H~II regions and 
thus high rms fluctuations at large scales and a very broad 
peak, with an almost constant value 
($\langle \delta T_b^2\rangle^{1/2}=10.39-10.94$) for a wide 
range of mean ionized fractions by mass, $x_m=0.21-0.58$. There 
is also no initial dip of the rms fluctuations, which normally
occurs when the highest density peaks are ionized, but the H~II 
regions are still much smaller than the smoothing beam size. In 
models S8 and S9 the ionized patches grow so fast that their 
typical sizes are of order or large than the beam at all times. 
Such a scenario therefore yields a signal which is both stronger 
and quite different from the others. The results for the lower 
minimum source mass cutoff case with same reionization history, 
S8, show similar properties to S9, namely a broad and relatively 
high rms peak and no initial dip. However, the peak value in 
this case, at $\sim8$~mK, is noticeably lower than for model S9 
($\sim11$~mK) and is more similar to the typical values for the 
majority of cases ($\sim5-8$~mK). The corresponding PDF 
distributions (not shown) are noticeably wider, with a long 
non-Gaussian tail at high differential brightness temperatures 
($\delta T_b-\bar{\delta T_b}>30-40$~mK) than in the fiducial 
self-regulation case, S1.

The corresponding 21-cm power spectra (Figure~\ref{21cm_power_uv_fig})
for cases S8 and S9 also show significantly higher signal on all 
scales compared to our fiducial case S1. During the early stages 
of reionization the power at large scales ($k\lesssim0.4\,\rm h/Mpc$) 
for S9 is almost an order of magnitude higher than for S1. During 
the late stages the difference decreases considerably, but still 
remains $\sim2$ on large ($k<2\,\rm h/Mpc$), as well as very small 
($k>8\,\rm h/Mpc$) scales. As could be expected, the case with lower 
minimum mass, S8, is intermediate between S1 and S9, but much closer 
to S9 throughout the evolution.

Our scenario S9 is similar to the high minimum source mass case, 
S4, considered in \citet{2007MNRAS.377.1043M}. These authors set 
the minimum source mass to $4\times10^{10}M_\odot$, somewhat higher 
than in S9. Their ionizing photon production is similarly set to 
reproduce, step-by-step, the one of their fiducial case. Their 
results are qualitatively similar to what we find. The rare, 
efficient and strongly clustered sources yielded 21-cm power 
spectra which were higher and flatter than in their fiducial case, 
with the difference decreasing over time 
\citep[cf. Fig.~17 in][]{2007MNRAS.377.1043M}.
However, some quantitative differences remain, due to the 
somewhat different approach we have taken, as well as some 
numerical and resolution differences. Apart from the higher
source mass cutoff adopted by \citet{2007MNRAS.377.1043M}, 
which results in a stronger source bias, other important 
differences include lower resolution of their N-body and 
radiative transfer simulations, and lack of Jeans mass 
filtering. Unlike our high-resolution simulations, which 
resolve all atomically-cooling halos ($M>10^8M_\odot$, the 
N-body structure formation simulations used by 
\citet{2007MNRAS.377.1043M} resolved only halos with mass 
above $10^9M_\odot$, with lower-mass sources included in some 
cases by sub-grid modelling. More importantly, their fiducial 
case (whose photon production per timestep was the basis for 
their high-mass cutoff case S4) yielded late overlap and 
included no Jeans mass filtering (several of their other 
simulations included it, but not this one). Therefore, their 
photon production per unit source mass was necessarily very 
low, making their fiducial case more similar to our low-efficiency
case S4 than to our fiducial simulation S1. Finally, we take 
account of peculiar velocity when calculating more precise 
21-cm power spectra (including redshift space distortions) 
\citep{2011arXiv1104.2094M}. Despite these differences, our 
results agree reasonably well on a qualitative level.  

\begin{figure*}
\begin{center}  
\includegraphics[width=3.2in]{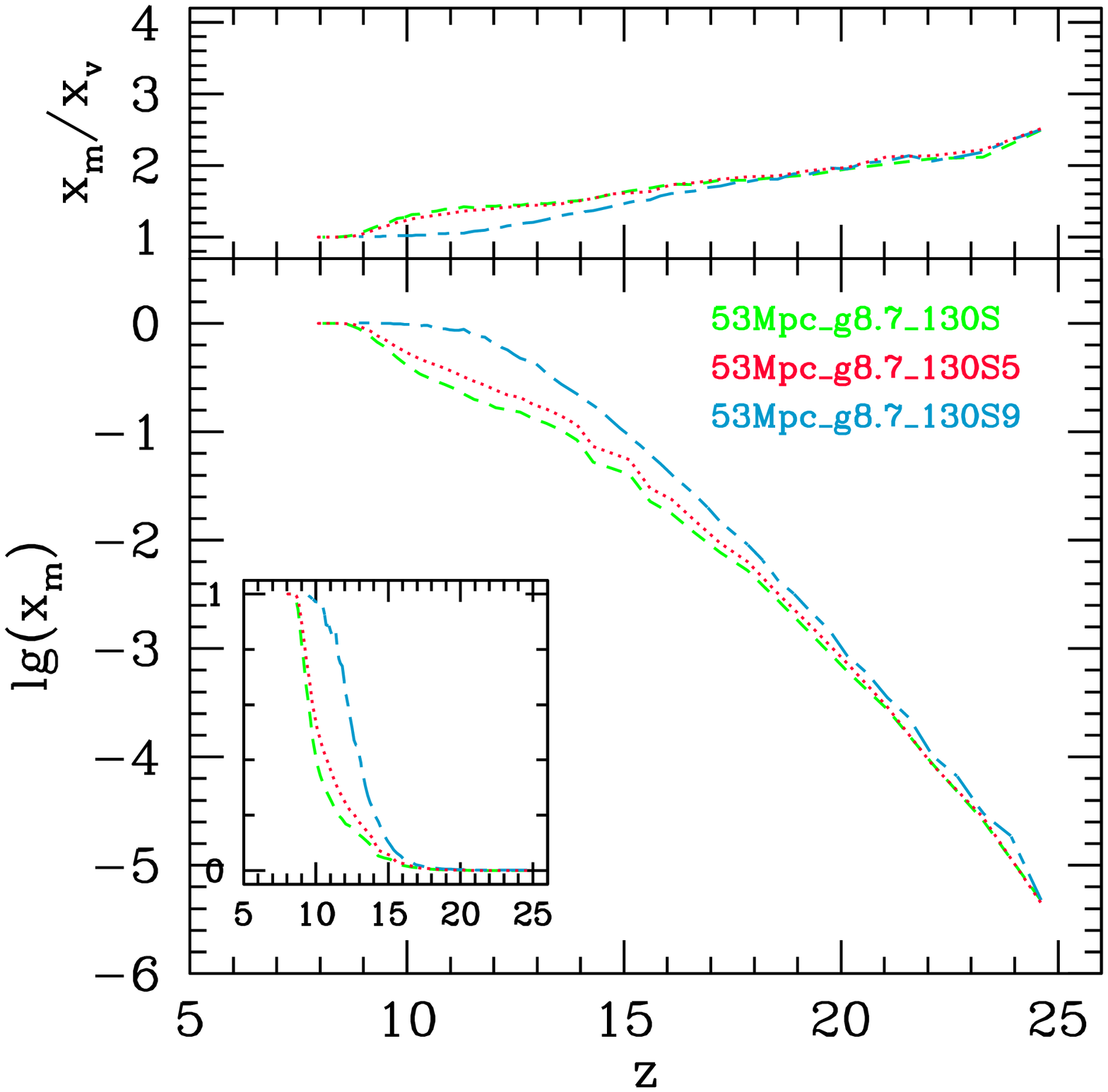} 
\includegraphics[width=3.2in]{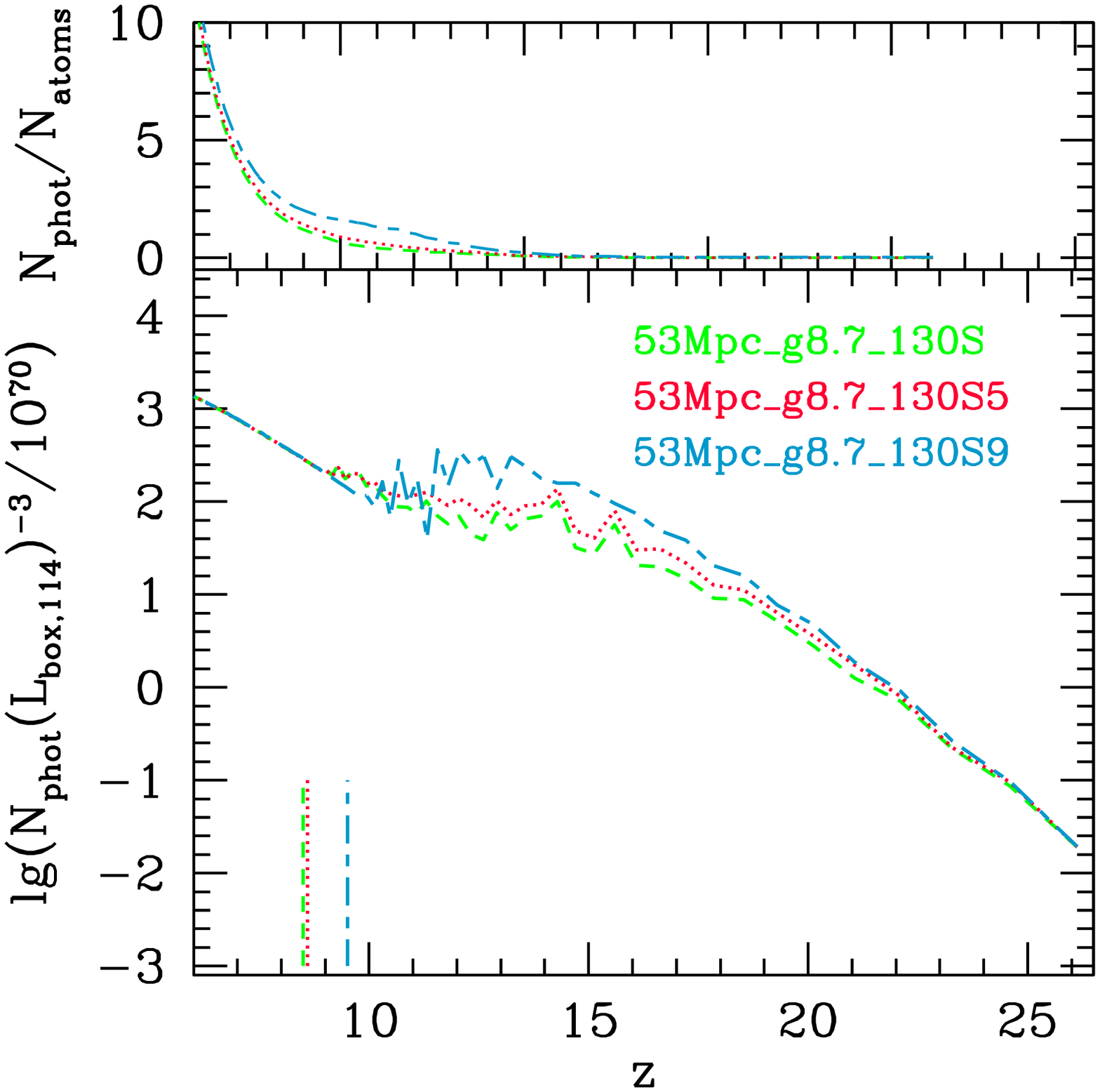} 
\caption{
\label{reion_hist_supp_fig}
(left) (bottom panel) Mass-weighted reionization histories for 
cases S1, S6 and S7, each with different Jeans mass filtering 
threshold. (top panel) Ratio of the corresponding mean 
mass-weighted and volume ionized fractions, $x_m/x_v$. 
(bottom panels) Number of ionizing photons emitted by all active 
sources (thick lines) in the computational volume per timestep and 
(top panels) cumulative number of photons per total gas atom released 
into the IGM. Vertical lines mark the overlap redshift in each case.
All curves on both left and right are labelled by color and line-type, 
as follows: S1 (green, short-dashed), S6 (light blue, long dash-short 
dash) and S7 (red, dotted).
}
%\vspace{-0.5cm}
\end{center}
\end{figure*}
  
We note that models S8 and S9 are rather unrealistic, as they 
assume unphysically high and time-variable luminosities, as well 
as the suppression of all sources with mass below $10^{10}M_\odot$ 
(or, less aggressively, $10^9M_\odot$ for case S8), 
for which no clear mechanism exists. We 
have included these models here primarily in order to demonstrate, 
under controlled circumstances, the effect of higher source-mass 
cutoff on the 21-cm observables. Such a higher source cutoff mass 
occurs numerically in simulations with large volumes and limited 
dynamic range \citep[e.g.][]{2009A&A...495..389B,2009MNRAS.393...32T}, 
and, therefore, it is important to evaluate the level of reliability 
of such models. Our results show that including only the high-mass 
sources can result in over-estimating the 21-cm rms fluctuations by up 
to a factor of 2, while $P(k)$ at small $k$ where the first generation
of observations will probe, could be over-estimated by as much as an 
order of magnitude at the 50\% ionized epoch. It can also yield quite 
a different evolution, even for the same boxsize, numerical resolution 
and the same integrated photon emissivity over time. One should 
therefore be aware of these potential pitfalls and adjust their 
modelling accordingly. A better simulation approach would be to 
add the lower-mass, unresolved sources by sub-grid modelling.

\section{Dependence on the Jeans suppression threshold}
\label{appendixB}

%S1 vs. S6 vs. S7

\begin{figure*}
  \begin{center}
  \includegraphics[width=2.2in]{simage_xy_9.690_37Mpc_f10_150S_x.5.ps}
  \includegraphics[width=2.2in]{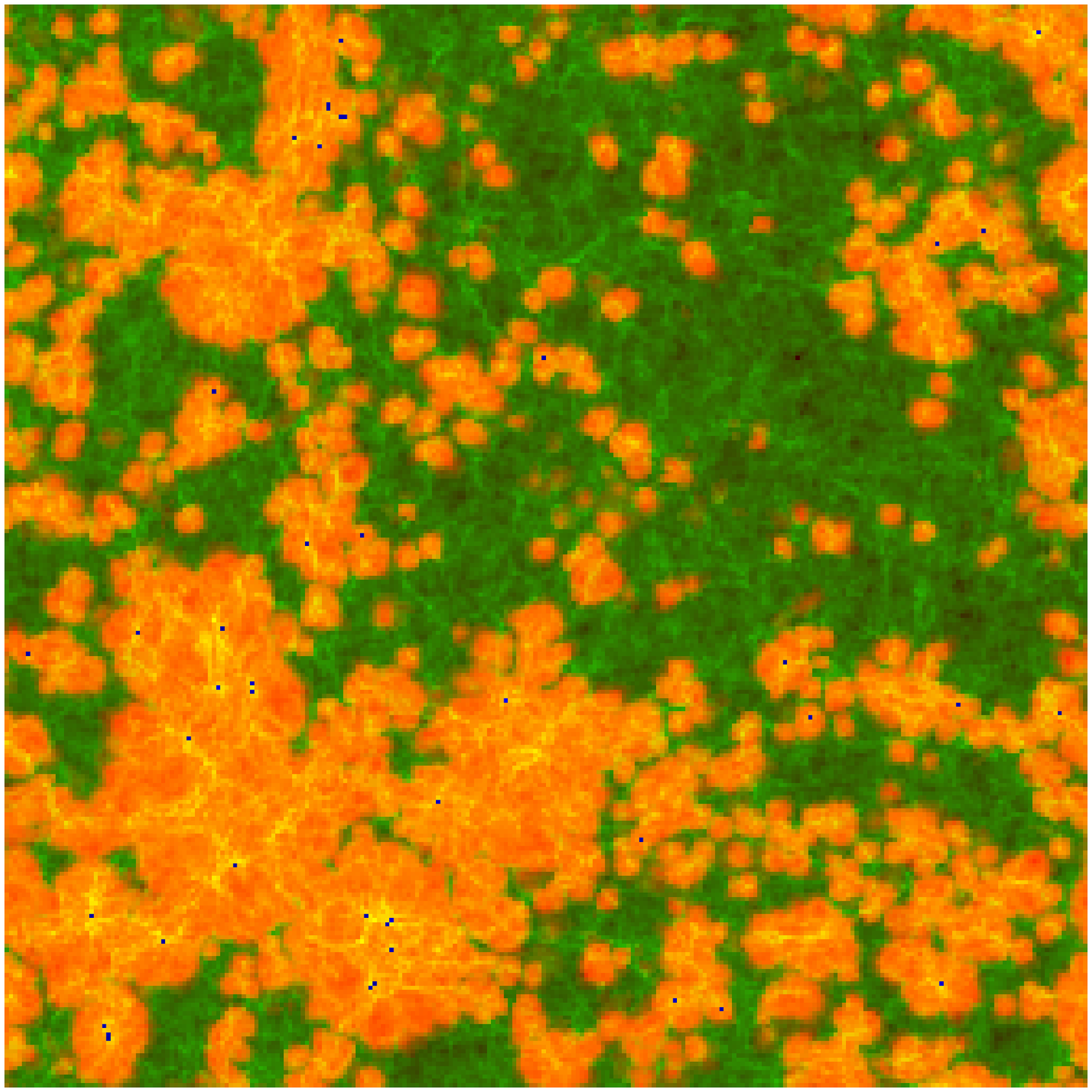}
  \includegraphics[width=2.2in]{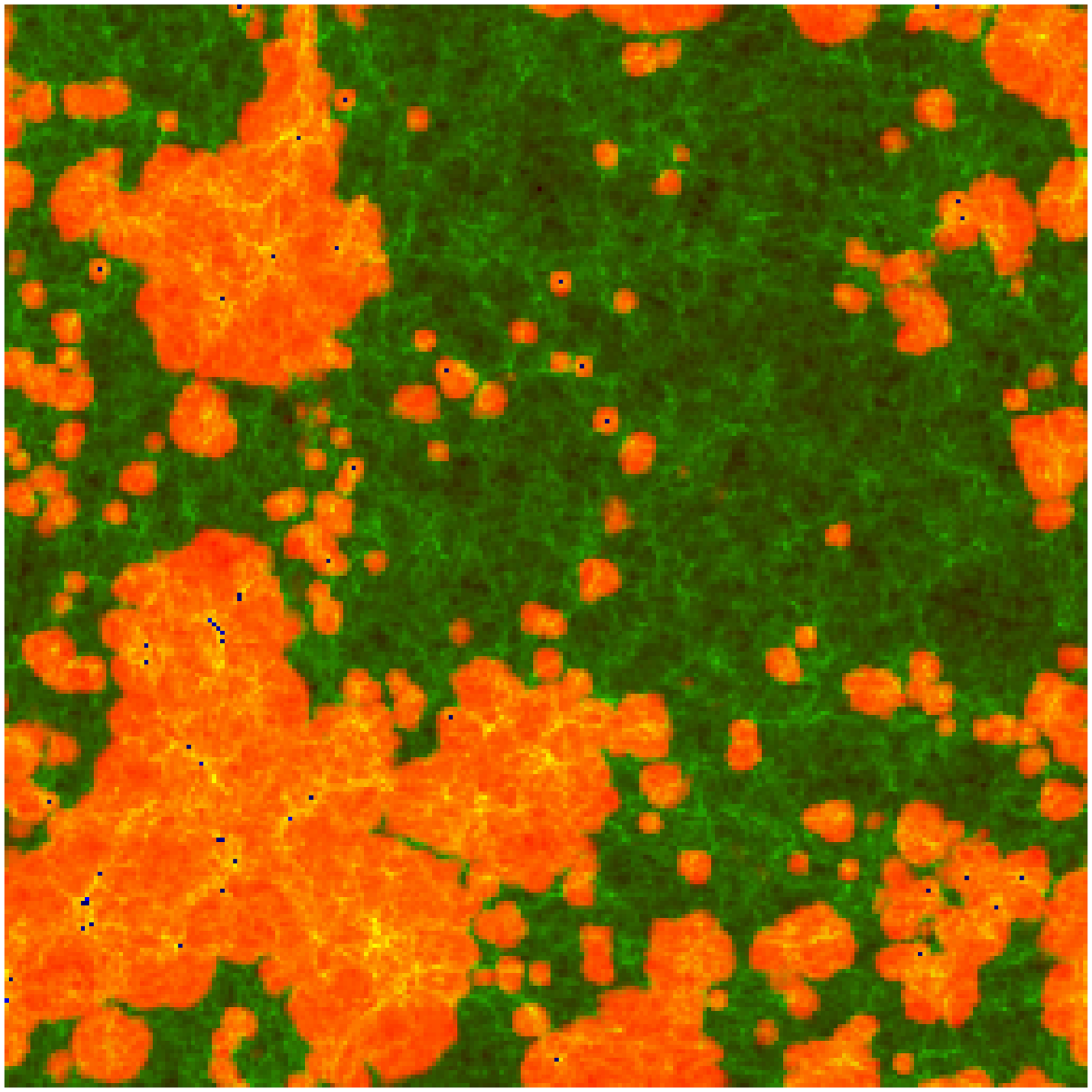}
    \vspace{-0.3cm}
  \end{center}
  \caption{Spatial slices of the ionized and neutral gas density 
    from our radiative transfer simulations with boxsize 
    $37\,h^{-1}$~Mpc , all at box-averaged ionized fraction by 
    mass $x_m\sim0.50$. Shown are the density field (green) 
    overlayed with the ionized fraction (red/orange/yellow) and 
    the cells containing sources (dark/blue). Shown are cases S1, 
    S7, and S6.
    \label{images_37Mpc_supp}}
\end{figure*}

\begin{figure*}
\begin{center}  
\includegraphics[width=2.3in]{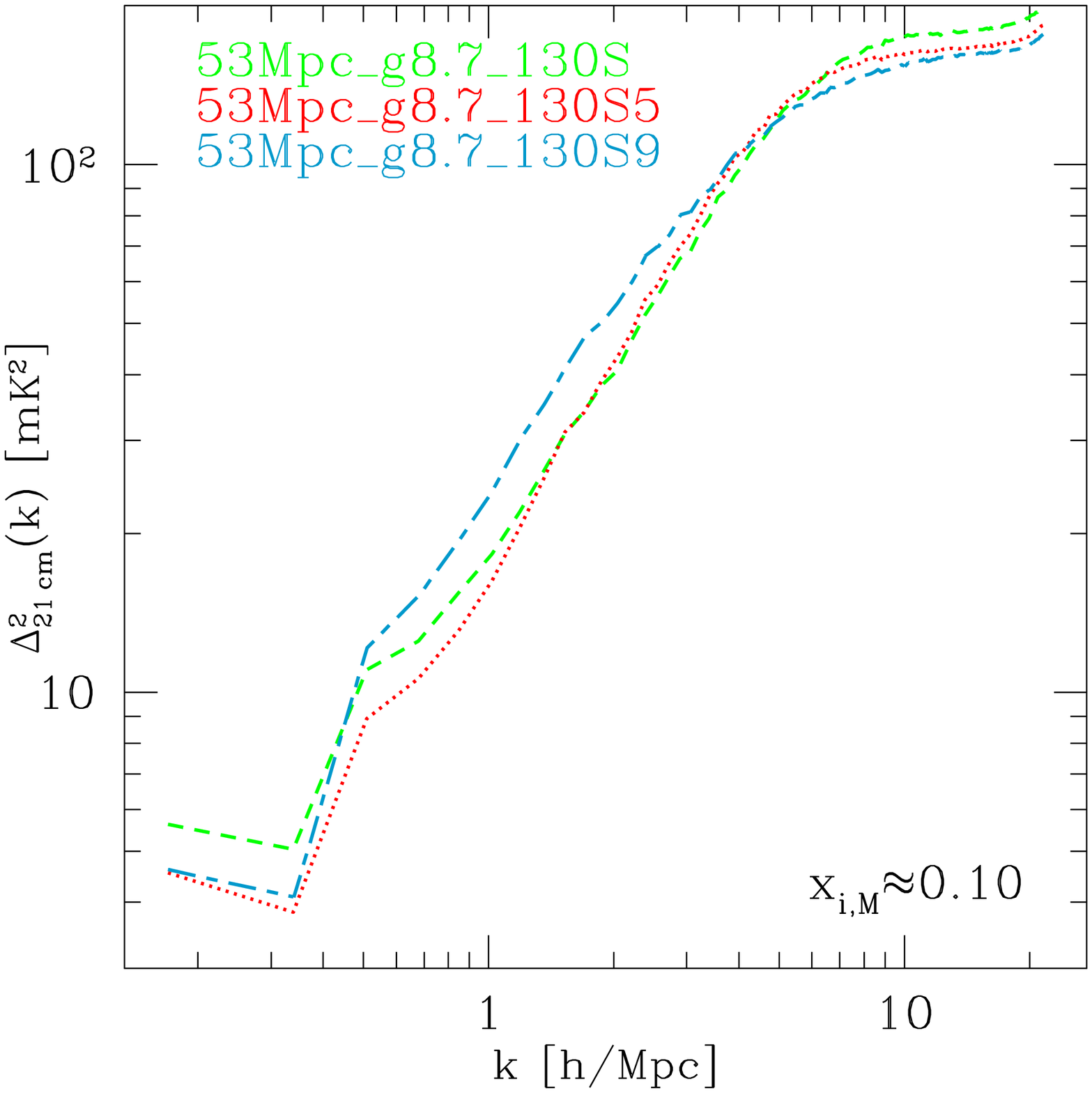} 
\includegraphics[width=2.3in]{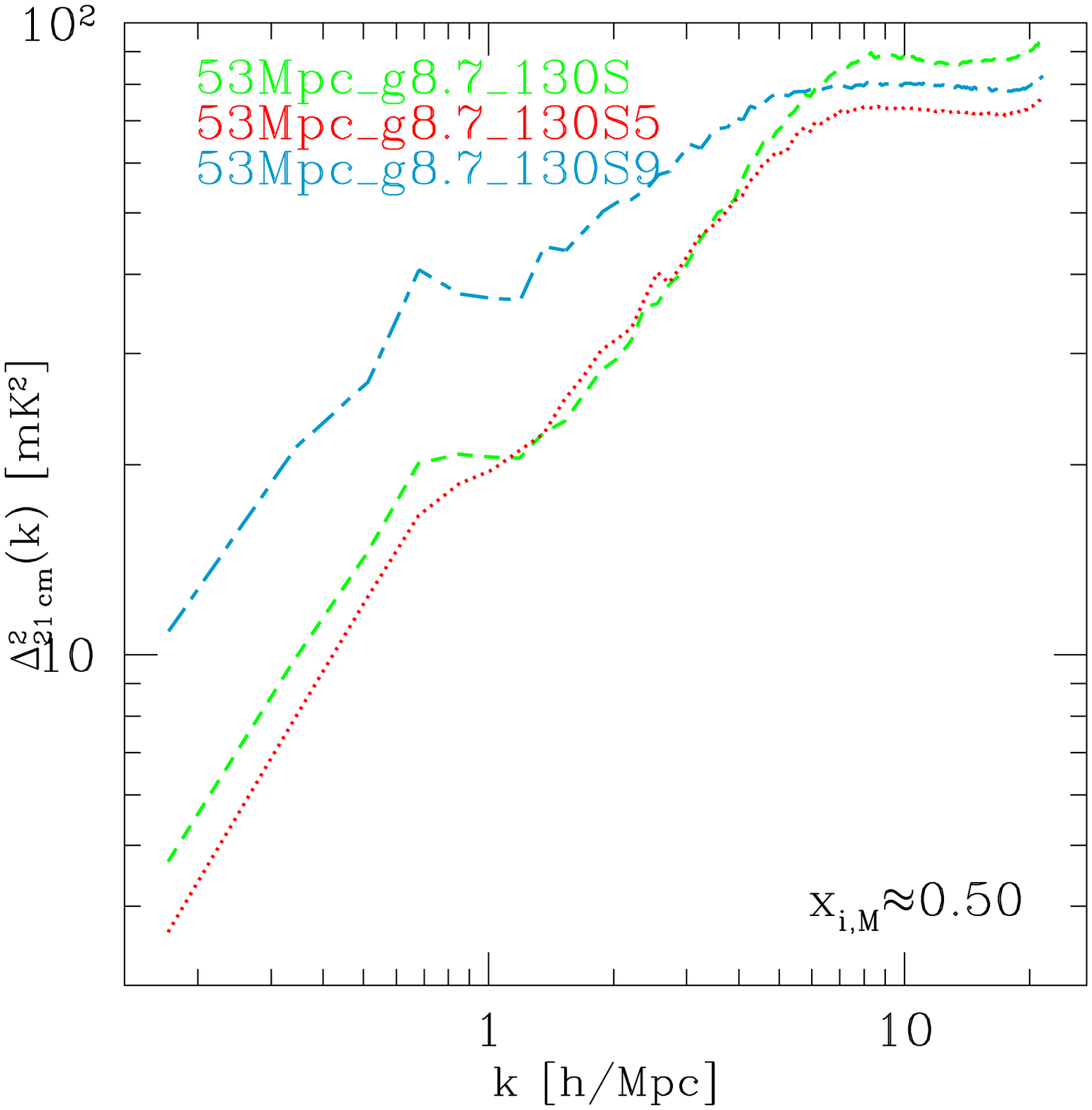} 
\includegraphics[width=2.3in]{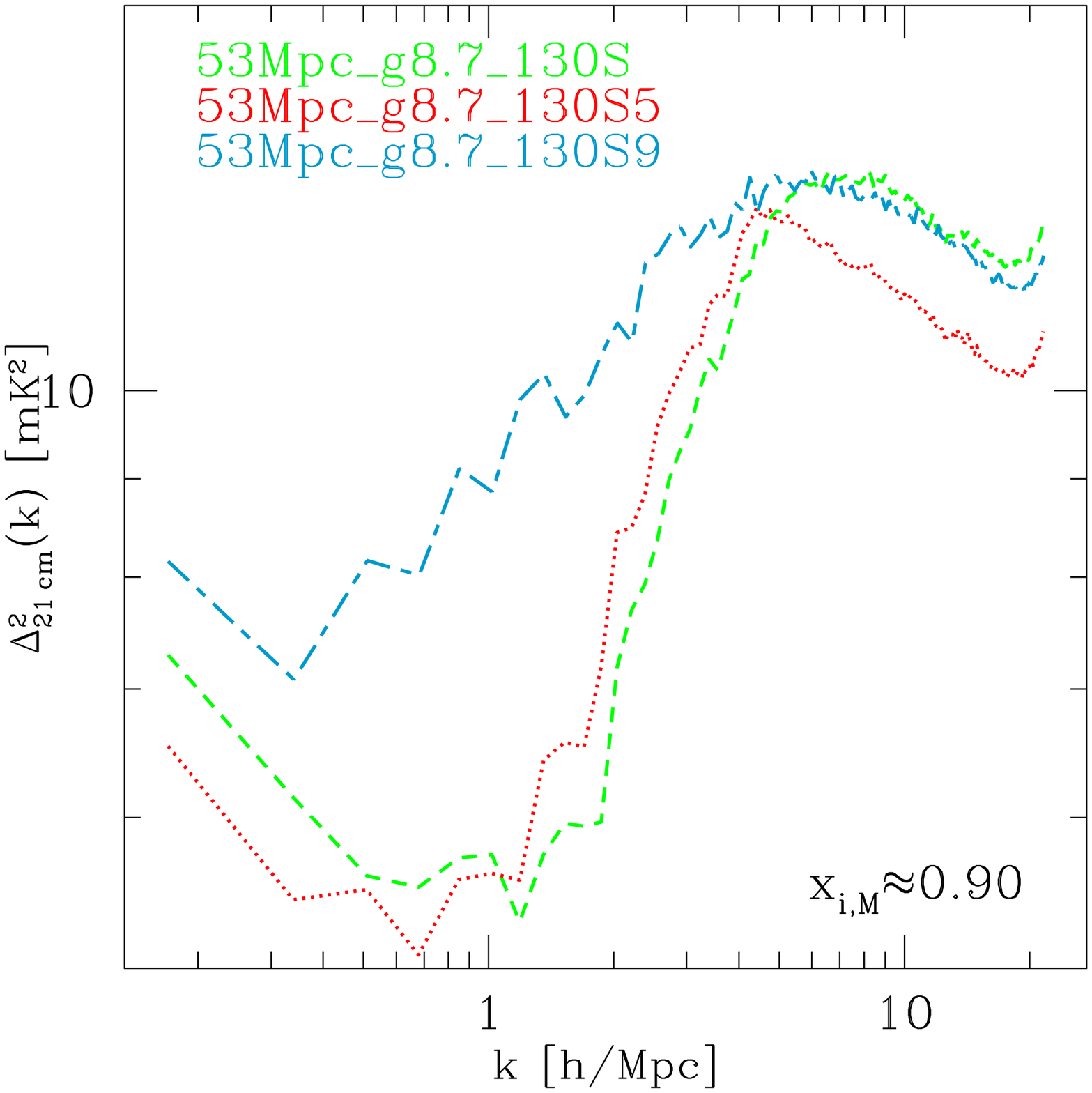} 
%\vspace{-1in}
\caption{
\label{21cm_power_sup_fig} 21-cm differential brightness temperature
fluctuation power spectra for varying source models. Shown are the 
epochs at which the ionized fractions are (left) $x_m=0.1$, (middle)  
$x_m=0.5$ and (right) $x_m=0.9$. All cases are labelled by color and 
line-type, as follows: S1 (green, short-dashed), S6 (light blue, 
long dash-short dash) and S7 (red, dotted).
}
%\vspace{-0.5cm}
\end{center}
\end{figure*}

\begin{figure*}
\begin{center}  
\includegraphics[width=3.2in]{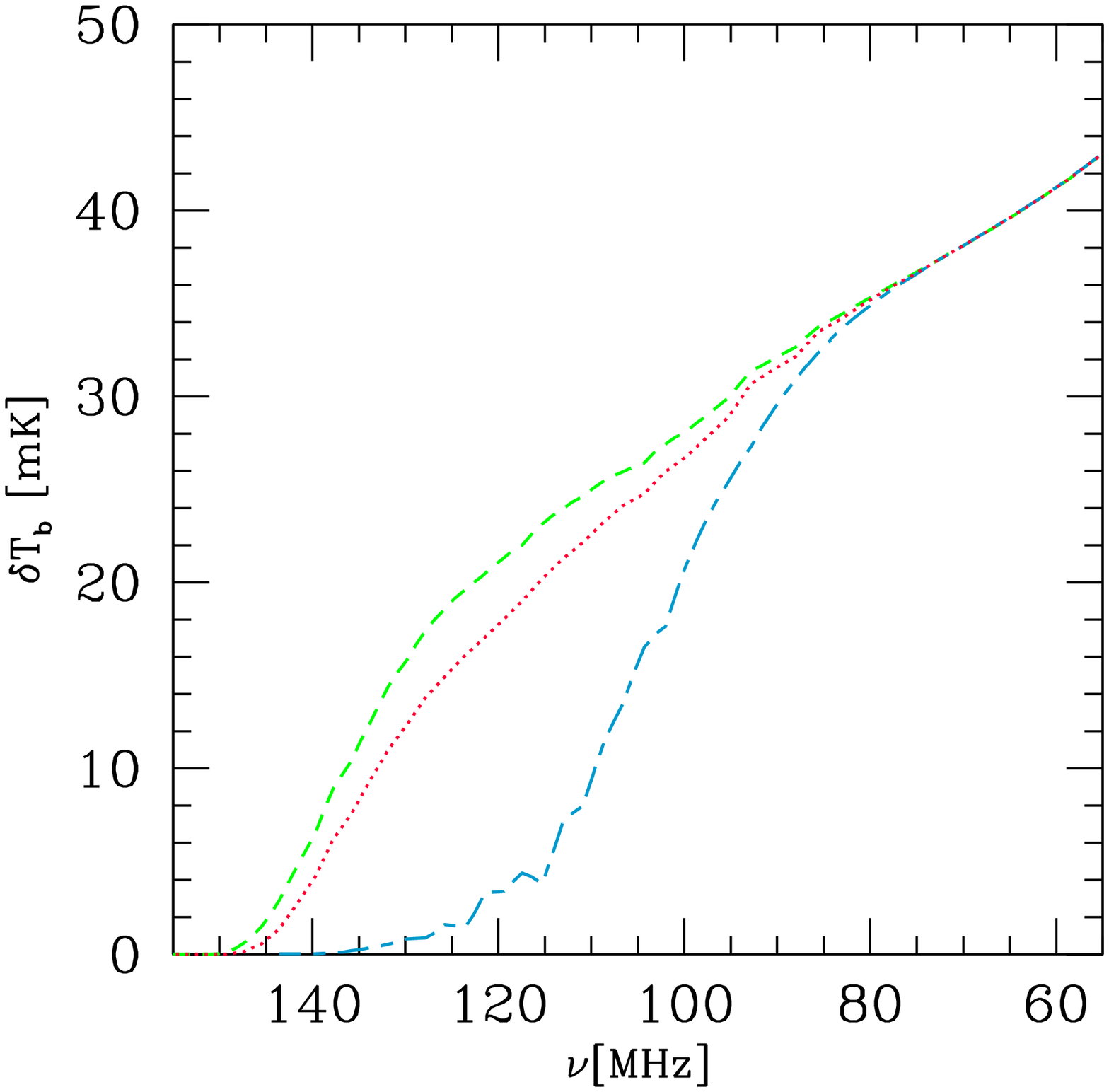} 
\includegraphics[width=3.2in]{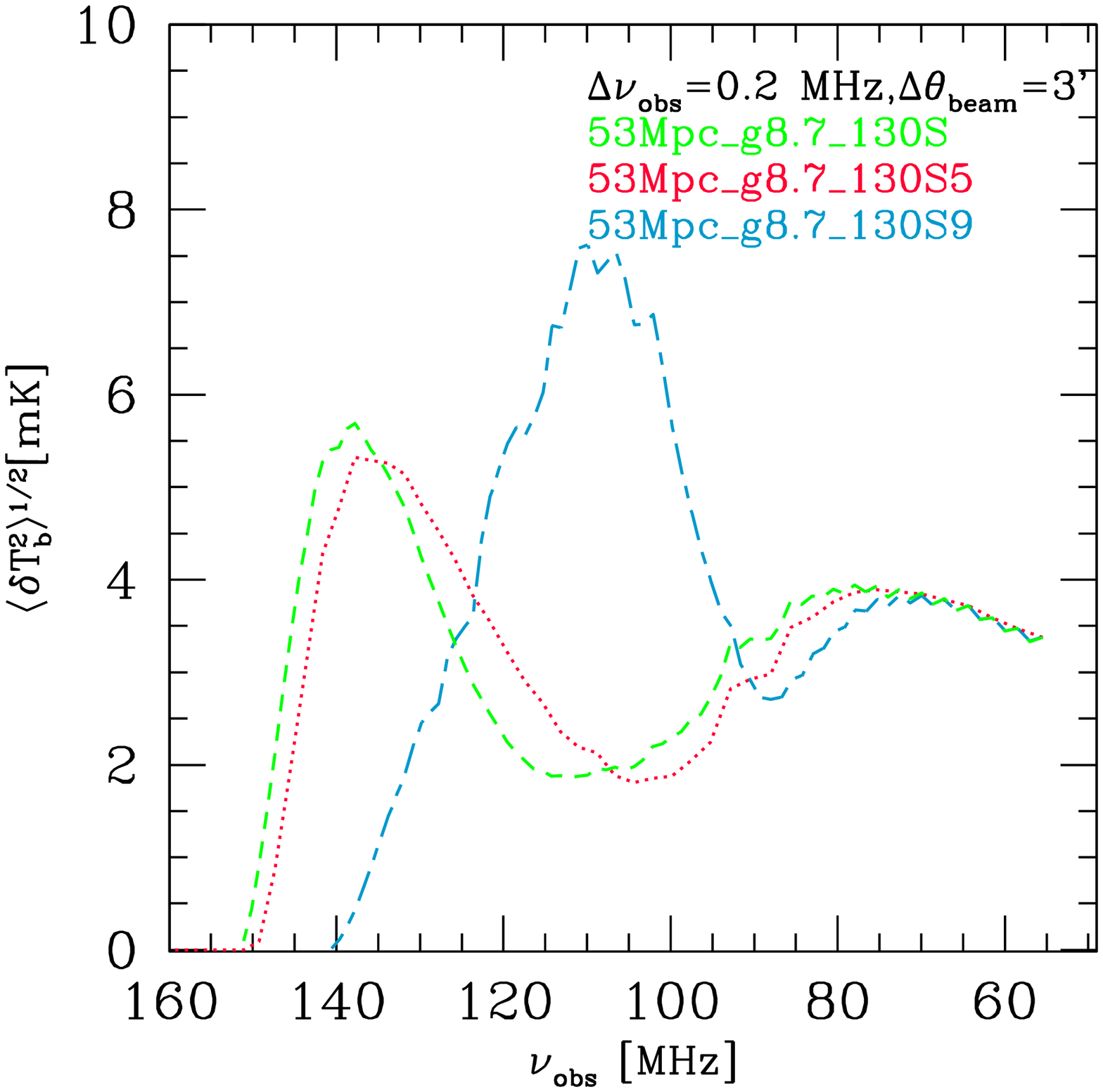}
\caption{The evolution of the mean 21-cm background (left) 
and its rms fluctuations for Gaussian beamsize $3'$ and 
bandwidth $0.2$~MHz and boxcar frequency filter (right) vs. 
observed 21-cm frequency. Shown are simulations (green, 
short-dashed), S6 (light blue, long dash-short dash) and 
S7 (red, dotted). 
\label{21cm_supp_fig}
}
%\vspace{-0.5cm}
\end{center}
\end{figure*}  

Here we consider variations of our source suppression threshold,
with the goal establishing the robustness and validity of our
fiducial model, S1, where we use $x_{\rm threshold}=0.1$. We have
ran two additional models, S6 and S7, whereby we raise this 
ionization threshold for low-mass source suppression to 
$x_{\rm threshold}=0.9$ and 0.5, respectively. In our fiducial case 
the suppression criterion for partially-ionized cells is more 
aggressive than in these new cases. The reasonable value to be 
adopted for this suppression threshold is still quite uncertain 
at present, thus it is important to check the sensitivity of our 
results to variations in its value. In fact, it is most likely 
that a sharp on-off condition like this is an oversimplification 
and in reality the suppression boundary is gradual, with full 
suppression of the smallest galaxies, partial one for 
intermediate-mass galaxies, up to no suppression at all for 
sufficiently massive galaxies. However, given the current 
uncertainties, the range of possible suppression models is very 
large and it is difficult to fully explore numerically. Instead, 
we have chosen to consider three very different cases covering 
the full range of the threshold value, in order to evaluate the 
effect of these uncertainties on the reionization history and 
observables. We note, however, that we consider our original 
source suppression criterion to be well motivated, for the 
following reasons. Although for numerical reasons our suppression 
criterion is defined in terms of ionized fraction, physically it 
is related to the temperature state of the IGM, for which the 
ionization state is used as a proxy. When a given region is 
photoionized, its temperature rises to $\sim10^4$~K, with the 
exact value dependent on the intensity and spectrum of the 
ionizing radiation, ranging from $\sim20,000$~K for Pop.~II 
stellar spectra and QSO's to $\sim30,000-40,000$~K for Pop.~III 
\citep{2004MNRAS.348..753S}. This rises the gas pressure and 
thus the Jeans mass, to $10^9M_\odot$ or more. In order for the 
low-mass halos to be able to re-form in a previously-ionized 
region its temperature should decrease to well below $10^4$~K. 
However, in the mostly metal-free gas during these early epochs
there is no efficient radiative coolant available and therefore 
the main cooling mechanisms are the local adiabatic expansion 
and Compton scattering of CMB photons. Since both of these 
processes are relatively slow and inefficient we expect that 
our fiducial more aggressive ionized fraction-based source 
suppression criterion is more physically realistic than the 
milder suppression of the new cases. However, given the 
significant uncertainties of the Jeans filtering process, which 
can only be properly modelled by hydrodynamical simulations 
with detailed and realistic microphysics, we consider all of 
these very different suppression criteria and study their 
consequences below.

The reionization histories and cumulative numbers of ionizing 
photons emitted derived for the three suppression criteria are 
shown in Figure~\ref{reion_hist_supp_fig} (left). Clearly, only 
a very weak suppression ($x_{\rm threshold}=0.9$, case S7) yields 
any significant differences. Compared to our fiducial case S1, 
many fewer low-mass sources are suppressed in S7, and of these 
a significant fraction are allowed to become active again 
shortly after suppression (since in absence of radiation 
recombinations quickly bring the neutral fraction back up above 
10\%). The evolution of the number of photons produced in 
simulation S6 is up to $\sim2$ higher in the middle stages of 
reionization, while the corresponding number for S7 is 
essentially the same as in S1 throughout the evolution.

The Jeans mass filtering nonetheless still has a significant 
effect, keeping the ionized fraction well below the corresponding 
one for the no-suppression case S3 (cf. Fig.~\ref{reion_hist_fig}, 
right panels). Eventually, the fully-ionized fraction of the 
volume becomes sufficiently large to suppress almost all low-mass 
sources even with this mild suppression criterion and the 
reionization process slows down until sufficient number of 
massive sources form and are able to finish this process and 
reach overlap. In contrast, the intermediate case, S7 
($x_{\rm threshold}=0.5$) shows only modest differences from the 
fiducial model S1, manifesting themselves mostly in bringing 
reionization forward by $\Delta z\sim0.4$, compared to 
$\Delta z\sim1.2-3.2$, and very different shape of the 
reionization history for case S6. Similarly, the integrated 
electron-scattering optical depth for the mild suppression 
case S6 is $\tau_{\rm es}=0.111$, much higher than in the 
fiducial case ($\tau_{\rm es}=0.080$), while for the intermediate 
suppression case the increase is much more modest, at 
$\tau_{\rm es}=0.089$. The cumulative number of photons per atom 
at overlap, $\sim2$, is very similar in all three cases.

The variations in the geometry of reionization 
(Fig.~\ref{images_37Mpc_supp}) are mostly found in the 
small-scale structures. There are significantly fewer such 
structures in the weak suppression case S6. The merged H~II 
regions are typically slightly larger, as well as rounder and 
with smoother boundaries compared to the fiducial simulation 
S1. Once again the intermediate case S7 is very similar to S1, 
with only minor differences in small-scale features. These 
visual impressions are further confirmed by comparing the 21-cm 
power spectra (Figure~\ref{21cm_power_sup_fig}). The weak 
suppression case S6 has significantly more power at intermediate 
and large scales ($k\lesssim5$) during all stages of reionization, 
more so at late times, but less power on small scales than our 
fiducial case S1. On the other hand, the intermediate model S7 
matches S1 fairly closely, except for having less power on very 
small scales.

The 21-cm mean differential brightness temperature 
(Fig.~\ref{21cm_supp_fig}, left) for the weak suppression 
case S7 shows an initial steep decline around $\nu\sim100$~MHz, 
followed by a sudden change of slope at $\nu\sim125$~MHz and 
a very slow decrease thereafter. This behaviour is quite 
different from models S1 and S6 (which again follow almost 
identical evolution), as well as from all other models 
discussed earlier. The only model with a similarly sharp 
decrease of the mean brightness temperature is the high 
efficiency, no suppression case S3, which however does not 
have the same long slow evolution tail at late times due to 
lack of suppression. 

Finally, the differential brightness temperature rms 
fluctuations (Fig.~\ref{21cm_supp_fig}, right) for the weak 
suppression case, S6, peak much earlier, at $\nu\approx110$~MHz 
(but still significantly later than the no suppression case S3, 
which underlines the importance of even a very weak low-mass 
source suppression) reaching 
$\langle \delta T_b^2\rangle^{1/2}\approx8$~mK. Uniquely, this model
exhibits a very long tail of slow decline of the rms fluctuations 
beyond the peak. This is related to its very different (and, as 
we argued earlier, possibly less physically realistic) suppression 
model. Case S6 also exhibits significant fluctuations in the 
differential brightness temperature fluctuations, once again 
indicating that this suppression model might be less physically 
realistic than our standard suppression model. The intermediate 
suppression model S6 follows the same evolution as the fiducial 
case S1, but shifted to slightly earlier time.

In summary, all results prove fairly insensitive to the precise 
value of the Jeans suppression threshold assumed, as long as it 
is not at the very weak suppression limit. Both $x_{\rm threshold}=0.1$
and 0.5 yield essentially the same evolution, apart from a slight 
offset in time. On the other hand, a very high suppression threshold
($x_{\rm threshold}=0.9$, i.e. weak suppression) results in a very
different (and somewhat unstable) evolution with several
characteristic observational signatures. We have, however, argued 
above that such a weak suppression is likely less realistic 
physically. Our suppression model therefore proves quite robust
to a threshold variation within the plausible range. 

\end{document}